\documentclass[11pt]{article}
\pdfoutput=1
\usepackage{amsmath,amssymb,amsthm,epsfig,amsfonts}
\usepackage{graphicx}
\usepackage{subfig}
\usepackage[usenames, dvipsnames]{color}
\usepackage{hyperref}
\usepackage{cite}
\usepackage{multirow}
\usepackage{verbatim} 
\usepackage{enumitem}
\usepackage{rotating}
\usepackage{xcolor}
\usepackage{multirow}
\usepackage{bm}
\usepackage[normalem]{ulem}
\usepackage{cleveref}
\usepackage{diagbox}
\usepackage{extarrows}

\graphicspath{ {figs/} }

\makeatletter

\theoremstyle{definition}
\newtheorem{definition}{Definition}[section]

\newcommand{\be}{\begin{equation}}
\newcommand{\ee}{\end{equation}}
\newcommand{\ba}{\begin{aligned}}
\newcommand{\ea}{\end{aligned}}

\newcommand{\bea}{\begin{eqnarray}}
\newcommand{\eea}{\end{eqnarray}}

\def\unit{{1\kern-.65ex {\rm l}}}
\def\1{{1\kern-.65ex {\rm l}}}

\def\CO{{\cal O}}

\def\bbP{{\mathbb{P}}}

\newcount\hour \newcount\minute
\hour=\time \divide \hour by 60
\minute=\time
\def\now{%
\ifnum \hour<13
  \ifnum \hour=0 \advance \hour by 12 \number\hour:\else \number\hour:\fi%
     \ifnum \minute<10 0\fi%
     \number\minute%
\ A.M.%
\else \advance \hour by -12 \number\hour:%
  \ifnum \minute<10 0\fi%
  \number\minute%
  \ P.M.%
\fi%
}

\makeatother

\def\mbf{\mathbf}
\def\mc{\mathcal}

\begin{document}

\baselineskip=18pt  
\numberwithin{equation}{section}  
\allowdisplaybreaks  


%
%


\thispagestyle{empty}

\vspace*{0.8cm} 
\begin{center}
{{\Huge {Fibers add Flavor, Part II}:\\ 

\bigskip 

\LARGE 5d SCFTs, Gauge Theories, and Dualities}}

 \vspace*{1.5cm}
Fabio Apruzzi$^1$, Craig Lawrie$^2$, Ling Lin$^2$,  Sakura Sch\"afer-Nameki$^1$, Yi-Nan Wang$^1$\\

 \vspace*{1.0cm} 
{\it ${}^1$ Mathematical Institute, University of Oxford, \\
Andrew-Wiles Building,  Woodstock Road, Oxford, OX2 6GG, UK}\\
\smallskip
{\it ${}^2$ Department of Physics and Astronomy, University of Pennsylvania, \\
Philadelphia, PA 19104, USA}\\

\vspace*{0.8cm}
\end{center}
\vspace*{.5cm}

\noindent
In \cite{Apruzzi:2019vpe, Apruzzi:2019opn} we proposed an approach based on graphs to characterize 5d
superconformal field theories (SCFTs), which arise as
compactifications of 6d $\mathcal{N}= (1,0)$ SCFTs. The graphs, so-called combined
fiber diagrams (CFDs), are derived using the realization of 5d SCFTs
via M-theory on a non-compact Calabi--Yau threefold
with a canonical singularity. In this paper we complement this
geometric approach by connecting the CFD of an SCFT to its weakly
coupled gauge theory or quiver descriptions
and demonstrate that the CFD as recovered from the gauge theory
approach is consistent with that as determined by geometry.
To each quiver description we also associate a graph, and
the embedding of this graph into the CFD that is associated
to an SCFT provides a systematic way to 
enumerate all possible consistent weakly coupled gauge
theory descriptions of this SCFT. Furthermore, different embeddings of
gauge theory graphs into a fixed CFD can give rise to new UV-dualities
for which we provide evidence through an analysis of the
prepotential, and which, for some examples, we substantiate by
constructing the M-theory geometry in which the dual quiver
descriptions are manifest.

\newpage

\tableofcontents


\section{Introduction}

Supersymmetric gauge theories are an ideal setup to explore strongly-coupled aspects of quantum field theories. In less than five dimensions they are renormalizable theories, whereas in higher (five and six) dimensions they can be effective descriptions at low energy. 
Thanks to the additional structure provided by supersymmetry, one can study features such as electric-magnetic dualities or renormalization group flows even in the absence of perturbative control at all energy scales.  
In practice, they can be used to probe strongly coupled regimes, giving insights about the non-perturbative dynamics of quantum field theories.

A particularly interesting class are five dimensional (5d) ${\cal N}= 1$ gauge theories, which can be low energy descriptions of superconformal field theories (SCFTs). More specifically, by studying the space of one-loop corrected couplings, parametrized by the Coulomb branch, one can argue necessary conditions for the existence of a strongly coupled ultraviolet (UV) fixed point \cite{Seiberg:1996bd}.
Another motivation to study these theories at present is recent progress in 6d SCFTs, where it is believed that a full classification of all UV-complete supersymmetric theories exists \cite{Witten:1995zh,Heckman:2013pva,Heckman:2015bfa,Bhardwaj:2015xxa}.
Indeed, recent works \cite{DelZotto:2017pti, Jefferson:2017ahm,Jefferson:2018irk, Closset:2018bjz, Apruzzi:2018nre, Bhardwaj:2018yhy, Bhardwaj:2018vuu, Apruzzi:2019vpe, Apruzzi:2019opn} have suggested that all 5d ${\cal N}=1$ UV-complete theories arise from appropriate circle reductions, possibly with holonomies for the global symmetries of 6d theories, thus conjecturing a classification of 5d theories.

Like their six dimensional cousins, 5d SCFTs are inherently strongly coupled.
In the absence of a Lagrangian description, methods from or inspired by string theory have proved to be invaluable in their studies \cite{Seiberg:1996bd, Aharony:1997ju,Aharony:1997bh,DeWolfe:1999hj,Brandhuber:1999np, Bergman:2012kr,Bergman:2015dpa, Zafrir:2015rga, Zafrir:2015ftn, Ohmori:2015tka, Hayashi:2017btw,Hayashi:2018bkd,Hayashi:2018lyv}. 
One of the important lessons we have learned from these methods is that many
different 5d gauge theories can have the same SCFT as UV-completion, thus
being \textit{UV-dual} (or, simply, dual) to each other.  Another crucial
aspect, which manifests itself at strong coupling, is that the
flavor symmetry of the gauge theory description can enhance
 at the UV-fixed point. This fact is due to the presence of
non-perturbative instanton operators, which quantum mechanically enhance the
classical flavor symmetry at the SCFT point. The state-of-the-art method to
calculate the SCFT's flavor symmetry typically involves a localization
computation in field theory or a description in terms of 5-brane webs
\cite{Kim:2012gu,Bergman:2013aca, Zafrir:2014ywa, Mitev:2014jza,
Hwang:2014uwa, Tachikawa:2015mha, Yonekura:2015ksa, Gaiotto:2015una,
Zafrir:2015uaa, Bergman:2016avc,Ferlito:2017xdq, Cabrera:2018jxt}.

In recent works \cite{Apruzzi:2019vpe, Apruzzi:2019opn} we proposed an alternative approach that arose out of the well-established geometric engineering via M-theory on a non-compact Calabi--Yau threefolds \cite{Morrison:1996xf, Intriligator:1997pq, Douglas:1996xp,DelZotto:2017pti, Xie:2017pfl,Jefferson:2017ahm, Jefferson:2018irk, Apruzzi:2018nre}.
One of the key insights of \cite{Apruzzi:2019vpe, Apruzzi:2019opn} is that there is a succinct description of the CFT data in terms of graphs, and transitions between graphs correspond to mass deformations and subsequent RG-flows. These graphs, the {\it combined fiber diagrams (CFDs)}, not only capture how 5d SCFTs are interconnected, but more importantly, they encode the strongly-coupled flavor symmetry of the UV fixed point SCFT, as well as the BPS states.

The central idea connecting the CFDs and 5d SCFTs is as follows: given a
marginal\footnote{In this paper, we will consider theories that are both
  marginal and have a 6d UV fixed point. As such we will use the notation
  interchangeably, however see \cite{Jefferson:2017ahm} for exceptions.} theory whose UV completion is a given 6d SCFT, all its descendant 5d SCFTs are obtained via mass deformations and RG-flows.
These field theoretic transitions can be encoded via simple graph-theoretic operations on the CFDs,
that is associated with each SCFT, and from which the complete tree of
descendants is obtained straightforwardly. 
The CFDs can be thought of as characterizing physically inequivalent M-theory geometries, which are in general non-flat resolutions (see \cite{Lawrie:2012gg} for an in-depth discussion) of the non-compact elliptic Calabi--Yau threefold underlying the F-theory realization of the given 6d SCFT.

The goal of the present paper is to put this into the context of a gauge theoretic description.
In particular, we connect the Coulomb branch phases of the effective theory \cite{Intriligator:1997pq}, described in terms of representation-theoretic graphs \cite{Hayashi:2014kca}, to the CFD-characterization of the SCFT limit.
The focus here is three-fold:
\begin{enumerate}
\item Constraining the possible weakly-coupled gauge theory descriptions of a 5d SCFT given in terms of a CFD, 
\item Derivation and constraints on UV-dualities using the CFD description,
\item Bootstrapping CFDs for marginal theories, in cases where no CFD-description is known, but weakly-coupled descriptions are available.
\end{enumerate}

\subsubsection*{Geometry and CFDs}

Before expanding on these points, let us briefly recapitulate the relation
between the geometry of non-compact elliptically fibered Calabi--Yau threefolds
with canonical singularities and 5d SCFTs.  The 5d SCFT arising from the
canonical singularity can be identified by virtue of the M-/F-theory duality
\cite{Vafa:1996xn,Morrison:1996na,Morrison:1996pp} with circle reductions of
the 6d theory realized in F-theory, including possible holonomies in the
flavor symmetry.  The resolutions of the canonical singularities consist of a
collection of intersecting compact surfaces, and, field theoretically, their
volumes parametrize the Coulomb branch of the theory. These surfaces shrink to
a point in the singular limit, which corresponds to the UV fixed point. When
these divisors are ruled ($\mathbb{P}^1$ fibered over a curve)
and intersect along sections of the rulings, the collection of surfaces may be collapsed to a 
curve of singularities after the ruling curves are collapsed to zero volume.
The additional light states appearing from M2-branes
wrapping the fibers of these rulings give rise to a gauge theory. Finally,
when a  bouquet of surfaces shrinks to a collection of intersecting curves of
singularities, the underlying low energy description is generically given by a
quiver gauge theory.

The starting point of our analysis is the so-called 5d marginal theory, which
is obtained by taking the 6d theory compactified on a circle (or alternatively
M-theory on the same elliptically fibered Calabi--Yau), without any holonomy
for the flavor symmetry.  This theory usually has an effective gauge theory
description, which has a 6d SCFT as its UV fixed point. Starting from the marginal
theory we can turn on mass deformations.  This procedure allows one to obtain
all descending 5d SCFTs corresponding to partial blow-downs of the fully
resolved geometry, and these descendants can be enumerated combinatorially.
From the gauge theory point of view this procedure corresponds to decoupling
matter hypermultiplets, whereas from a strongly coupled perspective, the
resulting descendants are the end-products after renormalization-group (RG)
flows that are triggered by the mass deformations. The set of descendant 5d
SCFTs linked by RG flow leads to a connected tree of theories. One of the main
advantages of our approach is that the complete tree of descendants is
obtained from the CFD associated to the marginal 5d theory by simple
operations on the graphs, and can be fully automated.

A complete classification of all 5d SCFTs that descend from 6d SCFTs by circle-reductions requires as input the set of all marginal theories, the associated CFDs (usually computed by resolving the geometry). From this the procedure determines the descendants uniquely.
The single gauge node marginal theories were determined in \cite{Jefferson:2017ahm}, and we will discuss this class of theories in the present paper. 
Another class that already featured prominently in \cite{Apruzzi:2019opn, Apruzzi:2019vpe} are 5d theories descending from 6d minimal conformal matter theories \cite{DelZotto:2014hpa}. 
One of the outputs of this paper are proposals for weakly coupled descriptions of these theories, as well as dualities among these. In many instances we can substantiate these weakly coupled descriptions as well as dualities by determining the associated rulings in the resolved elliptic Calabi--Yau geometry.

\subsubsection*{Gauge Theories, Coulomb Branches, Dualities and CFDs}

The strength of the approach that we proposed in \cite{Apruzzi:2019opn, Apruzzi:2019vpe} lies in  its combinatorial nature, which at the same time captures not only the network of 5d SCFTs that descend from a 6d theory, but also the flavor symmetry of the UV-fixed point.
While the latter is often enhanced compared to the classical gauge theory descriptions, our previous discussions were focused primarily on the SCFT itself.

In this work, we extend the scope of this approach by explicitly studying the effective gauge descriptions of the SCFTs. 

A central tool to achieve this is a representation-theoretic object, the \textit{box graph}, introduced in \cite{Hayashi:2014kca}, which captures all Coulomb branch phases of a given 5d gauge theory. 
The Coulomb branch on the other hand is intimately linked to the relative Mori cone of the elliptic Calabi--Yau threefold \cite{Hayashi:2014kca, Braun:2014kla, Braun:2015hkv, Lawrie:2015hia}\footnote{
This structure has played an important role also F-theory on elliptic
fourfolds and fivefolds in the context of $G_4$-fluxes and chiralitiy
\cite{Grimm:2011fx, Hayashi:2014kca, Braun:2014kla, Braun:2015hkv,
Lawrie:2015hia, Krause:2011xj, Marsano:2011hv, Schafer-Nameki:2016cfr}.}
The box graphs fully encode the sets of consistency conditions on the Coulomb
branch of a gauge theory with matter, where the matter classically transforms
under a flavor group, $G_{\rm F,  cl}$ as well as the gauge group
$G_{\text{gauge}}$. In particular, we couple the gauge theory to a non-trivial
background connection for the flavor symmetry, by weakly gauging it. This
leads to a set of cone inequalities not only for the Coulomb branch
parameters, but also to consistency conditions for the possible masses of the
flavor hypermultiplets. This description is very convenient, since the mass
deformations of the gauge theory are characterized in terms of simple
operations on the box graphs. In brief, a Coulomb branch phase is given in
terms of a representation graph (encoding the transformation of the matter
under both the gauge and classical flavor symmetries), as well as a
sign-assignment or decoration, which specifies the Coulomb branch phase. 

We will define a class of graphs, which characterize 5d gauge theories: they
encode the classical flavor symmetry of the gauge theory. These graphs, the {\it
box graph CFDs (BG-CFDs)}, encode equivalence classes of Coulomb branch phases,
which all carry the same classical flavor symmetry.  We first determine these
for all possible gauge groups and matter contents. From this we can then build
the corresponding BG-CFDs for quivers. 

We then use these to  constrain the possible weakly coupled gauge theory descriptions of a given CFD (starting with the CFD for a marginal 5d theory, but also for all its descendants), by embedding the BG-CFDs into the CFDs. This, for instance, implies that for rank two 5d SCFTs, the known weakly-coupled descriptions are a comprehensive list. More interestingly, however, we can predict new weakly coupled gauge theory or quiver descriptions for theories where only few such descriptions exist, such as the $(E_n, E_n)$ minimal conformal matter theories as well as $(E_8, SU(n))$, and $(E_7, SO(7))$ conformal matter. In all these cases a geometric derivation of the marginal CFD exists. 
Another implication of the relation between CFDs and BG-CFDs is that we can predict a large class of new dualities, i.e., gauge theories or quivers, which have the same UV fixed point.

There are 6d theories, where no known elliptic fibration in terms of a
Weierstrass model for the fully singular geometry exists. In such instances we
can turn the arguments around and use our approach to constrain the marginal
CFD, by using known gauge theory descriptions as well as flavor symmetry
enhancements of the 5d descendants.

The plan of the paper is as follows:  
To set the stage, we give a lightning review of 5d Coulomb branches in the language of box graphs in section \ref{sec:BG}.
We then propose how to use this approach to study 5d gauge/quiver theories
with matter and introduce the concept of {\it flavor equivalence classes of
Coulomb branch phases (or box graphs)} and the BG-CFDs in section
\ref{sec:somesingle}.  This is done for all types of gauge theories and matter
in 5d that have an  SCFT in the UV. 
In section \ref{sec:gauge_descriptions_from_CFDs} we use this to constrain the weakly-coupled descriptions of 
marginal theories for all rank two 5d theories, as well as the marginal theories associated to minimal conformal matter theories of type $(D_k, D_k)$, $(E_n, E_n)$, $(E_8, SU(n))$. 
For all these models, we computed the CFDs of the marginal theories from geometry. 
In section \ref{sec:Boot} we turn this around and discuss theories, which do not have a known description in terms of a fully singular Tate or Weierstrass model.
Nevertheless, we find that we can bootstrap the candidate marginal CFD using the information about 
known weakly coupled descriptions, and their flavor symmetry enhancements. 
Interestingly, these are precisely the theories that are relatively easily accessible using other methods (such as 5-brane webs), whereas for the models where we can determine the marginal CFD from geometry, the weakly coupled descriptions are often somewhat sparse (e.g., the $(E_n, E_n)$ conformal matter theories). 

Descendant 5d SCFTs and dualities among weakly coupled descriptions that can be infered from the CFDs are the topic of section \ref{sec:descquiv}. We first discuss two cases where the dualities have a geometric underpinning: the marginal theories from $(E_6, E_6)$ and $(E_7, SO(7))$ conformal matter and their descendants. We propose new quiver descriptions for these theories as well as the complete network of descendants and their gauge theory descriptions, whenever these exist. This is backed by a geometric analysis in appendix \ref{app:Resolutions}. 

Finally, in sections \ref{sec:FibPhas} and \ref{sec:CFDsfromBG} we return to geometry to tie up some loose ends, and show how all three strands of our analysis --- the resolved elliptic Calabi--Yau, the CFDs and the gauge theory Coulomb branch phases --- are connected. In particular we quantify how the gauge theory description needs to be supplemented to see, for instance, the full superconformal flavor symmetry manifest in geometry. We conclude with a summary and outlook in section \ref{sec:ConcOut}. 
In appendix \ref{app:RankTwoBG}, we summarize all gauge theory phases (and associated BG-CFDs) for the rank two 5d theories.
Appendix \ref{app:Resolutions} contains details of the resolutions for marginal and descendant theories.



\section{Coulomb Phases, Box Graphs, and 5d SCFTs}
\label{sec:BG}

In this section, we summarize some of the basic ingredients that will be
combined in this paper.  For starters, we discuss the structure of the Coulomb
branch of 5d gauge theories --- supplementing the material in Part I
\cite{Apruzzi:2019opn}, where some aspects of this were already discussed.
Here our focus will be to characterize the Coulomb branch of a 5d gauge theory
with matter, using the underlying representation-theoretic structure, based on
the classic \cite{Intriligator:1997pq} as well as the box graph description in
\cite{Hayashi:2014kca}.

\subsection{The Coulomb Branch of 5d Gauge Theories}

The Coulomb branch of a 5d $\mathcal{N}=1$ supersymmetric gauge theory coupled
to matter can have an intricate structure.  Let us consider a gauge theory
with reductive gauge group, $G_{\text{gauge}}$, written as a product
\begin{equation}
  G_{\text{gauge}}= \prod_i G_i^{r_i} \times U(1)^{r_A} \,,
\end{equation}
where $G_i$ are simple groups, the superscript indicates the rank, and
further 
\begin{equation}
  r_A = r - \sum_i r_i \,,
\end{equation}
is the rank of the abelian subgroup transverse to the Cartan subgroup of the
non-abelian factors. In this notation the Coulomb branch is isomorphic to 
\begin{equation}
  \left( \prod_i \mathbb{R}^{r_i} / W_{G_i} \right) \times
    \mathbb{R}^{r_A} \,,  
\end{equation}
where $W_{G_i}$ is the Weyl group of $G_i$. The quotient is the Weyl chamber,
defined by 
\begin{equation}
  \mathcal{C}_i = \mathbb{R}^{r_i} / W_{G_i} \,,  
\end{equation}
and it has the structure of a cone. Thus the grossest feature
of the Coulomb branch of 5d supersymmetric gauge theory is that it is a
collection of cones; this property comes only from considering the gauge group
itself, and this structure is further refined in a theory that also
incorporates matter \cite{Intriligator:1997pq}.

We can choose a basis such that the $\mathcal{C}_i$ are the fundamental Weyl
chambers of the $G_i$. Let $\alpha_j^{(i)}$ be the positive simple roots of
$G_i$, then we can write\footnote{Appropriate care must be taken here with
respect to weight and coweight lattices, which we are pairing between.}
\begin{equation}
  \mathcal{C}_i = \{ \phi \in \mathbb{R}^{r_i} \, | \, \langle \phi ,
  \alpha_j^{(i)}\rangle > 0
\text{ for all } j \} \,.
\end{equation}
Consider now a hypermultiplet, $H$, transforming in a representation $\bm{R}$
of $G$. On the Coulomb branch of the theory the gauge group is broken to
$U(1)^r$. The hypermultiplets transform as a collection of
$\text{dim}(\bm{R})$ hypermultiplets under the $U(1)^r$ in the
representation defined by the weights of $\bm{R}$. Let us, for the moment,
consider a representation $\bm{R}_i$ of $G_i$ and highlight the induced
structure on the Coulomb branch from the presence of these hypermultiplets. A
hypermultiplet carrying the charges under $U(1)^r$ corresponding to the weight
$\lambda$ of $\bm{R}_i$ becomes massless at the point in the Coulomb branch
where
\begin{equation}
  \langle \phi , \lambda \rangle = 0 \,.
\end{equation}
It is easy to see that for each $\lambda$ in $\bm{R}_i$ this gives rise to a
wall inside the Coulomb branch, and along this wall there exist additional
massless hypermultiplets. We can then describe the subchambers, or subwedges,
of $\mathcal{C}$ as defined by these walls. A phase of the gauge theory is
defined as a non-empty subwedge of the Coulomb branch such that each 
\begin{equation}
  \langle \phi , \lambda_I \rangle \text{ has a definite sign for each } \lambda_I \in \bm{R} \,.
\end{equation}
Determining the phase structure of the Coulomb branch involves determining
these subwedges, and the adjacency relations between them.

\subsection{Phases for 5d Gauge Theory via Box Graphs}

It is useful to formulate the problem of finding the Coulomb branch phases of
a 5d gauge theory in terms of so-called {\it Box Graphs}
\cite{Hayashi:2014kca}, which provide a succinct combinatorial way to list all
phases. 

The set of weights of each irreducible representation $\bm{R}$ of a group $G$
is generated by starting with a highest weight
and
from that highest weight one repeatedly subtracts positive simple roots,
following a simple prescription. That is, if $\lambda$ is a weight of
$\bm{R}$ then it can be written as the linear combination
\begin{equation}
  \lambda^{\text{hw}} - \sum_j n_j \alpha_j \,,
\end{equation}
where the $n_j$ are non-negative integers and $\lambda^{\text{hw}}$ is the
highest weight of $\bm{R}$. This action of generating all the weights of a
representation from the highest weight forms the weight diagram
\cite{MR1153249} of the representation. For an irreducible representation the
weight diagram is a connected directed graph where the nodes are the weights
of the representation and there exists an edge between two nodes if the two
weights differ only by a single positive simple root. We will use a particular
presentation of this weight diagram, as explained in the following
definition.

{\definition An \emph{undecorated box graph} is a graphical depiction of
the weight diagram \cite{MR1153249} for a representation $\bm{R}$ for a Lie
algebra $\mathfrak{g}$. Each weight of $\bm{R}$ is represented by a box, and if
two weights differ by the addition of a single simple positive root of
$\mathfrak{g}$ then their boxes are adjacent. }

As we have discussed above, each phase of the Coulomb branch of a 5d
$\mathcal{N}=1$ $G_\text{gauge}$ gauge theory with matter transforming in a
representation $\bm{R}$ of $G_\text{gauge}$ is specified by the signs of
$\langle \phi, \lambda \rangle$ for each $\lambda \in \bm{R}$. We mark the
sign assignment for the phase onto the undecorated box graph for the weight
diagram as in the following definition.

{\definition A \emph{(decorated) box graph} is an assignation of $\pm$
signs to each weight, $\lambda$, represented in an undecorated box graph such that 
\begin{equation}
  \{ \pm \langle \lambda , \phi \rangle > 0 \} \cap \{ \langle \alpha_i , \phi
  \rangle> 0 \} \,,
\end{equation}
has non-zero solutions for $\phi$. We will write $\lambda^\pm$ for the weight
appearing in the decorated box graph together with the assigned sign of
$\langle \lambda , \phi \rangle$.  In this way one can see that a decorated
box graph is defined such that it corresponds to a non-empty phase of the
Coulomb branch of a 5d gauge theory with gauge algebra $\mathfrak{g}$ and
matter transforming in the representation $\bm{R}$. }

In practice we will represent the positive and negative weights by blue/yellow
boxes. 

  It is clear to see that if the weight $\lambda$ is assigned the sign $+$ in
  the decorated box graph, then all weights $\widetilde{\lambda}$ such that
  \begin{equation}
    \lambda = \widetilde{\lambda} - n_j \alpha_j \,,
  \end{equation}
  for $n_j$ non-negative must also be assigned $+$. Explicitly, there only
  exists a non-zero value of $\phi$ solving
  \begin{equation}
    \langle \phi, \lambda \rangle = \langle \phi, \widetilde{\lambda} \rangle
    - n_j \langle \phi, \alpha_j \rangle \,,
  \end{equation}
  subject to the assumptions that
  \begin{equation}
    \langle \phi, \lambda \rangle > 0 \,, \quad \langle \phi, \alpha_j \rangle
    > 0 \,,
  \end{equation}
  if one also has 
  \begin{equation}
    \langle \phi, \widetilde{\lambda} \rangle > 0 \,.
  \end{equation}
  Similarly, one can show that if $\lambda$ is assigned a $-$ sign in the
  decorated box graph the all weights $\widetilde{\lambda}$ which satisfy
  \begin{equation}
    \widetilde{\lambda} = \lambda - n_j \alpha_j \,,
  \end{equation}
  again for $n_j$ non-negative, must also be assigned a $-$ sign.

  As we have just seen, not all the weights of the representation can be
  assigned signs independently, in fact it is usually the case that once the
  signs are associated to a few weights the signs of all the other weights
  follows of necessity. We will refer to these weights as \emph{extremal}, and
  they are defined as follows.

{\definition\label{def:extremal}
  A weight $\lambda^+$ in a decorated box graph is \emph{extremal} if
  there does not exist a weight $\widetilde{\lambda}^+$ in the decorated box
  graph such that
  \begin{equation}
   \widetilde{\lambda} = \lambda - n_j \alpha_j \,,
  \end{equation}
  for $n_j$ non-negative. Similarly a weight $\lambda^-$ is extremal if there
  does not exist a $\widetilde{\lambda}^-$ such that
  \begin{equation}
    \lambda = \widetilde{\lambda} - n_j \alpha_j \,,
  \end{equation}
  again for $n_j$ non-negative integers.

{\definition\label{def:split} In a decorated box graph a root, $\alpha_j$, of the gauge group $G$ is
  said to \emph{split} if we have two weights related in the box graph as
  \begin{equation}
    \widetilde{\lambda} = \lambda - \alpha_j \,,
  \end{equation}
  such that $\widetilde{\lambda}$ is assigned the sign $-$ and $\lambda$ is assigned
  $+$. For such split roots we can use
  \begin{equation}
    \langle \phi, \lambda \rangle - \langle \phi, \widetilde{\lambda} \rangle
    = \langle \phi, \alpha_j \rangle \,,
  \end{equation}
  to see that $\langle \phi, \alpha_j \rangle > 0$ is automatically satisfied
  by the sign assignment of $\lambda$ and $\widetilde{\lambda}$. 
  Generally the
  $\widetilde{\lambda}^+$ and $\lambda^-$ may be further rewritten as a
  positive linear combination of the extremal weights and the non-split roots.
  }

The motivation for the previous two definitions is as follows. The signs
associated to the non-extremal weights and the split roots are determined from
the signs associated to the extremal weights and the non-split roots. In this
manner each subwedge for a simple gauge group $G$ and matter in an irreducible
representation $\bm{R}$ can be minimally written as
\begin{equation}
  \mathcal{W}_i = \{ \phi \in \mathbb{R}^{\text{rank}(G)} \, | \, 
  \pm \langle \phi , \lambda^{\text{extremal}}_j\rangle > 0 
  \text{ and } 
  \langle \phi , \alpha^{\text{non-split}}_k\rangle > 0 \text{ for all } j, k \}
  \,,
\end{equation}
where the indices run over all extremal (the $\pm$ is given by the sign of the
extremal weight) and non-split roots in the particular
subwedge under consideration. In this way we can see that it is this
restricted set of weights and roots that are the ``irreducible`` objects
generating the subwedge.

In the pictorial decorated box graphs this leads to the following
definition of flow rules, which follow directly from the
above
consistency requirements for the sign assignments in the box graphs:  
{\definition The \emph{flow rules} state that if we assign the sign + to
a weight of an undecorated box graph then we must assign + also to every box
up and to the left of that weight. Similarly if we assign - to a particular
weight then we must assign - to every weight that is down and to the right.
This is captured graphically by 
\begin{equation}\label{FlowRules}
\includegraphics[width=4cm]{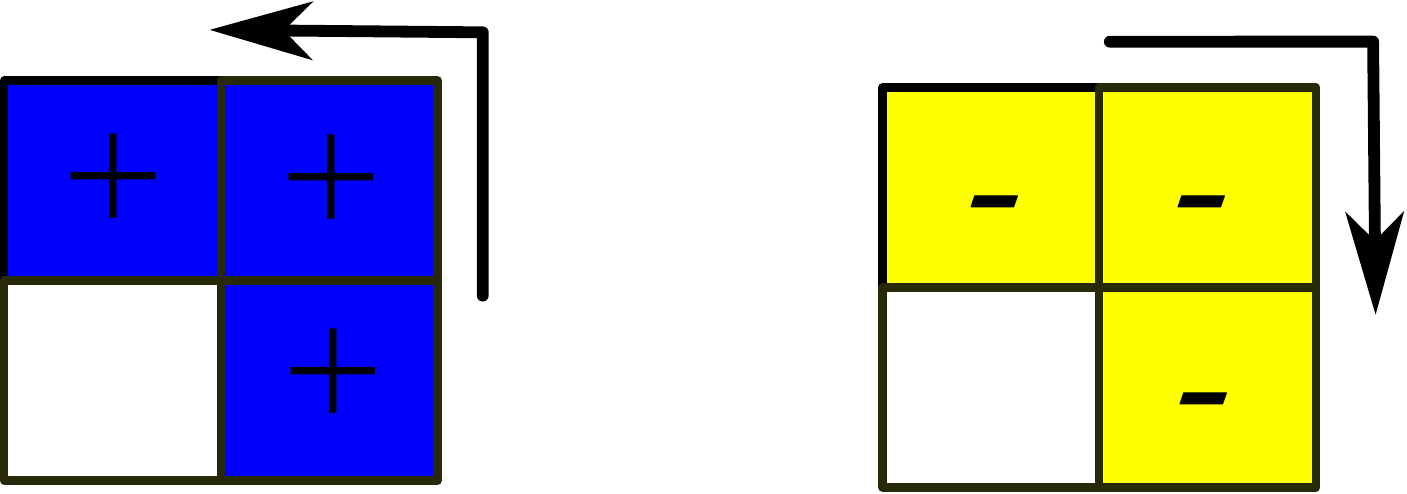} \,.
\end{equation}
}

{\definition 
A \emph{flop transition} exists between two decorated box graphs if a single
weight differs in assigned sign. These weights are necessarily extremal
weights, and a flop transition is changing precisely one sign assignment of an
extremal weight.  Generally we will be considering representations that
contain weights $\lambda$ and $-\lambda$, for example self-conjugate
representations, and when we say that a single weight differs in sign we mean
that the signs associated to $\lambda$ and $-\lambda$ are swapped. Two Coulomb
phases are adjacent inside of the Coulomb branch if the associated box graphs
are related by a flop transition.
}

For a more in depth discussion of box graph and Coulomb phases we refer the reader to 
\cite{Hayashi:2014kca} and \cite{Braun:2014kla, Braun:2015hkv, Lawrie:2015hia}. 

\subsection{Box Graphs and Flavor Symmetries}

Although box graphs are used to characterize the Coulomb branch phases of gauge theories in 5d (or 3d) with matter, we can equally apply them to determine the structure of the extended Coulomb branch, of a gauge theory with classical flavor symmetry $G_{\text{F, cl}}$. Consider a gauge theory with matter in $(\bm{R}, \bm{R}_{\text{BG}})$ of $G_{\text{gauge}} \times G_{\text{BG}}$. To determine which matter multiplets can be given masses and can be decoupled from the theory, recall that the prepotential has a contribution
\be
\ba
\mathcal{F}\supset -\frac{1}{12}  \sum_{{\bm{R}}_{\text{BG}}} 
\sum_{\lambda_{{\bm{R}_{\text{BG}}}
}} |\lambda_i \, \phi^i + m_f|^3\,,
\ea
\ee
where $\phi^i$ are the scalars in the vector multiplet, which are coordinates on the Coulomb branch, and $m_f$ are masses for hypermultiplets. The sum runs over the weights of the representation.  Promoting the masses $m_f$ as parameters of the Coulomb branch, corresponds to weakly gauging part of the flavor symmetry.
In practice, this is equivalent to studying the Coulomb branch, or box graphs, for bifundamental matter of $(\bm{R}_{\text{gauge}}, \bm{R}_{\text{BG}})$ of $G_{\text{gauge}} \times G_{\text{BG}}$. 

Our strategy will be to determine all phases of the extended Coulomb branch using box graphs, starting with a marginal 5d theory, i.e., the gauge theory description of a circle-reduction of a 6d SCFT. This determines all descendant gauge theories, that can be reached by successively decoupling hypermultiplets. As we shall see in section \ref{sec:somesingle}, equivalence classes of box graphs will then characterize all 5d SCFTs that admit a weakly coupled gauge theory description. 
We will illustrate the box graph approach in section \ref{sec:RankOne} with the rank one theories. This class of 5d theories, descend from a single marginal theory, which is the dimensional reduction of the rank one E-string.

We list for convenience the flavor group for all possible 5d gauge
  theories which can have a non-trivial UV fixed point\footnote{These are
    only the matter fields that we consider in this paper. In addition one can have the
    triple antisymmetric of $SU(n)$ and the spinor/conjugate spinor of $SO(n)$
  for low ranks of the gauge group; one can also include adjoint matter, which
may have a 6d fixed point with sixteen supercharges. See
\cite{Jefferson:2017ahm} for more details.}, following
  \cite{Intriligator:1997pq}.  For gauge theories with a simple gauge group,
  the data is summarized\footnote{We note that we rectify the
    typographical error in \cite{Intriligator:1997pq} whereby the classical
    flavor symmetry rotating $N_{27}$ fundamental hypermultiplets of $E_6$ was
    written as $SU(N_{27})$ rather than $U(N_{27})$.} in table \ref{tab:GTCFL}. 
  
\begin{table}
	\begin{align*}
	\renewcommand*{\arraystretch}{1.2}
		\begin{array}{c|c|c}
			\text{Gauge Group} & \text{Matter} & \text{Flavor Group }\, G_\text{F,cl} \\ \hline \hline
			& N_f \times \textbf{fund} & \\
			SU(n \geq 3) & + N_a \times \textbf{anti-sym} & U(N_f) \times U(N_a)
      \times U(N_s) \\
			& + N_s \times \textbf{sym} & \\ \hline
			\multirow{2}{*}{$Sp(n)$} & N_f \times \textbf{fund} & \multirow{2}{*}{$SO(2N_f) \times Sp(N_a)$} \\
			& + N_a \times \textbf{anti-sym} & \\ \hline
			SO(n) & N_v \times \textbf{vector} & Sp(N_v)   \\ \hline
			G_2 & N_7 \times {\bf 7} & Sp(N_7)\\ \hline
			F_4 & N_{26} \times {\bf 26} & Sp(N_{26}) \\ \hline
			E_6 & N_{27} \times {\bf 27} & U(N_{27}) \\ \hline
			E_7 & N_{56} \times {\bf 56} & SO(2N_{56})
		\end{array}	
	\end{align*}
	\caption{Flavor symmetries of 5d gauge theories with simple gauge groups. 
		The matter content has been restricted to representations that allow for the gauge theory to have an honest 5d SCFT limit.
		For this reason, there is no $E_8$ theory with non-trivial matter in this table.
		Note that for $SU(n\geq 3)$ gauge theories, the $N_i$ hypermultiplets transforming in the (anti-)fundamental of the $SU(N_i) \subset U(N_i)$ flavor factor have charges $(-)1$ under the baryonic $U(1) \subset U(N_i)$.
	}\label{tab:GTCFL}
\end{table}

For a quiver gauge theory, consisting of $Q$ gauge factors $G_I$, there are $Q-1$ hypermultiplets in the bifundamental of $G_I \times G_{I+1}$
\be
\includegraphics*[width=10cm]{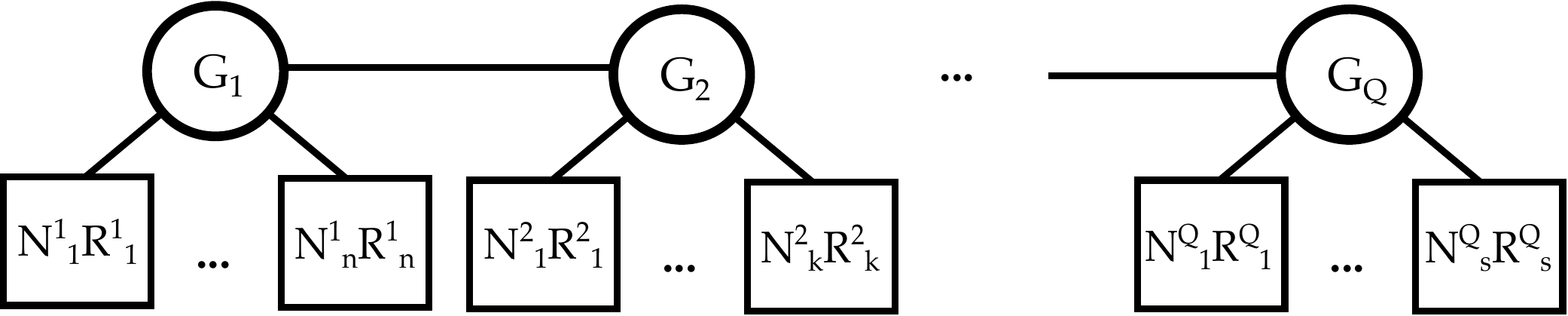}\,.
\ee
Furthermore, there can be $N_{f_I}$ hypermultiplets transforming in a representation $\bm{R}_{G_I}$ of the gauge group $G_I$.
Typically, one represents such a quiver as a set of nodes, each corresponding to one gauge factor $G_I$.
Bifundamental matter are showing as lines connecting two nodes, and additional hypermultiplets are indicated by lines attached to a single node.

The global symmetry group can be thought as coming from 3 different contribution:
\begin{itemize}
\item Each of the $Q$ gauge group nodes in the quiver has an associated $U(1)_T$. 
\item For each full hypermultiplet transforming in the bifundamental of $G_I \times G_{I+1}$, there will be a baryonic $U(1)_B$, which is an $SU(2)_B$ for an hypermultiplet in the fundamental of two $SU(2)$ gauge nodes.
\item The symmetry rotating the $N_{f_I}$ hypers can be read off from the single simple gauge group classical flavor symmetries.
\end{itemize}
The total global symmetry is a product of these factors.

\begin{figure}
  \centering
  \includegraphics[width=11cm]{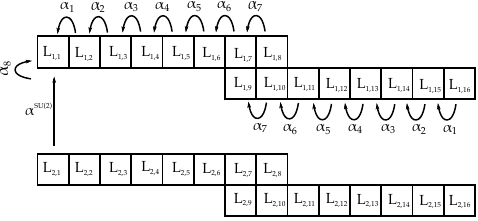}
  \caption{The representation graph (or undecorated box graph) for the $({\bf 2,16})$ representation of 
  $SU(2)_\text{gauge} \times SO(16)$. The simple roots of $SO(16)$  are $\alpha_i$, and for $SU(2)_\text{gauge}$ the simple root is $\alpha^{SU(2)}$. The arrows indicate how weights are mapped into each other under the addition of the roots. 
 The weights are $L_{i,j}= L_i^{\bf 2} +L_{j}^{\bf 16}$, where $L^{\bm R}_i$
 are the fundamental weights of the representation ${\bm R}$. The action of
 the roots is indicated by the arrows. Note that $L^{\bf 16}_{i+8}= -
 L^{\bf 16}_{9-i}$ for $i= 1, \ldots, 8$. 
  }\label{fig:SU2SO16bare}
\end{figure}


\subsection{Intermezzo: Gauge Theory Phases for Rank One 5d SCFTs}
\label{sec:RankOne}

To illustrate the inner workings of box graphs, let us first consider the
simplest example: the rank one theories in 5d, which e.g. arise as dimensional
reductions and mass deformations of the 6d rank one E-string theory.  The
marginal theory admits an $Sp(1) = SU(2)_\text{gauge}$ gauge theory
description with 8 fundamental flavors \cite{Seiberg:1996bd}
\be
SU(2) + 8 \bm{F} \,.
\ee
The flavor symmetry at weak coupling is then $SO(16) = G_{\text{BG}}$. 
In other words, the theory has matter in the $({\bf 2},{\bf 16})$ representation of the $SU(2)_\text{gauge} \times SO(16)$.
This induces a wedge structure on the Coulomb branch of the theory when we weakly gauge the
$SO(16)$, as explained earlier.

To study the different phases of this Coulomb branch, and the corresponding fiber structure, we denote the positive simple roots in the Cartan--Dynkin basis by \begin{equation}
  \begin{aligned}
	SO(16):\qquad &\left\{
	\ba
	 \alpha_1 &= (0; 2,-1,0,0,0,0,0,0)\,,\cr
	  \alpha_3& = (0; 0,-1,2,-1,0,0,0,0)\,,  \cr 
	   \alpha_5 &= (0; 0,0,0,-1,2,-1,0,0)  \,,\cr 
	  \alpha_7 &= (0; 0,0,0,0,0,-1,2,0)  \,,\cr 
	  \ea
	  	\ba
	\alpha_2 &= (0;-1,2,-1,0,0,0,0,0) \cr
	 \alpha_4& = (0;0,0,-1,2,-1,0,0,0) \cr 
	  \alpha_6& = (0;0,0,0,0,-1,2,-1,-1) \cr 
	  \alpha_8 &= (0; 0,0,0,0,0,-1,0,2) \cr 
	  \ea
\right. \cr 
SU(2) :\qquad & \alpha^{SU(2)} = (2;0,0,0,0,0,0,0,0) \, ,
	\end{aligned}
\end{equation}
and in this notation the highest weight of the ${(\bf 2, 16})$ is
\begin{equation}
  L_{1,1} = (1;1,0,0,0,0,0,0,0) \,.
\end{equation}
The undecorated box graph for this representation is given in figure \ref{fig:SU2SO16bare}. 
In particular we denote by 
\be
L_{i,j} = L_{i}^{\bf 2} + L_j^{\bf 16} \,,
\ee
the sum of fundamental weights of ${\bf 2}$ and ${\bf 16}$. The simple roots of $SO(16)$ can be written as 
\be
\alpha_i= L^{\bf 16}_i-L^{\bf 16}_{i+1}\,,\quad i = 1, \cdots , 7 \,,\qquad \alpha_8= L^{\bf 16}_7 + L^{\bf 16}_8 \,.
\ee
Starting with the marginal theory, we determine all the consistent phases using the box graphs. 
The marginal theory is such that all roots of the weakly gauged flavor symmetry $SO(16)$ are contained in the splitting of the $SU(2)$. 
In this case the two ${\bf 16}$ representation graphs that are part of the box graph in this case have the same coloring, i.e. coloring the $+/-$ sign assignments in blue/yellow, the decorated box graph associated to this phase is 
\be\label{EstringTopBox}
\includegraphics[width=4cm]{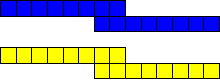} \,,
\ee
where each box corresponds to a weight as in figure \ref{fig:SU2SO16bare}. 
Consistency with the flow rules determine then all further phases, by applying flops. Lets illustrate this by performing one flop on the box graph (\ref{EstringTopBox}). The only extremal weights/boxes are $L_{1,16}$ and $L_{2,1}$ (recall that this representation is self-conjugate so each flop will require changing the color of one box in each of the two ${\bm{16}}$s). After the flop transition, the new box graph is 
\be
\includegraphics[width=4cm]{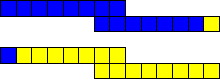} \,.
\ee
Continuing along these lines results in figure \ref{fig:Rank1BGALL}.

\begin{figure}
\centering
\includegraphics[width=6cm]{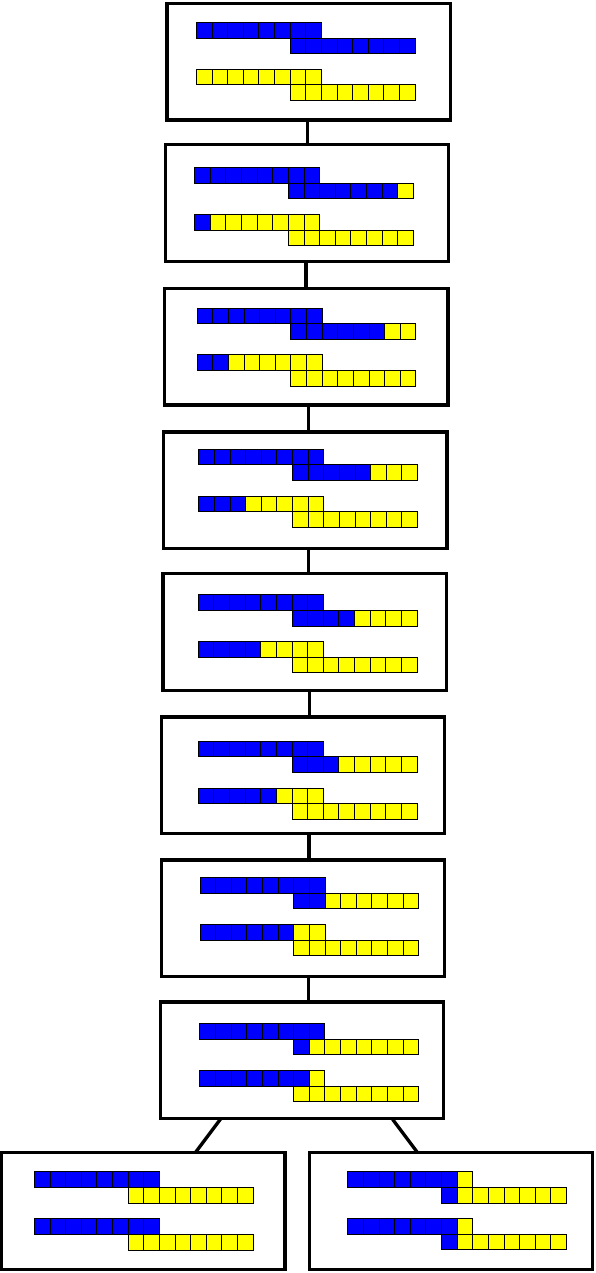}
\caption{5d rank one theories: Box graphs for $(\bf{2}, \bm{16})$ of
$SU(2)\times SO(16)$, for the marginal theory $SU(2)+ 8 \bm{F}$, which is
shown at the top of the tree.  Connections indicate `flop transitions',
which in the gauge theory correspond to different phases of the extended
Coulomb branch, and from the point of view of the $SU(2)$ gauge theory,
corresponds to decoupling fundamental hypermultiplets.
\label{fig:Rank1BGALL}}
\end{figure}

This chain is of course precisely the phases of the rank one theories in
  5d rank as described in the classic works of
  \cite{Seiberg:1996bd,Morrison:1996xf}, and each flop corresponds to
  decoupling a fundamental hypermultiplet.  In particular, the chain obtained
  from the box graphs also captures the two possible ways of decoupling the
  only hypermultiplet of an $SU(2) + 1 {\bm F}$ theory to flow either to an
  $SU(2)_{\theta = 0}$ or $SU(2)_{\theta = \pi}$ theory.  Indeed, as we will
  discuss in section \ref{sec:somesingle}, there is a natural way of
  interpreting the flop transitions as decoupling of matter multiplets in the
  limit where we restore the coupling of $G_\text{BG}$ to 0, which also
  establishes a \emph{subgroup} of $G_\text{BG}$ as the physical weakly
  coupled flavor
  symmetry after decoupling the matter.


\section{Gauge Theory Phases and Box Graphs for Arbitrary Quivers}
\label{sec:somesingle}

In this section we will determine the set of 5d gauge theories that arise as
mass deformations of a given theory with gauge group $G_\text{gauge}$.  We
will find that the structure of mass deformations amongst these form a tree of
theories with varying matter content charged under $G_\text{gauge}$.  These
theories will not necessarily be distinct, in that they may still admit, what
we will call, ``discrete dualities''; examples of such dualities are shifting
the Chern--Simons level, $k \rightarrow -k$, or shifts of $\theta$-angles in
such a way that the theories are identical.  More generally, we will also
discover such discrete dualities to incorporate simultaneous modification of
the number of hypermultiplets coming from different flavor nodes in a quiver. 

As we are interested in gauge theories with an SCFT limit, we will focus
either on theories where $G_\text{gauge}$ is a simple group that appears in
table \ref{tab:GTCFL}, or on quiver gauge theories where the nodes carry one
of these simple factors, together with some of the matter listed in the
aforementioned table. For low ranks of the gauge group there can be matter
fields transforming in more exotic representations, which still flow to an
interacting UV fixed point; one example is the triple anti-symmetric
representation of $SU(N)$, for sufficiently small values of $N$. Some of these
exceptional gauge theories are pointed out in \cite{Jefferson:2017ahm},
however we will not consider them further here, as the number of descendant
gauge theories can straightforwardly be determined by the methods explained
here. 

For the simple gauge theories in table \ref{tab:GTCFL} the phases of the
Coulomb branch has been often studied. For $SU(N)$ gauge theories the Coulomb
phases are enumerated for general $N$ in
\cite{Intriligator:1997pq,Hayashi:2014kca,Braun:2015hkv} and for specific low
values of $N$ in
\cite{Esole:2011sm,Marsano:2011hv,Hayashi:2013lra,Esole:2014bka,Esole:2014hya,Braun:2014kla,Lawrie:2015hia,Esole:2019asj},
for $Sp(N)$ in \cite{Intriligator:1997pq,Hayashi:2014kca,Esole:2017hlw}, for $SO(N)$ in
\cite{Intriligator:1997pq,Hayashi:2014kca,Esole:2017qeh,Esole:2018csl}, and
for the exceptional cases, $G_2$, $F_4$, $E_6$, and $E_7$ in
\cite{Esole:2017qeh,Esole:2018mqb}, \cite{Esole:2017rgz},
\cite{Hayashi:2014kca}, and \cite{Hayashi:2014kca,Esole:2018vnm},
respectively.  In this paper we will not require an understanding of the full
set of Coulomb phases, but, as we shall see momentarily, only of certain
equivalence classes of the extended Coulomb phases for the theory after weakly
gauging the classical flavor symmetry rotating the hypermultiplets, as it is
these that can be related each to a distinct descendant gauge theory. To
specify a Coulomb phase we shall use the object known as a box graph that was
introduced in \cite{Hayashi:2014kca}, and that has been summarized in section
\ref{sec:BG}, and for the equivalence classes that we shall define it is
necessary to know such box graphs for the fundamental or vector
reprensetations of $U(N)$, $Sp(N)$ and $SO(N)$, which are determined in the
aforementioned paper.

The procedure followed in this section to obtain the set of descendant gauge
theories is as follows. We will first weakly gauge the classical flavor
symmetry that rotates the hypermultiplets associated to a flavor node in a 5d
gauge theory quiver. This theory has a Coulomb branch,
$\mathcal{C}^\text{w.g.}$, and in the limit where we take the gauge coupling
of the weakly gauged flavor symmetry to zero, this Coulomb branch
fractionates. The result is a set of Coulomb branches of all of the descendant
gauge theories arising as mass deformations of the original gauge theory.
There may be redundancies in this description as, for instance, the same
Coulomb branch for a descendant can appear multiple times within this set. It
is vital, therefore, to, after determining in a redundant way all of the
descendant theories, identify those identical Coulomb branches as belonging to
the same theory. We refer to these identifications as ``discrete dualities'',
although this is something of a misnomer since they are often not dualities
but directly equivalent descriptions of the same theory. In this section we
will determine  the larger set of descendants,  where there are still these
redundencies.  This will first be carried out for the single node gauge
theories, as the logic therein will extend multiplicatively across arbitrary
quivers.

The key concepts that we will define and determine for all gauge theories, are
equivalence classes of box graphs called a flavor-equivalence class, and
associated graphs, the {\it Box Graph Combined Fiber Diagram} (BG-CFD), as a
collection of vertices and edges; these definitions appear in section
\ref{sec:3def}. In sections \ref{sec:3complex}, \ref{sec:3quat}, and
\ref{sec:3real} we will consider all of the gauge theoretic descendants for
single gauge node quivers of the form
\begin{equation}\label{fig:SingleNodeQuiver}
  \includegraphics[scale=2.0]{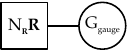} \,,
\end{equation}
where $G_\text{gauge}$ is a simple Lie group and the representation $\bm{R}$
is, respectively, complex, quaternionic, or real. Each of these descendants
will be in a one-to-one correspondence to a flavor-equivalence class, and will
have an associated BG-CFD. In section \ref{sec:3quiver} we show how this
analysis extends simply to determine all of the descendant gauge theories
associated to arbitrary quiver gauge theories with building blocks
(\ref{fig:SingleNodeQuiver}), by gluing together such gauge nodes with
bifundamental hypermultiplets, by attaching multiple flavor nodes to a given
gauge node, or combinations of both of these constructions.

\begin{sidewaystable}
  \centering
  \begin{tabular}{|c|c|c|c|c|c|}
    \hline
    & $G_\text{gauge} + N_{\bm{R}}\bm{R}$ & Quiver Descendants & Tree & $G_{\text{F,cl}}$& BG-CFD \cr\hline\hline
    $\mathbb{C}$ & $\begin{array}{c} 
    \cr 
      SU(N \geq 3) + N_{\bm{R}}\bm{F} \cr
      SU(N \geq 5) + N_{\bm{R}}\bm{AS} \cr
      SU(N \geq 3) + N_{\bm{R}}\bm{Sym} \cr
      E_6 + N_{\bm{R}}\bm{27} \cr 
      \cr 
    \end{array}$ 
    & 
    $\frac{1}{2}(N_{\bm{R}} + 1)(N_{\bm{R}} + 2)$ 
    &
    Figure \ref{fig:SUnUmFlopGraph} 
    &
    $U(N_{R})$
    & 
    $\begin{gathered}
      \includegraphics[width= 4cm]{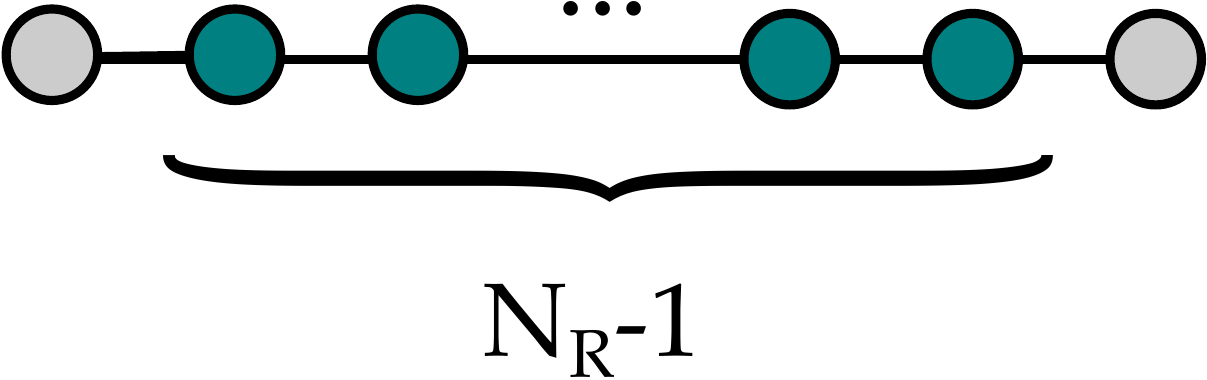}
    \end{gathered}$ \cr\hline
    $\mathbb{H}$ 
    & $\begin{array}{c}
      Sp(N \geq 1) + N_{\bm{R}}\bm{F}\,, \cr
      E_7 + N_{\bm{R}}\bm{56} \,, N_{\bm{R}} \in \mathbb{N}
    \end{array}$ 
    & 
    $\ba 
    N_{\bm{R}} = 0: &\  1  \cr
     N_{\bm{R}} \in \mathbb{N}_{>0}:&  N_{\bm{R}} + 2  \cr 
     \ea$
     & Figure   \ref{fig:SpNSO2Nftree} & 
    $SO(2N_{\bm{R}})$& 
    $\ba\cr 
      N_{\bm{R}} = 1:&\qquad  \includegraphics[width=1cm]{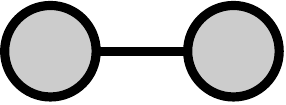}  \cr
     N_{\bm{R}} \in \mathbb{N}_{>1}&\qquad  \includegraphics[width= 4cm]{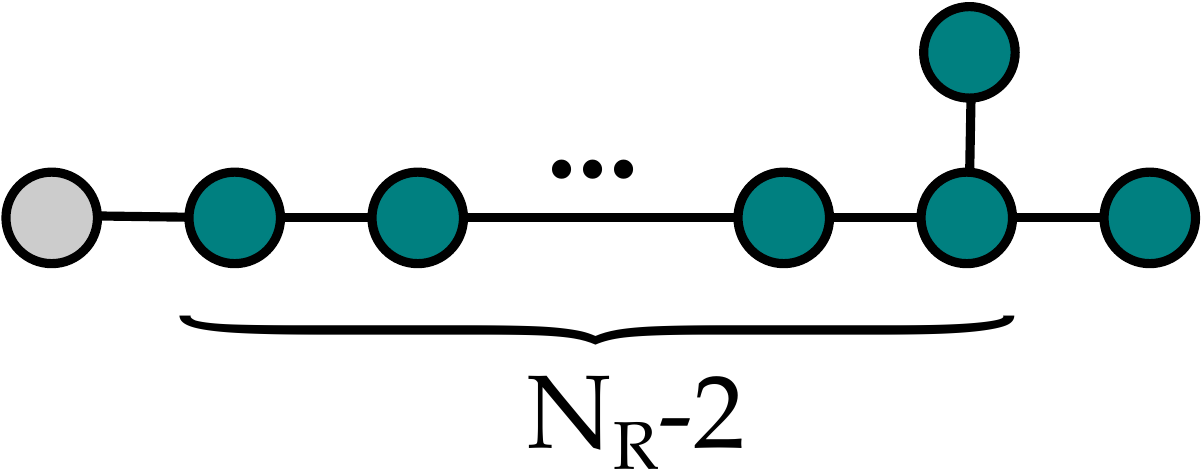} \cr 
    \ea$ \cr\hline
        $\mathbb{H}$ 
    & $\begin{array}{c}
      E_7 + N_{\bm{R}}\bm{56} \,, \cr 
      N_{\bm{R}} \text{ half-integral}
    \end{array}$ 
    & 
    $N_{\bm{R}} +\frac{1}{2}$
         & Figure   \ref{fig:SOoddFlop}
         & 
    $SO(2N_{\bm{R}} +1)$
    & 
	$\ba\cr 
     \includegraphics[width= 4cm]{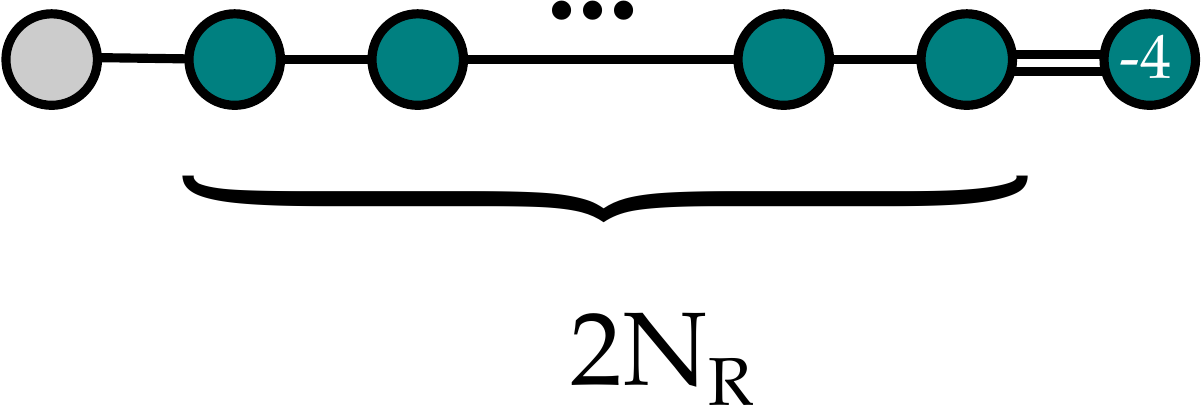}\cr 
     \ea $
  \cr\hline
    $\mathbb{R}$ & 
    $\begin{array}{c}\cr 
      Sp(N \geq 2) + N_{\bm{R}}\bm{AS} \cr
      SO(N \geq 5) + N_{\bm{R}}\bm{V} \cr
      SU(4) + N_{\bm{R}}\bm{AS} \cr
      G_2 + N_{\bm{R}}\bm{7} \cr
      F_4 + N_{\bm{R}}\bm{26} \cr 
      \cr 
    \end{array}$ & $N_{\bm{R}} + 1$ & Figure \ref{fig:SpNNatree} 
    & $Sp(N_{\bm{R}})$
    &
    $\ba
   \cr 
   \includegraphics[scale=0.2]{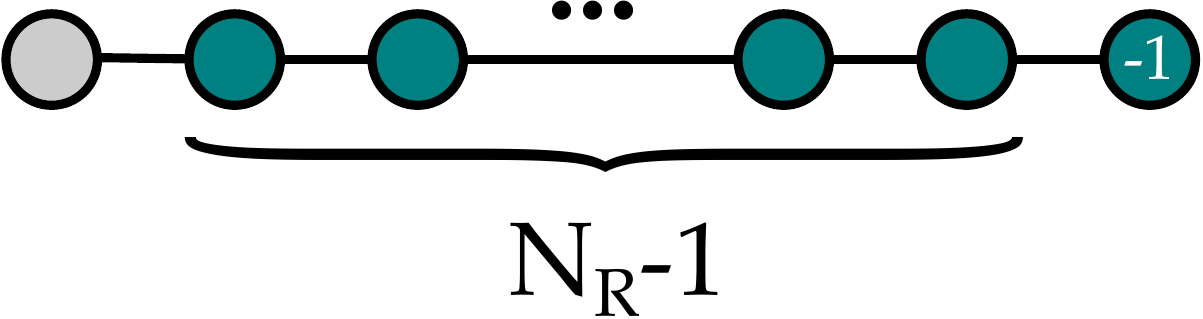}
 \ea$    \cr\hline
  \end{tabular}
  \caption{Classical flavor symmetries $G_{\text{F,cl}}$ and descendants as well as the corresponding   box-graph CFDs (BG-CFDs). The theories are arranged by the type of representation that the gauge group has (complex, quaternionic, real).
{Note that the number of quiver descendants include the starting point itself. }
   In the case of $E_7$ the number of hypermultiplets in the $\bm{56}$ can be half-integral, which gives classical flavor symmetries of $D$ or $B$ type. 
 Notation for the BG-CFDS: marked (blue)
  vertices are $(-2)$-vertices and grey vertices are $(-1)$-vertices of the CFD. The
  $Sp(n)$ diagram is non-simply-laced, which is indicated by a marked
  $(-1)$-vertex. \label{tab:BGCFDs}}
\end{sidewaystable}

\subsection{Flavor-equivalence Classes and Box Graph CFDs}\label{sec:3def}

In section \ref{sec:RankOne} we discussed the gauge theory phases for the
$SU(2)$ gauge theory with a weakly gauged $SO(16)$ flavor symmetry and matter
transforming in the $(\bm{2,16})$ representation. The set and structure of the
Coulomb phases was in a one-to-one correspondence with the set of $SU(2)$
gauge theories that arise from mass deformations of $SU(2) + 8\bm{F}$. For
more general gauge theories this will not always be the case, indeed the
Coulomb phases after weakly gauging the classical flavor symmetry rotating the
hypermultiplets will be in a many-to-one map onto the mass deformations of the
original gauge theory. The reason for this is that there will be distinct phases in the
weakly gauged theory where the distinction is only moving amongst the Coulomb
phases of the original gauge theory; we are not interested in the distinction
between these phases as they do not correspond to mass deformations of the
original theory, but instead to moving on its Coulomb branch. To remove this
redundancy and to restore a one-to-one relationship, we define an appropriate
equivalence class.

\begin{definition}{\it Flavor-equivalence Class of Box Graphs\\}
\label{DefFEC}
  Consider two box graphs associated to Coulomb phases of a gauge
  theory with symmetry groups $G_{\text{gauge}} \times G_\text{BG}$, where the
  $G_\text{BG}$ is considered as a weakly gauged flavor symmetry. Denote the
  simple roots of $G_{\text{gauge}}$ by $\alpha_i^{\text{gauge}}$ and those of
  $G_\text{BG}$ by $\alpha_j^{\text{BG}}$. 
  Then, these two box graphs are flavor-equivalent if the splitting of the $\alpha^{\text{gauge}}_j$ contains, in total, the same subset of roots $\alpha^{\text{BG}}_i$ of $G_{\text{BG}}$.  Furthermore, if none of the $\alpha^{\text{gauge}}_j$ split in the box graphs then the two box graphs are flavor-equivalent if and only if they are identical.
\end{definition}

It is easy to see that, for an irreducible representation $(\bm{R}_\text{gauge},
\bm{R}_\text{BG})$ of $G_{\text{gauge}} \times G_\text{BG}$, the
flavor-equivalence class is completely determined by the decoration of the
weights of $\bm{R}_\text{BG}$ associated to the highest and lowest weights of
$\bm{R}_\text{gauge}$, as they appear in the tensor product of the weights that
form the product representation $(\bm{R}_\text{gauge}, \bm{R}_\text{BG})$.
Furthermore, when there are multiple different matter representation of
$G_{\text{gauge}}$ there is a different flavor symmetry associated to each
matter field, and so the flavor-equivalence classes are multiplicative across
the different matter fields; a prominent example of this will appear in
section \ref{sec:3quiver}.

To determine the structure of the tree of mass deformations we need in
addition the following definitions. 

\begin{definition}{\it Flop Transitions for Flavor-equivalence
  Classes\\}\label{Def:Fflop}
Two flavor-equivalence classes of box graphs are
related by a \emph{flop transition} if there exist representatives of each
equivalence class which are related by a flop transition.
\end{definition}

\begin{definition}{\it F-extremal Weights\\}\label{Def:Fext}
  A weight is {\it F-extremal}, if it is an extremal weight which, when flopped, changes the flavor-equivalence class.
\end{definition}

\begin{definition}{\it F-extremal Weights Inside the Combined Roots of the Gauge Group\\}\label{def:curvesinsurface}
The flavor-equivalence class is associated to the splitting 
  \be
\ba\label{SplitBG}
  \sum_j \alpha_j^\text{gauge}
  \quad  \rightarrow\quad   \sum_\ell \alpha_{\ell}^\text{BG} + \sum_m \epsilon_m L^{\epsilon_m}_m \, ,
\ea
\ee
where $L_m^\epsilon$, with $\epsilon =\pm$ are, by definition, a subset of the F-extremal weights of this flavor-equivalence class. 
We refer to the $L_m^{\epsilon_m}$ that appear in \eqref{SplitBG} as F-extremal weights \emph{inside the combined roots of the gauge group}. 
\end{definition}

Moreover, this splitting associated with a reduced (flavor-equivalent) box
graph defines the {\it box graph CFD (BG-CFD)}. This is a sub-graph of the
full CFD \cite{Apruzzi:2019vpe} that is associated to the UV fixed point of
the flavor-equivalence class. 

\begin{definition}{\it Box Graph CFD (BG-CFD)\\}\label{def:BGCFD}
  Given a flavor-equivalence class the \emph{BG-CFD} is the intersection graph
  of the F-extremal weights inside the combined roots of the gauge group and the flavor roots $\alpha_\ell^\text{BG}$ that appear in \eqref{SplitBG}.
\end{definition}

An example of a BG-CFD is shown in figure \ref{fig:CFDInBG}.  The BG-CFD
encodes the part of the flavor symmetry of an SCFT that is manifest in the
weakly-coupled gauge theory description.  It forms generically a strict
subgraph of the CFD associated to said SCFT. A complete set of them is listed in table \ref{tab:BGCFDs}.

Given these definitions in the remainder of this section we are going to
determine the flop graph of all flavor-equivalence classes of box graphs for
an arbitrary gauge theory quiver built out of nodes corresponding to the
simple gauge theories listed in table \ref{tab:GTCFL}. Furthermore we will
determine the (disconnected) graphs, the BG-CFDs, associated to each quiver
gauge theory. We proceed by first determining the flavor-equivalence classes,
and the BG-CFDs, for the single node quivers listed in the aforementioned
table.

We summarize the results of this section, the number of flavor-equivalence
classes for the gauge theories with a simple gauge algebra as listed in table
\ref{tab:GTCFL}, and arbitrary quivers built therefrom. Similarly, for each possible classical flavor group, we
summarize the BG-CFDs in table \ref{tab:BGCFDs}.

\begin{figure}
\centering
\includegraphics*[width=8cm]{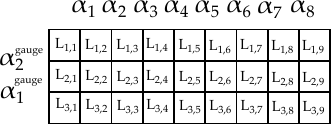}
\caption{Representation graph for $(\bm{3}, \bm{9})$ of $SU(3)_{\text{gauge}} \times U(9)_\text{BG}$. $L_{i,j}= L^{SU(3)}_i + L^{SU(9)}_j$, where $L_k$ indicates the fundamental weights of the respective groups. $\alpha_i$ are the simple roots of the classical flavor symmetry $U(9)$,  and $\alpha_i^{\text{gauge}}$ those of the gauge group $SU(3)$. \label{fig:RepGraphEx}}
\end{figure}

\begin{figure}
\centering
\includegraphics*[width=12cm]{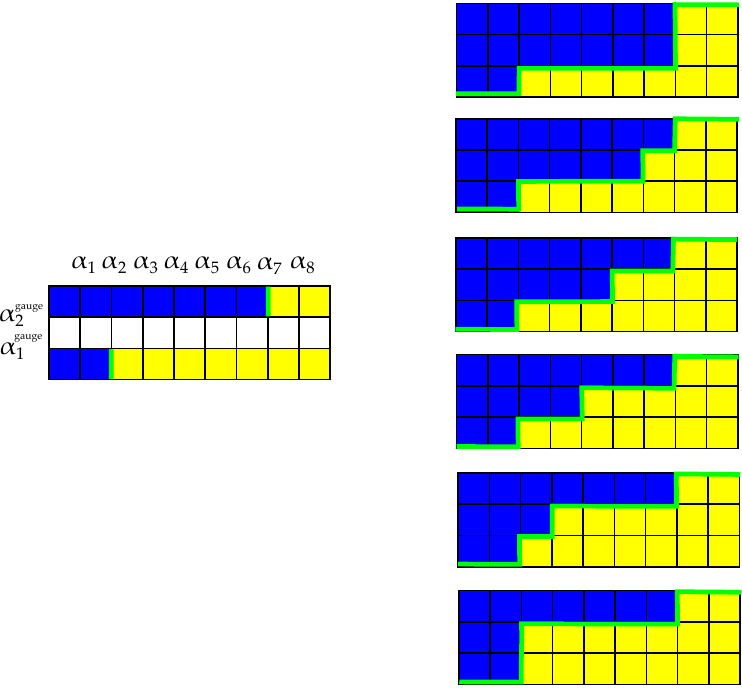}
\caption{An example of a flavor-equivalence class for $SU(3) + 9 \bm{F}$. The equivalence class is shown on the left-hand side --- there we indicate in green the key characteristics of the associated box graph: the splitting of both roots of the gauge group $\alpha_i^{\text{gauge}}$, $i=1,2$ combined will contain all the roots $\alpha_3, \alpha_4, \alpha_5, \alpha_6$. On the right-hand side, we show the complete set of box graphs that comprise the flavor-equivalence class. E.g., the top diagram corresponds to the case where the splitting of $\alpha_1^{\text{gauge}}$ contains all $\alpha_i$, $i=3,4,5,6$. In the second one, which is related by a flop to the top one, $\alpha_1^{\text{gauge}}$ does not contain $\alpha_6$, which is now part of the splitting of $\alpha_{2}^{\text{gauge}}$, etc. \label{fig:ExFEC}}
\end{figure}

\subsection[An Example: \texorpdfstring{$SU(3) + 9 \bm{F}$}{SU(3) + 9F}]{An Example: \boldmath{$SU(3) + 9 F$}}\label{sec:AnEx}

To fill these definitions with some life, before studying the single gauge node quivers comprehensively, we first work through an example in some detail. 
Consider the rank two theory $SU(3) + 9 \bm{F}$. The classical flavor symmetry is $U(9)$ and we consider box graphs for the $(\bm{3}, \bm{9})$ of $SU(3)_{\text{gauge}} \times U(9)_\text{BG}$. 
The representation graph is shown in figure \ref{fig:RepGraphEx}.
The complete set of flavor-equivalence classes for this theory are shown in figure \ref{fig:EstringSU3}.
To understand them in more detail, consider the flavor-equivalence class 
\be\label{eq:U9xSU3Example-FEC}
\includegraphics*[width=8cm]{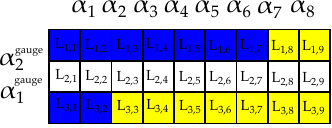} \,.
\ee
The  middle row has no sign assignment, as any consistent sign-assignment, which is subject to the flow rules, given the top and bottom rows, gives a representative of this equivalence class. 
It follows, e.g., immediately from the flow rules that the sign assignment for $L_{2,1}$ and $L_{2,2}$ has to be $+$ (blue), whereas $L_{2,8}$ and $L_{2,9}$ have $-$ (yellow).
The complete set of flavor-equivalent box graphs associated to this equivalence class are shown in figure \ref{fig:ExFEC}.

These are all flavor-equivalent in the sense that the splitting of the sum of roots of the gauge group contain the same roots of the flavor symmetry --- in this case $\alpha_3, \alpha_4, \alpha_5, \alpha_6$. 
Note that $\alpha_2$ and $\alpha_7$ split, into the sum of weights $L_{3,2} + (-L_{3,3})$ and $L_{1,7} + (-L_{1,8})$; these are F-extremal weights, cf.~definition \ref{Def:Fext}.
In each representative of the flavor-equivalence class, the splitting of $\alpha_i^\text{gauge}$ is different, however the sum of them always contain the same set of flavor roots. 
That is, the splitting (\ref{SplitBG}) for this flavor-equivalence class is
\be
\alpha_1^{\text{gauge}} +\alpha_2^{\text{gauge}}  \quad \rightarrow \quad 
\alpha_3+ \alpha_4+ \alpha_5+ \alpha_6 + (- L_{3,3}) + L_{1,7} \,.
\ee
Note that $- L_{3,3}$ and $L_{1,7}$ are the F-extremal weights inside the combined gauge roots, cf.~definition \ref{def:curvesinsurface}
As one can check using $\alpha_i = L_i - L_{i+1}$, this is indeed true, and follows directly from the sign assignments indicated in the box graph.
The different representatives in the flavor-equivalence class differ by those precisely the splitting is distributed among $\alpha_1^{\text{gauge}}$ and $\alpha_2^{\text{gauge}}$. 

Finally, the BG-CFD (see definition \ref{def:BGCFD}) is given by the chain of $-2$ vertices corresponding to the roots $\alpha_3, \alpha_4, \alpha_5, \alpha_6$, as well as the two F-extremal weights, which are $-1$ vertices. This is shown in figure \ref{fig:CFDInBG}.

\begin{figure}
\centering
\includegraphics*[width=7cm]{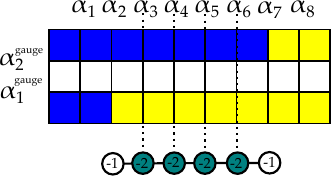}
\caption{Flavor-equivalence class and BG-CFD:
In this figure we show an example for a flavor-equivalence class of Coulomb branch phases for $SU(3)$ with $9{\bm{F}}$. The top graph is the box graph reduced to its flavor-equivalence class by omitting the sign assignments in the middle row: all sign assignments consistent with the standard box graph rules would correspond to the same combined gauge root splitting.
This is encoded in the CFD-subgraph that we refer to as BG-CFD: the roots $\alpha_i$, $i=3,4,5,6$ of $G_{\text{BG}}= U(9)$ that participate in the splitting of the roots $\alpha_i^{\text{gauge}}$ of $G_{\text{gauge}}= SU(3)$ are $(-2)$-vertices in the CFD --- shown at the bottom. The $(-1)$-vertices of the CFD correspond to the F-extremal weights $(-L_{3,3})$ and $L_{1,7}$. 
\label{fig:CFDInBG}}
\end{figure}

\subsection{From Box Graphs to 5d Gauge Theories and SCFTs}

We will now make the connection between box graphs, which capture the Coulomb
branch phases of 5d gauge theories, and the flavor-equivalence classes of
these box graphs, to five-dimensional superconformal field theories.  So far,
we characterized the Coulomb branch phases of a 5d $G_\text{gauge}$
theory, whose classical flavor symmetry $G_\text{BG}$ has been weakly gauged.
The key idea to relate these concepts to 5d gauge theories with gauge symmetry $G_\text{gauge}$ and their SCFT limits is to identify the $G_\text{BG}$ with the classical flavor symmetry of the marginal $G_\text{gauge}$ theory.

To obtain a theory that has a UV fixed point in 5d, we need to decouple
hypermultiplets from the marginal theory.  That is, we add mass terms to
matter charged under $G_\text{gauge}$ and formally send the mass to infinity.
In terms of the extended Coulomb branch, where we treat the classical flavor
symmetry $G_{\text{BG}}$ as a weakly gauged symmetry, this is achieved by
passing to a new flavor-equivalence class --- i.e., performing a flop
transition on the box graph.

More precisely, we can rephrase the key property of the marginal theory as
follows. Points in the extended Coulomb branch where $G_\text{gauge}$
is unbroken correspond to the marginal theory, where the manifest flavor symmetry
is $G_\text{BG}$, when the group $G_\text{BG}$ is also unbroken. 
In terms of the Coulomb branch parameters $\phi$, gauge enhancement to
$G_\text{gauge}$ occurs when\footnote{Note that when we write an expression
  like $\langle \phi, \alpha_j \rangle$ we are silently extending the root
  $\alpha_j$ of $G$ to the root lattice of the full semi-simple gauge group.}
  $\langle \phi, \alpha_i^\text{gauge} \rangle = 0$ for all gauge roots.
  Therefore, the subset of the extended Coulomb branch (with weakly
  gauged $G_\text{BG}$) describing the marginal theory in the above sense is
  one where we have 
\begin{align}
	\forall i : \, \langle \phi, \alpha_i^\text{gauge} \rangle = 0 \quad \text{
  and }\quad \forall j : \, \langle \phi, \alpha_j^\text{BG} \rangle =0 \,,
\end{align}
which is nothing other than the point in the extended Coulomb branch
\begin{equation}
  \phi = 0 \,.
\end{equation}

In terms of the box graphs, this condition applies exactly when the combined
splitting of the gauge roots \eqref{SplitBG} contain \emph{all} roots of $G_\text{BG}$. 
In this case, $\langle \phi, \alpha_i^\text{gauge} \rangle = 0$ implies $\langle \phi, \alpha_j^\text{BG} \rangle = \langle \phi, L_m \rangle = 0$. (Note that the signs in \eqref{SplitBG} are precisely such that $\langle \phi, \epsilon_m L_m^{\epsilon_m} \rangle \geq 0$.)
This means that also linear combinations $\lambda$ out of $L_m$ and $\alpha_j^\text{BG}$, which fill out representations of $G_\text{BG}$ \cite{Hayashi:2014kca, Braun:2014kla, Braun:2015hkv, Lawrie:2015hia}, have $\langle \phi, \lambda \rangle = 0$.
Physically, this characterizes the Coulomb phases of the marginal theory as those where, when the gauge symmetry $G_\text{gauge}$ is restored, the massless charged matter transforms under the full $G_\text{BG}$ --- which by definition was the classical flavor symmetry of the marginal $G_\text{gauge}$ theory.

In the example of the last subsection, $SU(3) + 9 \bm{F}$, the marginal theory is thus characterized by flavor-equivalence class represented by the box graph 
\be\label{eq:U9xSU3Marginal-FEC}
\includegraphics[width=8cm]{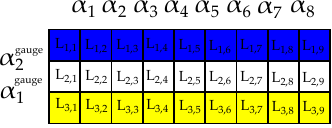}
\ee 
where the representatives of the equivalence class have any sign assignment that is consistent with the flow rules 
(\ref{FlowRules}). 
Indeed, the box graph rules stipulate that all roots $\alpha_i$ of the $SU(9) \subset U(9)_\text{BG}$ appear in the splitting (\ref{SplitBG}) of $\alpha_1^{\text{gauge}} + \alpha_2^{\text{gauge}}$.

A mass deformation that decouples a hypermultiplet requires a mass term that
remains non-zero (and can be sent to $\pm \infty$) when $\langle \phi,
\alpha_i^\text{gauge} \rangle = 0$.  This results in a smaller flavor symmetry
$G_\text{F,cl} \subset G_\text{BG}$, whose rank is lowered by one compared to
$G_\text{BG}$.  In the context of having weakly gauged $G_\text{BG}$, this
must therefore correspond to a phase on the extended Coulomb branch, where
there is one flavor root $\alpha_j^\text{BG}$ with $\langle \phi,
\alpha_j^\text{BG} \rangle \neq 0$ even when $\langle \phi, \alpha_i^\text{gauge} \rangle = 0$.  The associated box graph of such a phase
thus implies a combined splitting of the gauge roots which leaves out one
flavor root, whose mass may be identified with the non-zero Coulomb branch
parameter.

Starting from the flavor-equivalence class of the marginal theory, such a
phase is obtained from a flop, i.e., a change of sign assignment of an
F-extremal weight.  After the flop, the resulting $G_\text{gauge}$ theory has
less matter, and correspondingly a smaller classical flavor group
$G_\text{F,cl} \subset G_\text{BG}$, specified by the roots
$\alpha_j^\text{BG}$ which are still contained in the splitting of the gauge
roots. 

Returning to our $SU(3)$ example, we recognize $L_{1,9}$ and $-L_{3,1}$ to be
the two F-extremal weights in the box graph \eqref{eq:U9xSU3Marginal-FEC}.
For concreteness consider changing the sign assignment of $L_{1,9}$.  In the
field theory picture, without the gauging of the flavor symmetry, this
corresponds to the decoupling of a hypermultiplet associated with this weight,
under the flavor symmetry group.  After the flop the flavor-equivalence class
is 
\be
\includegraphics[width=8cm]{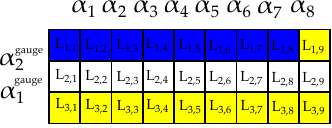}
\ee
The flow rules imply that the sign assignment for $L_{2,9}$ is $-$ as well,
and in the splitting of $\alpha_1^{\text{gauge}} + \alpha_2^{\text{gauge}}$,
only the roots $\alpha_1,\cdots, \alpha_8$ appear.  This means that we have
decoupled one fundamental flavor and ended up with an $SU(3) + 8 \bm{F}$
theory.  Consistently, the flavor symmetry of this descendant gauge theory is
$G_\text{F,cl} = U(8)$, whose roots are precisely $\alpha_1,\cdots, \alpha_7$.
If on the other hand we flop $-L_{3,1}$, then $L_{2,1}$ has fixed sign
assignment $+$ by the flow rules, and the flavor roots $\alpha_i$ that appear
in the splitting of the gauge roots are $\alpha_2, \cdots, \alpha_8$.
Continuing in this fashion, we can eventually reach the phase
\eqref{eq:U9xSU3Example-FEC} described earlier, after having flopped
$L_{1,9}$, $L_{1,8}$, $-L_{3,1}$ and $-L_{3,2}$.

Note that after flopping either $L_{1,9}$ or $-L_{3,1}$, both phases define a
$SU(3) + 8\bm{F}$ theory with classical $G_\text{F,cl} = U(8)$ flavor
symmetry.  On the other hand, the different \emph{embedding} of it into the
original $U(9)_\text{BG}$ flavor group will lead to a different superconformal
flavor enhancement as we approach the SCFT limits of these two descendant
gauge theories.  In fact, in case of $SU(3)$ (or any $SU(n)$) gauge theories
in 5d, the box graphs must be supplemented by the discrete Chern--Simons level
$k$, which we have neglected so far.  However, it is crucial that
$|k|=\frac32$ for $SU(3) + 9\bm{F}$ to be a marginal theory
\cite{Jefferson:2017ahm}.  The two different descendants, flopping either on
$L_{1,9}$ or $-L_{3,1}$, then correspond to $SU(3) + 8\bm{F}$ with $|k| = 1$
or $|k| = 2$, respectively.  The different superconformal flavor enhancements
of these theories cannot be described by the box graphs alone, but requires a
little geometric input related to M-theory realizations of 5d gauge theories,
see sections \ref{sec:FibPhas} and \ref{sec:CFDsfromBG}.  However, a more
succinct portrayal of this process can be developed using the embedding of the
BG-CFDs into the CFD description of 5d SCFTs developed in
\cite{Apruzzi:2019vpe,Apruzzi:2019opn}.  This will be one of the main themes of the
present work.

In summary, flops (or changes of sign assignments) in the $G_\text{gauge}
\times G_\text{BG}$ flavor-equivalence classes of box graphs provides an
alternative description of decoupling a matter hypermultiplet of a
$G_\text{gauge}$ gauge theory.  Starting with the weakly coupled marginal
description with gauge group $G_\text{gauge}$ and matter transforming under
the flavor symmetry $G_\text{BG}$, successive flop transitions of
flavor-equivalence classes map out all descendants with a $G_\text{gauge}$
description, while simultaneously keeping track of their classical flavor
symmetry $G_\text{F,cl} \subset G_\text{BG}$.  Each of these gauge theories
has a UV fixed point, and thus each flavor-equivalence class corresponds to a
5d SCFT. As noted before, there can be discrete identifications between the
flavor-equivalence classes, which then correspond to the same 5d SCFT; this
will be discussed later.

\subsection{Complex Representations}\label{sec:3complex}

In this section we will discuss the gauge theory descendants in terms of flavor-equivalence classes of box graphs
for single node quivers of the form (\ref{fig:SingleNodeQuiver}), where the
representation $\bm{R}$ is complex. We will be concerned with
the following theories
\begin{equation}\label{eqn:complexSN}
  \begin{array}{c|c}
    G_\text{gauge} & \bm{R} \cr\hline
    SU(N \geq 3) & \bm{F} \cr
    SU(N \geq 5) & \bm{AS} \cr
    SU(N \geq 3) & \bm{Sym} \cr
    E_6 & \bm{27} 
  \end{array} \,,
\end{equation}
where $\bm{F}$, $\bm{AS}$, and $\bm{Sym}$ refer to the  fundamental,
anti-symmetric, and symmetric representations, respectively.  We point out,
however, that the analysis herein applies to any such quiver where $\bm{R}$ is
complex, including, for example, the single node gauge theories with
exceptional matter that appear for low ranks of the gauge group in
\cite{Jefferson:2017ahm}.

In each of the above cases the classical flavor group that rotates the
hypermultiplets arising from the flavor node of the quiver is
\begin{equation}
  G_{\text{BG}} = U(N_{\bm{R}}) \,.
\end{equation}
After weakly gauging this global symmetry we are determining the structure of the
flavor-equivalences classes of the box graphs for the theory with gauge group
\begin{equation}
  G_\text{gauge} \times U(N_{\bm{R}}) \,,
\end{equation}
with matter transforming in the 
\begin{equation}
  (\bm{R}, \bm{F}) 
\end{equation}
representation, where $\bm{F}$ is the fundamental representation of
$U(N_{\bm{R}})$. As we have seen in section \ref{sec:3def}, the
flavor-equivalence classes are agnostic as to the particular $\bm{R}$ and
$G_\text{gauge}$ above, and thus they are completely determined by the
classical flavor group which is weakly gauged. Since all complex
representations have the same classical flavor group, the only parameter that
enters is the number of hypermultiplets on the flavor node, $N_{\bm{R}}$. 

Let us consider the illustrative example where we take $G_\text{gauge} =
SU(N)$ and $N_{\bm{R}}\bm{R} = N_f \bm{F}$. The flavor-equivalence classes
and the tree structure amongst them
will be identical for all of the other combinations of gauge groups and
matter appearing in (\ref{eqn:complexSN}). 

After weakly gauging the classical $U(N_f)$ flavor symmetry rotating the
hypermultiplets we are studying the product gauge theory
\begin{equation}
  SU(N) \times U(N_f) \quad \text{ with }\  (\bm{N}, {\bm{N_f}})_q \oplus
  (\overline{\bm{N}}, \overline{\bm{N_f}})_{-q} \,.
\end{equation}
For complex representations, $\bm{R} \oplus \bm{\overline{R}}$, it is necessary only to
determine the signs associated to the weights of $\bm{R}$, as this will
completely specify the signs associated to the weights of the
$\bm{\overline{R}}$. The positive simple roots for $SU(N)$ are 
\begin{equation}
  \begin{aligned}
    \alpha_1^n &= (2,-1,0,0,\cdots,0,0,0,0) \cr
    \alpha_2^n &= (-1,2,-1,0,\cdots,0,0,0,0) \cr
    &\cdots \cr
    \alpha_{n-2}^n &= (0,0,0,0,\cdots,0,-1,2,-1) \cr
    \alpha_{n-1}^n &= (0,0,0,0,\cdots,0,0,-1,2) \,,
  \end{aligned}
\end{equation}
and similarly for $SU(N_f)$, the semi-simple part of the $U(N_f)$. The highest
weight of the $(\bm{N,N_f})$ representation is given by
\begin{equation}
  L_{1,1} = (1,0,\cdots,0; 1,0, \cdots, 0) \,,
\end{equation}
where the semi-colon in the middle denotes the join between the highest weight
of the fundamental representation of each of the $SU(N)$ and $SU(N_f)$ factors.
The representation graph for this representation is represented in figure
\ref{fig:SUnUmBG}. This is the weight diagram for the $(\bm{N, N_f})$
representation displayed as a box graph. 

\begin{figure}
  \centering
  \subfloat[]{{\includegraphics[scale=1.5]{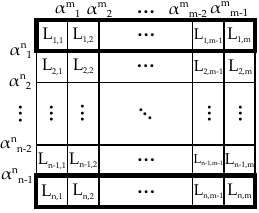}}}
  \qquad 
  \subfloat[]{{\includegraphics[scale=1.5]{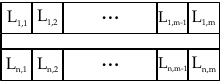}}}
  \caption{In (a) is the representation graph, or weight diagram, for $SU(N)_k \times
    U(N_f)$ with $(\bm{N,N_f})_q \oplus (\bm{\overline{N},
    \overline{N_f}})_{-q}$. Notice that we simply write the $(\bm{N,N_f})_q$ half
  of the representation, as the decoration thereon will imply the decoration
on the conjugate half. The Coulomb phases are given by all decorations of this
box graph subject to the flow rules. In (b) we give the subgraph (marked in
boldface in (a)) of the box
graph that contains only the weights whose signs are required to be specified
to determine the flavor-equivalence class. Any box graphs with the same
coloring for the weights in (b) are flavor-equivalent.}\label{fig:SUnUmBG}
\end{figure}

The phases are determined by all decorations of the box graph in figure
\ref{fig:SUnUmBG} with signs subject to the flow rules (\ref{FlowRules}).
Equivalently each phase can be characterized by a monotonic path between the
lower left and the upper right corners on the $N_f \times N$ grid that the
representation graph defines. An elementary computation reveals that the total number of 
such phases is
\begin{equation}
  \binom{N+N_f}{N} \,.
\end{equation}
Of course, the determination of the total number of gauge theory phases of
this weakly gauged product gauge theory is not the goal of this section. This
quantity will vary depending on the type of matter representation on the
flavor node; it will not just depend on $N_{\bm{R}}$ in the same manner for
all complex representations. 

We now turn to the sorting of these phases into flavor-equivalence classes of
 box graphs. It is clear from figure \ref{fig:SUnUmBG} that,
regardless of the coloring in the middle rows, the sum over all of the
$\alpha_\text{BG}$ can always be written as 
\begin{equation}
  \sum_{j=1}^{N-1} \alpha^{\text{gauge}}_j \rightarrow L_{1,k_u} - L_{N, k_l} +
  \sum_{i = k_l}^{k_u} \alpha^{\text{BG}}_i + \cdots \,,
\end{equation}
where the $\cdots$ represents weights that appear in the central $N-3$ rows of
the box graph, and where $L_{1,k_u}$ is the rightmost box on the upper row
decorated with a plus, and $L_{N,k_l}$ is the leftmost box on the lower row
decorated with a minus. It is clear that the set of $\alpha^\text{BG}_i$ that
are included in the splitting of the $\alpha^\text{gauge}_j$ depends only on the
$k_u$ and the $k_l$, and therefore the flavor-equivalence class depends only
on the choice of consistent decoration for the uppermost and lowermost rows in
the box graph\footnote{Note that even in the case where none of the
  $\alpha^\text{gauge}_j$ split, the values of $(k_u, k_l=k_u)$ specify the
flavor-equivalence class, as per the definition.} , corresponding to weights
that carry, respectively, the highest and the lowest weight of $SU(N)$. The
flavor-equivalence class is then completely defined by the consistent
decoration of the subdiagram of the box graph that is depicted in figure
\ref{fig:SUnUmBG} (b).

It is straightfoward to see that the total number of consistent decorations
that give the flavor-equivalence classes is
\begin{equation}
  \#\text{ flavor-equivalence classes for } SU(N) \text{ with } N_f \bm{F} =
  \frac{1}{2}(N_f + 1)(N_f + 2) \,,
\end{equation}
and furthermore one can study the flop transitions between these flavor-equivalence
classes to see that they form together in the tree structure that is depicted in figure
\ref{fig:SUnUmFlopGraph}. 

\begin{figure}
  \centering
  \includegraphics[width=\textwidth]{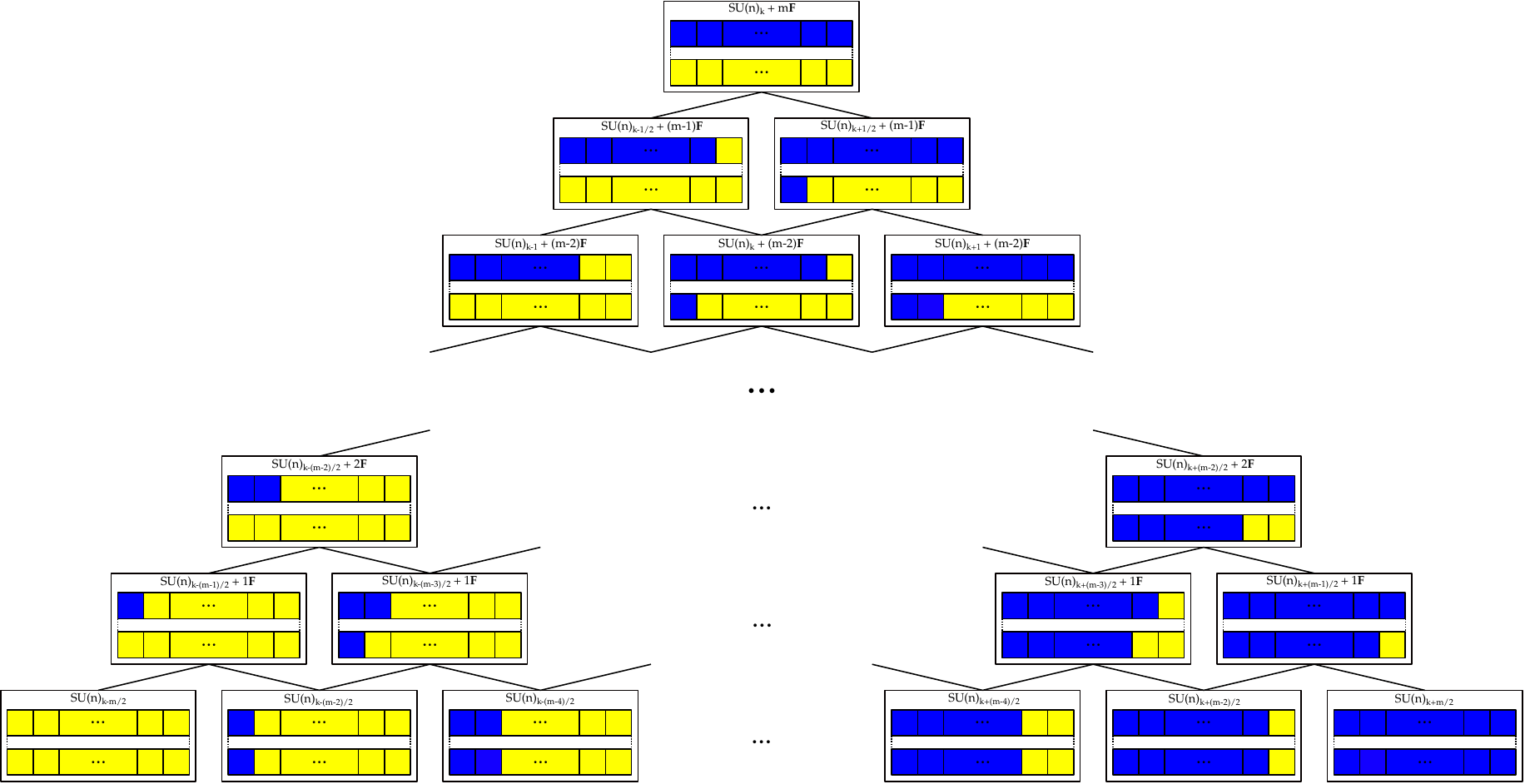}
  \caption{The flop graph for the flavor-equivalence classes of Coulomb phases
    associated to $SU(N)$ gauge theory with $\bm{N_f}$ fundmental
    hypermultiplets. This is identical to the flop graph for the Coulomb
  branches of all descendant gauge theories with matter in complex
  representations, as shown in (\ref{eqn:complexSN}).}\label{fig:SUnUmFlopGraph}
\end{figure}

\subsection{Quaternionic Representations}\label{sec:3quat}

In this section we consider quivers (\ref{fig:SingleNodeQuiver}), where
the hypermultiplets transform in a quaternionic, or pseudo-real,
representation of the gauge group $G_\text{gauge}$. There are only two such
kinds of quiver that we need to consider for the purposes of this paper, which are
\begin{equation}
  \begin{array}{c|c}
    G_\text{gauge} & \bm{R} \cr\hline
    Sp(N \geq 1) & \bm{F} \cr
    E_7 & \bm{56} 
  \end{array} \,,
\end{equation}

In this section we shall consider as an example the $Sp(N) + N_f\bm{F}$, and,
as in the case of $SU(N)$, the analysis shall also apply to $E_7 + N_f\bm{56}$
because the two representations are quaternionic. In the case of the $Sp(N)$
there is an anomaly that requires $N_f$ to be integer, so to say, that there
is an even number of half-hypermultiplets in the fundamental representation of
$Sp(N)$. For $E_7$ there is no such anomaly, and $N_f$ can be half-integer;
since $SO(\text{even})$ is particularly distinct from $SO(\text{odd})$ we
shall consider the former first, and then move on to the case of $E_7$ with an
odd number of half-hypermultiplets.  In either case the flavor-equivalence class will be
specified by the decorated subdiagram of the full box graph that
corresponds to the highest and the lowest weights of the representaton of the
$Sp(N)$ or $E_7$. 

The highest weight of the fundmental representation of
$Sp(N)$ is given by
\begin{equation}
  (1,0,\cdots,0) \,,
\end{equation}
in terms of the usual Cartan--Dynkin labels, and the lowest weight, as the
representation is self-conjugate, is given by
\begin{equation}
  (-1,0,\cdots,0) \,.
\end{equation}
To each of these two weights is associated a decoration of the vector
representation of the weakly gauged classical flavor group $SO(2N_f)$. The vector representation of $SO(2N_f)$  has highest weight
\begin{equation}
  (1,0,\cdots,0) \,,
\end{equation}
and is also a self-conjugate representation. Because of this self-conjugacy
there is a relation between the weights appearing in the representation
$(\bm{2N}, \bm{2N_f})$ of $Sp(N) \times SO(2N_f)$, in particular for the
weights that appear in the flavor-equivalence class, that is
\begin{equation}\label{eqn:SOdual}
  L_{1,j} = - L_{2N, 2N_f + 1 - j} \,.
\end{equation}
In figure \ref{fig:SpNSO2NF} we have drawn the subdiagram of the 
box graph that specifies the flavor-equivalence class, and further we have
decorated those weights appearing in the flavor-equivalence class that cannot
be consistently decorated in any other way, and all consistent decorations are
given by applying the flow rules (\ref{FlowRules}) and (\ref{eqn:SOdual}) to
this box graph. It is straightforward to see that this yields
\begin{equation}
  N_f + 2 \,,
\end{equation}
flavor-equivalence classes when $N_f > 0$, and when $N_f = 0$ there is exactly
$1$ flavor-equivalence class. This distinction is a consequence of the fact
that when an $Sp(N)$ gauge theory has no fundamental hypermultipelts it has an
additional physical discrete parameter, the $\theta$-angle, which must be
specified.  The tree structure generated by the flop transitions amongst these
flavor-equivalence classes is as given in figure \ref{fig:SpNSO2Nftree}.

\begin{figure}
  \centering
  \includegraphics[scale=1.5]{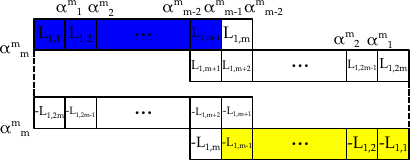}
  \caption{We show the subdiagram of the box graph for which the
    consistent decoration specifies the flavor-equivalence classes of $Sp(N)$
  with $m = N_f$ fundmental hypermultiplets. We have decorated all of the weights whose decoration is fixed by
(\ref{eqn:SOdual}) and the flow rules (\ref{FlowRules}).}\label{fig:SpNSO2NF}
\end{figure}

\begin{figure}
  \centering
  \includegraphics[scale=0.6]{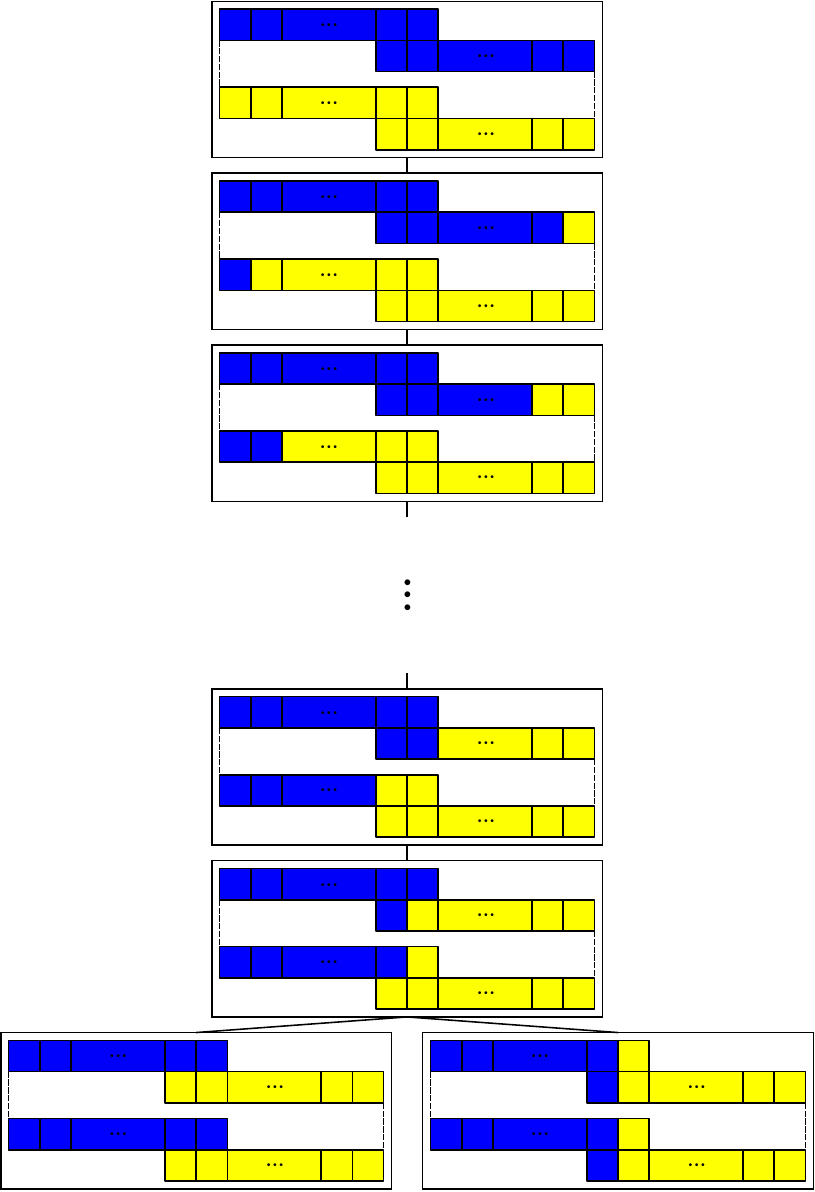}
  \caption{The tree of flop transitions of flavor-equivalence classes, and thus also of the
    descendant gauge theories, starting from $Sp(N) + N_f \bm{F}$. The tree
    structure is identical for the flavor-equivalence classes for $E_7 +
    N_{56}\bm{56}$, as they both have a special orthogonal group as flavor
  group.}\label{fig:SpNSO2Nftree}
\end{figure}

In fact, in this case the total number of Coulomb phases is straightforward to
determine, and we include the number here for the purposes of later making a
comparison between how the number Coulomb phases and the number of
flavor-equivalence classes of Coulomb phases scale when considering quivers
that combine such $Sp(N)$ single gauge nodes.  
The total number of phases for $Sp(N) \times SO(2N_f)$ with matter in the
$(\bm{2N, 2N_f})$ is given by
\begin{equation}
  \binom{N+N_f-1}{N_f-1} + \sum_{k=1}^N 2 \binom{N - k + N_f - 1}{N_f-1} =
  \frac{(2N+N_f)\Gamma(N+N_f)}{\Gamma(N + 1)\Gamma(N_f+1)} \,,
\end{equation}  
where $\Gamma$ is the Euler gamma function.

We now turn to the case where there are an odd number of half-hypermultiplets
transforming in a quaternionic representation. This can only occur, in the
cases we consider, for $E_7$ with matter in the $\bm{56}$ representation.
Since $2N_f$ is odd we can write it as $2k + 1$ and then we are considering the
classical flavor group $SO(2k+1)$ or $B_k$. The Cartan matrix of this algebra
is rank $k$ and looks like
\begin{equation}
  \begin{pmatrix}
    2 & -1 & 0 & \cdots & 0 & 0 & 0 \cr
    -1 & 2 & -1 & \cdots & 0 & 0 & 0 \cr
    0 & -1 & 2 & \cdots & 0 & 0 & 0 \cr
     & & & \cdots & & & \cr
    0 & 0 & 0 & \cdots & 2 & -1 & 0 \cr 
    0 & 0 & 0 & \cdots & -1 & 2 & -2 \cr 
    0 & 0 & 0 & \cdots & 0 & -1 & 2 
  \end{pmatrix} \,.
\end{equation}    
The highest weights of the fundmental representation of $SO(2k+1)$ is
\begin{equation}
  (1,0,\cdots,0) \,.
\end{equation}
This representation is depicted, in the usual way, in the undecorated
flavor-equivalence box graph that appears in figure \ref{fig:SOoddBG}.  Such a
representation has the novel feature that it contains a zero-weight, to which
a sign cannot be assigned -- the weights appearing in the flavor-equivalence
class box graph in figure \ref{fig:SOoddBG} with a cross through them are
exactly those such weights that are zero-weights under the weakly gauged
$SO(2n+1)$ factor.

\begin{figure}
  \centering
  \includegraphics[width=9cm]{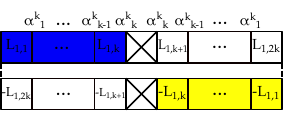}
  \caption{The flavor-equivalence classes associated to $E_7$ with
    $N_f\bm{56}$, where $N_f$ is a half-integer, involve the flavor-equivalence
    classes of $E_7 \times SO(2N_f)$ with matter in the $(\bm{56},
    \bm{2N_f})$. The flavor-equivalence classes depend only on the weights
  written here, and the sign associated to some of them is fixed, as shown.
The boxes marked with crosses correspond to the weights that are zero-weights
under the $\bm{2N_F}$ representation, which cannot be consistently assigned a
sign.}\label{fig:SOoddBG}
\end{figure}

As previously discussed, the $\bm{56}$ of $E_7$ is a self-conjugate
representation and so the weights appearing in the flavor-equivalence class of
box graphs are not independent, and thus cannot be assigned a sign
independently; this interdependence is shown in figure \ref{fig:SOoddBG},
where we also color the boxes for which the sign is fixed a priori, for all
phases, by this interdependence together with the flow rules.

The tree of descendants, or flop diagram for the flavor-equivalence classes,
is shown in figure \ref{fig:SOoddFlop}, and shows that descendants that arise
when decoupling one full hypermultiplet of the $\bm{56}$ at a time. One cannot consistently decouple an odd number of half-hypermultiplets, as there is no possible real mass term, see e.g. \cite{Bhardwaj:2013qia}.

\begin{figure}
  \centering
  \includegraphics[scale=1.2]{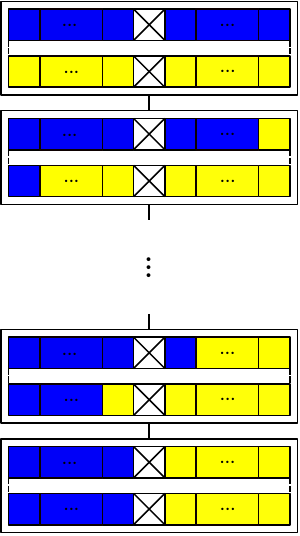}
  \caption{The tree of descendants for $E_7$ with $N_f\bm{56}$ where $N_f$ is a
half-integer. Decoupling one full hypermultiplet descends down the
tree.}\label{fig:SOoddFlop}
\end{figure}

\subsection{Real Representations}\label{sec:3real}

In this final case we consider the single gauge node quivers
(\ref{fig:SingleNodeQuiver}) where $\bm{R}$ is in a real representation of
$G_{\text{gauge}}$:
\begin{equation}
  \begin{array}{c|c}
    G_\text{gauge} & \bm{R} \cr\hline
    Sp(N \geq 2) & \bm{AS} \cr
    SO(N \geq 5) & \bm{V} \cr
    SU(4) & \bm{AS} \cr
    G_2 & \bm{7} \cr
    F_4 & \bm{26} 
  \end{array} \,,
\end{equation}
where we further add that the representaion $\bm{V}$ is the vector
representation. Such theories have a classical flavor group rotating the
$N_{\bm{R}}$ hypermultiplets being
\begin{equation}
 G_{\text{BG}}= Sp(N_{\bm{R}}) \,.
\end{equation}
As such, after weakly gauging this flavor group we are interested in
determining the flavor-equivalence classes of box graphs for the gauge theory
\begin{equation}
  G_{\text{gauge}} \times Sp(N_{\bm{R}}) \,,
\end{equation}
with matter transforming in the representation
\begin{equation}
  (\bm{R}, \bm{F}) \,,
\end{equation}
where $\bm{F}$ here denotes the fundamental representation of the
$Sp(N_{\bm{R}})$ rotation group of the hypermultiplets. 

The example that we will consider in this section, that will reveal the
structure of the flavor-equivalence classes when we have real representations
will be $Sp(N) + N_a\bm{AS}$.  These $N_a$ hypermultiplets are rotated by an
$Sp(N_a)$ flavor symmetry and thus we are considering the $(\bm{\Lambda^2 2N,
2N_a})$ representation of $Sp(N) \times Sp(N_a)$.  The highest and lowest
weights of the $\bm{\Lambda^2 2N}$ are given by
\begin{equation}
  (0,\pm 1, 0, \cdots, 0) \,,
\end{equation}
which is again a self-conjugate representation, similarly to the fundamental
represention of a symplectic group as has already been discussed. The
weight diagram which will capture all of the flavor-equivalence classes
for this gauge theory is shown in figure \ref{fig:SpNNaBG}.

\begin{figure}
  \centering
  \includegraphics[width=9cm]{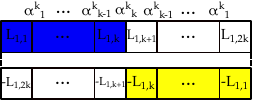}
  \caption{We show the box graph for the flavor-equivalence
    class of $Sp(N)$ with $k = N_a$ hypermultiplets transforming in the
    anti-symmetric representation. The boxes whose decoration is fixed
    by (\ref{eqn:SpNNadual}) are colored. All consistent colorings of the
    remaining boxes give rise to all consistent flavor-equivalence classes of
    this theory. Since the structure of the flavor-equivalence box graph is
    dependent on the $Sp(2N_a)$ symmetry rotating the $N_a$ hypermultiplets,
    and not on the particular highest and lowest weights of the anti-symmetric
    representation of the $Sp(N)$ but only that they are conjugate, 
    the same structure exists for all theories that have a sympletic symmetry
    group that acts as rotations on hypermultiplets that transform in a
  self-conjugate representation of the gauge group. This includes the $SO(N)$,
$G_2$, and $F_4$ theories of interest in this section.}\label{fig:SpNNaBG}
\end{figure}

Again, because the representation $(\bm{\Lambda^2 2N, 2N_a})$ is
self-conjugate there is a relationship amongst the weights of the
representation. For the weights relevant for the flavor-equivalence class this
is
\begin{equation}\label{eqn:SpNNadual}
  L_{1,i} = - L_{2N+1,2N_a + 1 - i} \,.
\end{equation}
Similarly to the case of fundamental matter, there are weights that can only
be consistently decorated with one particular sign due to
(\ref{eqn:SpNNadual}) combined with the flow rules (\ref{FlowRules}). These
weights are shown with their necessary decoration in figure \ref{fig:SpNNatree}, and
the rest of the flavor-equivalence classes come from the consistent decoration
of the remaining undecorated boxes. The total number of flavor-equivalence
classes is 
\begin{equation}
  N_a + 1 \,,
\end{equation}
and furthermore these flavor-equivalence classes arrange themselves, via flop
transitions, into the tree shown in figure \ref{fig:SpNNatree}. 

To give an explicit example, for the gauge theory $G_2 + 1 \times \bm{7}$ the
extended Coulomb phases were written down from a geometric realization in
\cite{Esole:2018mqb}, and one can see that the four phases found there sort
themselves into two flavor-equivalence classes with three and one
representatives, respectively. 

\begin{figure}
  \includegraphics[width=5cm]{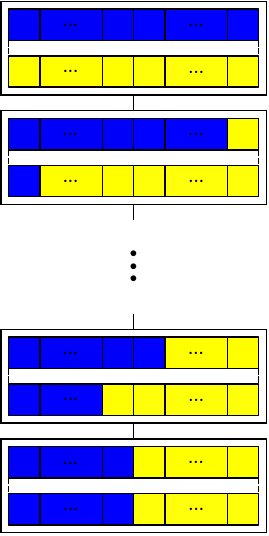}
  \centering
  \caption{The tree of flavor-equivalence classes for $Sp(N) + N_a\bm{AS}$,
  $SO(N) + N_V\bm{V}$, $G_2 + N_7\bm{7}$, and $F_4 + N_{26}\bm{26}$. 
 What is shown are the decorations/colorings of the undecorated part in figure \ref{fig:SpNNaBG}.
  All of these theories have a symplectic group acting as rotations on the
hypermultiplets, this is the key feature that controls the
flavor-equivalence classes, and thus each of these theories has the same
structure for their flavor-equivalence classes.}\label{fig:SpNNatree}
\end{figure}

\subsection{Flavor-equivalence Classes for Quiver Gauge
Theories}\label{sec:3quiver}

The quivers that we will consider are those that are built out of gauge nodes
that correspond to the gauge theories described in table \ref{tab:GTCFL}.
There are two ways to build such quivers out of the previously analyzed
quivers (\ref{fig:SingleNodeQuiver}). Either we chain together gauge nodes of
that form, potentially also without any associated flavor node, or else we add
more flavor nodes on to a given single gauge node.  The number of
flavor-equivalence classes is multiplicative across constructing quivers with
arbitrary numbers of gauge nodes, out of the single nodes in the table, via
gluing two gauge nodes together with bifundamental matter. Such bifundamental
matter is uncharged under any of the flavor symmetries, and thus the
flavor-equivalence classes, which are defined as those box graphs with the
same set of flavor roots contained inside of the gauge roots, are unchanged.
So the total number of weakly gauged phases will increase in an intricate way
upon gluing, but the flavor-equivalent phases will simply be all ways of
choosing one flavor-equivalence classes from the equivalence class associated
to each gauge node. The total number of flavor-equivalence classes attached to
a quiver, $Q$, is then 
\begin{equation}
  n_Q = \prod_G n_G \,,
\end{equation}
where $G$ runs over all the gauge nodes in the quiver, and $n_G$ is the number
of flavor-equivalence classes for that gauge node, as given above, and which
depends on the flavor nodes attached to that gauge node.

The quantity $n_G$ is determined above in the cases where the gauge node has
at most one flavor node attached. We will now show that, when multiple flavor
nodes are attached, which can only occur for $SU(N)$ or $Sp(N)$ gauge nodes if
we wish to have a interacting SCFT in the UV limit, the number of
flavor-equivalence classes, and thus the number of descendants (counting
redundantly) is multiplicative.

We will consider first the set of flavor-equivalence classes from
the 5d $\mathcal{N} = 1$ gauge theory with the following matter fields
\begin{equation}
  SU(N)_k \quad \text{ with } \quad N_f \bm{F} + N_a \bm{AS} + N_s \bm{Sym}
  \,,
\end{equation}
where $\bm{F}$, $\bm{AS}$, and $\bm{Sym}$, respectively, refer to
hypermultiplets that transform in the fundamental, anti-symmetric, and
symmetric representations of the $SU(N)$. For the purposes of the
flavor-equivalence classes the Chern--Simons level, $k$, will be immaterial,
as the box graph is not sensitive to such discrete data. The classical flavor
symmetry of this theory is 
\begin{equation}
  U(N_f) \times U(N_a) \times U(N_s)  \,.
\end{equation}
After weakly gauging the first three factors, which are the perturbative
flavor symmetry groups, the theory contains matter that transforms in the 
\begin{equation}\label{eqn:SUNreps}
  (\bm{N, N_f, 1, 1})_{q_f} \oplus (\bm{\Lambda^2 N, 1, N_a, 1})_{q_a} \oplus
  (\bm{Sym^2 N, 1, 1, N_s})_{q_s} \oplus \text{ c.c } \,,
\end{equation}
where the subscripts indicate the charges of the matter fields under the
$U(1)$ factor of the $U(N_{\bm{R}})$ symmetry that rotates the hypermultiplets
transforming in the representation $\bm{R}$. As before, we will assume that
$q_{\bm{R}} \neq 0$ as otherwise this would lead to the decoupling of the
$U(1)$ and thus a changing of the phase structure; this is of course true when
the $U(1)$ is part of a $U(k)$ global rotation group. It is with respect to
these representations that we must determine the flavor-equivalence classes. 

Each of the irreducible representations in (\ref{eqn:SUNreps}) are charged
under different gauge groups, after the weak gauging, and so the subsectors of
the Coulomb branch that capture moving the vacuum expectation values of the
matter fields of different irreducible representations are orthogonal to each
other. As such we can consider the fundamental, anti-symmetric, and symmetric
matter under the $SU(N)$ independently, and the number of flavor-equivalence
classes for each of these was determined in section \ref{sec:3complex}. 

Now we are considering the more general case, where $N_f$, $N_a$, and $N_s$ are
all, in principle, non-zero. As the Coulomb branch has the stucture of the
product of the four Coulomb branches give by the Weyl chambers of the $SU(N)$
and of the three weakly gauged flavor symmetries, and that the vevs 
 under consideration are orthogonal in this space, the total number of
flavor-equivalence classes (and indeed the number of Coulomb phases) is simply
the product of the total number from each irreducible matter representation.
The total number of flavor-equivalence classes is then given by the expression
\begin{equation}
  n_{SU(N)} = \frac{1}{8}(1+N_a)(2+N_a)(1+N_f)(2+N_f)(1+N_s)(2+N_s) \,.
\end{equation}
Each of these flavor-equivalence classes can be represented by a triplet of
consistently decorated diagrams as in figure \ref{fig:SUnUmBG} (b), where the
length of each is $N_f - 1$, $N_a - 1$, and $N_s - 1$. This simple structure
follows because each of the different kinds of $SU(N)$ matter all have an
$U(N_{\bm{R}})$ flavor symmetry which rotates the respective hypermultiplets
under their fundamental representation. The flop graph of these equivalence
classes then has the form of figure \ref{fig:SUnUmFlopGraph}, extended into a
space spanned by two additional transverse planes, which we do not attempt to
draw here.

For $Sp(N)$ gauge theories one can only have matter transforming in the
fundmamental and anti-symmetric representations if one wishes to have a
non-trivial interacting fixed point in the UV. We will consider such theories,
which we write as
\begin{equation}
  Sp(N) \quad \text{ with } \quad N_f \bm{F} + N_a \bm{AS} \,.
\end{equation}
We note that if $N_f = 0$ then we must, in addition, specify a discrete
$\theta$-parameter for the $Sp(N)$, being either $0$ or $\pi$.  The classical
flavor symmetry of the theory is 
\begin{equation}
  SO(2N_f) \times Sp(N_a)  \,,
\end{equation}
and when one weakly gauges the first two factors one has matter, which
determines the phase structure of the Coulomb branch, transforming in the 
\begin{equation}
  (\bm{N, 2N_f, 1}) \oplus (\bm{\Lambda^2 N, 1, 2N_a}) \,,
\end{equation}
representations of $Sp(N) \times SO(2N_f) \times Sp(N_a)$. Again, each of
these representations can be considered seperately for the purposes of the
flavor-equivalence classes and the result follows from sections
\ref{sec:3quat} and \ref{sec:3real}. 
Putting everything together we can determine that the total number of
flavor-equivalence classes for $Sp(N)$ gauge theories with arbitrary $N_f$ and
$N_a$ is
\begin{equation}
  n_{Sp(N)} = \begin{cases}
    (N_a + 1) \quad &\text{if} \quad N_f = 0 \cr
    (N_f + 2)(N_a + 1) \quad &\text{otherwise}
  \end{cases} \,.
\end{equation}

One example of a multi-node quiver which we will explore more
in appendix \ref{app:RankTwoBG} is when the gauge theory is given by the two-gauge-node quiver
\begin{equation}
  M_1\bm{F} - Sp(1) - Sp(1) - M_2\bm{F} \,,
\end{equation}
where $M_i \geq 1$. Such a theory has 
\begin{equation}
  (M_1 + 2)(M_2 + 2) \,,
\end{equation}
flavor-equivalence classes of phases. The total number of gauge theory phases is given by
\begin{equation}
  2(M_1 + 2)(M_2 + 2) \,,
\end{equation}
where the factor of $2$ comes from the two different phases of the
bifundamental of the two $Sp(1)$ factors. In this example we can see that
whilst determining the number of Coulomb phases for an arbitrary quiver may be
quite involved, the number of flavor-equivalence classes is obtained by a
simple combination of the number for each individual gauge node.


\section{Weakly-Coupled Descriptions from CFDs}
\label{sec:gauge_descriptions_from_CFDs}

In the previous section we have determined the set of descendants for a given
5d $\mathcal{N} = 1$ quiver gauge theory. In
\cite{Apruzzi:2019vpe,Apruzzi:2019opn} one determined a geometric object, a
graph known as a Combined Fiber Diagram, or CFD, that is associated in
principle to any 5d or 6d SCFT. Therein it was observed that, if one knows a weakly
coupled gauge theory description for a 5d SCFT, then any global symmetry
enhancement at the superconformal point, and furthermore the tree of all of
the descendants of that SCFT, and thus of the gauge theory, is captured in
the CFD.

In this section we will demonstrate that given a CFD the set of weakly coupled
quiver gauge theory descriptions that have the associated SCFT at the UV
fixed point are heavily constrained. Of particular interest will be to
constrain the marginal\footnote{The set of marginal 5d gauge theories and the
set of 5d gauge theories which have a UV fixed point that is a 6d SCFT are
closely overlapping but distinct sets \cite{Jefferson:2017ahm}. In this paper the
marginal theories that we consider will have 6d fixed points, and thus we will
utilize the adjective ``marginal'' without including the further qualification.} 5d quiver gauge theory descriptions of a given 6d SCFTs, as these
are conjectured to source all of the 5d SCFTs as descendants. One then has to
know the ``marginal CFD,'' which is the CFD associated to a 6d SCFT, of
which many interesting cases are known from \cite{Apruzzi:2019vpe,Apruzzi:2019opn}.

Let us briefly recap some of the salient details of 5d $\mathcal{N} = 1$
quiver gauge theories. A quiver consists of $n_G$ gauge nodes, each of which
supports some simple non-abelian\footnote{Gauge nodes carrying a $U(1)$ gauge
  group will not be considered, as quivers with such nodes cannot give rise to
  an interacting SCFT in the UV \cite{Intriligator:1997pq}.} gauge group
$G_i$, such that the total gauge algebra is
\begin{equation}
  G = \prod_{i=1}^{n_G} G_i \,.
\end{equation}
The rank of this gauge algebra will be denoted $r_G$.
Two gauge nodes can be connected by including matter transforming in the
bifundamental\footnote{Adding hypermultiplets charged under different
  non-trivial representations of the two gauge algebras, or indeed of any
  number of simple gauge factors, is a priori possible, however we will not
consider such quivers here.} representation of the two gauge algebras. We will
assume that any gauge nodes are connected in the most minimal way possible,
with the bifundamental matter being either a single hypermultiplet or a single
half-hypermultiplet, depending on what is least allowed. Furthermore, we will
assume that the quivers under discussion do not have loops. For most of the
analysis in this section these two assumptions will be unnecessary, and the
analysis is essentially unchanged by relaxing them. In addition a quiver can
have matter transforming in a representation $\bm{R}$ of a single gauge
factor; this is the matter captured in the flavor nodes of the quiver.

The global symmetry group of the quiver has three contributions, which can be
summarized by writing the rank of the flavor group as
\begin{equation}
  r_F = n_F + n_G + n_b \,.
\end{equation}
The most obvious contribution is from the number of gauge nodes, $n_G$; each
simple gauge factor has an associated topological symmetry, $U(1)_T$. The
other two factors, $n_F$ and $n_b$ come from the classical flavor group
rotating the charged hypermultiplets of the quiver. These rotation groups
depend on the type of representation under which the hypermultiplets
transform. They are:
\begin{equation}
  \begin{aligned}
    k \text{ hypermultiplets in a complex representation }
    &\Rightarrow U(k) \, , \cr 
    k \text{ hypermultiplets in a real representation }
    &\Rightarrow Sp(k) \, , \cr 
    k \text{ half-hypermultiplets in a quaternionic representation }
    &\Rightarrow SO(k) \, ,
  \end{aligned}
\end{equation}
where the hypermultiplets rotate under the fundamental representation of the
global symmetry group. We define $n_F$ to be the rank of this combined group
for all of the flavor nodes of the quiver. The last quantity, $n_b$, is
defined to be the rank of flavor group of the bifundamentals connecting the
gauge nodes; since $k = 1$ for such matter the contribution to $n_b$ is zero
when the bifundamental is real $\otimes$ quaternionic, and one in all other
cases.

The key thrust of this section lies in the fact that the flavor nodes of any marginal
quiver description are highly constrained by the structure of the CFD, as the BG-CFD, defined in \ref{def:BGCFD},  associated with classical flavor of the quiver must be a subgraph of the marginal CFD. Recall that all types of BG-CFDs these are listed in table \ref{tab:BGCFDs}.
The reason is that both graphs represent features of the same geometry underlying the M-theory realization.
Thus, a necessary condition for a gauge theory to be a consistent effective description of an SCFT is for the BG-CFD of the former to embed into the CFD of the latter.

The geometric details of this relationship will be spelled out in sections \ref{sec:FibPhas} and \ref{sec:CFDsfromBG}.
To get across our main points here, we will review the definition of the CFDs in section \ref{sec:CFDrecap},
followed by listing what constraints apply to the prospective quiver from a
known CFD in section \ref{sec:consts}; we will find that the possible flavor
nodes of any quiver are constrained to be one of a small finite list from the
embedding of the BG-CFD inside the CFD, further usage of the gauge rank and
flavor rank, together with the single node constraints of
\cite{Jefferson:2017ahm} will often allow one to specify the complete quiver
more restrictively. We determine the constraints on
the possible 5d quiver descriptions for many of the known marginal CFDs and
thus for their associated 6d SCFTs.

\subsection{Recap: CFDs}\label{sec:CFDrecap}

The CFD \cite{Apruzzi:2019vpe,Apruzzi:2019opn} is a marked undirected graph,
where each vertex $C_i$ is associated with two integers $(n_i,g_i)$ and each
edge between the two vertices $C_i$ and $C_j$ is marked with an integer
$m_{i,j}$. In the context of elliptic Calabi--Yau geometries, a CFD can
be interpreted as a flop equivalence class among a family of reducible complex
surfaces $\mathcal{S}$. Under this interpretation, each vertex $C_i$ is a
complex curve with self-intersection number $n_i$ and genus $g_i$, and the
integer $m_{i,j}$ is equal to the intersection number $C_i\cdot_\mathcal{S}
C_j$.

Qualitatively, the vertices can be classified into the following three classes:

\begin{enumerate}
\item{The marked vertices, which correspond to  flavor curves $F_i$, and are usually colored green. 
Typically, they have labels $(n_i,g_i)=(-2,0)$, and are called ``$(-2)$-vertices''. However, sometimes they are associated with $(n_i,g_i)=(-1,0)$ instead, see the $(E_8,SU(2))$ CFDs in \cite{Apruzzi:2019opn}. The subgraph of such vertices always form the Dynkin diagram of the flavor symmetry of the UV fixed point, $G_{\text{F}}$.

In the presence of some non-simply laced Lie algebra $G$ (such as the
$(E_7,SO(7))$ case in \eqref{E7SO7-topCFD}), the flavor curve corresponding to
the short root is a collection of $p$ green $(-2)$-vertices that are identical\footnote{Geometrically, there are $p$ curves with normal bundle $\mc{O}+\mc{O}(-2)$ that are homologous in the resolved Calabi-Yau threefold.}, where $p$ is the ratio between the length of the long root and the short root of the Lie algebra $G$. Specifically, for $G=B_k$, the single short root will be assigned to a reducible vertex, represented by with two $(-2)$-vertices that are encircled. For $G=G_2$, the short root will be assigned to a
reducible vertex with three $(-2)$-vertices that are encircled. For $G=C_k$, there is only a single long root,
along with $(k-1)$ short roots. In principle, we need to draw $(k-1)$ reducible vertices 
which are each containing two $(-2)$-vertices that are encircled, and a single vertex with
$(n_k,g_k)=(-2,0)$, while these vertices are connected with $m_{i,i+1}=2$
$(i=1,\dots,k-1)$. However, in practice we can simplify the CFD by taking
``half'' of this diagram, which ends up with $(k-1)$ vertices with
$(n_i,g_i))=(-2,0)$ $(i=1,\dots,k-1)$ and a single flavor vertex with
$(n_k,g_k)=(-1,0)$, while they are connected with $m_{i,i+1}=1$
$(i=1,\dots,k-1)$. This explains the convention of BG-CFDs for non-simply
laced $G_{\text{F,cl}}^{\rm 5d}$ in table~\ref{tab:BGCFDs}.

}

\item{The unmarked vertices with labels $(n_i,g_i)=(-1,0)$ will be denoted by ``$(-1)$-vertices'', and corresponds to an extremal curve in the geometry.  
A transitions between CFDs, and thus 5d SCFTs, is realized by removing such a curve. 
Certain extremal curves will correspond to the F-extremal weights in a gauge theory description of the SCFT.

Sometimes, there will be a reducible vertex comprised of multiple $(-1)$-vertices connected to a reducible vertex containing multiple $(-2)$-vertices, which each describe homologous curves in the Calabi-Yau threefold that have to be flopped simultaneously. In the CFD language, one has to remove all the $(-1)$-vertices in the reducible vertex at the same time.

}

\item{Other vertices with $n_i>0$, are unmarked, and are determined from the resolution of the singular geometry.  However, they cannot be directly seen from the gauge theory description.}

\end{enumerate}

We also list the rule of CFD transitions here. After the $(-1)$-vertex $C_i$ is removed, the new graph is constructed from the original CFD with the following rules:

\begin{enumerate}
\item{For any vertex $C_j$ with label $(n_j,g_j)$ that connects to $C_i$ ($m_{i,j}>0$), the updated vertex  $C_j'$ in the resulting CFD' has the following labels:
\be\ba
n_j'  = &n_j+m_{i,j}^2\cr 
 g_j' =& g_j+\frac{m_{i,j}^2-m_{i,j}}{2} \,.
\ea \ee
}

\item{For any two vertices $C_j,C_k$, $j\neq k$, that connect to $C_i$, the new label on the edge $(j,k)$ is given by
\be
m'_{j,k}=m_{j,k}+m_{i,j}m_{i,k}\,.
\ee
}

\item{If there are multiple $C_j$s connected to $C_i$, then the rule 2 applies for each pair of vertices.}
\end{enumerate}

The starting point of the CFD transitions is called a \textit{marginal CFD}, which corresponds to a 5d marginal theory that only has a UV fixed point in 6d. The flavor (marked) vertices in a marginal CFD can form affine Dynkin diagrams, but it is required that none of the affine Dynkin diagrams is present after any CFD transition is applied to the marginal CFD.

The 5d BPS states from the M2 brane wrapping modes can be read off from the linear combinations of the vertices in the CFD. For the 5d hypermultiplets, which can correspond to the matter fields in our gauge theory descriptions, they are read off from the unmarked vertices $C_i$ with $(n_i,g_i)=(-1,0)$.

If the SCFT has an effective gauge theory description, then its perturbative states are also formed by M2 branes wrapping certain curves that are encoded by the CFD.
As will become more apparent in sections \ref{sec:FibPhas} and \ref{sec:CFDsfromBG}, these curves precisely form the BG-CFD, which therefore must be contained inside the CFD.


\subsection{Constraints on Quiver Gauge Theories}\label{sec:consts}

To determine which quiver gauge theories are consistent with any marginal CFD
one can proceed in the following manner. Determine all possible embeddings of
(possibly disconnected) BG-CFDs into the marginal CFD as subgraphs.
These must be embedded in such a way that they are non-overlapping, and
furthermore such that the marked/flavor vertices in any connected BG-CFD are not
adjacent to the vertices of the embedding of any other connected BG-CFD. Such an
embedding is necessary if we want to obtain a quiver which has a consistent
classical flavor symmetry. 

Since the BG-CFDs are associated to the classical flavor symmetry rotating the
hypermultiplets at any flavor vertex, the BG-CFDs which can be embedded
gives immediately the set of flavor symmetry groups that can be realized as
rotation groups of the flavor verticess. This fixes the kinds of representations
that can be realized on the flavor vertex, whether they are complex, real, or
quaternionic, and also fixes the number of hypermultiplets that can appear
there. 

Since the CFD is, by construction, agnostic towards the details of the precise
configuration of surfaces in the geometry, and thus to the details of any
particular gauge sector that is disconnected from the global symmetry groups
that the CFD is sensitive to, we shall find that there is often a
pure-gauge\footnote{We remind the reader that by a ``pure-gauge'' quiver we
mean a quiver consisting only of gauge nodes --- there remain bifundamental
matter fields between these gauge nodes.} sub-quiver in any putative quiver
description. This $Q_s$ is generally unfixed, but of course
constrained\footnote{For low ranks these constraints will, in fact, generally
be enough to completely fix $Q_s$.}, however
the precise details of its structure are irrelevant for the tree of
descendant SCFTs, except for possible discrete dualities that depend on those
details.

In addition to this we further know that the flavor rank of the SCFT must be replicated in
the rank of the classical flavor symmetry of the quiver description of the
marginal theory. Similarly we know the gauge rank, $r_G$, required of any
prospective quiver from the SCFT which realizes the CFD in question. Thus we
have a further constraint on quiver gauge theory descriptions from knowledge
of the pair of ranks $(r_G, r_F)$. 

A further set of constraints, which we refer to as the ``constraints on the
number of hypermultiplets'', comes from the analysis of single gauge node
quivers in \cite{Jefferson:2017ahm}. In that paper it was shown that if a 5d
single gauge node quiver was to lead to an interacting SCFT in the UV then the
matter content (and where relevant the Chern--Simons level) must satisfy the
following constraints:\footnote{We will not consider here some of the outlier
  options for matter representations that can appear at low rank. These are
  the triple anti-symmetric representations of $SU(N)$ and the spinor and
  conjugate spinor representations of $SO(N)$. The methods given throughout
  this paper apply with little modification to these cases also, however to
  prevent a proliferation of subcases we do not write of them here.}
\begin{equation}\label{eqn:JKVZ1}
  \begin{aligned}
    &(\#\bm{Sym}, \#\bm{AS}, \#\bm{F}; k) \text{ of } SU(N) & &\leq (1,1,1;0), (1,0,N-2;0),
    (1,0,0;N/2), \cr
    & & &\qquad (0,2,8;0), (0,2,7;3/2), (0,1,N+6;0), \cr
    & & &\qquad (0,1,8; N/2), (0,0,2N + 4; 0) \cr
    &(\#\bm{AS},\#\bm{F}) \text{ of } Sp(N) & &\leq (1,8), (0, 2N+6) \cr
    &\#\bm{V} \text{ of } SO(N) & &\leq N - 2 \cr
    &\#\bm{7} \text{ of } G_2 & &\leq 6 \cr
    &\#\bm{26} \text{ of } F_4 & &\leq 3 \cr
    &\#\bm{27} \text{ of } E_6 & &\leq 4 \cr
    &\#\bm{56} \text{ of } E_7 & &\leq 3 \,. \cr
  \end{aligned}
\end{equation}
When we write $\leq$ in this context we mean that the possible data associated
to the gauge node must be that of a descendant of the gauge theory with the
data on the right-hand side. For almost all gauge groups except
$SU(N)$, this is equivalent to stating that there must be fewer
hypermultiplets transforming in one or more of the representations appearing
on the right. In the case of $SU(N)$, which has a non-trivial cubic Casimir,
one must also consider how the Chern--Simons level is shifted when decoupling
a hypermultiplet in a representation $\bm{R}$, 
\begin{equation}
  k \rightarrow k \pm A(\bm{R})/2 \,.
\end{equation}
For the representations of interest in this paper the quantities $A(\bm{R})$
are
\begin{equation}
  A(\bm{F}) = 1 \,, \quad A(\bm{AS}) = N - 4 \,, \quad A(\bm{Sym}) = N + 4 \,.
\end{equation}
In this way all of the descendant gauge theories of a single gauge node quiver
can be determined (one can also see them in section \ref{sec:somesingle}), and
each individual gauge node must have the data of a descendant of one of the
above right-hand side theories. In fact, for small values of the rank of the
gauge algebra there are additional options. For the representations that we
will consider these additional possibilities are exhausted by the following
\begin{equation}\label{eqn:JKVZ2}
  \begin{aligned}
    &(\#\bm{AS}, \#\bm{F}) \text{ of } Sp(2) & &\leq (3,0), (2,4) \cr
    &(\#\bm{AS}, \#\bm{F}) \text{ of } Sp(3) & &\leq (2,0) \cr
    &(\#\bm{F}; k) \text{ of } SU(3) & &\leq (6;4), (3;13/2), (0;9) \cr
    &(\#\bm{AS}, \#\bm{F}; k) \text{ of } SU(4) & &\leq (4,0;4), (3,4;2),
    (3,0;5),
    (2,0;6), \cr & & &\qquad (1,0;7),(0,8;3),(0,0;8) \cr
    &(\#\bm{AS}, \#\bm{F}; k) \text{ of } SU(5) & &\leq (3,3;0), (3,1;3),
    (3,2;3/2), (0,5;11/2) \cr
    &(\#\bm{AS}, \#\bm{F}; k) \text{ of } SU(6) & &\leq (3,0;3), (0,0;9) \,.
  \end{aligned}  
\end{equation}
Any other gauge algebras or matter outside of that satisfying the above will
not lead to an interacting SCFT as the UV fixed point of the gauge theory.
This is determined by studying the Coulomb branch of the gauge theory, and as
such these constraints will, in general, apply to any gauge node of an
arbitrary quiver. 

A further constraint is given by the positivity of the metric and the string tensions of the full quiver, which comes from study the second and first derivatives of the prepotential with respect to the Coulomb branch scalar vevs. The requirement that the metric is positive, provided that the tensions of the BPS strings do not change sign, sets sharper bounds on the possible quiver nodes and the amount of matter on them \cite{Jefferson:2017ahm}. In this paper we present quivers which, based on their classical flavor symmetry embeddings in the CFDs, are possible IR description of a given CFT, and in this sense we provide a set of necessary conditions for these quivers to exist as such. It would be interesting to further restrict these bounds by studying these Coulomb branch metric constraints. We plan to come back to this general analysis in the future.

\subsection{Consistent Quivers for the Rank One E-string}

To illustrate this procedure of constraining weakly coupled quiver gauge
theory descriptions for the SCFTs associated to a CFD, let us consider the CFD
that is associated to the 6d rank one E-string theory. 
By considering the marginal CFD that has a 6d, rather than 5d, SCFT as its fixed point
we are constraining the possible marginal 5d quiver gauge theories that flow to this 6d
theory in the UV. 
While we can of course determine possible quiver gauge theories for a non-marginal
CFD we shall not do so here; this is because if a gauge theory description
exists for a descendant SCFT then a similar description, with an increased
number of matter hypermultiplets, should exist for the marginal theory.

In this section we will show how the three different constraints laid out in
section \ref{sec:consts} leads to a unique possible weakly-coupled quiver
description for the rank one E-string. The marginal CFD for this theory has
been determined in \cite{Apruzzi:2019vpe}, and it is
\begin{equation}
  \includegraphics[scale=0.3]{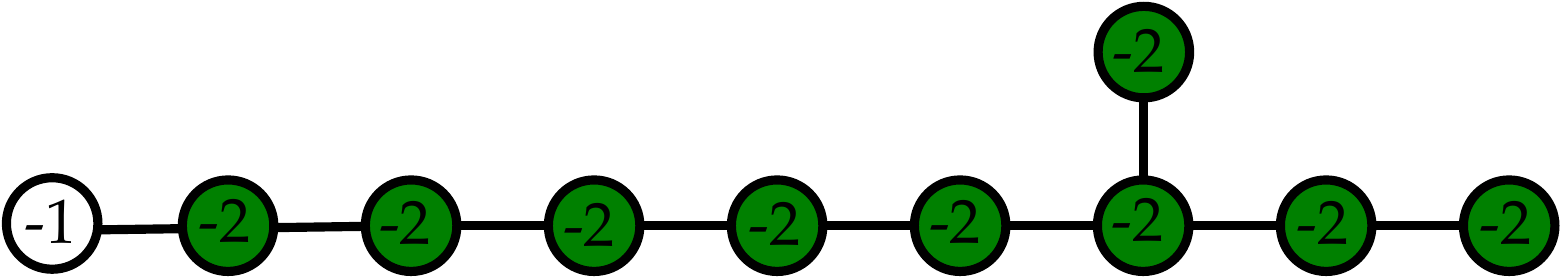} \,.
\end{equation}
We can see that there is only exactly one way to embed disjoint unions of the
BG-CFDs into the marginal CFD, that being the BG-CFD association to an
$SO(16)$ global symmetry group. This embedding can be drawn as
\begin{equation}
  \includegraphics[scale=0.3]{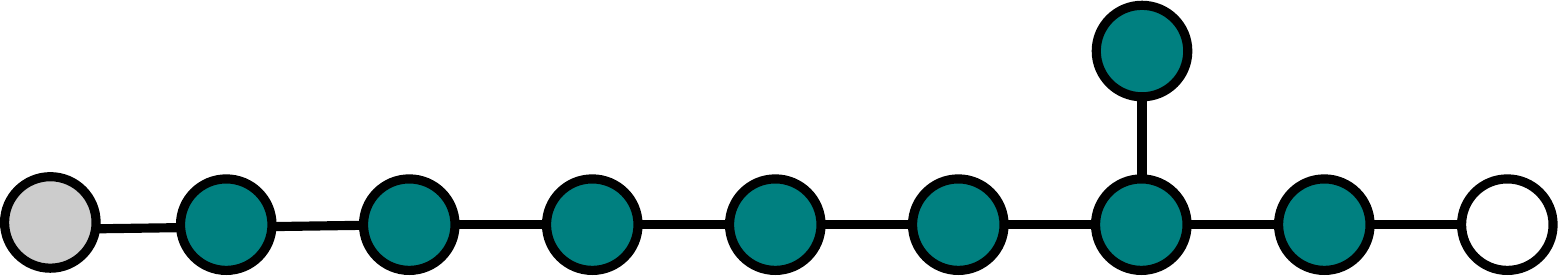} \, ,
\end{equation}
where, as in the previous section, the BG-CFD is the colored part, with turquoise denoting flavor $(-2)$- , and gray denoting $(-1)$-vertices.
This global symmetry group can only arise from a flavor vertex
associated to $m = 8$ full hypermultiplets transforming in either
the fundamental representation of $Sp(n)$ for some $n$, or the $\bm{56}$ of $E_7$. 
However, since the rank one E-string has a single tensor multiplet, and no gauge algebras on its tensor branch in 6d, its 5d descendants can only have a single vector multiplet coming from the reduction of the tensor.
Likewise, the rank $r_F$ of the global gauge symmetry group is $\text{rank}(G_{\text{6d}}) + 1 = 9$, where $G_{\text{6d}}$ is the global symmetry
group of the 6d SCFT. 
This leaves
only one option for the quiver, which is
\begin{equation}
  8\bm{F} - Sp(1) \,.
\end{equation}
The
final set of constraints, those coming from the number of hypermultiplets
allowed for each gauge node, are not required to nail down this quiver,
however we can see that the constraint
\begin{equation}
  \# \bm{F} \text{ of } Sp(n) \leq 2n + 6 \,,
\end{equation}
is saturated for this quiver. 
This is, of course, nothing other than $SU(2)$
with eight fundamental hypermultiplets which has long since been known to have
the rank one E-string as a UV fixed point \cite{Seiberg:1996bd}. 
Furthermore
what we show here, which is also long since known, is that this is the only
possible 5d quiver gauge theory description to have the rank one E-string as a
UV fixed point.

In the next section we shall perform the same analysis
for a variety of other, much more non-trivial, 6d SCFT starting points, and
determine their heretofore unknown consistent 5d quiver gauge theory descriptions.

\subsection[Consistent Quivers for \texorpdfstring{$(D_k, D_k)$}{(Dk,Dk)}
Minimal Conformal Matter]{Consistent Quivers for \boldmath{$(D_k, D_k)$} Minimal Conformal Matter}

In \cite{Apruzzi:2019vpe} we considered the descendants of the 6d SCFT known
as minimal $(D_k, D_k)$ conformal matter. Based on the known (quiver) gauge
theory descriptions for theories that have this SCFT at their UV fixed point,
we were able to determine a host of superconformal flavor symmetry
enhancements for said theories. 

In this section we will show that these quiver gauge theories descriptions discussed therein, and previously known, are found as consistent quivers
satisfying the three following constraints, as laid out previously, 
\begin{itemize}
  \item the BG-CFDs for the quiver can be embedded into the marginal CFD in a
    non-adjacent way,
  \item the gauge and flavor ranks of the quiver match $(r_G, r_F) = (k-3,
    2k+1)$,
  \item each gauge node of the quiver satisfies the constraints on the number
    of hypermultiplets as written in \cite{Jefferson:2017ahm} and summarized
    in (\ref{eqn:JKVZ1}) and (\ref{eqn:JKVZ2}).
\end{itemize} 
Furthermore, we will see that, up to a caveat with a reordering of the
``pure-gauge'' part of the quiver that we will discuss anon, the known
theories saturate the options for marginal theories consistent with the above
constraints.  
The marginal CFD, for arbitrary $k \geq 5$,\footnote{The smallest $k$ such that there is a 6d SCFT is $k=4$, which however is an alternative realization of the rank one E-string discussed above.
} was determined in
\cite{Apruzzi:2019vpe} to be
\begin{equation}
  \includegraphics[scale=0.3]{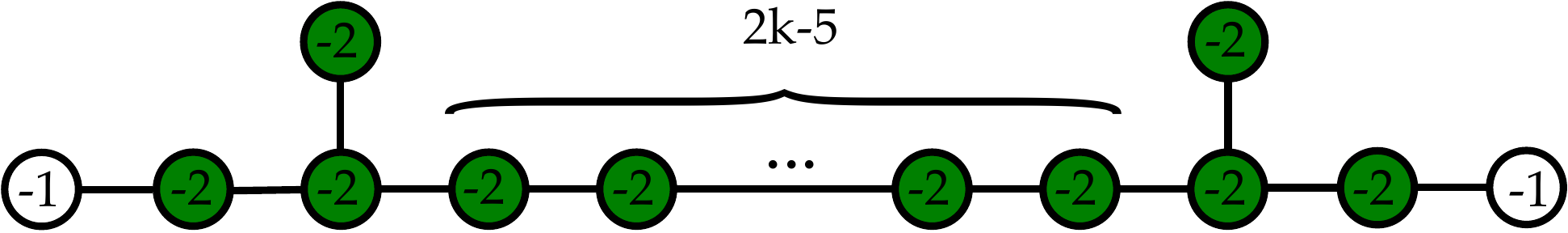} \,.
\end{equation}
We must determine all of the possible embeddings of the BG-CFDs into this CFD.
We write, for each possible embedding, the flavor symmetry group that rotates the hypermultiplets at each
flavor node, for which the BG-CFD must embed, together with the subquiver
associated to that embedded BG-CFD in the following table, 
\begin{equation}\label{eqn:DkDkqH}
  \begin{array}{c|c}
    \text{Subgroup of $G_\text{F,cl}$} & \text{(Sub)quivers} \cr\hline
    U(2k) & 2k\bm{F} - SU(k-2) \cr\hline
    SO(4k) & 2k\bm{F} - Sp(k-2) \cr\hline
    SO(8) \times SO(8) & \bigg( 4\bm{F} - Sp(n_1) \bigg) \oplus \bigg( 4\bm{F}
    - Sp(n_2) \bigg)\cr\hline
    SO(8) & 4\bm{F} - Sp(n) \cr\hline
  \end{array} \,.
\end{equation}
In this table we write only the part of the quiver that is directly implied by
the existence of an embedding of the BG-CFD into the marginal CFD. We have
also used the constraints on the number of hypermultiplets associated to any
gauge node to rule out, for example, $4 \times \bm{56} - E_7$ from appearing. The
$\oplus$ means that these are two subquivers which are part of whatever the
complete quiver is that would describe this marginal theory. 

Let us consider each of these possible subquivers in turns and determine
whether or not there exist complete quivers, satisfying all of the
constraints, for which these are subquivers. The first example is
\begin{equation}
  2k\bm{F} - SU(n) \,,
\end{equation}
for which we can see that it is not possible to add any further gauge nodes
without violating the constraint on the flavor rank, $r_F = (2k+1)$. As such
the gauge rank constraint, $r_G = k-3$ fixes $n$, and we can see that the
quiver
\begin{equation}
  2k\bm{F} - SU(k-2) \,,
\end{equation}
is a consistent quiver for a marginal theory which has, as UV fixed point, the
minimal $(D_k, D_k)$ conformal matter theory. The second possible subquiver is
\begin{equation}
  2k\bm{F} - Sp(n) \,,
\end{equation}
which we can see, by exactly the same arguments as applied for the first
example, that the only possible quiver containing this subquiver and
satisfying all of the constraints is
\begin{equation}
    2k\bm{F} - Sp(k-2) \,.
\end{equation}
We will now study all of the remaining cases simultaneously. Let us consider a
subquiver, $Q_s$, of the full prospective quiver, where the full quiver is
formed by adding either one or two flavor nodes, as in (\ref{eqn:DkDkqH}), to
$Q_s$. We can see that $Q_s$ must satisfy the following properties
\begin{equation}
  r_G(Q_s) = r_G = k - 3 \,, \quad r_F(Q_s) \geq r_F - 8 = 2k - 7 \,.
\end{equation}
The latter inequality follows as the maximal number of hypermultiplets
associated to the flavor nodes is eight, as one can see from the third line of
(\ref{eqn:DkDkqH}). For any quiver without loops and where each gauge node is
connected to other gauge nodes by only a single (half-)hypermultiplet in the
bifundamental representation, the rank of the subgroup of the global
symmetry group that rotates these bifundamentals is bounded by the number of
gauge nodes,
\begin{equation}
  n_b(Q_s) \leq n_G(Q_s) - 1 \,.
\end{equation}
Furthermore, the bound on the rank of the gauge algebra also bounds the total
number of gauge nodes of the quiver
\begin{equation}
  n_G(Q_s) \leq k - 3 \,.
\end{equation}
Putting all this together we find that there are two inequalities that $Q_s$
must satisfy. These are
\begin{equation}
  \begin{aligned}
    r_F(Q_s) = n_G(Q_s) + n_b(Q_s) &\leq 2k - 7 \, ,\cr
    r_F(Q_s) &\geq 2k - 7 \,,
  \end{aligned}
\end{equation}
for which there is only one solution:
Any such $Q_s$ must be a quiver formed
out of $k-3$ gauge nodes each carrying gauge group $Sp(1)$. The total quiver
which may be a marginal theory for the minimal $(D_k, D_k)$ conformal matter
theory is then any such $Q_s$ where the flavor nodes $4\bm{F}$ and $4\bm{F}$
are attached to any two distinct $Sp(1)$ gauge nodes. 
These must be distinct
nodes as otherwise the constraint that any $Sp(1)$ gauge node must have at
most $8$ total fundamental hypermultiplets is violated. As we can see this
gives a great variety of potential quiver gauge theory descriptions for the
marginal theory, based on all the different configuration of the $Sp(1)$ nodes
in the ``pure-gauge'' quiver $Q_s$. That the quiver is not unqiue, but this is
expected, as the CFD, by construction, does not contain the information about
the gauge algebra; the algebra itself depends on the details of the
decomposition of the reducible surface of the CFD into irreducible surfaces.
This goes above and beyond the purpose of the CFD, and as such we can see that
the CFD itself constrains only which flavor nodes (and thus the attached gauge
nodes) can appear, and combined with the rank constraints and the constraints
from the number of hypermultiplets, one can determine a set of possible
quivers. This is highly restrictive, but it remains a superset of the 5d
quiver gauge theory descriptions which are marginal descriptions of the
particular 6d SCFT under consideration.

\subsection{Consistent Quivers for Rank Two ``Model 3''}

The rank two theory that was referred to as ``Model 3'' in
\cite{Apruzzi:2019opn}, i.e. $SU(3)$ on a $(-1)$ curve with 12 hypers,  was found to have a marginal CFD being
\begin{equation}\label{eqn:mCFDModel3}
  \includegraphics[scale=0.25]{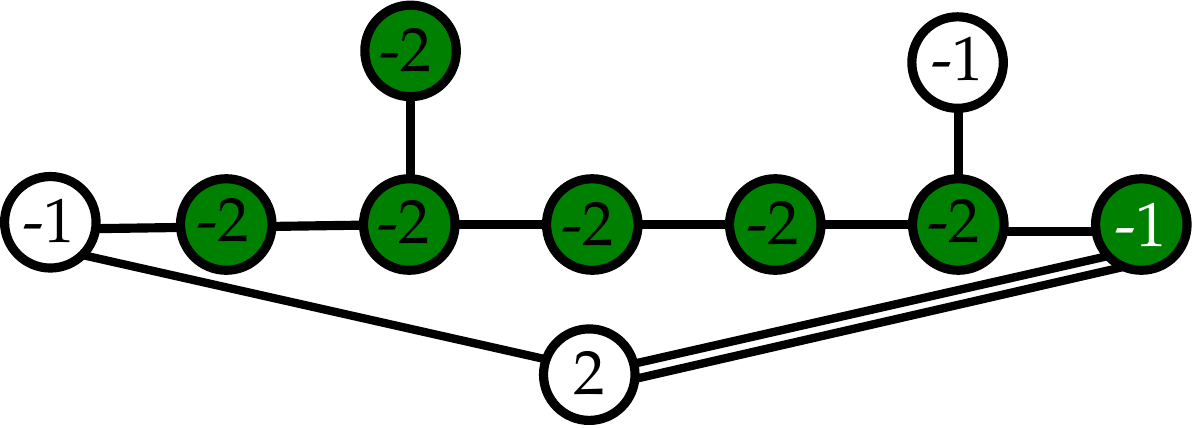} \,,
\end{equation}
and the ranks of the gauge and classical flavor groups for any 5d gauge theory
description of this theory are known to be
\begin{equation}
  (r_G, r_F) = (2, 7) \,.
\end{equation}
\begin{figure}
\centering
\includegraphics[scale=0.2]{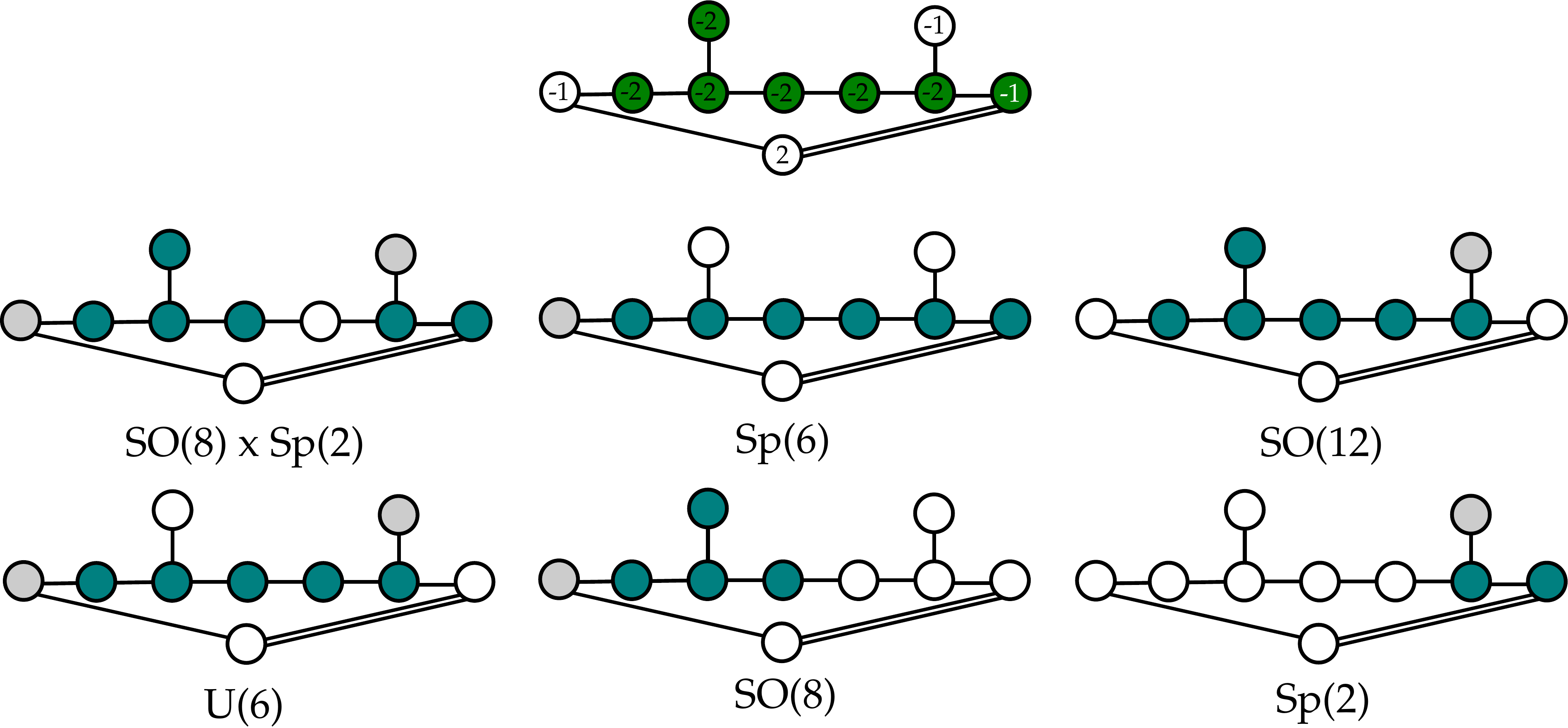}
\caption{Embedding of BG-CFDs into the marginal CFDs of Model 3 of the rank 2
  classification. From the embedding the classical flavor symmetry, and
  thereby $G_{\text{BG}}$ can be read off.\label{fig:Model3embed}}
\end{figure}
We again carry out the same procedure of determine possible weakly coupled
gauge theory descriptions for this theory: we study all possible ways of
embedding the BG-CFDs into (\ref{eqn:mCFDModel3}). There are eight possible
combinations of embeddings, and these are depicted in figure
\ref{fig:Model3embed}. First of all, let us just utilize the gauge rank,
$r_G$, and the embedding of the BG-CFDs inside of the marginal CFD to write
down all of the potential quivers that may be marginal descriptions of this 6d
SCFT. These are
\begin{equation}
  \begin{aligned}
    &6\bm{F} - SU(3)_k \, , & & \cr
    &m_1\bm{F} - Sp(2) - m_2\bm{AS}\, , &\quad &m_1 = 0,4 \,, m_2 = 0, 2 \, , \cr
    &6\bm{F} - Sp(2) \, , & & \cr
    &m\bm{7} - G_2 \, , &\quad &m = 2,6 \, , \cr
    &m\bm{F} - Sp(1) - Sp(1)_\theta \, , &\quad &m = 4,6 \,.
  \end{aligned}
\end{equation}
Compatibility with the fixed flavor rank, $r_F = 7$, leaves only five possible
quivers
\begin{equation}
  \begin{aligned}
    &6\bm{F} - SU(3)_k \, , \cr
    &4\bm{F} - Sp(2) - 2\bm{AS} \, , \cr
    &6\times \bm{7} - G_2 \, , \cr
    &6\bm{F} - Sp(2) \, , \cr
    &4\bm{F} - Sp(1) - Sp(1)_\theta \,.
  \end{aligned}
\end{equation}
Of these five theories the latter two are known to be descendants of
\begin{equation}
  10\bm{F} - Sp(2) \quad \text{ and } \quad  4\bm{F} - Sp(1) - Sp(1) - 4\bm{F} \,,
\end{equation}
respectively, which are both marginal gauge theory descriptions of minimal $(D_5, D_5)$
conformal matter. As descendants of marginal theories they
cannot, in themselves, be marginal, and thus we must rule them out of the set
of possible marginal quiver descriptions associated to the
Model 3 CFD. This leaves only the three previously known gauge theory
descriptions as the complete set of options, 
\begin{equation}
  \begin{aligned}
    &6\bm{F} - SU(3)_k \, , \cr
    &4\bm{F} - Sp(2) - 2\bm{AS} \, , \cr
    &6 \times \bm{7} - G_2 \,.
  \end{aligned}
\end{equation}
Furthermore, for the $SU(3)$ description the Chern--Simons level can be fixed
to $k = 4$ by compatibility with the constraints on the number of
hypermultipets as given previously.

\subsection{Consistent Quivers for Rank Two ``Model 4''}

Continuing with our study of the possible rank two quiver gauge theory descriptions of
6d SCFTs, we now turn to the 6d SCFT starting point referred to as ``Model 4''
in \cite{Apruzzi:2019opn}, i.e. $SU(3)$ theory on a $(-2)$-curve with 6 hypers. There, the marginal CFD for such a CFD was
determined to be
\begin{equation}
  \includegraphics[scale=0.25]{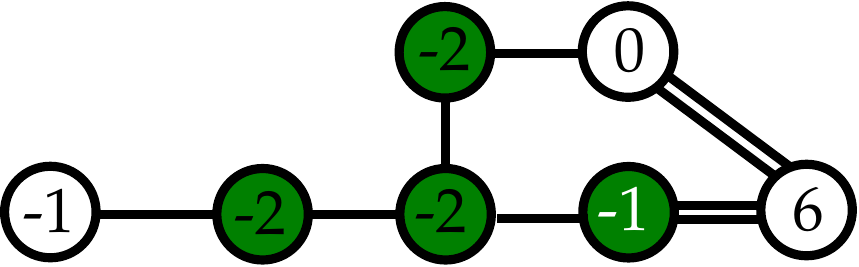}\,,
\end{equation}
and the gauge and flavor ranks of any 5d quiver gauge theory description are
required to be
\begin{equation}
  (r_G, r_F) = (2, 4) \,.
\end{equation}
We can see that, after looking only at the embedding of possible BG-CFDs inside of
the marginal CFD there is only one option for the embedded BG-CFD,
corresponding to
\begin{equation}
  \includegraphics[scale=0.25]{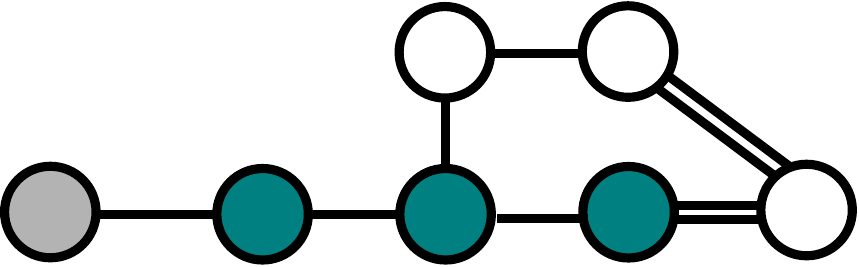} \,,
\end{equation}
which comes from an $Sp(3)$ flavor group.  Restricting such that the
ranks match those given above, the only possible 5d quiver gauge theory
descriptions of this marginal CFT are
\begin{equation}
  \begin{aligned}
    &3\bm{AS} - Sp(2) \, ,\cr
    &3 \times \bm{7} - G_2 \, . \cr
  \end{aligned}
\end{equation}
The latter quiver is a descendant of $6 \times \bm{7} - G_2$, which is a known
marginal theory describing Model 3.
As such it cannot be associated to a marginal theory, and thus our analysis
leaves the only possible quiver description $3\bm{AS} - Sp(2)$ for Model 4.

\subsection[Consistent Quivers for \texorpdfstring{$(E_6, E_6)$}{(E6,E6)} Minimal Conformal
Matter]{Consistent Quivers for \boldmath{$(E_6, E_6)$} Minimal Conformal
Matter}\label{sec:E6E6qH}

Let us now consider the case of minimal $(E_6, E_6)$ conformal matter. As
before, to determine possible quiver descriptions of the marginal theory it is
of the first importance to determine which possible BG-CFDs associated to any
classical flavor group can be embedded into the marginal CFD. As it turns out,
there are a limited number of options. Since the marginal CFD is simply-laced,
in the sense that all of the marked green vertices have label $n = -2$, the
only types of connected BG-CFDs that can be embedded are those associated to
$U(k)$ and $SO(2k)$ flavor factors.

The marginal CFD was determined in \cite{Apruzzi:2019vpe,Apruzzi:2019opn} from
geometric considerations, and it is
\begin{equation}\label{eqn:E6E6MCFD}
  \includegraphics[scale=0.25]{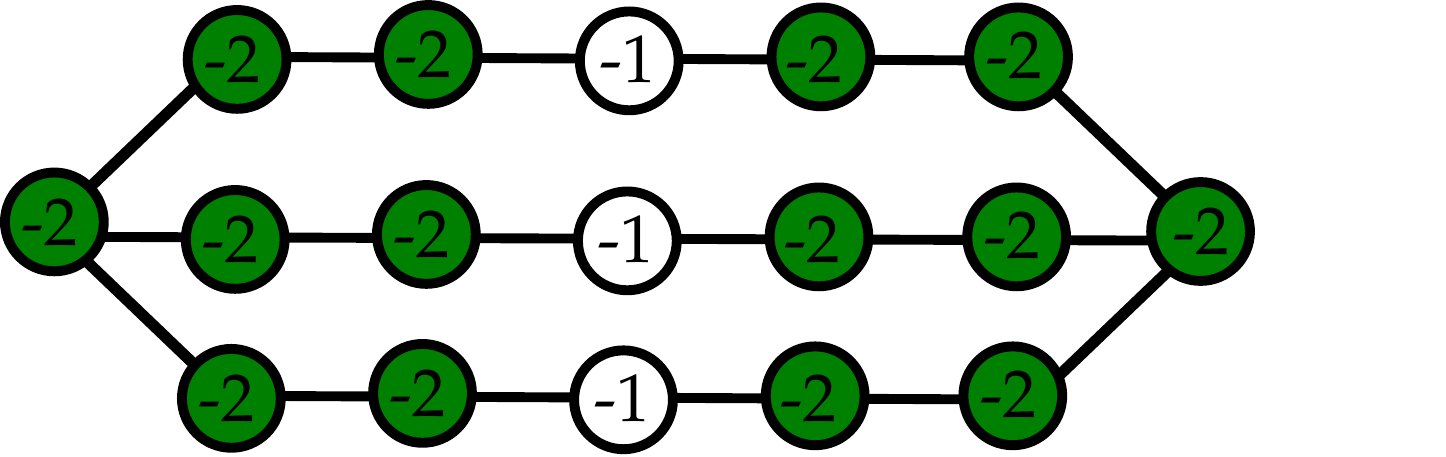} \,,
\end{equation}
and only allows gauge descriptions with gauge and flavor ranks given by
\begin{equation}
  (r_G, r_F) = (5, 13) \,.
\end{equation}
All of the different possible embeddings are shown in figure
\ref{fig:E6E6embed}. We see immediately that if there is a $U(k)$ factor in
the classical flavor group then there can only be one such factor, and
furthermore it must have $k = 6$; it is apparent even from this immediate
result that the structure of the marginal CFD powerfully constrains possible
quiver gauge theories that have minimal $(E_6, E_6)$ conformal matter as a UV
fixed point.

\begin{figure}
\centering
\includegraphics*[width=\textwidth]{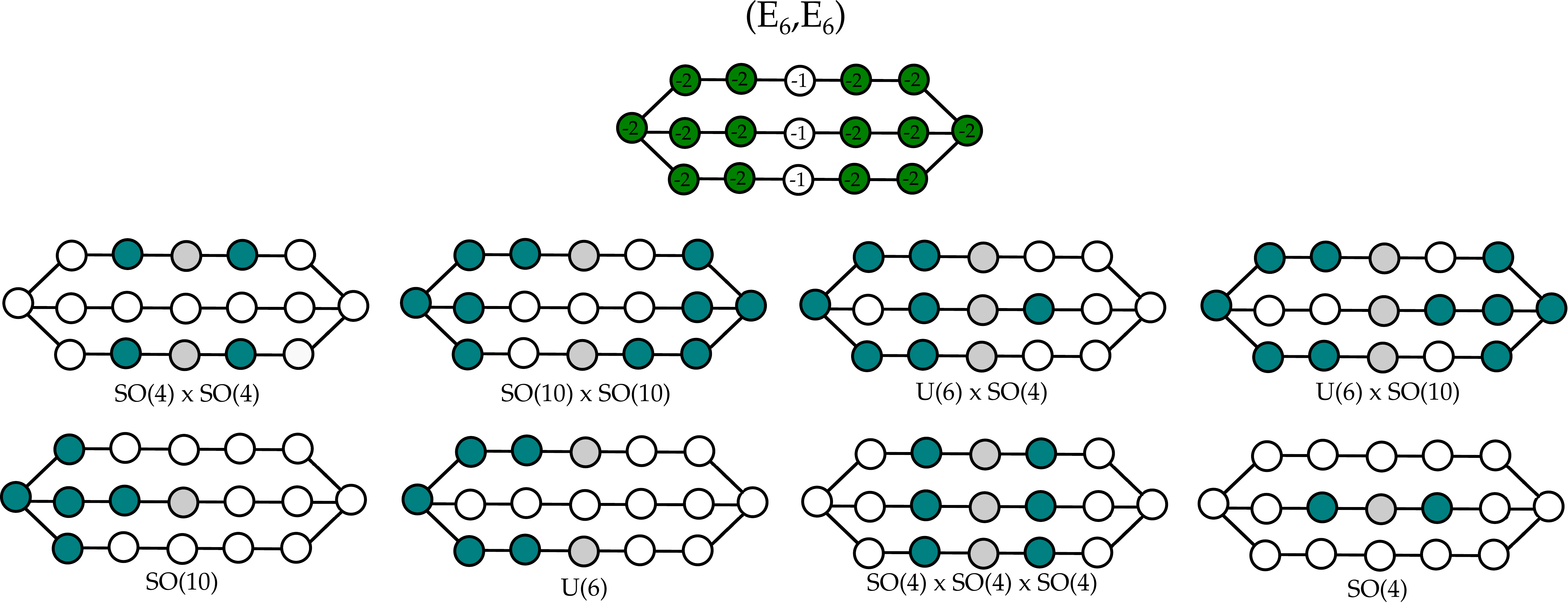}
\caption{Embedding of potential BG-CFDs into the $(E_6, E_6)$ marginal
CFD. Below each embedding of the BG-CFD into the marginal CFD we list the classical flavor symmetry. }\label{fig:E6E6embed}
\end{figure}

Let us now combine this analysis with the full set of constraints that were
previously described. The possible flavor nodes of any quiver description are
one of the following
\begin{equation}
  \begin{aligned}
    &6\bm{F} - SU(n) \, , & & \cr
    &\bigg(6\bm{F} - SU(n_1)\bigg) \oplus \bigg(m\bm{F} - Sp(n_2)\bigg) \, , &\quad & m = 2,5 \, , \cr
    &m\bm{F} - Sp(n) \, ,&\quad & m = 2,5 \, , \cr
    &\bigg(m\bm{F} - Sp(n_1)\bigg) \oplus \bigg(m\bm{F} - Sp(n_2)\bigg) \, , &\quad
    &m = 2,5 \, ,\cr
    &\bigg(2\bm{F} - Sp(n_1)\bigg) \oplus \bigg(2\bm{F} - Sp(n_2)\bigg) \oplus
    \bigg(2\bm{F} - Sp(n_3)\bigg) \, , & \cr
  \end{aligned}
\end{equation}

We will now write down some explicit quivers with the maximal rank of the
global symmetry group rotating the hypermultiplets at each flavor nodes. Such
a quiver will extend down the maximal depth inside of the CFD tree, and thus
have a wide variety of 5d SCFTs at the UV fixed points of its descendants. It
is immediately clear that the subquiver corresponding to
\begin{equation}
  \bigg(6\bm{F} - SU(n_1)\bigg) \oplus \bigg(5\bm{F} - Sp(n_2)\bigg)\,,
\end{equation}
cannot be combined into a quiver satisfying $r_F = 13$. The maximal quivers
that we can attempt to determine must be constructed from the subquiver option
\begin{equation}
  \bigg(5\bm{F} - Sp(n_1)\bigg) \oplus \bigg(5\bm{F} - Sp(n_2)\bigg) \,.
\end{equation}
One option is to consider the quiver
\begin{equation}
  5\bm{F} - Sp(n_1) - Sp(n_2) - 5\bm{F} \,,
\end{equation}
which has the correct global symmetry group rank. The constraints that such a
quiver is consistent with the gauge rank and with the constraints on the
number of hypermultiplets are
\begin{equation}
  \begin{aligned}
    &\#\bm{F} \text{ of } Sp(n_1): &\quad &2n_2 + 5 \leq 2n_1 + 6 \cr
    &\#\bm{F} \text{ of } Sp(n_2): &\quad &2n_1 + 5 \leq 2n_2 + 6 \cr
    &r_G = 5: &\quad &n_1 + n_2 = 5 \,,
  \end{aligned}
\end{equation}
for which it is straightforward to see that there is no solution. The second
possible quiver is
\begin{equation}
  5\bm{F} - Sp(n_1) - G - Sp(n_2) - 5\bm{F} \,,
\end{equation}
where $G$ has a real (for consistency with $r_F = 13$) even-dimensional (for
consistency with the anomaly requiring $Sp$ gauge algebras to have an even
number of fundamental half-hypermultiplets) fundamental representation.
Furthermore, the rank of $G$ must be rank$(G) \leq 3)$ for consistency with
$r_G = 5$. This leaves precisely one option, which is indeed a consistent
quiver description satisfying all of the constraints,
\begin{equation}
  5\bm{F} - Sp(1) - SO(6) - Sp(1) - 5\bm{F} \,.
\end{equation}
This is the only possible maximal depth quiver which may be a marginal theory
for $(E_6, E_6)$ minimal conformal matter. 

\subsection[Consistent Quivers for \texorpdfstring{$(E_7, E_7)$}{(E7,E7)} Minimal Conformal
Matter]{Consistent Quivers for \boldmath{$(E_7, E_7)$} Minimal Conformal
Matter}\label{sec:E7E7qH}

In this subsection we will consider potential marginal quiver gauge theories
that flow to 6d minimal $(E_7, E_7)$ conformal matter at the UV fixed point.
Any such theory must satisfy
\begin{equation}
  (r_G, r_F) = (10, 15) \,,
\end{equation}
and the marginal CFD for this CFT is known to be \cite{Apruzzi:2019opn}
\begin{equation}\label{eqn:E7E7M}
  \includegraphics[scale=0.25]{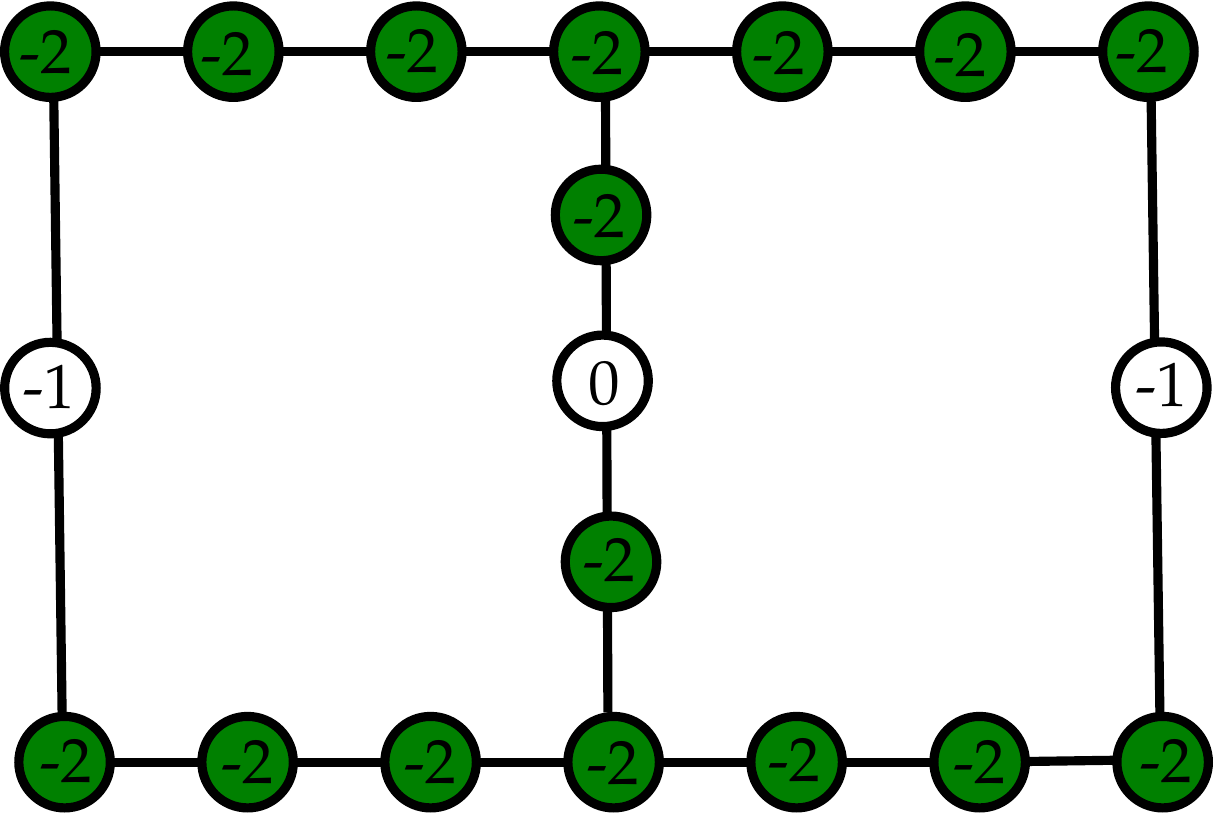} \,.
\end{equation}
All possible combinations of embedding the BG-CFDs into this marginal CFD are
shown in figure \ref{fig:E7E7embed}.
\begin{figure}
  \centering
  \includegraphics[scale=0.18]{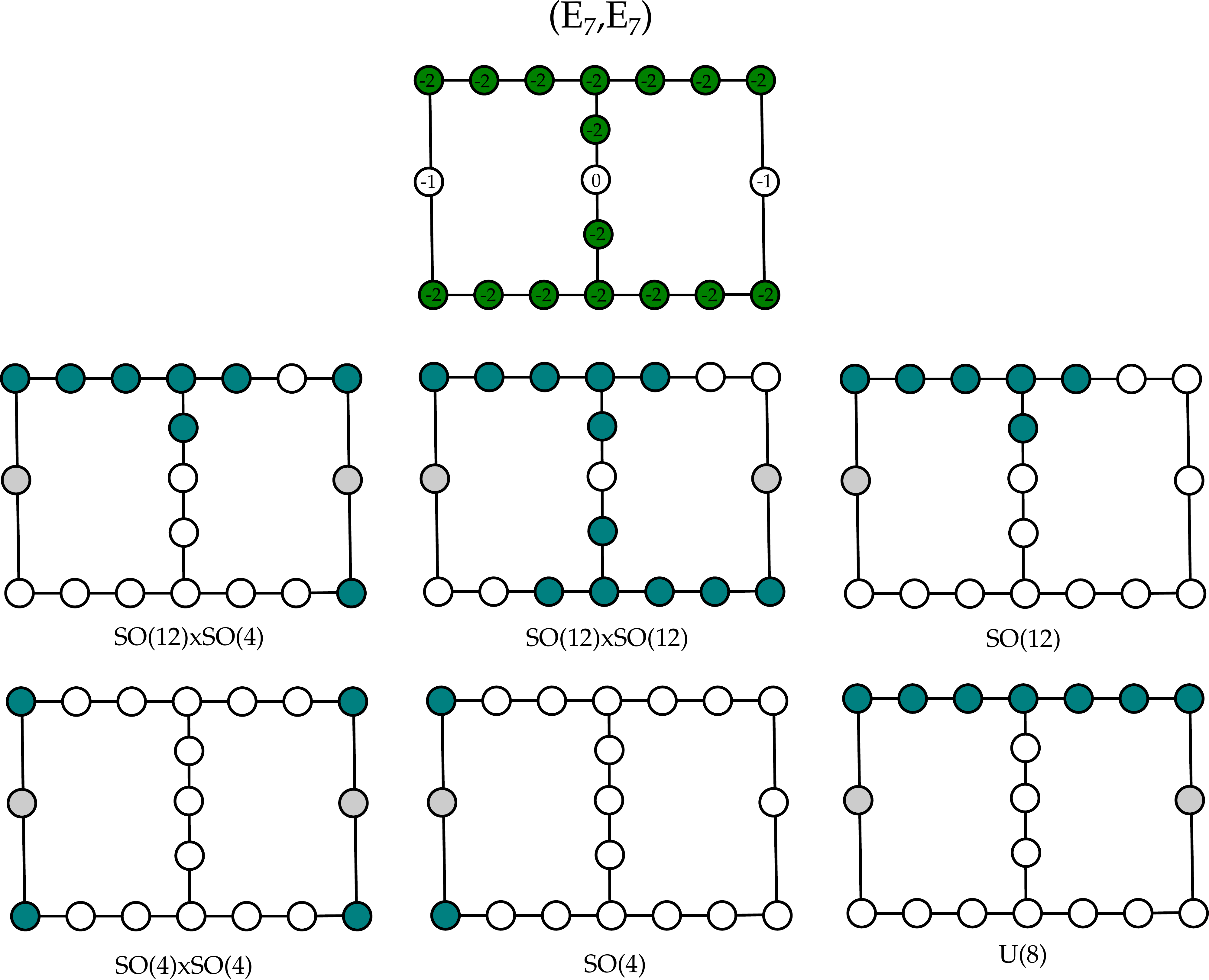}
  \caption{Embedding of potential BG-CFDs into the $(E_7, E_7)$ Marginal
CFD.}\label{fig:E7E7embed}
\end{figure}
The possible flavor nodes of the quivers can be completely classified by the
possible embeddings of the BG-CFDs into the marginal CFD, and the following
options for the flavor nodes are found:
\begin{equation}
  \begin{aligned}
    &8\bm{F} - SU(n) \, , & & \cr
    &m\bm{F} - Sp(n) \, , &\quad &m=2,6 \, , \cr
    &\bigg(m_1\bm{F} - Sp(n_1)\bigg) \oplus \bigg(m_2\bm{F} - Sp(n_2)\bigg) \, , &\quad &m_i \in \{2,6\} \, , \cr
    &2\times\bm{56} - E_7 \, , &\quad & \cr
    &\bigg(m\bm{F} - Sp(n_1)\bigg) \oplus \bigg(2\times\bm{56} - E_7\bigg) \, , &\quad &m = 2,6 \,.
  \end{aligned}
\end{equation}
There are many quivers which can be found with these as the flavor nodes,
which also satisfy the constraints imposed by $r_G$, $r_F$, and those
described earlier as the constraints on the number of hypermultiplets. The
enumeration of all such quivers is unenlightening, however, we can determine
some interesting potential quiver descriptions. One kind of description of
interest, as in the $(E_6, E_6)$ case just discussed, is a quiver with the
maximal number of matter hypermultiplets associated to flavor nodes. Here this
would be any quivers associated to
\begin{equation}
  \bigg(6\bm{F} - Sp(n_1)\bigg) \oplus \bigg(6\bm{F} - Sp(n_2)\bigg) \,.
\end{equation}
There are only two possibilities to connect these two subquivers together
while satisfying the constraint from $r_F$ and that is either to write
\begin{equation}\label{eqn:E7E7max1}
  6\bm{F} - Sp(n_1) - Sp(n_2) - 6\bm{F} \,,
\end{equation}
or
\begin{equation}\label{eqn:E7E7max2}
  6\bm{F} - Sp(n_1) - G - Sp(n_2) - 6\bm{F} \,,
\end{equation}
where $G$ has a real fundamental representation.
We consider the case of (\ref{eqn:E7E7max1}) first. The constraints on
the number of hypermultiplets require that
\begin{equation}
  \begin{aligned}
    &2n_2 + 6 \leq 2 n_1 + 6 \cr
    &2n_1 + 6 \leq 2 n_2 + 6 \,,
  \end{aligned}
\end{equation}
and, together with the gauge rank constraint that
\begin{equation}
  n_1 + n_2 = 10 \,,
\end{equation}
one finds that the only possible solution is
\begin{equation}
  6\bm{F} - Sp(5) - Sp(5) - 6\bm{F} \,.
\end{equation}

Now we turn to maximal depth quivers of the form (\ref{eqn:E7E7max2}). It is
immediate to see that $G$ cannot be either of $SO(2k+1)$ or $G_2$ as these
groups have odd-dimensional fundamental representations, which is incompatible
with the anomaly requirement that $Sp(n)$ gauge groups must come with an even
number of fundamental hypermultiplets. If we take $G = F_4$ then the
constraints on the number of hypermultiplets become
\begin{equation}
  \begin{aligned}
    &\# \bm{F} \text{ of } Sp(n_1): &\quad 6 + \frac{1}{2}\times 26 &\leq 2n_1 + 6 \cr
    &\# \bm{F} \text{ of } Sp(n_2): &\quad 6 + \frac{1}{2}\times 26 &\leq 2n_2 + 6 \cr
    &\# \bm{26} \text{ of } F_4: &\quad \frac{1}{2} \times (2n_1 + 2n_2) &\leq 3 \,,
  \end{aligned}
\end{equation}
which, together with the gauge rank constraint that $n_1 + n_2 = 6$, clearly
has no solutions. If we instead consider the case where $G = SO(2r)$ then the
set of constraints is
\begin{equation}
  \begin{aligned}
    &\# \bm{F} \text{ of } Sp(n_1): &\quad &6 + r \leq 2n_1 + 6 \cr
    &\# \bm{F} \text{ of } Sp(n_2): &\quad &6 + r \leq 2n_2 + 6 \cr
    &\# \bm{F} \text{ of } SO(2r): &\quad &n_1 + n_2 \leq 2r - 2 \cr
    &r_G = 10 : &\quad &n_1 + n_2 + r = 10 \,.
  \end{aligned}
\end{equation}
This system of equations has only two solutions
\begin{equation}
  (n_1, n_2, r) = (3,3,4) \quad \text{ or } \quad (2,4,4) \,.
\end{equation}
In short, we deduce that there are three potential marginal quiver gauge
theories for minimal $(E_7, E_7)$ conformal matter which have the maximum rank
of the flavor nodes. These are
\begin{equation}
  \begin{aligned}
    &6\bm{F} - Sp(5) - Sp(5) - 6\bm{F} \cr
    &6\bm{F} - Sp(3) - SO(8) - Sp(3) - 6\bm{F} \cr
    &6\bm{F} - Sp(2) - SO(8) - Sp(4) - 6\bm{F} \,.
  \end{aligned}
\end{equation}

\subsection[Consistent Quivers for \texorpdfstring{$(E_8, E_8)$}{(E8,E8)}
Minimal Conformal Matter]{Consistent Quivers for \boldmath{$(E_8, E_8)$} Minimal Conformal Matter}

Minimal $(E_8, E_8)$ conformal matter is a 6d SCFT with associated marginal
CFD \cite{Apruzzi:2019opn}
\begin{equation}
  \includegraphics[scale=0.25]{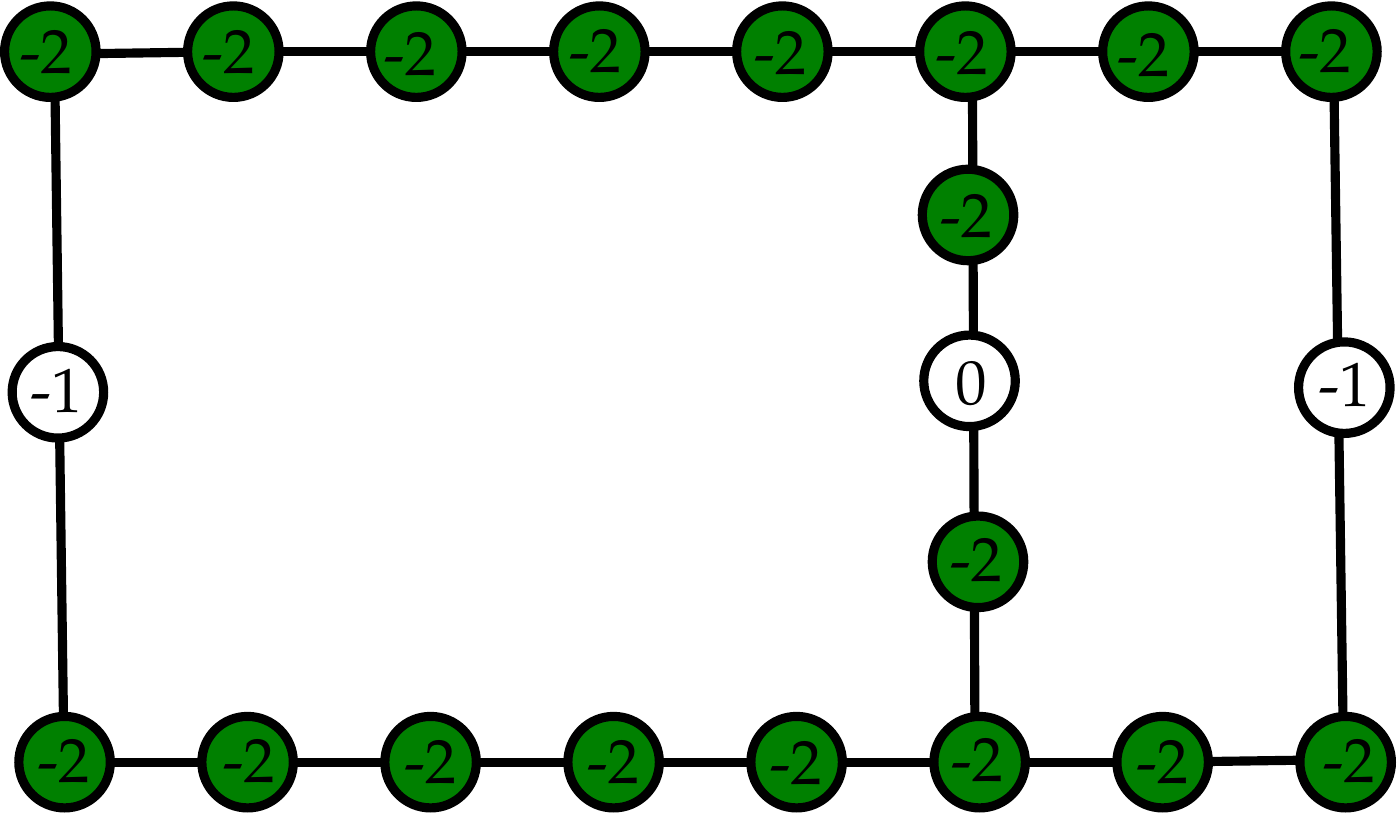} \,.
\end{equation}
Any 5d quiver gauge theory that realizes minimal $(E_8, E_8)$ conformal matter
at its UV fixed point must have the rank of the gauge algebra and the rank of
the classical flavor symmetry being
\begin{equation}
  (r_G, r_F) = (21, 17) \,.
\end{equation}
\begin{figure}
  \centering
  \includegraphics*[width=\textwidth]{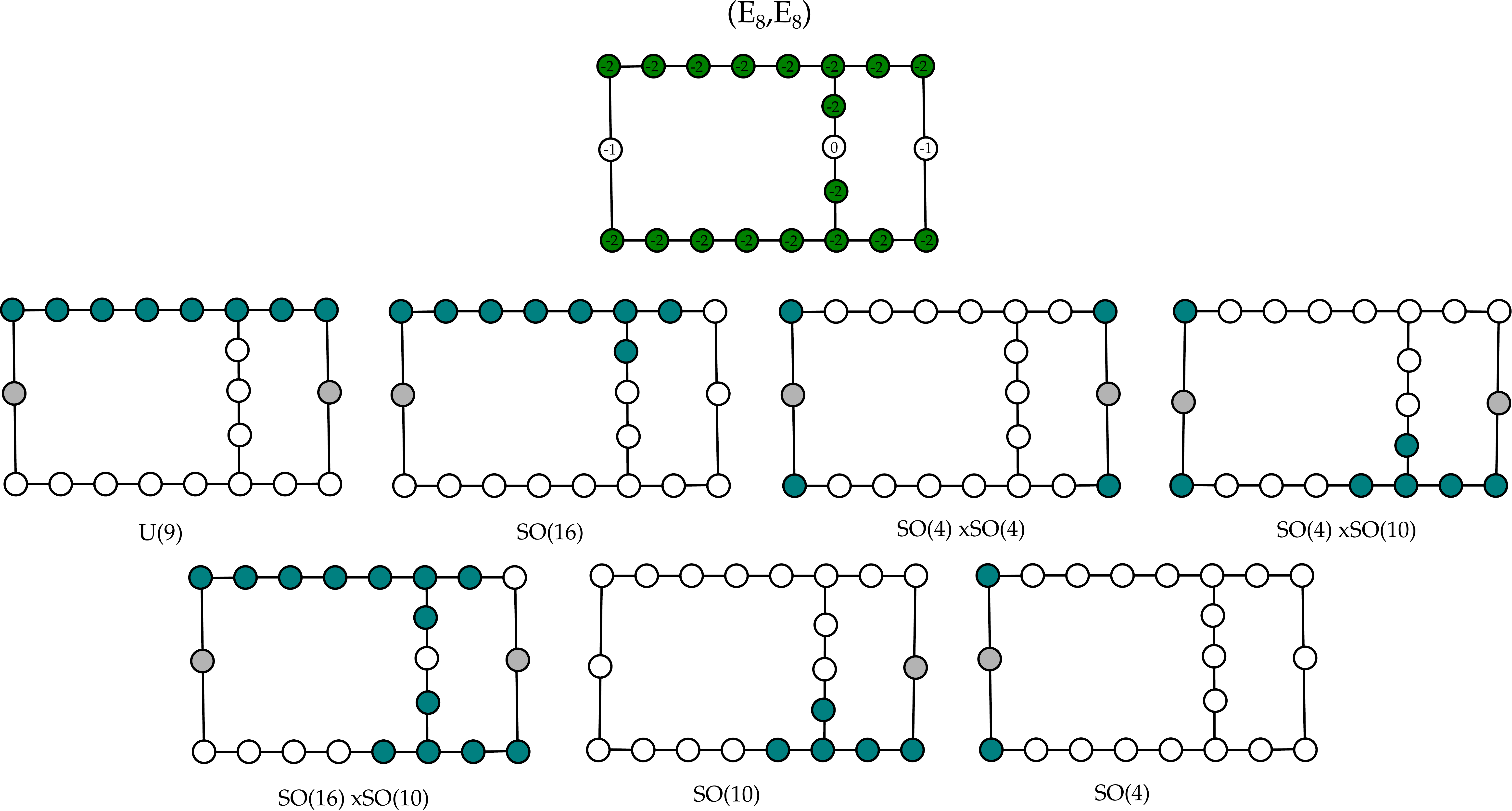}
  \caption{Embedding of potential BG-CFDs into the $(E_8, E_8)$ Marginal
CFD.}\label{fig:E8E8embed}
\end{figure}
All of the possible embeddings of the BG-CFDs into this marginal CFD are given
in figure \ref{fig:E8E8embed}, and from this one can determine that the
possible flavor nodes of any quiver that may be a marginal theory for minimal
$(E_8, E_8)$ conformal matter are the following
\begin{equation}
  \begin{aligned}
    &9\bm{F} - SU(n) \, , &\quad & \cr
    &m\bm{F} - Sp(n) \, ,&\quad &m=2,5,8 \, ,\cr
    &2 \times \bm{56} - E_7 \, , &\quad & \cr
    &\bigg(m_1\bm{F} - Sp(n_1)\bigg) \oplus \bigg(m_2\bm{F} - Sp(n_2)\bigg)\, , &\quad &(m_1, m_2) = (2,2), (2,5), (5,8) \, , \cr
    &\bigg(m\bm{F} - Sp(n_1)\bigg) \oplus \bigg(2\times\bm{56} - E_7\bigg) \, ,&\quad &m
    = 2,5 \, ,\cr
    &\bigg(2\times\bm{56} - E_7\bigg) \oplus \bigg(2\times\bm{56} - E_7\bigg) \, .&\quad
    & 
  \end{aligned}
\end{equation}
As in the other cases of $(E_n, E_n)$ conformal matter, we will be interested
in determining precise possible quivers for the theories which have the
deepest descendants; that is, the quivers for which a gauge theory description
exists farthest down the CFD-tree. In this case such a quiver is one with
flavor nodes
\begin{equation}
    \bigg(8\bm{F} - Sp(n_1)\bigg) \oplus \bigg(5\bm{F} - Sp(n_2)\bigg) \,.
\end{equation}
There are two possible options to connect this into a complete quiver
satisfying that $r_F = 17$. The first option is to connect the two $Sp(n)$
factors via a bifundamental hypermultiplet, and add an additional gauge node
connecting to one of the $Sp(n)$ in such a way that the bifundamental gluing
those two nodes is quaternionic, and thus does not provide any additional
classical flavor symmetry. A straightforward analysis from the constraints on
the number of hypermultiplets shows that such a quiver does not have an
interacting 5d SCFT fixed point. 

There are then three possible options if we consider only quivers without
loops, and where each gauge node is glued to another gauge node by only a
single bifundamental (half-)hypermultiplet. One of these is
\begin{equation}
  \begin{gathered}
    G_{(2)} \cr
    | \cr
    8\bm{F} - Sp(n_1) - G_{(1)} - Sp(n_2) - 5\bm{F}
  \end{gathered} \,.
\end{equation}
The other options involve attaching the group $G_{(2)}$ to one of the
$Sp(n_i)$ factors instead of $G_{(1)}$, however, for brevity, we shall not consider those
options here. $G_{(1)}$ is required to be a group with a real,
even-dimensional, fundamental
representation and $G_{(2)}$ one with a quaternionic fundamental
representation. It is straightforward to see that neither of these groups can
consistently be exceptional groups, and the only option is 
\begin{equation}
  (G_{(1)}, G_{(2)}) = (SO(2r_1), Sp(r_2)) \,.
\end{equation}
Such a quiver satisfies the constraints coming from $r_F = 17$ and the
remaining constraints are
\begin{equation}
  \begin{aligned}
    &\# \bm{F} \text{ of } Sp(n_1): &\quad &8 + r_1 \leq 2n_1 + 6 \, , \cr
    &\# \bm{F} \text{ of } Sp(n_2): &\quad &5 + r_1 \leq 2n_2 + 6 \, , \cr
    &\# \bm{F} \text{ of } SO(2r_1): &\quad &n_1 + n_2 + r_2 \leq 2r_1 - 2 \, , \cr
    &\# \bm{F} \text{ of } Sp(r_2): &\quad &r_1 \leq 2r_2 + 6 \, , \cr
    &r_G = 21 : &\quad & n_1 + n_2 + r_1 + r_2 = 21 \,.
  \end{aligned}
\end{equation}
There are many explicit quivers that exist as solutions of these constraints.
Therefore there are potential maximal depth quivers for minimal $(E_8, E_8)$
conformal matter of the form
\begin{equation}
  \begin{gathered}
    Sp(r_2) \cr
    | \cr
    8\bm{F} - Sp(n_1) - SO(2r_1) - Sp(n_2) - 5\bm{F}
  \end{gathered} \,,
\end{equation}
where
\begin{equation}
  \begin{aligned}
    (n_1, n_2, r_1, r_2) = \,\, &(5,4,8,4) \,, (5,5,8,3) \,, (5,7,8,1) \,,
    (5,6,8,2) \,, \cr
    \quad &(6,4,8,3) \,, (6,5,8,2) \,, (6,6,8,1) \,, (7,4,8,2) \,, \cr
    \quad &(7,5,8,1) \,, (8,4,8,1) \,, (6,4,9,2) \,.
  \end{aligned}
\end{equation}

\subsection[Consistent Quivers for \texorpdfstring{$(E_8, SU(2k+1))$}{(E8,
SU(2k+1))} Minimal Conformal Matter]{Consistent Quivers for \boldmath{$(E_8, SU(2k+1))$} Minimal Conformal Matter}

The marginal CFD for $(E_8, SU(2k+1))$ minimal conformal matter, which is a
theory with
\begin{equation}
  (r_G, r_F) = (2k^2+k+1, 2k+9) \,.
\end{equation}
was determined in Appendix D of \cite{Apruzzi:2019opn}. There it was found to
be
\begin{equation}
  \includegraphics[scale=0.25]{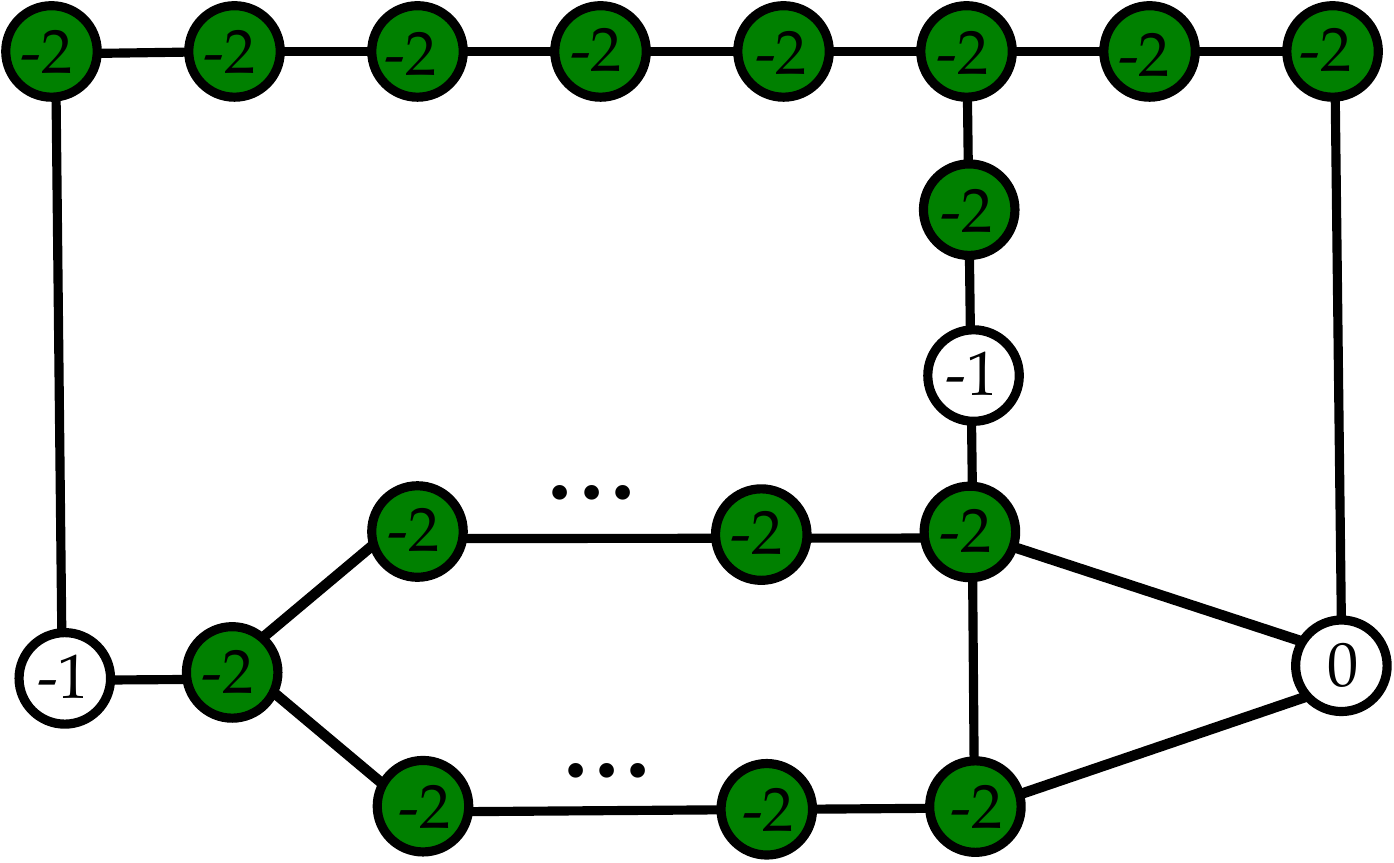} \,.
\end{equation}
By studying the embeddings of the BG-CFDs in this marginal CFD one can again
determine that the possible flavor nodes of any quiver are
\begin{equation}
  \begin{aligned}
    &m\bm{F} - SU(n) \, , &\quad &m = 8, k+2, k+3 \, , \cr
    &m\bm{AS} - SU(n) \, , &\quad &m = k+2, k+3 \,, \quad (\text{if } k = 1,2) \, , \cr
    &m\bm{27} - E_6 \, , &\quad &m = k+2, k+3 \,,\quad k \leq 2 \, ,\cr
    &m\bm{F} - Sp(n) \, ,&\quad &m = 2,3,4,8 \,, \quad (m=3 \text{ only if } k > 1) \, ,\cr
    &m\bm{56} - E_7 \, ,&\quad &m = 2,3 \,, \quad (m=3 \text{ only if } k > 1)\, ,\cr
    &\bigg(m_1\bm{F} - Sp(n_1)\bigg) \oplus \bigg(m_2\bm{F} - Sp(n_2)\bigg) \, ,&\quad &m_1 \in \{2,3\}
    \,,\, m_2 \in \{2,3,4\} \, ,\cr
    &\bigg(m\bm{F} - Sp(n_1)\bigg) \oplus \bigg(2\times\bm{56} - E_7\bigg) \, ,&\quad &m
    = 2,3,4 \, ,\cr
    &\bigg(2 \times \bm{56} - E_7\bigg) \oplus \bigg(2\times \bm{56} - E_7\bigg) \, .&\quad & 
  \end{aligned}
\end{equation}
In the latter three quivers the general values of $m_i$ given are only
potential options for sufficiently large $k$. We find that the following are
not allowed
\begin{equation}
  \begin{aligned}
    (m_1, m_2) \text{ or } (m_2, m_1) \neq &(3,4)\,, (2,2) &\quad &\text{ when } k \leq 1
   \, , \cr
     &(2,3) &\quad &\text{ when } k \leq 2 \, , \cr
     &(3,3) &\quad &\text{ when } k \leq 3 \,.
  \end{aligned}
\end{equation}
Let us now write down some explicit potential quivers for the marginal theory
associated to $(E_8, SU(2k+1))$ minimal conformal matter for some particular
small values of $k$. We will take $k = 1$, which was discussed from the
geometric point of view in \cite{Apruzzi:2019opn}. In fact 6d theory has a
further enhancement of the superconformal flavor symmetry, to $E_8 \times
G_2$, and thus to determine the full superconformal flavor symmetry of the
descendants it was useful to introduce a different marginal CFD that captured
this enhancement. Of course, by studying the $(E_8, SU(3))$ marginal CFD and
its descendants one can still determine a non-trivial enhancement of the
superconformal flavor symmetry directly from the CFD, and when computing the
BPS spectrum, as in \cite{Apruzzi:2019opn}, one observes that the states
organize into representations of the larger flavor symmetry group. This is to
say, a potential quiver derived from the $(E_8, SU(3))$ marginal CFD is a
necessary condition for the quiver to describe the marginal theory, regardless
of the further symmetry enhancement. 

We are considering a theory with 
\begin{equation}
  (r_G, r_F) = (4, 11) \,.
\end{equation}
We now attempt to determine a quiver with the maximal depth of descendants.
After little reflection one can see that any quiver with a flavor node charged under
an $SU(n \geq 3)$ gauge group cannot be consistent with the ranks and the
constraints on the number of hypermultiplets. The maximal quiver would then
involve an $Sp(n)$ gauge node with $8\bm{F}$, and it is straightforward to see
that there is only one such possible quiver, being
\begin{equation}
  8\bm{F} - Sp(3) - Sp(1)_\theta \,.
\end{equation}
Furthermore, if one is interested in quivers corresponding to the flavor nodes
\begin{equation}
  \bigg(m_1\bm{F} - Sp(n_1)\bigg) \oplus \bigg(m_2\bm{F} - Sp(n_2)\bigg) \,,
\end{equation}
then one can determine that the only options are when $m_1 = 4$ and $m_2 = 2$,
for which there are precisely nine different complete quivers satisfying all
of the consistency requirements.

\subsection[Consistent Quivers for \texorpdfstring{$(E_8, SU(2k))$}{(E8,
SU(2k))} Minimal Conformal Matter]{Consistent Quivers for \boldmath{$(E_8, SU(2k))$} Minimal Conformal Matter}

The marginal CFD for minimal $(E_8, SU(2k))$ conformal matter is
\cite{Apruzzi:2019opn}
\begin{equation}
  \includegraphics[scale=0.25]{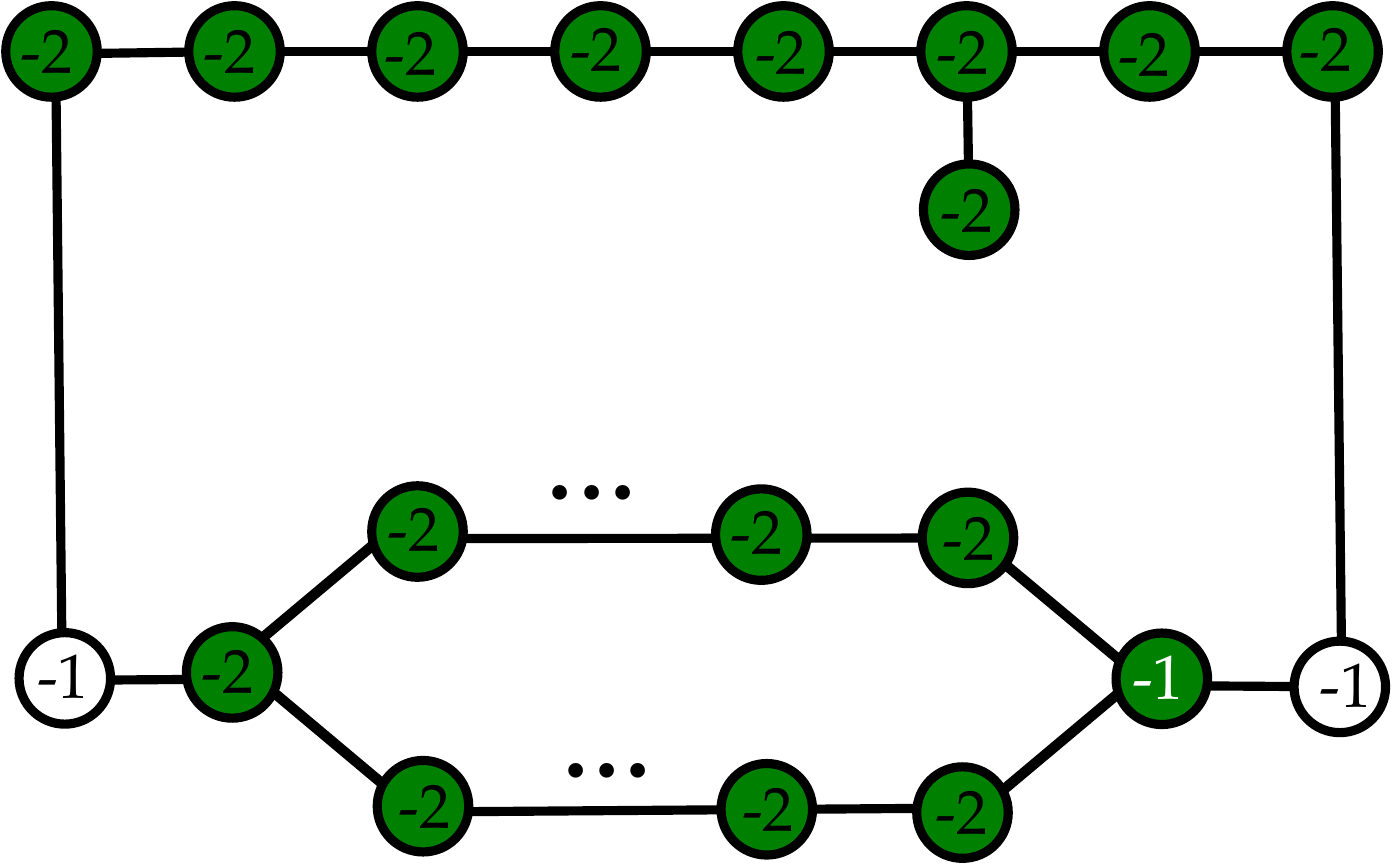} \,.
\end{equation}
Any 5d quiver gauge theory description that flows in the UV to this 6d SCFT
must have
\begin{equation}
  (r_G, r_F) = (2k^2-k+1, 2k+8) \,.
\end{equation}
By determining all of the possible embeddings of BG-CFDs into the marginal CFD
one finds the following set of possible flavor nodes for 5d quiver
descriptions of this 6d SCFT
\begin{equation}\label{eqn:E8SU2kqH}
  \begin{aligned}
    &9\bm{F} - SU(n)  \, , & & \cr
    &m\bm{F} - Sp(n)  \, , &\quad &m =  2,3,5,8 \, , \cr
    &m\bm{56} - E_7 \, , &\quad &m =  2,3 \, , \cr
    &m\bm{F} - Sp(n) - 1\bm{AS} \, , &\quad &m =  2,3,8 \, , \cr
    &\bigg(m\bm{F} - Sp(n)\bigg) \oplus \bigg(1\bm{AS} - Sp(r)\bigg) \, , &\quad &m =  2,3,8 \, , \cr
    &\bigg(m\bm{F} - Sp(n)\bigg) \oplus \bigg(1\bm{V} - SO(r)\bigg) \, , &\quad &m = 2,3,8 \, , \cr
    &\bigg(m\bm{F} - Sp(n)\bigg) \oplus \bigg(1\times \bm{7} - G_2\bigg) \, , &\quad &m = 2,3,8 \, , \cr
    &\bigg(m\bm{F} - Sp(n)\bigg) \oplus \bigg(1\times \bm{26} - F_4\bigg) \, , &\quad &m = 2,3,8 \, , \cr 
    &\bigg(2\times \bm{56} - E_7\bigg) \oplus \bigg(1\bm{AS} - Sp(r)\bigg) \, , &\quad & \cr
    &\bigg(2\times \bm{56} - E_7\bigg) \oplus \bigg(1\bm{V} - SO(r)\bigg) \, , &\quad & \cr
    &\bigg(2\times \bm{56} - E_7\bigg) \oplus \bigg(1\times \bm{7} - G_2\bigg) \, , &\quad & \cr
    &\bigg(2\times \bm{56} - E_7\bigg) \oplus \bigg(1\times \bm{26} - F_4\bigg) \, , &\quad & \cr
    &k\bm{AS} - Sp(r) \, , & & \cr
    &k\bm{V} - SO(r) \, , & & \cr
    &k\bm{7} - G_2 \, , & & \cr
    &k\bm{26} - F_4 \, , & & \cr
    &\bigg(5\bm{F} - Sp(n_1)\bigg) \oplus \bigg(m\bm{F} - Sp(n_2)\bigg) \, , &\quad &m = 2,3 \, , \cr
    &\bigg(5\bm{F} - Sp(n_1)\bigg) \oplus \bigg(2\times\bm{56} - E_7\bigg) \, . &\quad &  \cr
  \end{aligned}
\end{equation}
Furthermore, quivers with classical flavor symmetry 
\begin{equation}
  SO(10) \times SO(6) \quad \text{
  or } \quad SO(6) \quad \text{ or } \quad  SO(4) \times Sp(1) \,,
\end{equation}  
rotating the hypermultiplets attached to the flavor nodes requires one to have
\begin{equation}
  k > 1 \,,
\end{equation}
and similarly we find that 
\begin{equation}
  SO(6) \times Sp(1) \quad \text{ requires } \quad k > 2 \,.
\end{equation}

When $k=1$ this 6d $\mathcal{N} = (1,0)$ SCFT is the rank two E-string theory,
for which 5d quiver gauge theory descriptions are known. We can now reproduce
this set of descriptions from our approach, and thus show that all
possibilities are realized. It is immediate on checking all options in
(\ref{eqn:E8SU2kqH}) that the only potential quiver descriptions, consistent
with the gauge and flavor ranks, together with the constraints on the number
of hypermultiplets are
\begin{equation}
  \begin{aligned}
    &9\bm{F} - SU(3) \cr 
    &8\bm{F} - Sp(2) - 1\bm{AS} \cr 
    &5\bm{F} - Sp(1) - Sp(1) - 2\bm{F} \,.
  \end{aligned}
\end{equation}
These are exactly the set of known quiver gauge theory descriptions that have
the rank two E-string as their UV fixed point.

We stress that while all potential quivers are realized for ranks one and two
6d SCFTs, we do not expect this to generalize to higher ranks --- it is
essentially an accident that the low rank combined with the restrictions on
the flavor nodes is exceptionally constraining. As we have seen, for higher
ranks in general there are many more potential quivers than there are known
gauge theory descriptions. It remains to determine which of these quivers are,
in fact, realized, however we do not expect that a pure CFD approach is
capable of answering this question. The CFD is, by definition, defined in
terms of a reducible surface, and the details of the ``pure gauge'' part of
any quiver description is contained precisely inside of the details of how
that reducible surface is glued together from irreducible surfaces. However,
at all ranks we can see that the embedding of the BG-CFDs inside of the
marginal CFD is extremely constraining on what possible flavor nodes can
appear in any quiver description.

\section{Bootstrapping CFDs}

\subsection{Constraining Marginal CFDs of Single Gauge Node Theories}
\label{sec:Boot}

In this section, we demonstrate the power of BG-CFDs as an alternative approach to constrain and ``derive'' the actual CFDs. In certain instances, there is no known geometric realization of the marginal theory in 6d F-theory language. In such instances, we can nevertheless `bootstrap' the marginal CFD using consistency requirements with known properties. 

\begin{enumerate}
\item The marginal CFD has a marked subgraph, which is given by the Dynkin diagram of the 6d superconformal flavor symmetry (generically these will be affine Dynkin diagrams). 
\item The rules for constructing CFD-descendants need to work in parallel with the mass deformations. Along with the known superconformal flavor symmetries, this condition will largely fix the location of $(-1)$-vertices in the marginal CFD.
 \item The classical flavor symmetry determines a set of BG-CFDs, which have to be embeddable into the CFDs. If there are multiple non-Abelian factors, the corresponding BG-CFDs cannot intersect each other. This rule also applies to the marginal CFD. 
 \item Applying any mass deformation (i.e. CFD-transition to any $(-1)$-vertex) to the marginal CFD has to result in a CFD, whose marked sub-graph is a Dynkin diagram (without any affine marked subgraphs).
\end{enumerate}
These conditions do not necessarily constrain the CFDs entirely and uniquely, but they give at worst a subgraph that encodes a subset of mass-deformations. Nonetheless, we will try to reconstruct the conjectural marginal CFD in this way. 

In this section, we will focus on the subclass of marginal theories which have a description in terms of a gauge theory with a simple gauge algebra $SU(N)$, $(N\geq 4)$, which has the matter contents of $2\bm{AS}+8\bm{F}$ and the following 6d tensor branch  \cite{Jefferson:2017ahm}:
\begin{equation}
\label{single-node-tensor-branch}
 \ba
  N \text{ odd:}\quad   & [SO(16)]-\overset{\mathfrak{sp}(1)}{1}-\overset{\mathfrak{su}(2)}{2}-\dots-\overset{\mathfrak{su}(2)}{2}-[SU(2)]\,,\cr 
  N\geq 6 \text{ even:}\quad & [E_7]-1-\underset{[SU(2)]}{\overset{\mathfrak{su}(2)}{2}}-\overset{\mathfrak{su}(2)}{2}-\dots-\overset{\mathfrak{su}(2)}{2}-[SU(2)]
 \,, \cr 
  N=4: \quad & [E_7]-1-\overset{\mathfrak{su}(2)}{2}-[SO(7)]\,.\cr
\ea
\end{equation}

The 6d flavor symmetries are \cite{Mekareeya:2017jgc, Jefferson:2017ahm}
\begin{equation}
\ba
  N\geq 5 \text{ odd:}\quad & SO(16) \times SU(2)^2\,,\cr 
  N\geq 6 \text{ even:} \quad & E_7 \times SU(2)^3\,,
  \cr 
  N=4: \quad & E_7\times SO(7)\,.\cr
\ea
\end{equation}
For $N\geq 5$, there is an extra $SU(2)$ flavor symmetry at the 6d superconformal point comparing to the tensor branches in (\ref{single-node-tensor-branch}). This is related to the unique linear combination of the baryonic $SU(2)$s, which remains non-anomalous. 

The $N=4$ case corresponds to the $(E_7,SO(7))$ conformal matter theory. For $N>4$, they do not have a known singular Weiertrass model in the 6d F-theory description, and we will apply the bootstrap methodology to get a conjectural marginal CFD. Finally, we see that the resulting marginal CFD has in general more descendants than those realized by known gauge theory descriptions, which are indicators for dual gauge/quiver descriptions. Such quiver gauge theory descriptions and dualities are extensively discussed in section \ref{sec:descquiv}.


\subsection[\texorpdfstring{$SU(2n)_0 + 2\bm{AS} + 8\bm{F}$}{SU(2n)0 + 2AS + 8F}]{\boldmath{$SU(2n)_0+ 2AS + 8{ F}$}}
\label{sec:SU2n2AS8F}

\begin{figure}
\centering
\includegraphics*[height=0.9\textheight]{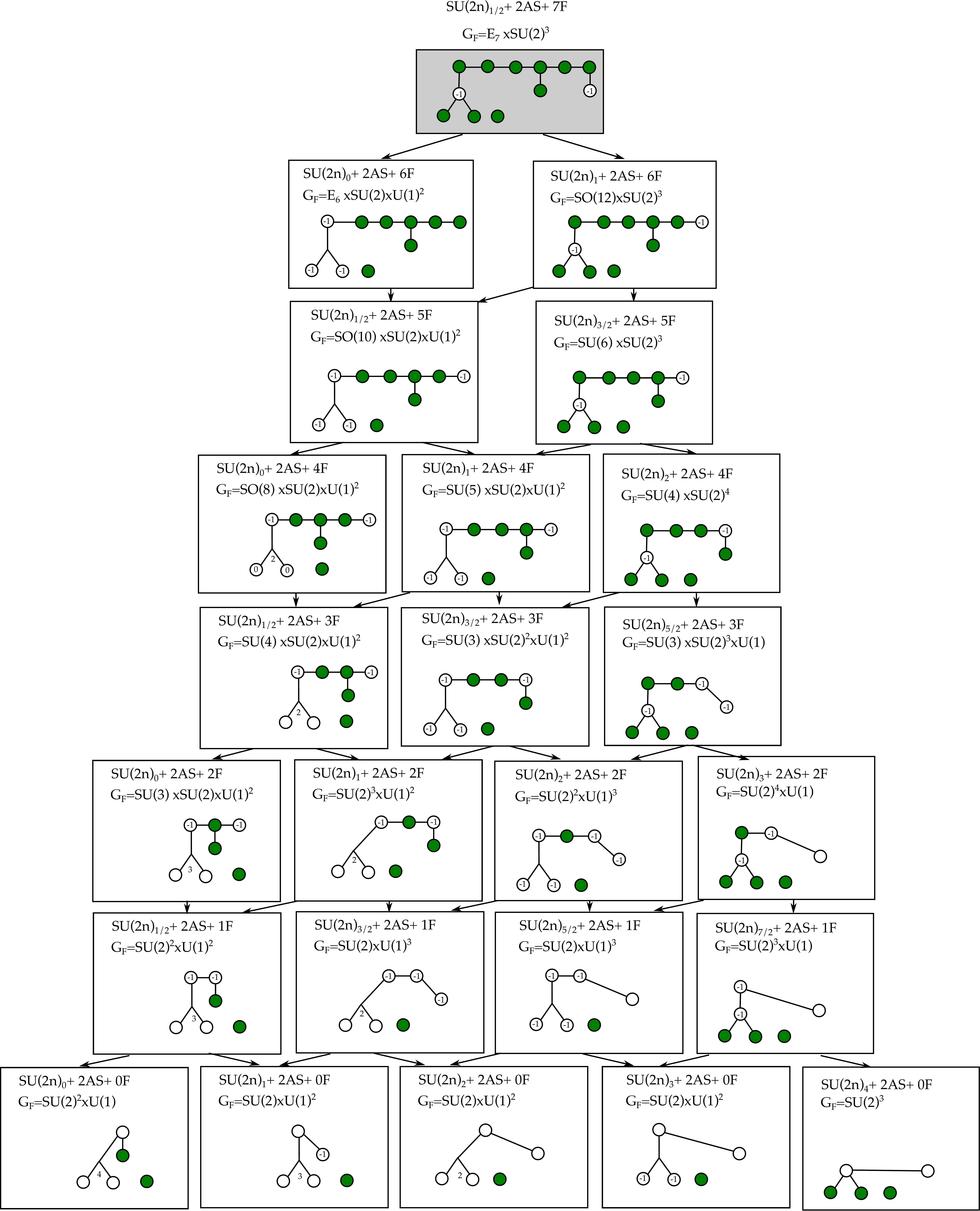}
\caption{CFD-tree consistent with the decoupling of fundamental flavors of $SU(2n)_0+ 2\bm{AS} + 8\bm{F}$. The gauge theory descriptions and superconformal flavor symmetry $G_\text{F}$ are labeled.\label{fig:SUn2ASLessF}}
\end{figure}

Consider the marginal theory 
\be
SU(2n)_0+ 2\bm{AS} + 8\bm{F} \,,\qquad n>2\,. 
\ee
The classical flavor symmetry is $U(2)\times U(8)$, and the associated BG-CFD needs to embed into the marginal CFD. 

In the following it will be useful to recall some of the known flavor symmetry enhancement for 5d  $SU(N)$ gauge theories with $N_A \bm{AS} + N_f \bm{F}$ matter fields at the their UV fixed points. For $N_A= 2$ and $N_f \leq 8$, the flavor symmetry enhancements were determined in the appendix of \cite{Zafrir:2015rga}. For the cases of $N_A=1$, the UV flavor symmetry enhancements are implicitly given in \cite{Yonekura:2015ksa}, and we summarize them here explicitly in appendix \ref{app:FlavorEnha}. Finally, for the cases of $N_A=0$, the superconformal flavor symmetries were determined in \cite{Apruzzi:2019vpe, Cabrera:2018jxt}.

We will first constrain the CFDs by fitting the flavor symmetries of the descendant CFDs with the known ones after decoupling the fundamental flavors. Comparing with the flavor symmetry enhancements predicted in \cite{Zafrir:2015rga}, we find that the following graph should be a subgraph of the actual marginal CFD:
\be\label{2AS7F}
\includegraphics*[width=5cm]{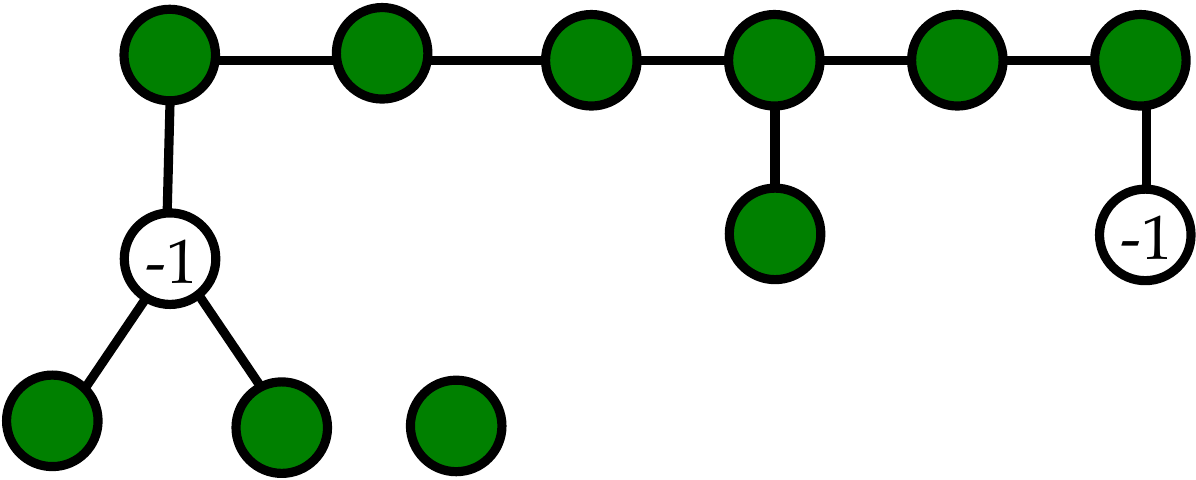}\,.
\ee
The CFD tree generated by decoupling fundamental flavors is shown in figure~\ref{fig:SUn2ASLessF}.

To see that this is the only way the $SU(2)$ nodes can attach, consider the first descendant. From the enhancements of the flavor symmetry \cite{Zafrir:2015rga} there are two enhancement patterns: 
\be
\ba
SU(2n)_{0} + 2 \bm{AS} + 6 \bm{F} :\qquad & G_F = E_6 \times SU(2) \times U(1)^2 \cr 
SU(2n)_{1} + 2 \bm{AS} + 6 \bm{F} :\qquad & G_F = SO(12) \times SU(2)^3  \,.
\ea
\ee
To get the first, it is clear that two $SU(2)$ nodes have to attached with a single $(-1)$ vertex to the longer tail of $E_7$. 
From the latter it is clear that another $(-1)$ vertex has to be attached to the short tail of the $E_7$ diagram. 
Note that decoupling only the fundamental matter retains an $SU(2)$ factor in the flavor symmetry. We will see that additional constraints on this node follow from the decoupling of the anti-symmetric matter. The subtree that corresponds to the models with a fixed number of 2$\bm{AS}$ and decoupling $\bm{F}$ is consistent indeed with all the known flavor symmetry enhancements.

On the other hand, if we decouple the anti-symmetric hypermultiplets, we obtain the following gauge  theories: 
\be 
SU(2n)_{0} + 2 \bm{AS} + 8 \bm{F} \quad \rightarrow \quad 
SU(2n)_{\pm (n-2)} + 1 \bm{AS} + 8 \bm{F} \quad \rightarrow \quad 
\left\{ 
\ba
&SU(2n)_{2n-4} + 0 \bm{AS} + 8 \bm{F} \cr 
& SU(2n)_{0} + 0 \bm{AS} + 8 \bm{F} \cr 
&SU(2n)_{-2n+4} + 0 \bm{AS} + 8 \bm{F} \cr 
\ea\right.
\ee
Further decoupling a fundamental flavor, we should get:
\be
\ba
SU(2n)_{2n-4}  + 8 \bm{F} \quad \rightarrow\quad  &SU(2n)_{2n-4\pm1/2}  + 7 \bm{F}\cr 
 SU(2n)_{0}  + 8 \bm{F}  \quad \rightarrow \quad & SU(2n)_{\pm 1/2}  + 7 \bm{F}\cr 
SU(2n)_{-2n+4}  + 8 \bm{F} \quad \rightarrow\quad  &SU(2n)_{-2n+4\pm 1/2}  + 7 \bm{F} \,.\cr 
\ea
\ee
According to \cite{Cabrera:2018jxt, Apruzzi:2019vpe}, the flavor symmetry enhancements for the theories without $\bm{AS}$ are 
\be
\ba
SU(2n)_{\pm(2n-4)}  + 8 \bm{F}: \qquad &G_\text{F}= U(8) \times SU(2) \cr 
SU(2n)_{0}  + 8 \bm{F}: \qquad &G_\text{F}= U(8) \times U(1) \cr  
SU(2n)_{\pm\left(2n-4+\frac{1}{2}\right)}  + 7 \bm{F}: \qquad &G_\text{F}= U(7) \times SU(2) \cr 
SU(2n)_{\pm\left(2n-4-\frac{1}{2}\right)}  + 7 \bm{F}: \qquad &G_\text{F}= U(7) \times U(1) \cr 
SU(2n)_{\pm \frac{1}{2}}  + 7 \bm{F}: \qquad &G_\text{F}= U(7) \times U(1) 
\ea
\ee
With 1$\bm{AS}$, the superconformal flavor symmetries are \cite{Yonekura:2015ksa}
\be
\ba
SU(2n)_{\pm(n-2)}  + 1 \bm{AS} + 8 \bm{F}: \qquad &G_\text{F}= U(8) \times SU(2)\times U(1)  \cr 
SU(2n)_{0} + 1 \bm{AS}  + 8 \bm{F}: \qquad &G_\text{F}= U(8) \times U(1) \times U(1)   \cr  
SU(2n)_{\pm\left(n-2+\frac{1}{2}\right)} + 1 \bm{AS}  + 7 \bm{F}: \qquad &G_\text{F}= U(7) \times SU(2) \times U(1)  \cr 
SU(2n)_{\pm\left(n-2-\frac{1}{2}\right)}+ 1 \bm{AS}  + 7 \bm{F}: \qquad &G_\text{F}= U(7) \times U(1) \times U(1)  \cr 
SU(2n)_{\pm \frac{1}{2}} + 1 \bm{AS} + 7 \bm{F}: \qquad &G_\text{F}= U(7) \times U(1) \times U(1)  
\ea
\ee

\begin{figure}
\centering
\includegraphics*[width=\textwidth]{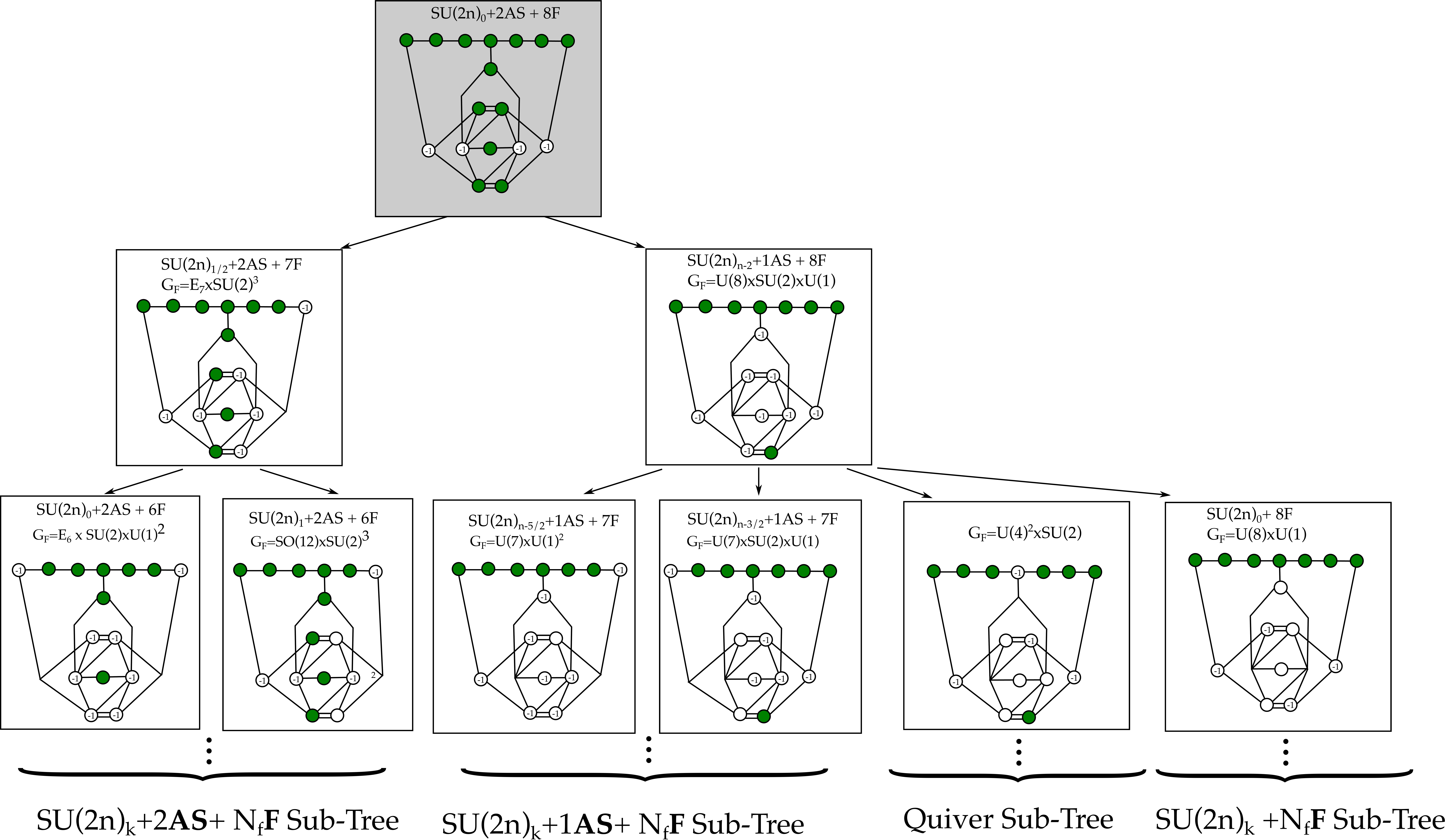}
\caption{CFD-tree consistent with the decoupling of antisymmetric flavors of $SU(2n)_0+ 2\bm{AS} + 8\bm{F}$. On the LHS, the decoupling of fundamental flavors results in the sub-tree shown in figure \ref{fig:SUn2ASLessF}. 
The mass deformations removing the anti-symmetric representations is consistent with the enhancement of flavor symmetries. The models with $1\bm{AS}$ are consistent with the gauge theory description in tables \ref{tab:GFsu2n1AS2} and \ref{tab:GFsu2n1AS3}.
\label{fig:modifiedSU2n2AS7F}}
\end{figure}

Besides the consistency requirements with the known flavor symmetry enhancements, the other constraints for this marginal CFD are:
\begin{itemize}
\item Embedding of the BG-CFDs for the classical global symmetry $U(8)\times U(2)\times U(1)_T$.
\item The marginal CFD contains the Dynkin diagrams for $\widehat{E}_7 \times \widehat{SU(2)}^2 \times SU(2)$ that was observed in \cite{Jefferson:2017ahm}. 
\item Mass deforming the marginal CFD (i.e., CFD-transitioning on any of the $-1$-vertices) results in marked subdiagram that is a collection of non-affine Dynkin diagrams. 
\end{itemize}
Finally, the resulting marginal CFD is highly constrained to be the following
\be\label{Bambi}
\includegraphics*[width=5.0cm]{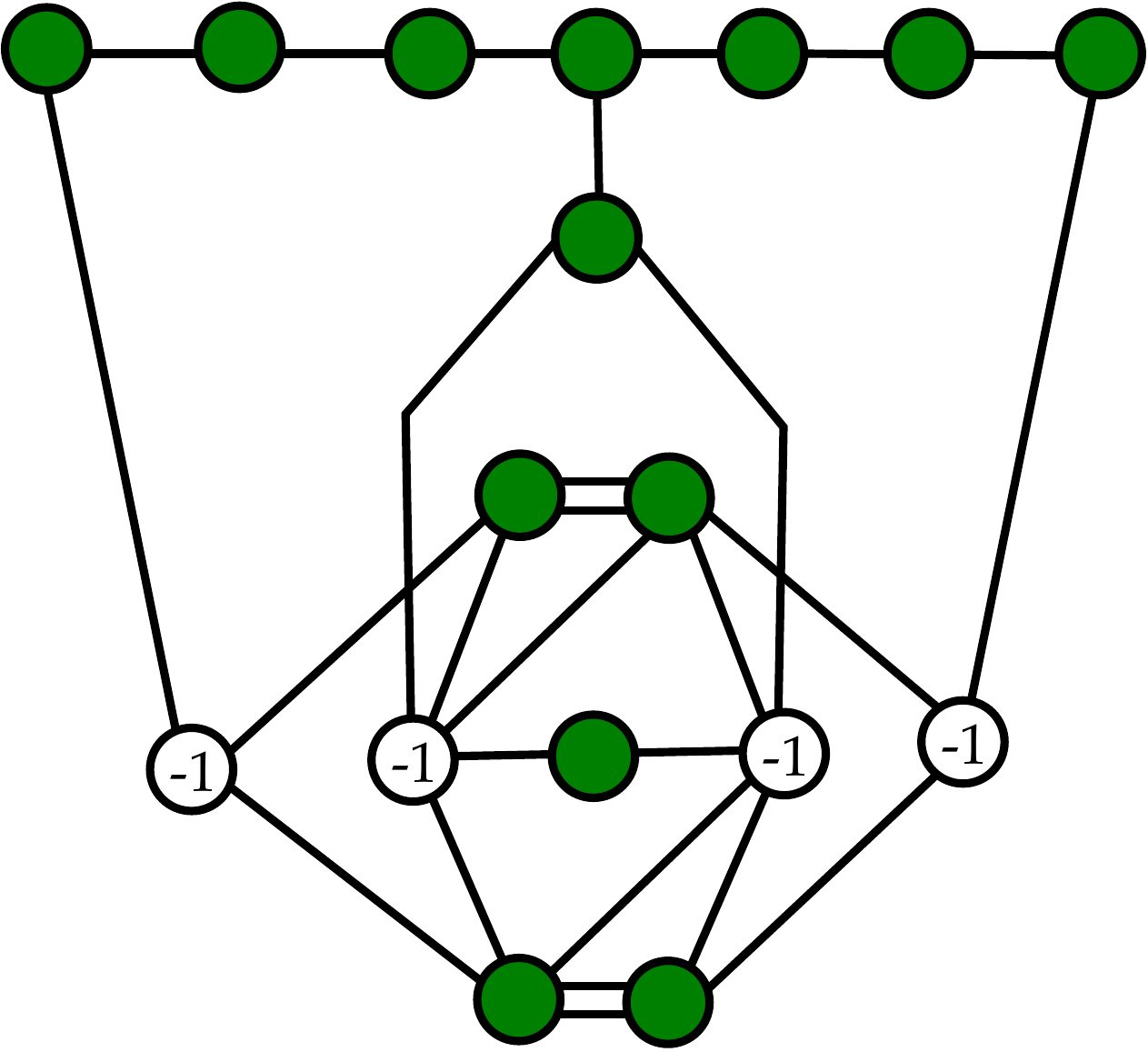}
\ee

There are four $(-1)$-vertices in this graph. From left to the right, removing these vertices will correspond to the following four different ways of decoupling matter fields:
\begin{enumerate}
\item{Decoupling $1\bm{F}$, shifting the CS level $k$ by $\frac{1}{2}$.}
\item{Decoupling $1\bm{AS}$, shifting the CS level $k$ by $(n-2)$.}
\item{Decoupling $1\bm{AS}$, shifting the CS level $k$ by $-(n-2)$.}
\item{Decoupling $1\bm{F}$, shifting the CS level $k$ by $-\frac{1}{2}$.}
\end{enumerate}

We list all the possible maximal embedding of BG-CFDs into this marginal CFD in figure~\ref{fig:FavoriteBGCFDS}. Together with the classical flavor symmetry shown in table \ref{tab:BGCFDs}, this put constraints on the possible quiver gauge theory descriptions. We will return to this and the resulting 
dualities in section \ref{sec:descquiv}. 

\begin{figure}
\centering
\includegraphics*[width=\textwidth]{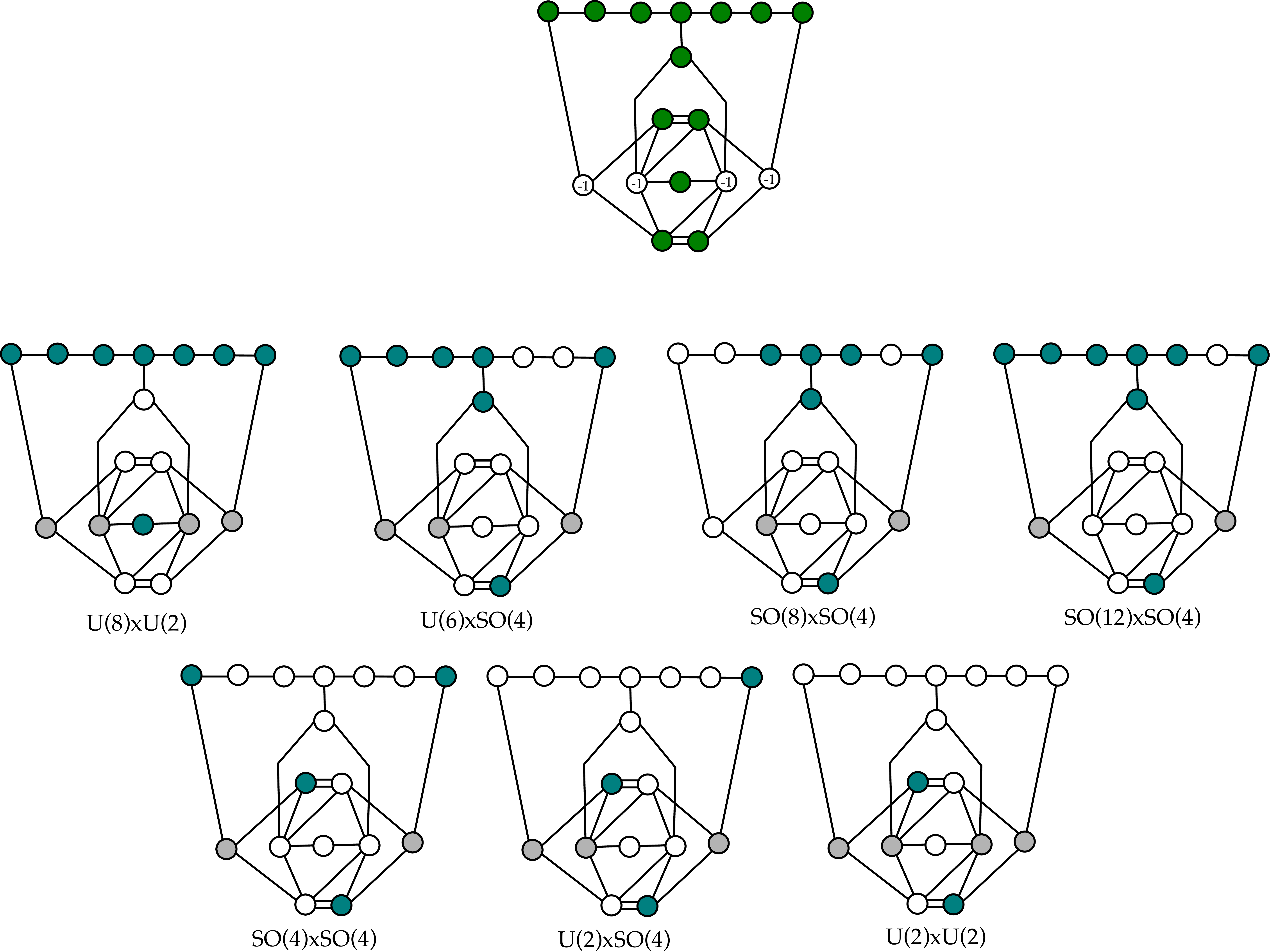}
\caption{At the top we show the CFD for the marginal theory $SU(2n) + 2 \bm{AS}+ 8\bm{F}$, and below, using the notation of section \ref{sec:gauge_descriptions_from_CFDs} the possible embeddings of the BG-CFDs into the marginal CFD. Below the diagrams we note the classical flavor symmetry of the corresponding weakly coupled description.  
\label{fig:FavoriteBGCFDS}}
\end{figure}

\begin{figure}
\centering
\includegraphics*[height=0.9\textheight]{CFD-BG-CFDs-SU2n-1AS-8F-SubTree.pdf}
\caption{Sub-trees of the CFD-tree figure \ref{fig:modifiedSU2n2AS7F}:  
$SU(2n)_0+ 1\bm{AS} + N_f \bm{F}$, $N_f\leq 7$ decoupling the fundamental flavors results in the sub-tree shown in this figure. The enhancements are consistent with the ones.  
\label{fig:modifiedSU2n1AS7F}}
\end{figure}

From this proposed marginal CFD, we obtain the tree of descendants, part of which is shown in figure \ref{fig:modifiedSU2n2AS7F}. The sub-tree with gauge theory descriptions $SU(2n)_k+1\bm{AS}+m\bm{F}$ is shown in figure \ref{fig:modifiedSU2n1AS7F}. Besides the $SU(2n)$ gauge theory descriptions, in figure \ref{fig:modifiedSU2n2AS7F} there is already a descendant theory with $G_\text{F}=U(4)^2\times SU(2)$ with a different quiver gauge theory description.

\subsection[\texorpdfstring{$SU(2n+1)_0 + 2\bm{AS} + 8\bm{F}$}{SU(2n+1)0 + 2AS + 8F}]{\boldmath{$SU(2n+1)_0+ 2{AS} + 8{F}$}}
\label{sec:SU2nplus12AS8F}

We can apply similar logic to the case with $SU(N=2n+1)_0$ gauge group and $2\bm{AS} + 8 \bm{F}$.
Note that in this case the marginal theory in 5d has flavor symmetry \cite{Jefferson:2017ahm}
\be
G_\text{F}^{\text{marginal}}=\widehat{SO(16)}\times \widehat{SU(2)} \times SU(2)\,.
\ee
Using the known UV flavor symmetry enhancements as before, we conjecture the marginal CFD for this class of models to be 
\be
\includegraphics*[width=6cm]{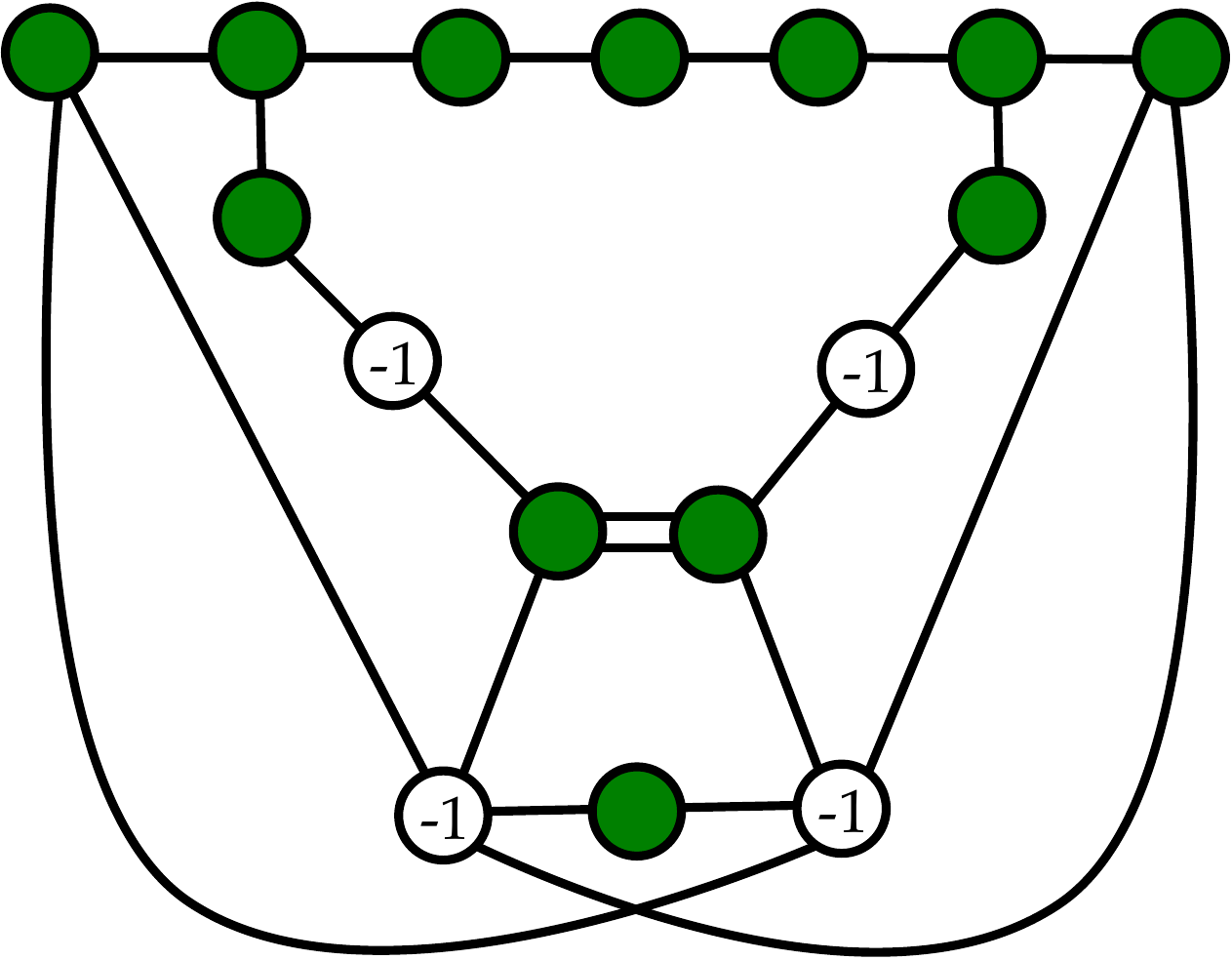} \,.
\ee
It contains the BG-CFDs for the classical flavor symmetries $U(8)$ and  $U(2)$, and furthermore the marked vertices in the CFD realize $G_\text{F}^{\text{marginal}}$. There are four $(-1)$-vertices in this graph. From upper left to the bottom right, removing these vertices will correspond to:
\begin{enumerate}
\item{Decoupling $1\bm{F}$, shifting the CS level $k$ by $\frac{1}{2}$.}
\item{Decoupling $1\bm{F}$, shifting the CS level $k$ by $-\frac{1}{2}$.}
\item{Decoupling $1\bm{AS}$, shifting the CS level $k$ by $(n-\frac{3}{2})$.}
\item{Decoupling $1\bm{AS}$, shifting the CS level $k$ by $-(n-\frac{3}{2})$.}
\end{enumerate}

The descendants are in agreement with the $2\bm{AS} + N_f \bm{F}$ flavor symmetry enhancements in \cite{Zafrir:2015rga} and the ones with one $\bm{AS}$ (as listed in appendix \ref{app:FlavorEnha}) and no $\bm{AS}$ in \cite{Yonekura:2015ksa, Cabrera:2018jxt, Apruzzi:2019vpe}. The CFD-descendants that model the theories with 1 or 2 $\bm{AS}$ are shown in figure \ref{fig:FavoriteODDTree}. 
Note that there are more descendants, which will correspond to other gauge theory descriptions. 
The possible BG-CFD embeddings,  are shown in figure \ref{fig:FavoriteBGCFDSOdd}, from which we can determine alternative weakly coupled descriptions, which in particular should model some of the other descendants.

\begin{sidewaysfigure}
\centering
\includegraphics*[width=\textwidth]{CFD-BG-CFDs-SU2nplus1-0AS-8F-TREE-NEW.pdf}
\caption{The CFD-subtrees that correspond to the gauge theory descriptions $SU(2n+1)_0 + 2 \bm{AS}+ 8\bm{F}$, where on the RHS we decouple first one $\bm{AS}$ and then the fundamental flavors, whereas the LHS is the subtree with $2 \bm{AS}$. The resulting flavor symmetries for the SCFTs are listed above each diagram, and agree with the known enhancements. This is a strict subtree of the CFD descendant tree, as is obvious from the configuration of $(-1)$ curves. 
\label{fig:FavoriteODDTree}}
\end{sidewaysfigure}

\begin{figure}
\centering
\includegraphics*[width=\textwidth]{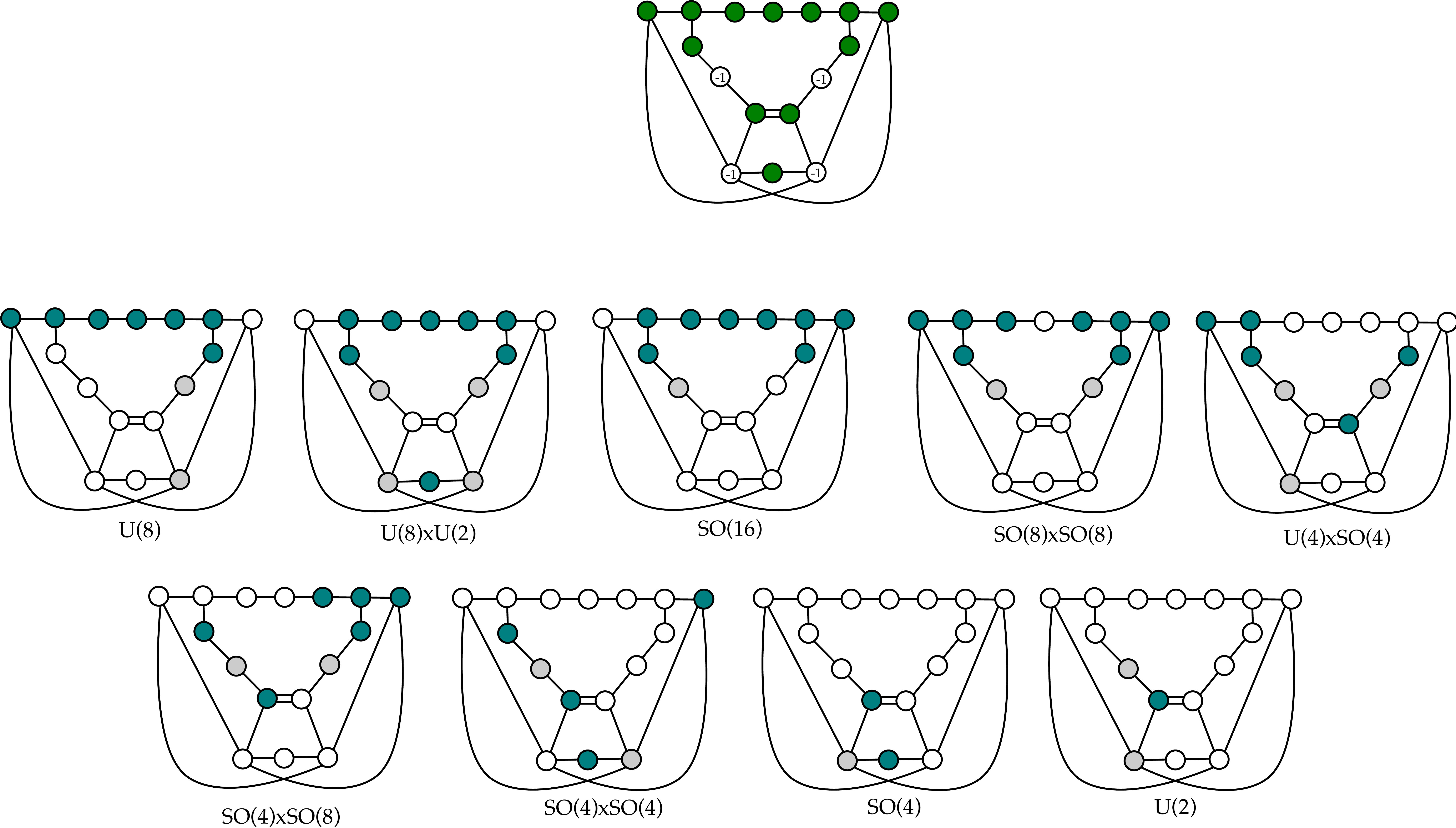}
\caption{The possible embeddings of the BG-CFDs into the marginal CFD, which is shown at the top, for 
$SU(2n+1)+  2 \bm{AS}+ 8\bm{F}$. Below the diagrams we note the classical flavor symmetry for the putative weakly coupled description. 
\label{fig:FavoriteBGCFDSOdd}}
\end{figure}


\subsection[\texorpdfstring{$SU(4)_0 + 2\bm{AS} + 8\bm{F}$}{SU(4)0 + 2AS + 8F} and \texorpdfstring{$(E_7, SO(7))$}{(E7,SO(7))} Conformal Matter]{\boldmath{$SU(4)_0 + 2 { AS} + 8{ F}$} and \boldmath{$(E_7, SO(7))$} Conformal Matter} 
\label{sec:E7SO7}

A special case of the theories discussed in section \ref{sec:SU2n2AS8F} is $n=2$:
$SU(4)_0 + 2 \bm{AS} + 8\bm{F}$. 
In this case there is an enhanced superconformal symmetry $G_\text{F}^\text{marginal} = \widehat{E}_7 \times \widehat{SO(7)}$ for the marginal theory, which has its origin in the 6d realization in terms of the $(E_7, SO(7))$ minimal conformal matter.

The marginal CFD can in this case in fact be computed directly using a geometric resolution from the conformal matter description, which is done in appendix \ref{app:E7SO7}. 
From this we determine the marginal CFD to be:
\be
\centering
\label{E7SO7-topCFD}
\includegraphics[width=5.5cm]{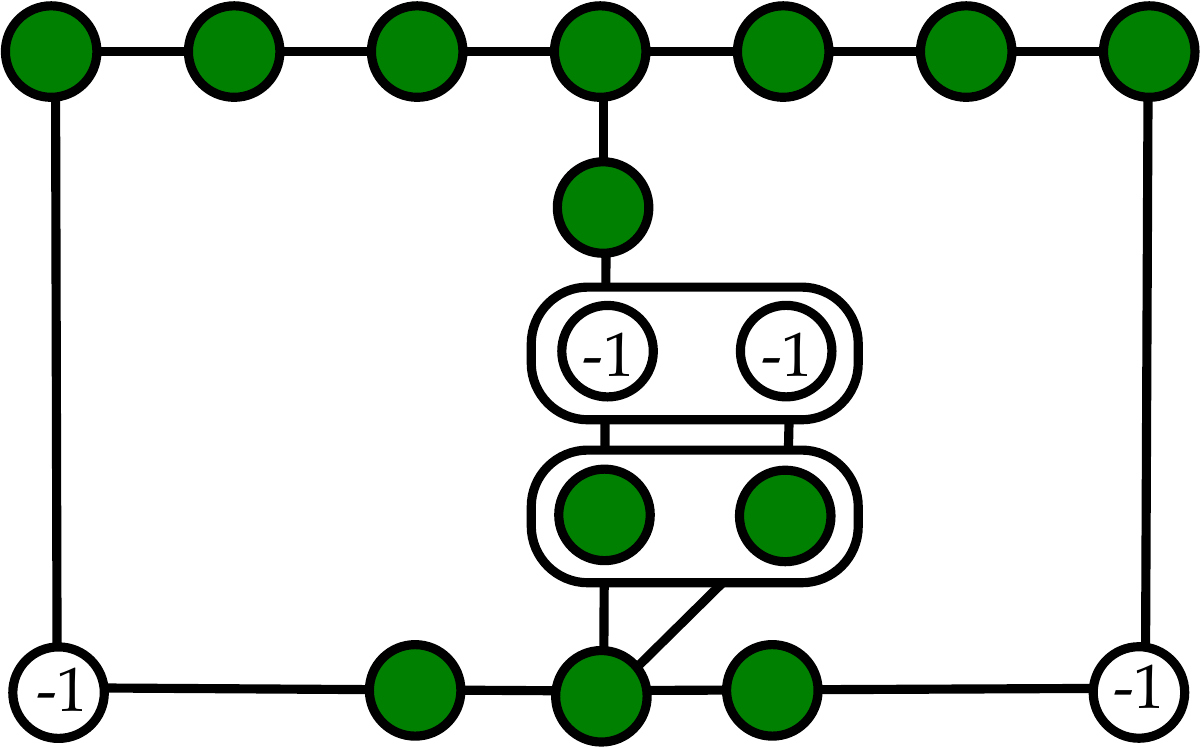} \,.
\ee
Note that the in the middle of the graph, there is a reducible vertex containing two $(-1)$-vertices, which are encircle to indicate that they are homologous, and  that they have to be removed simultaneously\footnote{In the resolved Calabi--Yau threefold geometry, they correspond to two curves with normal bundle $\mc{O}(-1)+\mc{O}(-1)$ that are homologous in the Calabi--Yau threefold but not on the surface components.}. Likewise, the reducible vertex below it contains two $(-2)$-vertices that are encircled and homologous curves in the geometry, and are part of the $SO(7)$ affine Dynkin diagram. 

This graph indeed contains the BG-CFDs for $8\bm{F}$ with classical flavor symmetry $U(8)$ and $2\bm{AS}$ with classical flavor symmetry $Sp(2)$, see table \ref{tab:BGCFDs}:
\be
\centering
\includegraphics[width=5.5cm]{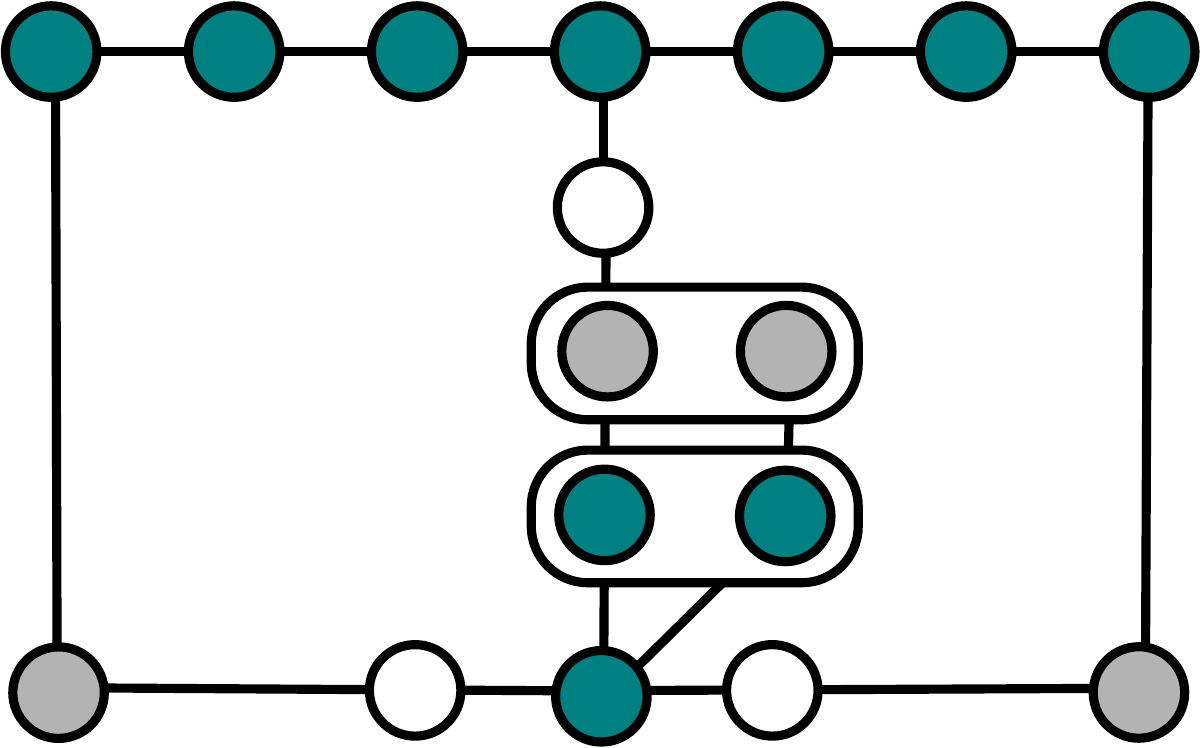} \,.
\ee
Thus if we were to bootstrap this theory as we did in the general case of $SU(2n)$, we would arrive at precisely this diagram. Consistency with the known flavor symmetry enhancements can also be checked.

On the other hand, there is another way of embedding the BG-CFD with classical flavor symmetry $SO(12)$ and $SO(4)$: 
\be
\centering
\includegraphics[width=5.5cm]{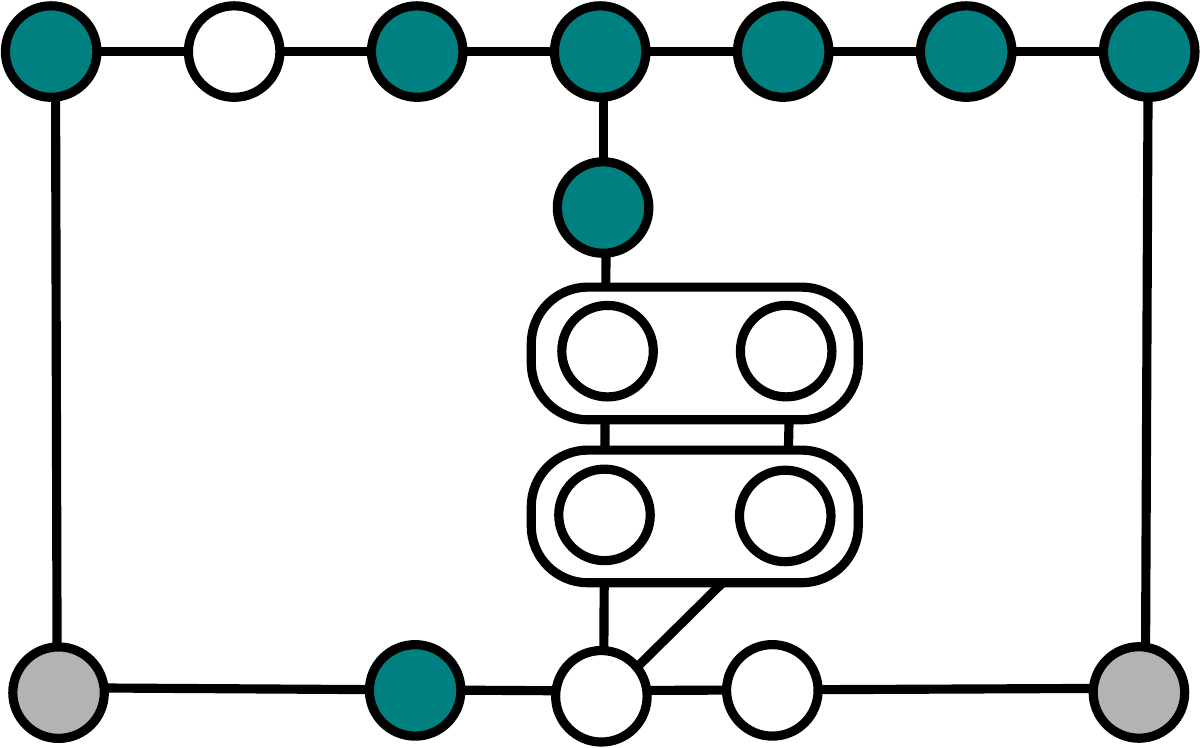} \,,
\ee
which corresponds to the quiver gauge theory $6\bm{F}-Sp(2)-Sp(1)-2\bm{F}$, see section \ref{sec:descquiv} for more details.

Some of the descendants of the $(E_7,SO(7))$ marginal CFD are shown in figure~\ref{f:E7SO7-CFDtree}. For example, after removing the reducible vertex containing the two $(-1)$-vertices, we arrive at the theory (3), where the reducible vertex below it will contain two $(-1)$-vertices and the vertex above it will become a $(-1)$-vertex. Then if we remove the reducible vertex containing two $(-1)$-vertices in the graph (3) to get graph (8), the reducible vertex below it will become a $0$-vertex because of the double connection in (3). 

We list the $SU(4)$ gauge theory descriptions of these theories in table~\ref{t:E7SO7-gauge}, which can be compared with the flavor symmetry enhancements in the literature~\cite{Zafrir:2015rga,Yonekura:2015ksa}. For the CFDs (2), (4) and (6) with $2\bm{AS}$, the enhanced flavor symmetry matches the table 3 in~\cite{Zafrir:2015rga}. For the cases (3) and (7) with $1\bm{AS}$, the flavor symmetry was correctly predicted in~\cite{Yonekura:2015ksa}. For the case 8, which does not have any anti-symmetric matter, the CFD matches the descendant from $(D_6,D_6)$ marginal CFD~\cite{Apruzzi:2019vpe} with flavor symmetry $SU(8)\times SU(2)\times SU(2)$. Especially, there should not be any additional extremal $(-1)$-curve in the middle part of the picture. Finally, for the case (5), the CFD transition from case (2) does not correspond to the decoupling of a matter multiplet in the $SU(4)$ gauge theory. Hence this theory with $G_\text{F}=E_7\times SU(2)\times SU(2)$ is not expected to have an $SU(4)$ gauge theory description. On the other hand, from the quiver gauge theory $6\bm{F}-Sp(2)-Sp(1)-2\bm{F}$ of the marginal theory, the CFD (5) is generated by decoupling the $2\bm{F}$ of the $Sp(1)$ gauge group. Hence we expect this theory to have a $6\bm{F}-Sp(2)-Sp(1)$ quiver gauge theory description.

\begin{figure}
\centering
\includegraphics[width=\textwidth]{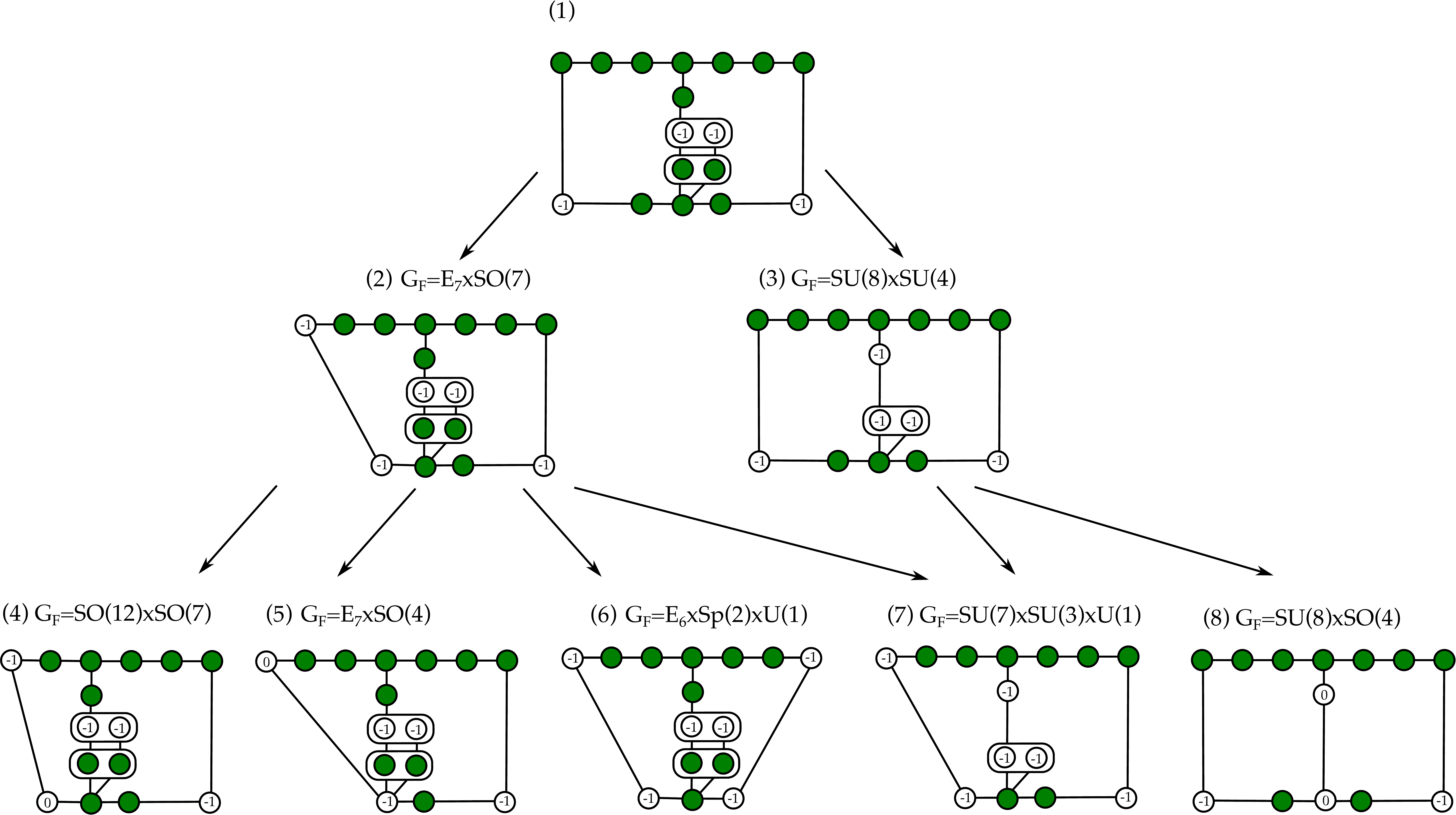}
\caption{$(E_7, SO(7))$: the figure shows the  first few descendants of the marginal CFD (\ref{E7SO7-topCFD}) for the $(E_7,SO(7))$ theory, including the enhanced superconformal flavor symmetries $G_\text{F}$. \label{f:E7SO7-CFDtree}}
\end{figure}

\begin{table}
\centering
\begin{tabular}{|c|c|c|c|}
\hline
CFD No. & Matter fields & $\kappa$ & $G_\text{F}$\\\hline
\hline
(1) & $2\bm{AS}+8\bm{F}$ & 0 & -\\
(2) & $2\bm{AS}+7\bm{F}$ & $1/2$ & $E_7\times SO(7)$\\
(3) & $1\bm{AS}+8\bm{F}$ & 0 & $SU(4)\times SU(8)$\\
(4) & $2\bm{AS}+6\bm{F}$ & $1$ & $SO(7)\times SO(12)$\\
(5) & - & - & $E_7\times SU(2)\times SU(2)$\\
(6) & $2\bm{AS}+6\bm{F}$ & $0$ & $Sp(2)\times E_6\times U(1)$\\
(7) & $1\bm{AS}+7\bm{F}$ & $1/2$ & $SU(7)\times SU(3)\times U(1)$\\
(8) & $8\bm{F}$ & 0 & $SU(8)\times SU(2)\times SU(2)$\\
\hline
\end{tabular}
\caption{The $SU(4)_\kappa$ gauge theory description and superconformal flavor symmetry $G_\text{F}$ of the descendant theories from the $(E_7,SO(7))$ marginal CFD. The CFD numbering corresponds to that  in figure~\ref{f:E7SO7-CFDtree}. Model $(5)$ does not have a description in terms of an $SU(4)$ gauge theory, but it has a $6\bm{F}-Sp(2)-Sp(1)$ quiver gauge theory description instead.}
\label{t:E7SO7-gauge}
\end{table}


\newpage


\section{Descendants and Dualities}
\label{sec:descquiv}

A 5d SCFT can be effectively described by multiple gauge theories at
low-energy, equivalently different gauge theories can
have the same UV-fixed point. Such theories can be viewed as dual effective
descriptions of the same UV 5d SCFT. In particular, because of the IR
effective nature of 5d gauge theories, these are called UV-dualities.

In this section we study UV-dualities among quiver gauge theories from the
point of view of the CFDs. More precisely, by embedding the BG-CFDs into the
marginal CFDs we are able to predict possible dual gauge theory phases, which
UV-complete to a 6d SCFT. We then check if these gauge theories are consistent
geometrically, by studying the ruling of the resolutions corresponding to
different descendants. From these resolutions, we also consistently blow up
the collection of surfaces to get a candidate resolved geometry for the
marginal theory. We then compute the triple intersection numbers and check
them against the prepotential computed from the gauge theory. 

Having set the dual effective descriptions for the marginal theory we can
immediately predict many novel dualities for the descendant theories. In  many
cases, we support these by explicitly finding the corresponding rulings in the
resolved Calabi--Yau threefolds engineering the 5d theories. In addition, we
also compute the prepotential of the candidate duals. For many theories, which
are not distinguished by non-trivial physical theta angles, we check that the
prepotentials match in some region of the Coulomb branches. This, together
with the prediction of the superconformal flavor symmetries supplied by the
CFD and BG-CFD embeddings, provides a good test for these novel proposed
UV-dualities. 

\subsection[Descendants and Dualities for Minimal \texorpdfstring{$(E_6, E_6)$}{(E6,E6)} Conformal
Matter]{Descendants and Dualities for Minimal \boldmath{$(E_6, E_6)$} Conformal
Matter}

As we showed in \cite{Apruzzi:2019opn}, the known weakly coupled quiver description of the $(E_6, E_6)$ minimal conformal matter theory only captures a very small subset of descendant SCFTs in 5d. Starting with the CFD, we have seen that there are multiple ways that BG-CFDs can be embedded. This results in new dualities, both for the marginal theory as well as the descendants. In this class of theories, all dualities can be checked by comparing with the geometry and finding the corresponding ``dual'' rulings of the surfaces.

\subsubsection{An Asymmetric Quiver}\label{sec:E6asym}

In this section we will consider quiver gauge theories that are descendants of
the quiver
\begin{equation}\label{eqn:diego}
  6\bm{F} - SU(4)_0 - Sp(1)_0 - Sp(1) - 2\bm{F} \,.
\end{equation}
This quiver is derived from geometric considerations, in appendix
\ref{app:E6E6resol} and shown to be marginal, and to have as 6d UV fixed point the
minimal $(E_6, E_6)$ conformal matter theory. 
While the $\theta$-angle of the
central $Sp(1)$ gauge node is not directly fixed by the geometry\footnote{If we assume the following empirical evidence that a trivial theta angle leads to an enhancement of flavor symmetry at strong coupling, whereas a non-trivial one leads to an abelian factor, we can actually present a criterion to compute the theta angle at least for a $SU(2)$ gauge theory factor: if the surface is ruled over a curve corresponding to a green node in the CFD, then $\theta=0$, otherwise $\theta=\pi$.}, the superconformal
flavor symmetries for the alternate case where $\theta = \pi$ are not
consistent with the quiver being a descendant of $(E_6, E_6)$ conformal matter
\cite{ZafrirPC}, and thus we are led inexorably to the conclusion that the
only option is $\theta = 0$. 

The descendants of this quiver have the form
\begin{equation}\label{eqn:arthur}
 (m_1, k, m_2):\qquad  m_1 \bm{F} - SU(4)_k - Sp(1)_0 - Sp(1) - m_2 \bm{F} \,.
\end{equation}
If $m_2 = 0$ then we must specify a $\theta$-angle for the rightmost $Sp(1)$
factor, we shall, without ambiguity, use the shorthand $m_2 = 0, \pi$ for
these two options. 

The classical flavor group of the marginal theory is $U(6) \times SO(4) \times
SU(2) \times U(1)^4$, where the first two factors are the global symmetry
groups rotating the hypermultiplets on each of the two flavor nodes. The
BG-CFDs associated to these two flavor groups must then be embedded into the
marginal $(E_6, E_6)$ CFD, which was given in (\ref{eqn:E6E6MCFD}). Up to the symmetry of
the CFD there is a unique form of such an embedding, which is
\begin{equation}\label{eqn:boots}
  \includegraphics[width=5cm]{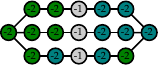} \,.
\end{equation}

Quivers of the form (\ref{eqn:arthur}) admit dualities amongst theories with
different $(m_1, k, m_2)$, and these can be observed from the symmetry of the
descandent CFDs. Let us consider the first descendant of the marginal CFD as
an illustrative example; despite the fact that there are three $(-1)$-curves,
each of which gives a CFD-transition to a descendant CFD, all of these three
descendants are the same CFD up to a reordering in the 2d-plane in which we
draw the image. However if we consider the CFD together with the marked
BG-CFDs, as in (\ref{eqn:boots}), then there would appear to be three distinct
gauge theory descendants, for which the embeddings of the BG-CFDs are
\begin{equation}
  \includegraphics[width= \textwidth]{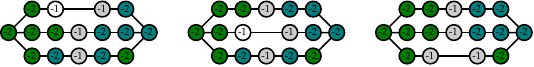} \,.
\end{equation}
These correspond to the three gauge theory descriptions
\begin{equation}
  (m_1, k, m_2) = (5, + 1/2, 2) \,,\, (5, -1/2, 2) \,,\, (6, 0, 1) \,.
\end{equation}
Since the underlying CFD for each of these theories is the same then these
theories have the same interacting SCFT as their UV fixed point, and as such
they are all dual to each other. To be extremely explicit, the CFD implies
that the three quiver gauge theories
\begin{equation}
  \begin{aligned}
    &5\bm{F} - SU(4)_{1/2} - Sp(1)_0 - Sp(1) - 2\bm{F} \cr
    &5\bm{F} - SU(4)_{-1/2} - Sp(1)_0 - Sp(1) - 2\bm{F} \cr
    &6\bm{F} - SU(4)_{0} - Sp(1)_0 - Sp(1) - 1\bm{F} \,,
  \end{aligned}
\end{equation}
flow to the same 5d $\mathcal{N} = 1$ SCFT in the UV.  In this way one can see
that novel dualities between quiver gauge theories, with the same gauge
algebra, can be observed from the CFD. We determine the superconformal flavor
symmetry for every descendant of the quiver (\ref{eqn:diego}), and furthermore
we determine for what values of $(m_1, k, m_2)$ the descendant quivers
(\ref{eqn:arthur}) are dual. 

More generally, if we compute the prepotential for the two sets of quiver specified by $(m_1, k, m_2)$ and $(\widetilde m_1, \widetilde k,\widetilde  m_2)$, it matches in a non-trivial region of the two gauge theories respectively, if the following conditions are satisfied,
\begin{equation} \label{eq:su4su2su2dual}
\widetilde m_1 =  1 - k+ m_2+ \frac{m_1}{2} ,  \qquad  \widetilde m_2 = \frac{1}{2}(m_1+2k-2)  \qquad \widetilde k = \frac{1}{4}|m_1-2m_2-2-2k|
\end{equation}
Since the prepotenial analysis is not sensitive to the $\theta$-angles these
expressions involve some care when $m_2 = 0,\pi$. For $m_2 = 0$ they capture
the dualities as determined from the CFD.  These results for the descendants
of the marginal quiver with $(m_1, k, m_2)=(6,0,2)$ are summarized in table
\ref{tbl:E6E6asym}.

\begin{table}
  \centering
  \footnotesize
  \begin{tabular}{|c|c|}
    \hline
    $(m_1, k, m_2)$ & Superconformal Flavor Symmetry \cr\hline\hline
    $(5, \pm 1/2, 2)$, $(6, 0, 1)$ & $E_6 \times E_6$ \cr\hline
    $(4, \pm 1, 2)$, $(6, 0, 0)$, $(6, 0, \pi)$ & $E_6 \times SU(6)$ \cr\hline
    $(5, \pm 1/2, 1)$, $(4, 0, 2)$ & $SO(10)^2 \times U(1)$ \cr\hline
    $(3, \pm 3/2, 2)$ & $E_6 \times SU(3)^2$ \cr\hline
    $(3, \pm 1/2, 2)$, $(4, \pm 1, 1)$, $(5, \pm 1/2, 0)$, $(5,\pm 1/2,
    \pi)$ & $SO(10) \times SU(5) \times U(1)$ \cr\hline
    $(4, 0, 1)$ & $SO(8)^2 \times U(1)^2$ \cr\hline
  $(2, \pm 2, 2)$ & $E_6 \times SU(2)^2 \times U(1)$ \cr\hline
  $(2, \pm 1, 2)$, $(3, \pm 3/2, 1)$ & $SO(10) \times SU(3) \times SU(2)
  \times U(1)$ \cr\hline
  $(2,0,2)$, $(4, \pm 1, 0)$ & $SO(10) \times SU(4) \times U(1)$ \cr\hline
  $(4, \pm 1, \pi)$ & $SU(5)^2 \times U(1)$ \cr\hline
  $(3, \pm 1/2, 1)$, $(4, 0, 0)$, $(4, 0, \pi)$ & $SO(8) \times SU(4) \times U(1)^2$ \cr\hline
  $(1, \pm 5/2, 2)$ & $E_6 \times SU(2) \times U(1)$ \cr\hline
  $(1, \pm 3/2, 2)$, $(2, \pm 2, 1)$ & $SO(10) \times SU(2) \times U(1)^2$
  \cr\hline
  $(1, \pm 1/2, 2)$ & $SO(10) \times SU(3) \times U(1)$ \cr\hline
  $(2, \pm 1, 1)$ & $SO(8) \times SU(2)^2 \times U(1)^2$ \cr\hline
  $(2, 0, 1)$, $(3, \pm 1/2, 0)$ & $SO(8) \times SU(3) \times U(1)^2$ \cr\hline
  $(3, \pm 3/2, 0)$ & $SO(10) \times SU(3) \times U(1)$ \cr\hline
  $(3, \pm 3/2, \pi)$ & $SU(5) \times SU(3) \times SU(2) \times U(1)$ \cr\hline
  $(3, \pm 1/2, \pi)$ & $SU(4)^2 \times U(1)^2$ \cr\hline
  $(0, \pm 3, 2)$ & $E_6 \times SU(2)$ \cr\hline
  $(0, \pm 2, 2)$ & $SO(10) \times SU(2) \times U(1)$ \cr\hline
  $(1, \pm 5/2, 1)$ & $SO(10) \times U(1)^2$ \cr\hline
  $(0, \pm 1, 2)$, $(2, \pm 2, 0)$ & $SO(10) \times SU(2) \times U(1)$ \cr\hline
  $(0, \pm 3/2, 1)$ & $SO(8) \times U(1)^3$ \cr\hline
  $(2, \pm 2, \pi)$ & $SU(5) \times SU(2) \times U(1)^2$ \cr\hline
  $(1, \pm 1/2, 1)$, $(2, \pm 1, 0)$ & $SO(8) \times SU(2) \times U(1)^2$
  \cr\hline
  $(0, 0, 2)$ & $SO(10) \times SU(3)$ \cr\hline
  $(2, \pm 1, \pi)$ & $SU(4) \times SU(2)^2 \times U(1)^2$ \cr\hline
  $(2, 0, 0)$ & $SO(8) \times SU(2) \times U(1)^2$ \cr\hline
  $(2, 0, \pi)$ & $SU(4) \times SU(3) \times U(1)^2$ \cr\hline
  $(0, \pm 3, 1)$ & $SO(10) \times U(1)$ \cr\hline
  $(0, \pm 2, 1)$ & $SO(8) \times U(1)^2$ \cr\hline
  $(0, \pm 1, 1)$, $(1, \pm 3/2, 0)$ & $SO(8) \times U(1)^2$ \cr\hline
  $(0, 0, 1)$ & $SO(8) \times SU(2) \times U(1)$ \cr\hline
  $(1, \pm 5/2, 0)$ & $SO(10) \times U(1)$ \cr\hline
  $(1, \pm 5/2, \pi)$ & $SU(5) \times U(1)^2$ \cr\hline
  $(1, \pm 3/2, \pi)$ & $SU(4) \times U(1)^3$ \cr\hline
  $(1, \pm 1/2, 0)$ & $SO(8) \times U(1)^2$ \cr\hline
  $(1, \pm 1/2, \pi)$ & $SU(4) \times SU(2) \times U(1)^2$ \cr\hline
  $(0, \pm 3, 0)$ & $SO(10)$ \cr\hline
  $(0, \pm 2, 0)$ & $SO(8) \times U(1)$ \cr\hline
  $(0, \pm 1, 0)$ & $SO(8) \times U(1)$ \cr\hline
  $(0, 0, 0)$ & $SO(8) \times U(1)$ \cr\hline
  $(0, \pm 3, \pi)$ & $SU(5) \times U(1)$ \cr\hline
  $(0, \pm 2, \pi)$ & $SU(4) \times U(1)^2$ \cr\hline
  $(0, \pm 1, \pi)$ & $SU(4) \times U(1)^2$ \cr\hline
  $(0, 0, \pi)$ & $SU(4) \times SU(2) \times U(1)$ \cr\hline
  \end{tabular}\caption{Dualities among and superconformal flavor symmetries
    for quivers of the form $m_1\bm{F} - SU(4)_{k} - Sp(1)_0 - Sp(1) -
    m_2\bm{F}$ that are descendants of the quiver $6\bm{F} -
    SU(4)_{0} - Sp(1)_0 - Sp(1) -
        2\bm{F}$. We abuse notation and capture the $\theta$-angle of the
  rightmost $Sp(1)$ when it has no fundamental hypermultiplets by allowing
$m_2 = 0, \pi$.}\label{tbl:E6E6asym}
\end{table}

\subsubsection{Maximal Quivers}\label{sec:E6maxi}
As a proof of principle we present here a prospective dual marginal description of $(E_6, E_6)$ conformal matter on a circle and its descendants, which satisfy the classical flavor symmetry embedding in the CFD diagrams in figure \ref{fig:E6E6embed}. This IR effective description was already introduced in \ref{sec:E6E6qH}, and it has ten
full hypermultiplets, the maximal number possible. This feature makes this possible description very interesting, since it would provide the IR effective gauge theories for as many descendants as possible. It reads
\begin{equation}
  5\bm{F} - Sp(1) - SO(6) - Sp(1) - 5\bm{F} \,.
\end{equation}
While there is no evidence from geometry that this is indeed a quiver
description of $(E_6, E_6)$ conformal matter, we will, in this section,
determine the descendants and their superconformal flavor symmetries under the
assumption that this quiver is indeed a realized description. It would be interesting to study the Coulomb branch metric and the BPS string tensions of the full quiver in order to verify the validity of this effective theory. 

To express the global symmetries in a concise way we first introduce a helpful
notation for the $E$-type exceptional simple groups
\begin{equation}\label{eqn:specialE}
  E^{(k)} = \begin{cases} 
    E_{k+1} \quad &\text{ for } k = 0, \cdots, 7 \cr
    \widetilde{E}_1 = U(1) \quad &\text{ for } k = \pi
  \end{cases} \,.
\end{equation}

The descendant quivers have the form
\begin{equation}
  m_1\bm{F} - Sp(1) - SO(6) - Sp(1) - m_2 \bm{F} \,,
\end{equation}
where we again allow the abuse of notation to write $m_1, m_2 = 0, \pi$ to
describe the $\theta$-angle when the respective flavor nodes become trivial.
If we write the tuple $(m_1, m_2)$ to describe one of these descendant quivers
we find, from an analysis of the CFD and its descendants, that the
superconformal flavor symmetries are, 
\begin{equation}
  \begin{aligned}
    (4, 5) \,&: \quad E_6 \times E_6 & & \cr
    (3, 5) \,&: \quad SU(6) \times E_6 & &  \cr
    (2, 5) \,&: \quad SU(3)^2 \times E_6 & & \cr
    (1, 5) \,&: \quad SU(2)^2 \times U(1) \times E_6 & & \cr
    (\theta, 5) \,&: \quad SU(2) \times U(1) \times E_6 &\qquad &\text{ for }
    \theta = 0, \pi \cr
    (m_1, m_2) \,&: \quad E^{(m_1)} \times E^{(m_2)} \times U(1) &\qquad
    &\text{ for } m_1, m_2 = 0, \pi, 1, 2, 3, 4 \,.
  \end{aligned}
\end{equation}

\subsubsection{Dualities Between Different Quivers}

We have two known quiver gauge theories that are marginal for the minimal
$(E_6, E_6)$ conformal matter theory. These are
\begin{equation}\label{eqn:E6MQ1}
 [2\bm{F}]-SU(2)-\overset{%
\begin{array}
[c]{c}%
[2\bm{F}]\\
|\\
SU(2)\\
|
\end{array}
}{SU(3)_0}-SU(2)-[2\bm{F}] \,.
\end{equation}
and
\begin{equation}\label{eqn:E6MQ2}
  6\bm{F} - SU(4)_0 - Sp(1)_0 - Sp(1) - 2\bm{F} \,.
\end{equation}
The first gauge theory description comes from string dualities, namely the
circle compactification of an M5-brane probing an $\mathbb R^5/E_6$ singularity.
We also find these two gauge theory descriptions geometrically by studying
some resolutions and their rulings, as well as matching the triple intersection
numbers with the putative quiver gauge theory prepotential, see appendix
\ref{app:Resolutions}.  Furthermore there is a prospective description in terms of the
quiver
\begin{equation}\label{eqn:E6MQ3}
  5\bm{F} - Sp(1) - SO(6) - Sp(1) - 5\bm{F} \,,
\end{equation}
for which we stress that there is no geometric underpinning. Curiously, this
quiver can be obtained by gluing two marginal theories of the rank one E-string,
by gauging a diagonal $SO(6)$, 
\begin{equation}
 \ba
 5\bm{F} - Sp(1) - 3\bm{F}  \qquad 
  \xlongleftrightarrow{\rm gauge\; diag.\; SO(6)}
 \qquad & 3\bm{F} - Sp(1) - 5\bm{F} \cr 
& =  5\bm{F} - Sp(1) - SO(6) - Sp(1) - 5\bm{F} \, .
\ea
\end{equation}
 For \eqref{eqn:E6MQ2} and \eqref{eqn:E6MQ3} we have discussed the descendants
 and the dualities amongst those descendants in sections \ref{sec:E6asym} and
 \ref{sec:E6maxi}, and for \eqref{eqn:E6MQ1} these were discussed in
 \cite{Apruzzi:2019opn}. In this section we will determine, again from the
 structure of the CFDs in the CFD-tree, the dualities not only amongst
 descendants of the same marginal quiver, but also the dualities amongst the
 descendants of all three marginal quivers. As we have previously stated,
 descendants of (\ref{eqn:E6MQ1}) are captured by the tuple $(m_1, m_2, m_3)$,
 those of (\ref{eqn:E6MQ2}) by $(m_1, k, m_2)$, and finally the data of the
 descendants of the quiver (\ref{eqn:E6MQ3}) can be specified by $(m_1, m_2)$.
 For the first two class of quiver gauge theories, we further find that the
 prepotentials agree in some region of the respective Coulomb branches. The
 dualities amongst all these descendants are shown in table
 \ref{tbl:E6E6alldualities}.

\begin{table}
  \centering
  \begin{tabular}{|c|c|c|}
    \hline
    $(m_1, |k|, m_2)$ & $(m_1, m_2, m_3)$ & $(m_1, m_2)$ \cr\hline
    $(6, 0, 2)$ & $(2,2,2)$ & $(5,5)$ \cr
    $(5, 1/2, 2), (6, 0, 1)$ & $(1,2,2)$ & $(4,5)$ \cr
    $(5, 1/2, 1), (4, 0, 2)$ & $(1,1,2)$ & $(4,4)$  \cr
    $(4,1,2), (6,0,0), (6,0,\pi)$ & $(0,2,2), (\pi,2,2)$ & $(3,5)$ \cr
    $(4,0,1)$ & $(1,1,1)$ & -- \cr
    $(3, 3/2, 2)$ & -- & $(2,5)$ \cr
    $(3,1/2,2), (4,1,1), (5,1/2,0), (5,1/2,\pi)$ & $(0,1,2),(\pi,1,2)$ &
    $(3,4)$ \cr
    $(2,0,2), (4,0,1)$ & $(0,0,2), (\pi,\pi,2)$ & -- \cr
    $(4,1,\pi)$ & $(0,\pi,2)$ & $(3,3)$ \cr
    $(3,1/2,1), (4,0,0), (4,0,\pi)$ & $(0,1,1), (\pi,1,1)$ & -- \cr
    $(3, 3/2, 1)$ & -- & $(2,4)$ \cr
    $(2,2,2)$ & -- & $(1,5)$ \cr
    $(2,0,1), (3, 1/2, 0)$ & $(0,0,1), (\pi,\pi,1)$ & -- \cr
    $(3,1/2,\pi)$ & $(0,\pi,1)$ & -- \cr
    $(3, 3/2, \pi)$ & -- & $(2,3)$ \cr
    $(2,0,0)$ & $(0,0,0), (\pi,\pi,\pi)$ & -- \cr
    $(2,0,\pi)$ & $(0,0,\pi), (0,\pi,\pi)$ & -- \cr
    $(2,1,2), (3,3/2,1)$ & -- & -- \cr
    $(1,3/2,2), (2,2,1)$ & -- & $(1,4)$ \cr
    $(1,5/2,1)$ & -- & $(0,4)$ \cr
    $(0,2,2)$ & -- & $(\pi,4)$ \cr
    $(0,1,2), (2,2,0)$ & -- & -- \cr
    $(2,2,\pi)$ & -- & $(1,3)$ \cr
    $(1,1/2,1), (2,1,0)$ & -- & -- \cr
    $(1, 5/2, \pi)$ & -- & $(0, 3)$ \cr
    $(0,1,1), (1,3/2,0)$ & -- & -- \cr\hline
  \end{tabular}\caption{Dualities amongst marginal quiver gauge theory descriptions of
  minimal $(E_6, E_6)$ conformal matter. 
  The three columns correspond to the quiver gauge theories (\ref{eqn:E6MQ2}), (\ref{eqn:E6MQ1}) and (\ref{eqn:E6MQ3}), respectively.
  We do not write explicitly the obvious duality
between $(m_1, k, m_2)$ and $(m_1, -k, m_2)$, and simply write the
Chern--Simons level in terms of an absolute
value.}\label{tbl:E6E6alldualities}
\end{table}

\subsection[Descendants and Dualities for Minimal \texorpdfstring{$(E_7, E_7)$}{(E7,E7)} Conformal Matter]{Descendants and Dualities for Minimal \boldmath{$(E_7, E_7)$} Conformal Matter}

In section \ref{sec:E7E7qH} we speculated as to maximal quivers that may be marginal
theories for minimal $(E_7, E_7)$ conformal matter. We found, from the
embeddings of the BG-CFDs into the marginal CFD, that there are precisely two
symmetric options. These were
\begin{equation}
\ba
&    6\bm{F} - Sp(5) - Sp(5) - 6\bm{F} \cr
 &   6\bm{F} - Sp(3) - SO(8) - Sp(3) - 6\bm{F} \,.
  \ea
  \end{equation}
In this section we will first show that the former quiver is, in fact,
inconsistent with the CFD-tree, and thus cannot be a marginal description of
the $(E_7, E_7)$ theory. Secondly, we shall consider the latter quiver, which
is consistent, and determine the superconformal flavor symmetry of its
descendant quiver gauge theories. 

First we rule out the quiver containing two $Sp(5)$ gauge nodes. Since the
bifundamental representation between the two $Sp(5)$ nodes is real then there
is an $SU(2)$ classical flavor symmetry factor rotating that bifundamental.
The descendant quiver
\begin{equation}
  5\bm{F} - Sp(5) - Sp(5) - 5\bm{F} \,,
\end{equation}
has classical flavor symmetry
\begin{equation}
  SO(10)^2 \times U(1)^2 \times SU(2) \,,
\end{equation}
and a study of the descendants of the marginal CFD informs that the
superconformal flavor symmetry, if the quiver were indeed a description of
minimal $(E_7, E_7)$ conformal matter, would be
\begin{equation}
  E_6^2 \times U(1) \,.
\end{equation}
However, there does not exist an inclusion of this classical global symmetry
group into the superconformal flavor symmetry group, and thus we find a
contradiction with the assumption that the quiver $6\bm{F} - Sp(5) - Sp(5) -
6\bm{F}$ does have minimal $(E_7, E_7)$ conformal matter at its UV fixed
point. As such, this quiver must be ruled out as a possibility.

The potential marginal quiver $6\bm{F} - Sp(3) - SO(8) - Sp(3) - 6\bm{F}$ does
not have such an issue as there are no real bifundamental hypermultiplets
which would have an $SU(2)$ rotation group.  From the CFD, (\ref{eqn:E7E7M}),
one finds that quivers of the form
\begin{equation}
  m_1 \bm{F} - Sp(3) - SO(8) - Sp(3) - m_2\bm{F} \,,
\end{equation}
are descendants of this putative marginal quiver. For each of the following
parameters $(m_1, m_2)$, they have superconformal flavor symmetries, which enhance
from the classical flavor symmetry\footnote{To make sense of this expression
  we use the shorthand that the groups $SO(0)$ and $SO(\pi)$ are trivial.}
\begin{equation}
  SO(2m_1) \times SO(2m_2) \times U(1)^3 \,,
\end{equation}
to the superconformally enhanced values for $(m_1, m_2)$ given by
\begin{equation}
  \begin{aligned}
    (5, 6) \,&: \quad E_7 \times E_7 & & \cr
    (4, 6) \,&: \quad SO(12) \times E_7 & & \cr
    (3, 6) \,&: \quad SU(6) \times E_7 & &  \cr
    (2, 6) \,&: \quad SU(4) \times SU(2) \times E_7 & & \cr
    (1, 6) \,&: \quad SU(3) \times U(1) \times E_7 & & \cr
    (0, 6) \,&: \quad SU(3) \times U(1) \times E_7 & & \cr
    (\pi, 6) \,&: \quad SU(2) \times U(1) \times E_7 & & \cr
    (m_1, m_2) \,&: \quad E^{(m_1)} \times E^{(m_2)} \times U(1) &\qquad
    &\text{ for } m_1, m_2 = 0, \pi, 1, 2, 3, 4, 5 \,.
  \end{aligned}
\end{equation}
Note that we have again used the $E^{(k)}$ shorthand for the exceptional
groups, as defined in (\ref{eqn:specialE}). We stress that these dualities are
only determined under the assumption that the putative marginal quiver is, in
fact, a quiver gauge theory which realizes minimal $(E_7, E_7)$ conformal
matter at its UV fixed point. 
One approach to verify these dualities from first principles would be to find the associated rulings in resolutions of the $(E_7, E_7)$ geometry.

\subsection[Descendants and Dualities for Minimal \texorpdfstring{$(E_7, SO(7))$}{(E7,SO(7))} Conformal
Matter]{Descendants and Dualities for Minimal \boldmath{$(E_7, SO(7))$} Conformal
Matter}

As shown in \cite{Zafrir:2015rga}, the marginal theory for $(E_7, SO(7))$ minimal conformal matter
theory has a single gauge node description of the form $SU(4)_0 +
2\bm{AS} + 8\bm{F}$. Here we propose a dual description as the quiver, which will be supported geometrically as well as from the CFD and by matching the prepotential, given by
\begin{equation}
  6\bm{F} - Sp(2) - Sp(1) - 2\bm{F} \,.
\end{equation}
In this section we use the CFD to determine the superconformal flavor
symmetries of and the dualities amongst the descendants of this quiver. Such
quivers take the form
\begin{equation} \label{eq:quive7so7}
  m_1\bm{F} - Sp(2) - Sp(1) - m_2\bm{F} \,.
\end{equation}
This gauge theory description (as well as the $SU(4)_0 + 2\bm{AS} + 8\bm{F}$)
is supported by the rulings of the geometric resolutions as shown in 
appendix \ref{app:Resolutions}, and the triple intersection numbers match the putative
gauge theory prepotential. As a further outcome of the prepotential analysis,
we notice that there are regions of the Coulomb branch of  $SU(4)_k + 2\bm{AS} + m
\bm{F}$ and \eqref{eq:quive7so7}, where the prepotential coincide, provided
that
\begin{equation}
m_1 =2\pm k+\frac{m}{2}, \qquad m_2= \mp k-2+\frac{m}{2}\, .
\end{equation}
The superconformal flavor symmetries and the dualites for such gauge theories
are\footnote{Recall once more our convention that $SO(x)$ for $x$ not a
positive integer is the trivial group.}
\begin{equation}
  \begin{array}{|c|c|c|}
  \hline
      (m,|k|) & (m_1, m_2) &G_\text{F}  \cr \hline \hline
    (7, \frac{1}{2}) & (5, 2), (6, 1)  &SO(7) \times E_7  \cr \hline
 (6,1) & (4, 2), (6, 0) &SO(7) \times SO(12)  \cr\hline
  (6,0) & (5,1) & SO(5) \times E_6 \times U(1)  \cr\hline
 (5,\frac{3}{2}) &  (3, 2) &SO(7) \times SU(6)  \cr\hline
  (4,2)  & (2, 2) &SO(7) \times SU(4) \times SU(2)  \cr\hline
    (3,\frac{5}{2}) &(1, 2) &SO(7) \times SU(3) \times U(1)  \cr\hline
    -&(0, 2) &SO(7) \times SU(3)  \cr\hline
  (2,3)  &(\pi, 2) &SO(7) \times SU(2) \times U(1)  \cr\hline
   (5, \frac{1}{2}) &(4, 1), (5, 0) &SO(5) \times SO(10) \times U(1)  \cr\hline
       (1+m,\frac{5-m}{2}) &(m, 1) &SO(5) \times E_{m+1} \times U(1) \quad m =  1, 2,
    3 \cr\hline
       (1,\frac{5}{2}) &(\pi, 1) &SO(5) \times U(1) \times U(1)  \cr\hline
       - &(0 , 1) &SO(5) \times SU(2) \times U(1)  \cr\hline
    (m,\frac{4-m}{2})&(m, 0) &SO(5) \times SO(2m) \times U(1) \quad m = 1,
    \cdots, 4\cr\hline
        (0, 2)& (\pi, 0) &SO(5)  \times U(1) \cr\hline
       - & (0, 0) &SO(5)  \times U(1) \cr\hline
   - &(6, \pi) &SU(2)^2 \times E_7  \cr \hline
   -&(m, \pi) &SU(2) \times E^{(m)} \times U(1) \quad m = 0, \pi, 1,
    \cdots, 5 \cr  \hline
  \end{array}
\end{equation}

\subsection[Dualities for the Marginal Theory: \texorpdfstring{$SU(2n+1)_0 + 2 \bm{AS}+ 8 \bm{F}$}{SU(2n+1)0 + 2AS + 8F}]{Dualities for the Marginal Theory of \boldmath{$SU(2n+1)_0 + 2 {AS}+ 8 {F}$}}

The gauge theory, $SU(2n+1)_0 + 2 \bm{AS}+ 8 \bm{F}$, and its 6d origin have been discussed in section \ref{sec:SU2nplus12AS8F}, where we have also shown all the possible embeddings of classical flavor symmetries for putative dual theories, see figure \ref{fig:FavoriteBGCFDSOdd}. To test examples in this infinite class of dualities via checks of the prepotential, we specialize to $n=2$ and compare the proposed dual theories.
In fact, the prepotential matches in non-trivial regions of the respective Coulomb branches for the following quivers:
\be
\ba
&4\bm{F}  - Sp(2) - Sp(1)_{\theta} - Sp(1) - 2\bm{F}  \\
&4\bm{F}  - SU(3)_2 - Sp(1)_{\theta} - Sp(1) - 2\bm{F} \\
&8\bm{F}  - Sp(3) - Sp(1)_{\theta}   \\
&8\bm{F}  - SU(4)_1 - Sp(1)_{\theta} \\
&4\bm{F}  - Sp(2) - Sp(2) - 4\bm{F} \\
&2\bm{F}  - Sp(1) - Sp(1)_{\theta} - \overset{%
\begin{array}
[c]{c}%
[2\bm{F}]\\
|\end{array}}{Sp(1)} - Sp(1)_{\theta}\\
\ea
\ee
The dualities should hold for specific values of the $\theta$-angles, even if the prepotential matching is not sufficient to resolve this ambiguity. 
This gives strong evidence that our method of embedding the BG-CFDs of a 5d gauge theory with classical flavor symmetry into the CFD is particularly useful and efficient in predicting candidate 5d UV-dual theories.


\subsection[Dualities for the Marginal Theory: \texorpdfstring{$SU(2n)_0 + 2 \bm{AS}+ 8 \bm{F}$}{SU(2n+1)0 + 2AS + 8F}]{Dualities for the Marginal Theory of \boldmath{$SU(2n)_0 + 2 {AS}+ 8 {F}$}}
Similarly to the previous section, we now focus on the following theory $SU(2n)_0 + 2 \bm{AS}+ 8 \bm{F}$, which has been already discussed in section \ref{sec:SU2n2AS8F}.  In particular, we test here the predictions for the existence of alternative effective gauge theory descriptions, which come from embedding the classical flavor symmetry into the CFD, \eqref{Bambi}. These embeddings have been proposed in figure \ref{fig:FavoriteBGCFDS}. Specializing to the case $n=3$, we can explicitly compute the prepotential, and, for instance, we find that the prepotential for $SU(6)_0 + 2 \bm{AS}+ 8 \bm{F}$ is consistent with the following quivers:
\be
\ba
&6\bm{F}  - SU(4)_1  - Sp(1)_{\theta} - Sp(1)_{\theta} \\
&6\bm{F}  - SU(4)_1 - Sp(2) -  2\bm{F}\\
&2\bm{F}  - Sp(2) - Sp(3) -  6\bm{F} \, .
\ea
\ee
Again, the dualities will hold for specific values of the $\theta$-angles, which have to be determined by alternative methods. 
We observe here, that the proposed duals match with the  embeddings of the classical flavor symmetry as dictacted by the embedding of BG-CFDs into the CFDs as shown in figure \ref{fig:FavoriteBGCFDS}.


\section{Fibers from Coulomb Branch Phases}
\label{sec:FibPhas}

In the previous sections, we have heavily employed the description in terms of the
BG-CFDs, which are graphs associated to gauge theories, in order to study their SCFT limits in terms of CFDs.  The
connection between these two related concepts is based on the M-theory
description underlying both.  In particular, we will discuss in this section
how the box graphs encode the geometry for the BG-CFDs, and how, with minimal
additional input, it also determines the superconformal flavor symmetry enhancement.

To begin with, we briefly recall the necessary background material on
non-minimal singularities in elliptic fibrations, their (possibly non-flat)
resolutions, and their role in engineering 5d SCFTs, and refer to
\cite{Apruzzi:2018nre,Apruzzi:2019opn} for more details.

\subsection{5d SCFTs and M-theory  on Elliptic Calabi--Yau Threefolds}

We can associate a singular, non-compact elliptic Calabi--Yau threefold $\pi:
Y \rightarrow B$ to each marginal theory, whose 6d UV-completion is described
by F-theory compactified on $Y$.  Different mass deformations of the marginal
theory, which results in different 5d SCFTs pushed onto their Coulomb
branches, correspond to different crepant resolutions $\widehat{Y} \rightarrow
Y$.  The resolution introduces compact and non-compact surfaces/divisors,
$S_j$ and $D^{(\nu)}_i$, respectively.  In fact, the compact reducible surface
${\cal S} = \bigcup_j S_j$, $j=1,\cdots,r$, fully determines the
(local) geometry and completely characterizes the 5d SCFT.  In particular, the
gauge group $G_\text{gauge}$ is determined by how each surface $S_j$ is ruled,
i.e., fibered by $\bbP^1_j \equiv f_j$
\cite{Intriligator:1997pq}.\footnote{See also section 2.2 of
  \cite{Apruzzi:2019opn} for a summary in the same notation as here.} There
  are generally multiple compatible rulings, denoted by $f_j^{(n)}
  \hookrightarrow S_j$, which correspond to different effective gauge theories
  with the same SCFT limit, i.e., that are UV-dual to each other.

The smooth geometry $\widehat{Y}$ corresponds to a generic point on the
Coulomb branch, where the effective description is just a $U(1)^r$ theory with
no charged light states.  By partially blowing down all rulings $f_j$ of the
compact surfaces, thus shrinking $S_j$ to a curve, the gauge symmetry enhances
to the full non-abelian group $G_\text{gauge}$.  M2-branes wrapping
holomorphic curves that collapse in this limit give rise to massless charged
hypermultiplets and the W-bosons of $G_\text{gauge}$.  By further collapsing $S_j$
to a point, the theory becomes a strongly coupled 5d SCFT, as the volume is inversely proportional to the 
gauge coupling.

At each stage of the two-step collapse, it can happen that non-compact
divisors $D_i^{(\nu)}$ are also forced to shrink, leading to canonical surface singularities
along a non-compact curve in the threefold, whose singularity type encodes the
flavor symmetry \cite{Xie:2017pfl}.  In particular, the singularity type can
at most become worse when $S_j$ is collapsed first to a curve and then to a
point.  This reflects the field theory intuition on the enhancement
$G_\text{F,cl} \subseteq G_\text{F}$ from classical to superconformal flavor
symmetry.

The key proposal of \cite{Apruzzi:2018nre} to read off the flavor symmetry
geometrically, which was systematized and condensed into CFDs in
\cite{Apruzzi:2019vpe,Apruzzi:2019opn}, is to track how the \emph{non-compact}
divisors $D^{(\nu)}_i$ intersect ${\cal S}$.  More precisely, these divisors
are $\bbP^1$-fibered over a non-compact curve $W_\nu \subset B$.  Over a
generic point on $W_\nu$, the fibers $F^{(\nu)}_i$ intersect in an affine
Dynkin diagram $\widehat{G}_{\text{F}, \nu}^\text{(6d)}$.  In F-theory, these
determine the 6d superconformal flavor symmetry $G_\text{F}^\text{(6d)} =
\prod_\nu G_{\text{F}, \nu}^\text{(6d)}$.  By construction, a non-compact
divisor $D^{(\nu)}_i$ intersects the compact surfaces $\bigcup_j S_j = {\cal
S}$ over isolated points $p \in W_\nu$.  Depending on the resolution, the
$\bbP^1$ fiber $F^{(\nu)}_i$ may or may not be contained in $\cal S$.  Those
$F^{(\nu)}_i$ that are contained form the \emph{non-affine} Dynkin diagram of
$G_{\text{F}, \nu}$.  Since these fibers shrink \emph{everywhere} over $W_\nu$
when ${\cal S}$ is collapsed to a point, they determine the non-abelian part
of the 5d superconformal flavor symmetry, $G_\text{F,na} = \prod_\nu
G_{\text{F}, \nu}$.  We will refer to these curves, as in
\cite{Apruzzi:2019opn}, as {\it flavor curves}.

Abelian factors of the full 5d superconformal flavor symmetry $G_\text{F}$ come
from non-compact divisors $D_i^{(\nu)}$ whose fibers are not fully contained
in ${\cal S}$, but nevertheless intersects ${\cal S}$ in curves.  In general,
there can be linear redundancies amongst different such divisors, which can be
inferred from the intersection numbers of all curves inside ${\cal S}$ and all
divisors, see \cite{Apruzzi:2018nre,Apruzzi:2019opn}.  
In practice, we know on
general grounds the full rank of $G_\text{F}$ from the classical flavor
symmetry $G_\text{F,cl}$ and the number $N$ of topological $U(1)$s, 
\begin{align}\label{eq:rank_counting_G_F}
  \text{rank}(G_\text{F}) = \text{rank}(G_\text{F,cl}) + N \, ,
\end{align}
so that the number of $U(1)$ factors is determined as $\text{rank}(G_\text{F}) - \text{rank}(G_\text{F,cl})$.

\subsection{Fibers from Box Graphs}

Consider now a marginal theory with symmetry 
$G_\text{gauge} \times G_\text{BG}$ (where $G_{\text{BG}}$ is the classical flavor symmetry of the 
marginal theory).
We now determine from the box graphs, that are associated to each descendant theory, the intersection structure amongst the divisors $S_j$ and
$D^{(\nu)}_i$, which in particular specifies which codimension one fibers
$F^{(\nu)}_i$ are contained in $\cal S$, i.e., which are flavor curves. Both
$S_j$ and $D^{(\nu)}_i$ are ruled surfaces, compact and non-compact,
respectively. The fibers are denoted by $f_j$ and $F^{(\nu)}_i$, respectively and we introduce the 
notation
\be
\mathcal{D}_\ell \in \{S_j, D^{(\nu)}_i \}\,,\qquad \mathcal{F}_\ell \in \{f_j, F_j^{(\nu)}\} 
\ee
for all divisors and fibral curves.
Next, recall the relationship between the representation theory of a Lie group $G$
and intersection theory in M-theory on a smooth Calabi--Yau threefold
$\widehat{Y}$: to each ruled surface ${\cal F}_\ell \hookrightarrow {\cal
D}_\ell$, we associate a simple root $\alpha_\ell$ to the curve ${\cal
F}_\ell$ (with normal bundle $\CO \oplus \CO(-2)$ inside $\widehat{Y}$), and
its coroot $\alpha_\ell^\vee$ to the divisor ${\cal D}_\ell$, such that
\begin{align}
	\mathcal{C}_{\ell \kappa}^G \equiv \langle \alpha_\kappa^\vee , \alpha_\ell \rangle = - {\cal D}_{\kappa} \cdot {\cal F}_\ell \, ,
\end{align}
where $\mathcal{C}_{\ell \kappa}^G$ is the Cartan matrix of $G$.  In our
setup, the (co-)roots of $G_\text{gauge}$ arise from ${\cal D}_\ell = S_j$,
whereas those of $G_\text{BG}$ come from ${\cal D}_\ell = D_i$.

The pairing between weights and coweights is identified with the intersection
pairing between divisors and curves.  Consider a box graph associated to a
representation of $G_{\text{gauge}} \times G_{\text{BG}}$.  To an extremal
weight $L$ in the box graph (see definition \ref{def:extremal}) we associate a
curve $C$ with normal bundle $\CO(-1) \oplus \CO(-1)$, and refer to such a
curve as an {\it extremal curve}.  Denote by $L^{\pm}= \pm L$, where in a given
box graph only one of these is in the cone defining the Coulomb branch, and associated to that the curve $C^\pm = \pm C$ (which is an effective curve if the corresponding weight is in the cone), related by
\begin{align}\label{eq:weight_vector_intersection_number}
  (L^\pm)_\ell \equiv  \pm  \langle \alpha_\ell^\vee, L \rangle = \mp {\cal D}_{\ell} \cdot C^\pm \, .
\end{align}
All other weights are then realized as the linear combination of extremal
curves and $\mathcal{F}_\ell$.

\begin{figure}
\centering
\includegraphics*[width=8cm]{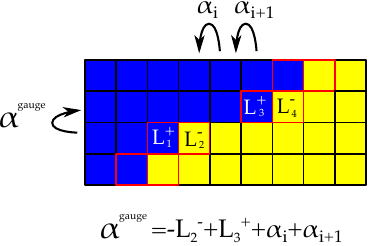} 
\caption{Example box graph for $SU(n)_{\text{gauge}}\times U(m)_{\text{BG}}$: the root  $\alpha^{\text{gauge}}$  of the gauge algebra $SU(n)$ splits. The extremal weights are boxed in red. The splitting of $\alpha^{\text{gauge}}$  into the extremal weights $L_2^-$ and $L_3^+$, as well as the roots of $G_{\text{BG}}$, $\alpha_i$, is determined from the box graph. The arrows indicate, as usual for box graphs, addition of roots.
In the geometry, we associate to $\alpha^\text{gauge}$ the curves $f_i$ and to $\alpha_j$ the curves $F_j$.
\label{fig:SplitEx}}
\end{figure}

A curve $\mathcal{F}$ associated to a root, which can be either a codimension
one fiber $F^{(\nu)}_i$ or a ruling $f_j$ of the surface $S_j$, is reducible
in codimension two, if the associated box graph indicates a splitting 
\begin{align}\label{eq:fiber_splitting_general}
  {\cal F} = \sum_a n_a C^{\epsilon_a}_a + \sum_{\nu} \sum_i \omega^{(\nu)}_i F^{(\nu)}_i + \sum_j \eta_j f_j \, , \quad \omega_i , \eta_j \geq 0 \, .
\end{align}
The first sum is over all extremal curves $C^{\epsilon_a}_a$, where $\epsilon_a = \pm$. An example is shown in figure \ref{fig:SplitEx}.

If ${\cal F} = f_j$ is the ruling of a compact divisor $S_j$, then we can
immediately deduce that all curves appearing on the right-hand side of
\eqref{eq:fiber_splitting_general} must be contained in $S_j$, and thus in
${\cal S}$.  In particular, the simple roots of the classical flavor symmetry
$G_\text{F,cl}$ are precisely those $F^{(\nu)}_i$ into which an $f_j$ splits
via \eqref{eq:fiber_splitting_general}. In the example of figure
\ref{fig:SplitEx} these are the roots associated to $F_i$ and $F_{i+1}$.

As for the superconformal flavor symmetry, recall that the Dynkin diagram of
$G_\text{BG}$ --- which is also the classical flavor symmetry of the marginal
theory --- fills only a subpart of the affine Dynkin diagram
$\widehat{G}_\text{F}^\text{(6d)} \equiv \prod_\nu \widehat{G}_{\text{F},
\nu}^\text{(6d)}$.  We denote divisors that are not captured in the box graph,
but are also ruled non-compact surfaces, by
\be 
F_{\Phi_l} \hookrightarrow D_{\Phi_l} \,.
\ee
In order to distinguish them from the roots/Cartans of $G_\text{BG}$, these
will be denoted by $F_i \hookrightarrow D_i$ in the following.  To
encode in the box graph approach the full superconformal flavor symmetry
requires determining how these additional nodes are attached to the part of
the fiber that we reconstruct from the box graph. 

Physically, these ``missing'' divisors either correspond to abelian factors of
the classical flavor symmetry (and hence have no roots that appear in the box
graph), or even the non-perturbative instanton $U(1)$s associated with each
simple factor of $G_\text{gauge}$.  Nevertheless, they can be fully contained
in $\cal S$ (as (multi-)sections of the rulings), and thus lead to an enhanced
superconformal flavor group.

The precise nature of the curves $F_{\Phi_l}$ (e.g., the information on how
they split, or in which surface $S_j$ they are contained) is dictated by
consistency conditions from intersection theory.  For that, we need one piece
of information which is not contained in the box graphs, namely the
intersection numbers $D_{\Phi_l} \cdot C$, where $C$ is an extremal curve.
These numbers depend on the resolution phase, as the extremal curves $C$ are
different in each phase.  However, what remains invariant throughout all
phases is $D_{\Phi_l} \cdot C(L)$, where $C(L)$ is the curve (possibly
reducible) associated to a particular weight $L$ in the box graph.  This is
because the linear combinations of divisors corresponding to abelian flavor
symmetries must give rise to well-defined charges for all weights $L \in
\bm{R}$ of the box graph.  These linear combinations do not change across
different resolutions, as they are --- similar to the Shioda-map for $U(1)$s
in F-theory --- divisorial, i.e., codimension one data.  Correspondingly, the
charges of the individual weights $\bm{R}$ under each $D_{\Phi_l}$ must
remain invariant.  In practice, we therefore compute $D_{\Phi_l} \cdot C(L)$
for all $l$ and weights $L$.  This has to be done in one specific resolution,
e.g., for the marginal theory having a $G_\text{gauge}$ gauge description, or
indeed for any other resolution. 

This data can also be phrased representation-theoretically.  We can extend the
(co-)weight lattice by additional (co-)roots $\Phi^\vee_l$ and $\Phi_l$, such
that every weight $L_m$ carries additional charges given by the pairing 
\begin{align}\label{eq:pairing_extra_roots}
	\langle \Phi^\vee_l , L_m \rangle  \, .
\end{align}
The holomorphic curve $C(L^{\epsilon_m}_m)$ that corresponds to a decoration
$\epsilon_m = \pm 1$ of $L_m$ then has, as in
\eqref{eq:weight_vector_intersection_number}, the intersection
\begin{align}
  \mathcal{E}_{l,m} = D_{\Phi_l} \cdot C_m^{\epsilon_m} = - \epsilon_m \langle \Phi^\vee_l , L_m \rangle \, .
\end{align} 
In concrete examples, we will provide the numbers ${\cal E}_{l,m}$, as
determined by any one specific resolution, for the associated undecorated box
graph.

With these numbers, the fiber geometry can be deduced from well-known
intersection properties of elliptically fibered threefolds.  Firstly, we know
that the intersection numbers
\begin{align}\label{eq:affine_Cartan_matrix}
  \widehat{\mathcal{C}}_{\ell \kappa}^\text{(6d)} = -{\cal F}_\ell \cdot {\cal D}_\kappa \, , \quad \text{with } {\cal F}_\ell \in \{F_i, F_{\Phi_l}\} \, \, \text{and} \, \, {\cal D}_\kappa \in \{D_i, D_{\Phi_l}\} \, ,
\end{align}
form the affine Cartan matrix of $\widehat{G}^\text{(6d)}_\text{F}$.
Secondly, we have
\begin{align}\label{eq:zero-intersection_S.F}
  S_j \cdot {\cal F}_\ell = 0 \quad \text{for } {\cal F}_\ell \in \{F_i, F_{\Phi_l}\} \, .
\end{align}
Finally, we know from the factorization $G_\text{gauge} \times G_\text{BG}$ that
\begin{align}\label{eq:zero-intersection_product_group}
  D_i \cdot f_j = S_j \cdot F_i = 0 \quad \text{for all } i,j \, .
\end{align}

With the above information, we can apply in each phase the following rules from intersection theory to determine the fiber structure:
\begin{enumerate}
  \item If an irreducible curve has negative intersection with a divisor, $C
    \cdot {\cal D} < 0$, then $C$ must be contained in ${\cal D}$.
    
  \item If a ruling of a surface (compact or not) splits as in
    \eqref{eq:fiber_splitting_general}, then all curves into which it splits
    must form a connected curve.

  \item If a codimension one fiber ${\cal F}_\ell$ does \emph{not} split, then
    ${\cal D}_\ell \cdot C \geq 0$ for any curve $C \neq {\cal F}_\ell$,
      such that $C$ is not contained inside of $\mathcal{D}_\ell$.  If
    ${\cal D}_\ell \cdot C = n > 0$, then $C$ and ${\cal F}_\ell$ intersect in
    $n$ points.

  \item If a non-splitting codimension one fiber ${\cal F}$ intersects a curve $C$ contained in a compact surface $S_j$, then ${\cal F} \subset S_j$, and is a flavor curve. 
\end{enumerate}
While the first three points follow from basic algebraic geometry, the last
point is due to the intersection number \eqref{eq:zero-intersection_S.F}: if
${\cal F}$ intersects a curve $C \subset S$, the only way to preserve this
intersection number is if ${\cal F}$ is also contained.  By these rules, it is
straightforward to reconstruct the configurations of the compact surfaces
$S_j$ and the relative positions of the codimension one fibers ${\cal F} =
F_i, F_{\Phi_l}$.

As a last comment, note that the intersection pattern of extremal curves and
the roots $F_i$ of $G_\text{BG}$ contained in ${\cal S}$ precisely form the
BG-CFD.  These curves shrink when we blow down the surfaces $S_j$ to curves in
order to have non-abelian gauge enhancement.  Conversely, if a geometry is
supposed to contain a specific gauge description, then it must contain the
BG-CFD as a subset of curves.

In summary our strategy will be as follows: 

\noindent
For the marginal 5d theories, we determine the geometric resolution. The
flavor symmetry $G_{\text{F}}^\text{(6d)}$ of the parent 6d theory will be
manifest in this description.  Different rulings of the surfaces yield
different weakly coupled gauge theory descriptions with
$G_{\text{gauge}}\times G_{\text{BG}}$. The geometric resolution provides the
following data:
\begin{itemize}
\item The embedding of the BG-CFD of $G_{\text{BG}}$ into $G_{\text{F}}^\text{(6d)}$ (i.e., the information about how the curves $F_{\Phi_l}$ are attached to the curves associated to the roots of $G_{\text{BG}}$),
\item Pairings $\langle \Phi_l^\vee, L \rangle$ for all weights $L$.
\end{itemize}
To determine the descendant 5d SCFTs with an effective $G_\text{gauge}$ description, we first construct all (flavor
equivalence classes of) Coulomb branch phases. 
For each descendant, self-consistency of the box
graphs and intersections of $\Phi_\ell$ fixes the fiber, and thus the
full superconformal flavor symmetry enhancement in the SCFT limit. 

\subsection{Fiber Reconstruction for Rank One SCFTs}
\label{sec:EstringBG}

In the following, we will first discuss in detail how these methods apply to
5d SCFTs of rank one.  The marginal geometry, descending from the 6d rank one
E-string, has an $SU(2)_\text{gauge} \times SO(16)_\text{F}$ gauge theory
description.  The corresponding $SU(2)_\text{gauge} \times SO(16)_\text{BG}$
box graphs have been presented previously in section \ref{sec:RankOne}.

To translate them into geometry, we first associate to the simple roots the rational curves
\begin{align}
\begin{split}
  SO(16)_\text{BG} : \quad F_i \quad & \leftrightarrow \quad -\alpha_i \,,\quad i=1,\cdots, 8 \, , \\
  SU(2)_\text{gauge}:\quad f \quad & \leftrightarrow \quad -\alpha_{SU(2)} \, ,
\end{split}
\end{align}
where $f \hookrightarrow S$ is the ruling of the compact surface introduced in
the resolution of the non-minimal singularity of the elliptic threefold at the
$(E_8, I_1)$ collision (see \cite{Apruzzi:2019opn} for more details).  The $F_i$ rule non-compact divisors $D_{i}$ resolving
the codimension one $E_8$ singularity and intersect in the non-affine Dynkin
diagram of $SO(16)$.  They are embedded into the affine $E_8$ fiber as
follows:
\be\label{eq:SO16_in_affine_E8}
 \includegraphics[width=7cm]{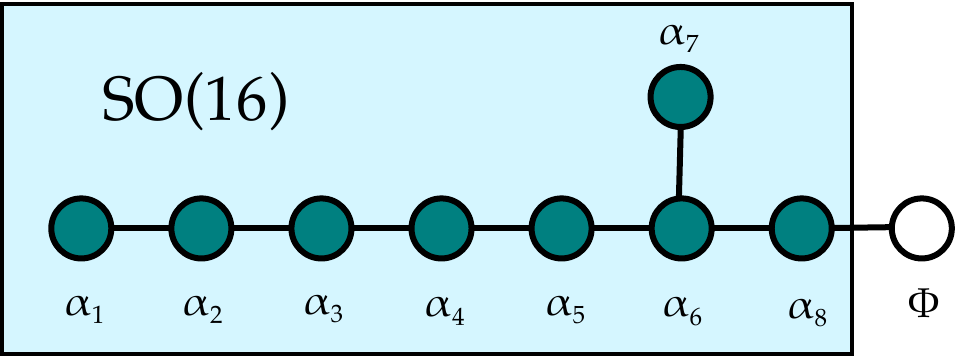}
\ee

The intersection numbers $(D_{i} \cdot F_j)_{ij}$ give the negative Cartan
matrix of $SO(16)$.  The additional node $F_\Phi \hookrightarrow D_{\Phi}$
corresponds to the ``extra'' root of $E_8$ missed by the $SO(16)_\text{BG}$
embedding (see \eqref{eq:SO16_in_affine_E8}) and has intersection numbers
\begin{align}
  D_\Phi \cdot F_\Phi = -2, \, \quad D_{i} \cdot F_\Phi = D_{\Phi} \cdot F_i = \delta_{i 8} \, .
\end{align}
These intersection numbers are independent of the resolution phase, as they
pertain to the codimension one fibers.  Likewise, we have for every phase
\begin{align}
  D_{i} \cdot f = S \cdot F_i = S \cdot F_\Phi =0 \, , \qquad i = 1 , \cdots, 8 \,.
\end{align}

The different phases are given by the decorated box graphs in figure
\ref{fig:Rank1BGALL}.  In the following, we construct the fibers for these
phases.  The result, including the flavor group $G_\text{F}$ of the SCFT (if
existent), is listed in table \ref{tab:rank1BG}, together with the
decorated box graphs for completeness.

\begin{table}\centering
\begin{tabular}{|c|c|c|c|c|}\hline
Phase &Topology of $S$&  Codim.~2 Fiber  & Box Graph  & $G_\text{F}$\cr \hline 
I & gdP$_9$& \includegraphics[width=4cm]{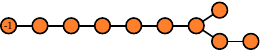} &  \includegraphics[width=3.5cm]{SO16xSU2BGI.pdf} &\cr \hline 
II &gdP$_8$&  \includegraphics[width=4cm]{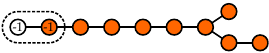} & \includegraphics[width=3.5cm]{SO16xSU2BGII.pdf} & $E_8$ \cr \hline 
III &gdP$_7$&  \includegraphics[width=4cm]{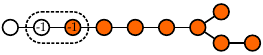} &  \includegraphics[width=3.5cm]{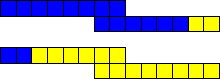} &$E_7$ \cr \hline 
IV& gdP$_6$&  \includegraphics[width=4cm]{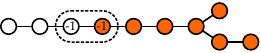} &  \includegraphics[width=3.5cm]{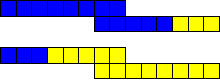} &$E_6$ \cr \hline 
V&gdP$_5$&  \includegraphics[width=4cm]{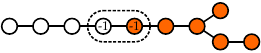} &  \includegraphics[width=3.5cm]{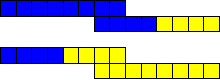} &$SO(10)$ \cr \hline 
VI&gdP$_4$&  \includegraphics[width=4cm]{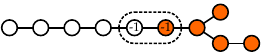} &  \includegraphics[width=3.5cm]{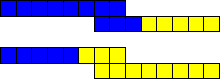} &$ SU(5)$ \cr \hline 
VII& gdP$_3$&  \includegraphics[width=4cm]{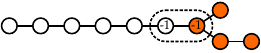} & \includegraphics[width=3.5cm]{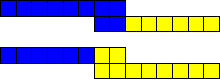} & $SU(2)\times SU(3)$ \cr \hline 
VIII & gdP$_2$&  \includegraphics[width=4cm]{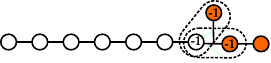} &  \includegraphics[width=3.5cm]{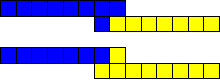} &$SU(2)\times U(1)$ \cr \hline 
IX& dP$_1$&  \includegraphics[width=4cm]{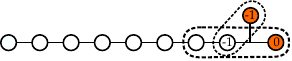} &  \includegraphics[width=3.5cm]{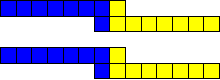} &$ U(1)$ \cr \hline 
X& gdP$_1 \cong \mathbb{F}_2$
&\includegraphics[width=4cm]{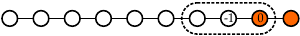} & \includegraphics[width=3.5cm]{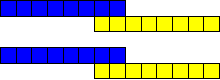} & $ SU(2)$ \cr \hline 
\end{tabular}
\caption{
Box graphs and codimension two fibers corresponding to the rank one 5d SCFTs. 
The surface $S$ with given topology, which in an M-theory realization would supply the $SU(2)$ gauge theory description, contains the codimension two fiber that is shown.
Obtained in \cite{Apruzzi:2019opn} from non-flat resolutions, all curves have self-intersection $-2$ inside $S$, except when otherwise noted. 
The orange colored rational curves are those contained in the surface component $S$. 
The flavor curves are the fully contained $(-2)$-curves (colored and unlabeled). 
We furthermore list the box graph/Coulomb branch phase for the gauge theory description of $SU(2)_{\text{gauge}}\times SO(16)_{\text{BG}}$ with matter in the $(\bm{2}, \bm{16})$. The last column contains the enhanced flavor symmetry $G_\text{F}$ of the 5d SCFT obtained from shrinking $S$.}\label{tab:rank1BG}
\end{table}


\paragraph{\bf Phase I:}
The box graph for phase I is
\be\label{eq:rank_one_decorated_BG_phase_I}
\includegraphics[width=4cm]{SO16xSU2BGI.pdf} \,,
\ee
The only curve that splits in codimension two is the root of the $SU(2)$,
i.e., $f$, which maps from the lower representation graph of the $\bm{16}$ to
the upper.  There are two extremal weights, which are identified due to the
pseudo-reality of the representation
\be
	L_{1,16}^+ = L_{2,1}^- \, .
\ee
For definiteness, we will work with $L_{2,1}^-$, and associate a minus-sign with the corresponding extremal curve $C_{2,1}^- \equiv C^-$.

The box graph dictates the following splitting for $f$ 
\begin{equation}\label{eqn:Ifsplit}
  f \rightarrow 2 C^- + 2 F_1 + 2 F_2 + 2 F_3 + 2 F_4 + 2 F_5 + 2 F_6 + F_7 + F_8 \, ,
\end{equation}
where the only non-zero intersections of $C^-$ are
\begin{equation}
  \langle \alpha_1^\vee, L_{2,1} \rangle = D_{1} \cdot C^- = 1 \,, \quad \langle \alpha_{SU(2)}^\vee, L_{2,1} \rangle = S \cdot C^- = -1 \,.
\end{equation}
Thus, the extremal curve $C^-$ is attached to $F_1$, and is contained in $S$.
By rule 4 in the previous subsection, all $F_i$ must be contained in $S$ as
well.

It remains to analyze the fate of $F_\Phi$ --- the extra curve, that is not
contained in the data of the gauge theory description.  Without any further
input, we have to consider the explicit resolution presented in Part I
\cite{Apruzzi:2019opn}, where the geometry was determined to be gdP$_9$.  The
geometry implies that $F_\Phi$ is irreducible and is fully contained in $S$.
The corresponding fiber is depicted at the top of table \ref{tab:rank1BG}.

With this  geometric input, we can compute the  intersections $\langle
\Phi^\vee , L_{i,j} \rangle$ for all weights $L_{i,j}$ of $SU(2)_\text{gauge}
\times SO(16)_\text{BG}$, which we can use for the subsequent phases.  First,
there are sixteen effective curve classes corresponding to the weights $L^-_{2,j}$,
$j=1,\ldots,16$ (see figure \ref{fig:SU2SO16bare}) corresponding to the eight
$SU(2)_\text{gauge}$ flavors which transform as a half-hypermultiplet in the
bifundamental of $SU(2)_\text{gauge} \times SO(16)_\text{BG}$.  In the
decorated box graph \eqref{eq:rank_one_decorated_BG_phase_I}, they are:
\begin{align}\label{eq:weight_curves_rank_one_phase_I}
\begin{split}
  & L^-_{2,1} \ \leftrightarrow\  C^- \cr 
  & L^-_{2,2} \ \leftrightarrow\  C^- + F_1 \cr 
  & \qquad\vdots \cr 
  & L^-_{2,8} \ \leftrightarrow\  C^- + \sum_{i=1}^7 F_i  \cr 
  & L^-_{2,9} \ \leftrightarrow\  C^- + \sum_{i=1}^6 F_i + F_8 \cr 
   & L^-_{2,10} \ \leftrightarrow\  C^- + \sum_{i=1}^6 F_i + 2F_7 + F_8 \cr 
  & L^-_{2,11} \ \leftrightarrow\  C^- + \sum_{i=1}^5 F_i + \sum_{i=6}^7 2F_i + F_8 \cr 
  & \qquad \vdots \cr 
  & L^-_{2,16} \ \leftrightarrow\  C^- + \sum_{i=1}^7 2 F_i + F_8 \, .
\end{split}
\end{align}
From the codimension one affine $E_8$
fiber, we know that $D_\Phi \cdot F_8 =1$ and $D_\Phi
\cdot F_i = 0$ for $i \neq 8$, and furthermore we determined that $D_\Phi
\cdot C^- = 0$.  Thus, the curves in
\eqref{eq:weight_curves_rank_one_phase_I} associated to the weights $L_{2,j}$
of the box graphs have intersections with the additional divisor $D_\Phi$ as
follows
\begin{align}\label{eq:Phi-charges_rank_one}
  D_\Phi \cdot C(L^-_{2,j}) = \langle \Phi^\vee, L_{2,j} \rangle = 
  \begin{cases}
    0 \, , \quad \text{if } j < 9 \, ,\\
    1 \, , \quad \text{if } j \geq 9 \, .
  \end{cases}
\end{align}
Finally, the box graphs also determine the charges of the conjugate states via
$L_{1, \, j+8} = -L_{2, \, 9-j}$.

\paragraph{\bf Phases II--VII:} 
Denote the phase given by the Roman numeral corresponding to $n + 1$ by the
Arabic numeral $n$, i.e., phase II corresponds to $n=1$, phase III to $n=2$,
etc.  Then the splitting dictated by the box graph is
\begin{equation}
  \begin{aligned}
    F_n &\ \rightarrow\  C_{1,16-n}^+ + C_{1,17-n}^- \, , \cr
    f &\ \rightarrow\ 2 C_{1,16-n}^+ + F_7 + F_8 + \sum_{j=n+1}^6 2 F_j \,.
  \end{aligned}
\end{equation}
Recall $F_i$ are the curves associated to the roots of the
$SO(16)_{\text{BG}}$, which act along the $\bm{16}$ representation, and $f$
the one of the $SU(2)_{\rm gauge}$.  The intersections of the curves with the
Cartan divisors $D_{i}$ and $S$ can be read off from the associated weight in
the box graph.  From figure \ref{fig:SU2SO16bare}, the non-zero numbers are
\begin{align}
\begin{split}
  & (D_{{n-1}} , D_{n}, D_{{n+1}} , S) \cdot C^+_{1,16-n} = (0, -1, 1, -1) \, , \\
  & (D_{{n-1}} , D_{n}, D_{{n+1}} , S) \cdot C^-_{1,17-n} = (1, -1, 0, 1)  \,.
\end{split}
\end{align}
For  $n = 1$  there is of course no $D_{0}$, and these terms are thus ignored.
Furthermore, \eqref{eq:Phi-charges_rank_one} implies $D_\Phi \cdot
C^+_{1,16-n} = D_\Phi \cdot C^-_{1,17-n} =0$ for $1 \leq n \leq 6$.  This
means $F_\Phi$ does not split into any of these components.  Similarly as in
the case for Phase I, the curve $F_8$ is contained inside of $S$, and thus
$F_\Phi$, since it does not split, must also be contained. The fibers and box
graphs are shown in table  \ref{tab:rank1BG}.

The weakly coupled 5d gauge theory of these phases is $SU(2)_\text{gauge} +
(8-n) {\bf F}$.  Therefore the weakly coupled flavor group is $G_\text{F,cl} =
SO(16-2n)$, which including the $U(1)_T$ gives rank $M = 9-n$ for
$G_\text{F}$.  In these phases, all ranks are accounted for by the shrinking
simple roots including $\Phi$, so there is no abelian factor and one has
$G_\text{F} = E_{9-n}$.

\paragraph{\bf Phase VIII:} In this phase the splitting is 
\begin{equation}
  \begin{aligned}
    F_7 &\rightarrow C_{1,9}^+ + C_{1,10}^- \, ,\cr
    F_8 &\rightarrow C_{1,8}^+ + C_{1,10}^- \,, \cr
    f &\rightarrow C_{1,8}^+ + C_{1,9}^+ \,,
  \end{aligned}
\end{equation}
with non-trivial intersection numbers
\begin{align}
  \begin{split}
    (D_{6}, D_{7}, D_{8}, S) \cdot C^+_{1,8} & = (0, 1,-1,-1) \, , \\
    (D_{6}, D_{7}, D_{8}, S) \cdot C^+_{1,9} & = (0, -1,1,-1) \, , \\
    (D_{6}, D_{7}, D_{8}, S) \cdot C^-_{1,10} & = (1,-1,-1,1) \, .
  \end{split}
\end{align}
By \eqref{eq:Phi-charges_rank_one}, we then have
\begin{equation}
  D_{\Phi} \cdot C_{1,9}^+ = D_{\Phi} \cdot C_{1,10}^- = 0 \,, \quad D_{\Phi} \cdot C_{1,8}^+ = 1 \,.
\end{equation}
These intersection numbers are again consistent with a non-splitting of $F_\Phi$,
which intersects the split curve $F_8$ at the component $C^+_{1,8}$.  Because
$C^+_{1,8}$ is contained in $S$, so $F_\Phi$ must be contained (see table
\ref{tab:rank1BG}).

At weak coupling, this phase has a 5d $SU(2)_\text{gauge} + 1{\bf F}$
description with flavor symmetry $G_\text{F,cl} = SO(2) \cong U(1)$ as well as the $U(1)_T$ symmetry,
consistent with no $F_i$ fully wrapped.  However, by passing to strong
coupling, there is a non-trivial enhancement induced by the $U(1)_T$, leading
to a non-abelian $SU(2)_\text{F}$ part indicated by $F_\Phi$ being wrapped.
To preserve rank, we must then have $G_\text{F} = SU(2) \times U(1)$.

\paragraph{\bf Phase IX:} The box graph for phase IX implies the splitting 
\begin{equation}
  F_7 \rightarrow 2 C_{1,8}^- + F_8 + f \,.
\end{equation}
These three curves arrange as
\begin{align}
    \ldots - F_6 - F_8 - C^-_{1,8} - f \, ,
\end{align}
in order to comply with the intersection numbers $S\cdot F_7 = S \cdot F_8 = D_{7} \cdot f =0$, as well as the weights of the curve $C^-_{1,8}$:
\begin{align}
  (D_{7}, D_{8}, S) \cdot C^-_{1,8} = (-1,1,1) \, .
\end{align}
These intersection numbers in turn determine, as we know from
(\ref{eq:Phi-charges_rank_one}),
\begin{align}
  D_{\Phi} \cdot C^-_{1,8} = -1 \, .
\end{align}
This means that $D_\Phi$ contains $C^-_{1,8}$, and hence the fiber component $F_\Phi$ must split,
\begin{equation}\label{eq:Phi_splitting_phase_IX}
  F_\Phi \rightarrow C_{1,8}^- + \Gamma \,.
\end{equation}
Since we know $S \cdot F_\Phi = 0$, we can compute the intersection numbers of
the new curve $\Gamma$,
\begin{equation}
  (D_{1}, \cdots, D_{8}, D_{\Phi}, S) \cdot \Gamma = (0,0,0,0,0,0,1,0,-1,-1) \,.
\end{equation}
Thus $\Gamma \subset S$.  Since $\Gamma$ is not a curve appearing in the box
graph, it has to be a (multi-)section of the ruling, and hence $\Gamma \cdot_S f \geq 1$\footnote{
	In the explicit resolution, cf.~Part I \cite{Apruzzi:2019opn}, one can explicitly show $\Gamma \cdot_S f = 1$.}.
	
On the other hand, $\Gamma$ and $C^-_{1,8}$ must be also attached due to the
splitting \eqref{eq:Phi_splitting_phase_IX}.

Naively, it would appear as if $C^-_{1,8}$, which
lies outside of $S$, had two different intersection points with each $f$ and
$\Gamma$ that are in $S$. 
This is clearly in violation of $S \cdot C^-_{1,8}
= 1$.  The resolution of this puzzle is that the point in which $S$ and
$C^-_{1,8}$ intersect is also an intersection point of $f$ and $\Gamma$ inside $S$.
The resulting fiber picture is depicted in table \ref{tab:rank1BG}.  Since the
weak coupling $SU(2)_\text{gauge}$ description has no matter, there is only
the topological $U(1)_T$.  As $F_\Phi$ splits off $C_{1,8}^-$ which is not
wrapped, there is no non-abelian enhancement, and we have $G_\text{F} = U(1)$.

Note that because the $(-1)$-curve $\Gamma$ is not part of the box graph, the
gauge theory description of the geometry does not see the transition
corresponding to flopping this curve.  Nevertheless, to consistently embed the
$SO(16)_\text{F}$ box graph splittings into the affine $E_8$ Dynkin diagram
requires the existence of this curve, which we can now flop.  The resulting
{\bf Phase XI} has no gauge theory description and no flavor symmetry.
Obviously, this phase is not visible in the gauge theoretic approach, but
nevertheless part of the geometric classification, see
\cite{Morrison:1996xf,Jefferson:2018irk,Apruzzi:2018nre,Apruzzi:2019opn}.

\paragraph{\bf Phase X:} The splitting is
\begin{equation}
  F_8 \ \rightarrow \ 2 C_{1,9}^- + F_7 + f \,.
\end{equation}
From the intersections, 
\begin{align}
  (D_{7}, D_{8}, S) \cdot C^-_{1,9} = (1,-1,1) \, ,
\end{align}
it follows that $C_{1,9}^-$ is not contained inside the surface $S$.
Furthermore, we have $D_{7} \cdot C^-_{1,9} = 1$ and $D_{7} \cdot f =0$,
corresponding to a fiber with
\begin{align}
  \ldots - F_6 - F_7 - C^-_{1,9} - f - \ldots  \, .
\end{align}

Now, because we have $D_\Phi \cdot C^-_{1,9} = D_\Phi \cdot F_7 = 0$, $F_\Phi$
does not split.  However, since $D_\Phi \cdot F_8 = 1$, we see that $F_\Phi$
must attach to the above chain at the curve $f$ to preserve the intersection
number.  Then, because $f$ is contained in $S$ as its ruling, $F_\Phi$ must as
well be a curve in $S$ to satisfy $S \cdot F_\Phi = 0$.  We can see the
structure in table \ref{tab:rank1BG}.

The wrapped $(-2)$-curve $F_\Phi$ gives the $SU(2)_\text{F}$ flavor group in
the SCFT limit of the pure $SU(2)_\text{gauge}$ gauge theory.  Note that in
this case, we see that the surface $S$ does not contain any $(-1)$-curves.
This not only explains the absence of any massless charged matter at weak
coupling, but also why this phase cannot be further flopped geometrically,
i.e., the SCFT does not have any further mass deformations.


\subsection{Classification of Rank 2 Theories from Box Graphs}

From the Coulomb branch phases/box graphs of the rank one 5d gauge theories we
learned two things: the box graphs give a succinct representation-theoretic
description of all the Coulomb branch phases --- and thereby characterization
of all 5d SCFTs with a weakly-coupled gauge theory description, as well as the
mass deformations and RG-flows connecting them.  Secondly, the box graphs
reconstruct the geometry, in particular curves that are contained in the
compact divisors of the M-theory realization. These in turn determine flavor
symmetries of the 5d UV fixed point theories. The only additional input that
is necessary is the embedding of the weakly-coupled flavor symmetry of
the marginal theory $G_\text{BG}$, which is fixed with one geometric input (in rank
one, this is the embedding of $SO(16)$ into $\widehat{E}_8$, which determines how the
additional curve $F_\Phi$ is attached).  This approach also provides a gauge
theoretic counterpart to the geometric classification and properties of the
rank two 5d SCFTs \cite{Jefferson:2018irk,Apruzzi:2019opn}.  The corresponding
marginal theories have weakly-coupled gauge theory descriptions as summarized
in Part I, appendix A \cite{Apruzzi:2019opn}.

\subsubsection{Marginal Theories and Box Graphs}

We present all rank two gauge/quiver descriptions, and determine their box
graphs and descendant trees in appendix \ref{app:RankTwoBG}.  Let us briefly
summarize the theories here: 

The marginal 5d theory arising from the rank two E-string theory on $S^1$ has
the following weakly coupled gauge theory descriptions: 
\begin{itemize}
\item $SU(3)_{\frac32} + 9 {\bm F}$ ,
\item $Sp(2) + 8{\bm F}+ 1 \bm{AS} $ ,
\item $5 {\bm F} - SU(2) - SU(2) - 2{\bm F}$ .
\end{itemize}
The box graphs for these theories are shown in figures \ref{fig:EstringSU3},
\ref{fig:EstringSp2}, and \ref{fig:EstringSU2SU2}, respectively. 

For the minimal $(D_5, D_5)$ conformal matter theory on $S^1$, there are also
three weakly coupled descriptions of the marginal theory: 
\begin{itemize}
\item $SU(3)_{0} + 10 {\bm F}$ ,
\item $Sp(2) + 10{\bm F}$ ,
\item $4 {\bm F} - SU(2) - SU(2) - 4{\bm F}$ .
\end{itemize}
The box graphs for these, and the descendant theories are shown in figures
\ref{fig:D5D5SU3}, \ref{fig:D5D5Sp2}, and \ref{fig:D5D5SU2SU2}, respectively. 

There are additional 5d marginal gauge theories which act as starting points
for RG-flows, which are discussed in appendix \ref{app:RankTwoBG}. There are a
few observations to be made: the tree structure matches that of the geometric
transitions/flops as well as CFD-transitions for rank two SCFTs. Furthermore,
the theories match precisely with those that are known to have a gauge theory
description. This is strong evidence that the flavor-equivalence classes of
box graphs captures these theories correctly. 

To make further use of these Coulomb branch descriptions, we need to add, much
like in the rank one case, the information about the embedding of the flavor
symmetry of the marginal theory into the 6d flavor symmetry.  Once we have
supplemented the box graphs with this information, the superconformal flavor
symmetries of all descendants can be read off as well --- this is already
included in the figures in appendix \ref{app:RankTwoBG}.

\subsubsection{Phases to Fibers}

We exemplify this now by studying the rank two E-string and $(D_5, D_5)$
minimal conformal matter theories, that have both a description in terms of a
marginal $SU(3)$ gauge theory.

Recall that geometrically, the theories descending from the rank two E-string
are obtained from M-theory on an elliptically fibered Calabi--Yau threefold,
with a non-minimal singularity from an $(E_8, SU(2))$ collision.  Let us
denote the affine $E_8$ fiber components and corresponding non-compact
divisors by $F^{E_8}_i \hookrightarrow D^{E_8}_i$, $i=1,\cdots,9$ as shown in
\eqref{U9IntoE8}\footnote{We choose this slightly non-standard enumeration as
this is more natural when identifying the embedding of the $U(9)$ flavor
symmetry.}, and the two $SU(2)$ components by $F^{SU(2)}_i \hookrightarrow
D^{SU(2)}_i$, $i=1,2$.

The 5d theories descending from circle compactifications of this 6d model have rank two.
From an explicit blow-up resolution (see appendix \ref{sec:explicit_res_E8SU2}) one can read off the three different 5d gauge theory descriptions listed above. 
In the following we will discuss the fiber reconstruction from the box graphs from the $SU(3)$ gauge theory description. 

For the rank two E-string on $S^1$, the marginal theory is $SU(3) +  9 \bm{F}$, and the descendants are characterized in terms of box graphs for $SU(3) \times U(9)_\text{BG}$. 
There is essentially one unique way to embed the eight roots $F_i$ of $SU(9) \subset U(9)$ into the codimension one fibers of $E_8 \times SU(2)$, namely, into the connected chain of eight nodes inside the affine $E_8$ diagram.
We fix the ambiguity of ordering by the identification:
\be\label{U9IntoE8}
\includegraphics*[width = 8.5cm]{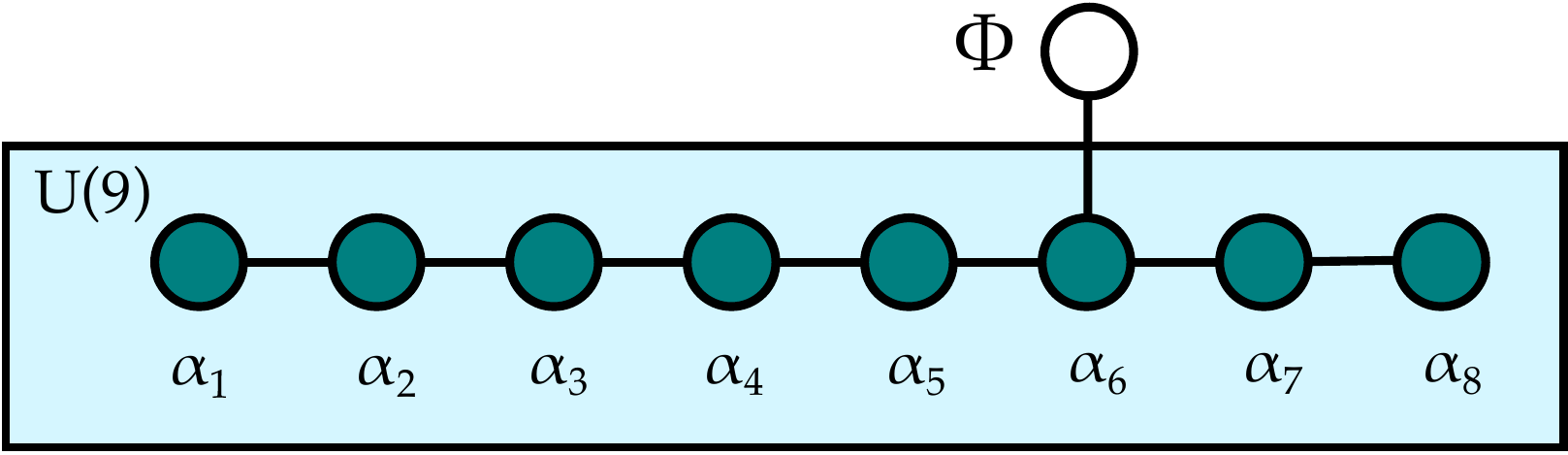} \, ,
\ee
where $F_\Phi \equiv F^{E_8}_9$ is the additional node that the gauge theory phase does not capture. 
This leaves codimension one fibers $F^{E_8}_\Phi \hookrightarrow D_\Phi^{E_8}$ and $F^{SU(2)}_{1,2} \hookrightarrow D_{1,2}^{SU(2)}$ out of the box graphs, which may be interpreted as additional coroots $\Phi^\vee \equiv \Phi_0^\vee$ and $\Phi^\vee_{1,2}$, respectively.
As in the rank one cases, we first determine the intersections $\langle \Phi_l^\vee, L_m \rangle$ of the box graph weights $L_m$.
We do this in the explicit resolution detailed in appendix \ref{sec:explicit_res_E8SU2}.
From the intersection numbers \eqref{eq:int_numbers_extremal_curves_E8SU2_res} between the extremal curves associated with these sign assignments and the divisors $D^{E_8}_\Phi$ and $D_{1,2}^{SU(2)}$, we can then infer the intersections of all the curves.
We collect this information in the representation graph in figure \ref{fig:intersection_rep_graph_SU9xSU3}.

\begin{figure}
  \centering
  \includegraphics[width=.7\hsize]{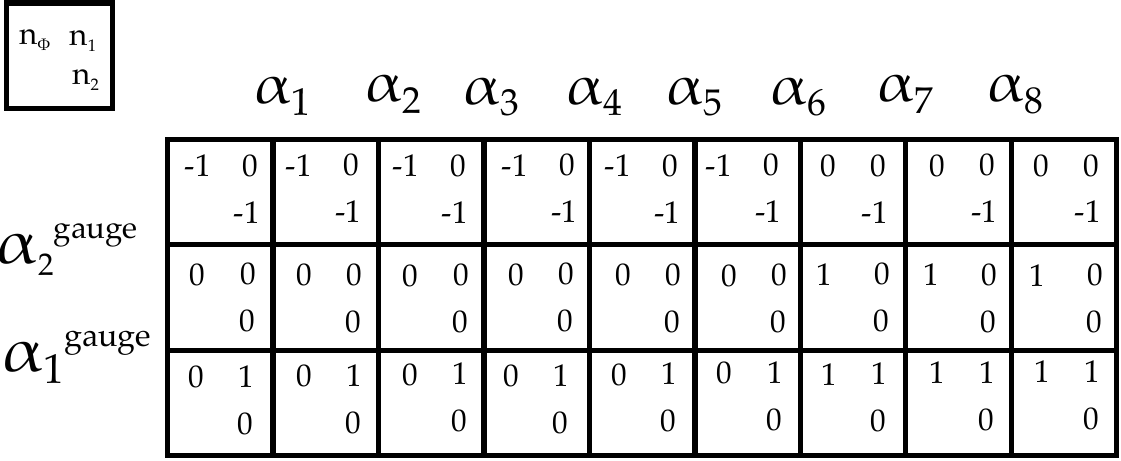}
  \caption{The pairings $n_l = \langle \alpha_l^\vee, L \rangle$ between the $({\bf 3,9})$-weights $L$ and the coroots $\alpha_l^\vee$ that are not part of $U(9)_\text{BG}$.
  For any decoration of this box graph, these numbers encode via \eqref{eq:weight_vector_intersection_number} the geometric intersections with additional non-compact divisors $D_{\Phi_l}$ over the affine $E_8$ and $SU(2)$ fibers.
  Specifically, $n_\Phi$ denotes the pairing with the coroot $D_\Phi$ (the additional $E_8$ coroot), and $n_{1,2}$ those with $D^{SU(2)}_{1,2}$.
  }\label{fig:intersection_rep_graph_SU9xSU3}
\end{figure}
To clarify the process, we provide three concrete examples in figure \ref{fig:U9SU3Examples}. The fibers are precisely the ones we discussed in Part I \cite{Apruzzi:2019opn} from a direct resolution computation of non-minimal singularity.

\begin{figure}
\centering
  \subfloat[$SU(3)\times U(9)$ Box Graph.]{\includegraphics*[width= 6cm]{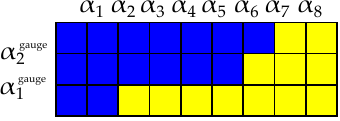}}
  \qquad  \qquad
  \subfloat[Rank 2 E-string codim 2 fiber for (a) (BU$_1^{(E_8, SU(2))}$).]{{\includegraphics[width=7cm]{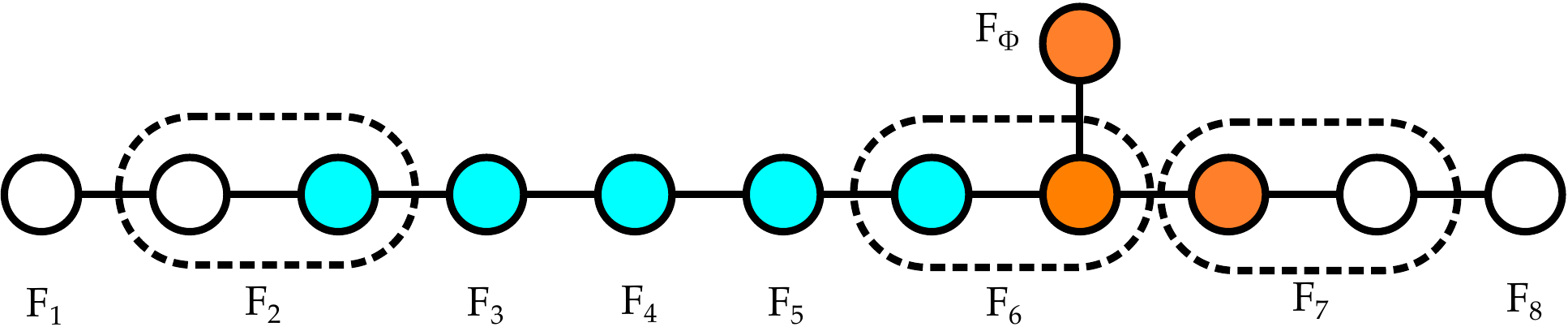}}}\\

 \subfloat[$SU(3)\times U(9)$ Box Graph.]{\includegraphics*[width= 6cm]{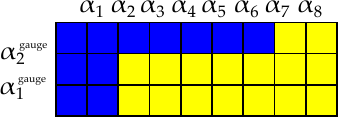}}  \qquad  \qquad
  \subfloat[Rank 2 E-string codim 2 fiber for (c) (BU$_2^{(E_8, SU(2))}$).]{{\includegraphics[width=7cm]{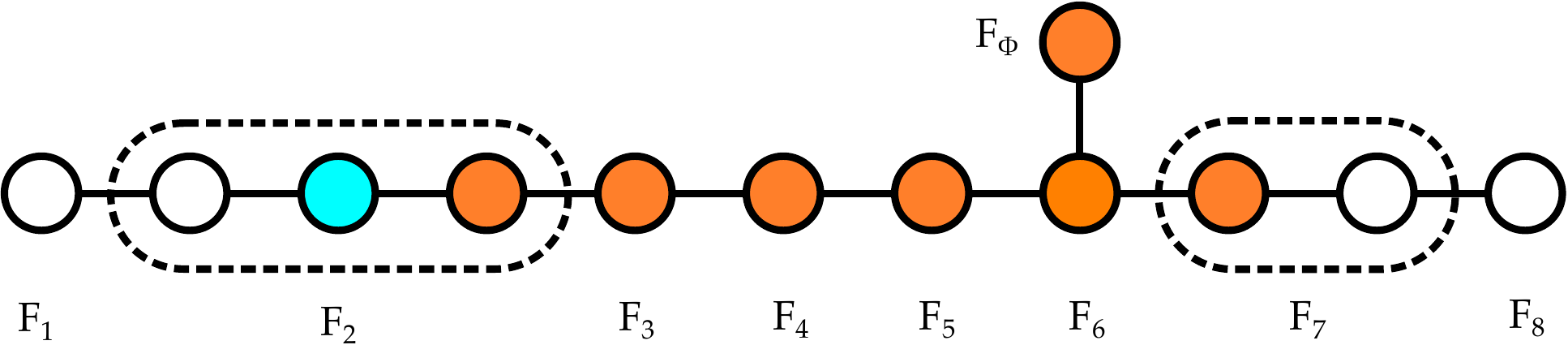}}}\\

 \subfloat[$SU(3)\times U(9)$ Box Graph.]{\includegraphics*[width= 6cm]{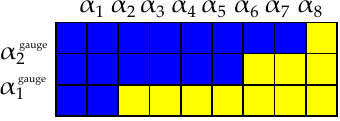}}
  \qquad  \qquad
  \subfloat[Rank 2 E-string Codim 2 fiber for (e) (BU$_3^{(E_8, SU(2))}$).]{{\includegraphics[width=7cm]{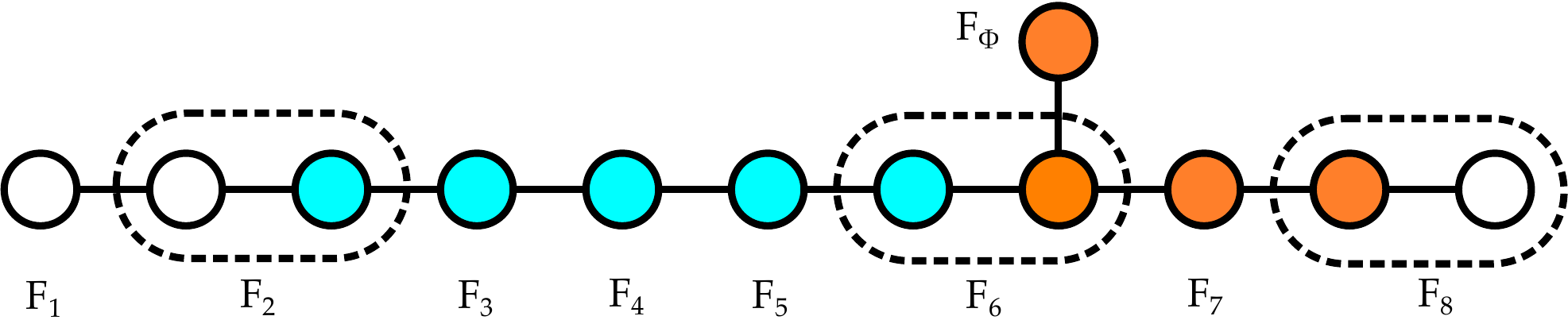}}}\\
  
 \subfloat[$SU(3)\times U(10)$ Box Graph.]{\includegraphics*[width= 6cm]{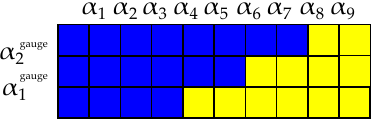}}\qquad \qquad 
  \subfloat[$(D_5, D_5)$ Codim 2 fiber for (g)(BU$_1^{(D_{10}, I_1)}$).]{{
  \includegraphics[width=7cm]{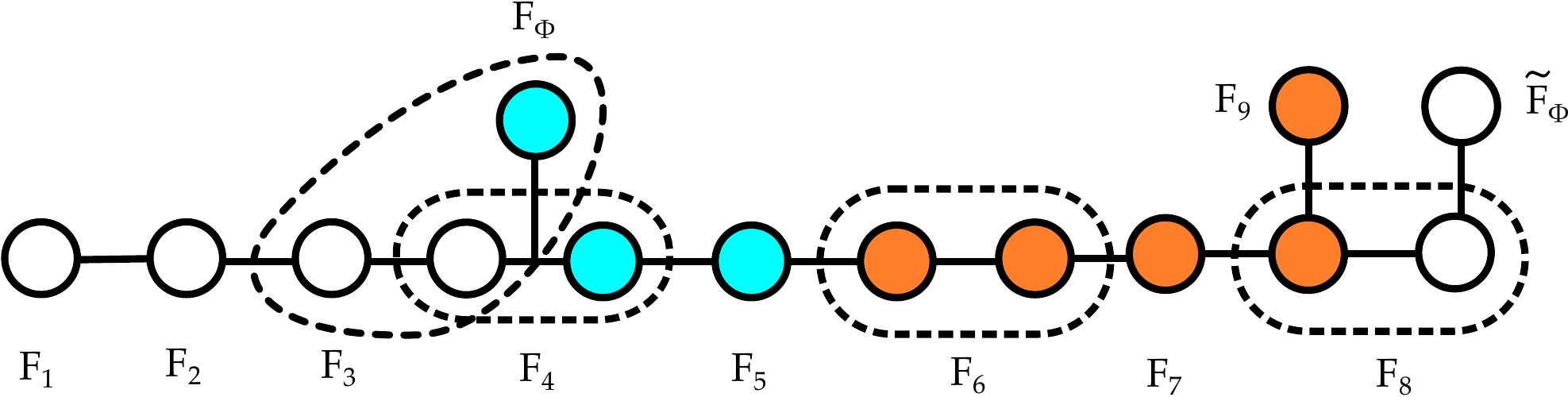}}}\\

   \caption{Examples, (a) to (f), of box graphs and the associated codimension two fibers for $SU(3)_\text{gauge} \times U(9)_\text{BG}$ phases of the rank two E-string, as well as one example, (g) and (h), for $(D_5, D_5)$ conformal matter; these examples have appeared in Part I, \cite{Apruzzi:2019opn}, in terms of explicit resolutions.
   Note that (a) and (c) are in the same flavor equivalence class. 
   The roots for $G_{\text{BG}}$ are denoted $\alpha_i$ and associated curves $F_i$. The roots of the gauge group are $\alpha_i^{\text{gauge}}$, which are dual to the compact surfaces $S_i$, $i=1,2$.
   The colors cyan/orange on the RHS indicate which codimension one curves $F_i$ are contained in which $S_i$.
Note that for the rank two E-string the $SU(2)$ part of the fiber also splits, but does not contribute in these examples to the flavor symmetry and we omit to draw it.}
  \label{fig:U9SU3Examples}
\end{figure}

We can repeat the exercise for the marginal $SU(3) + 10{\bm F}$ gauge description of the $S^1$-reduction of the $(D_5, D_5)$ minimal conformal matter theory.
 The box graphs for all descendants are shown in figure \ref{fig:D5D5SU3}.
To reconstruct the fiber in these cases, we consider the embedding of the classical flavor symmetry $U(10)_\text{BG}$ into 6d superconformal flavor symmetry, i.e., the affine $SO(20)$ Dynkin diagram
\be
\includegraphics*[width=9cm]{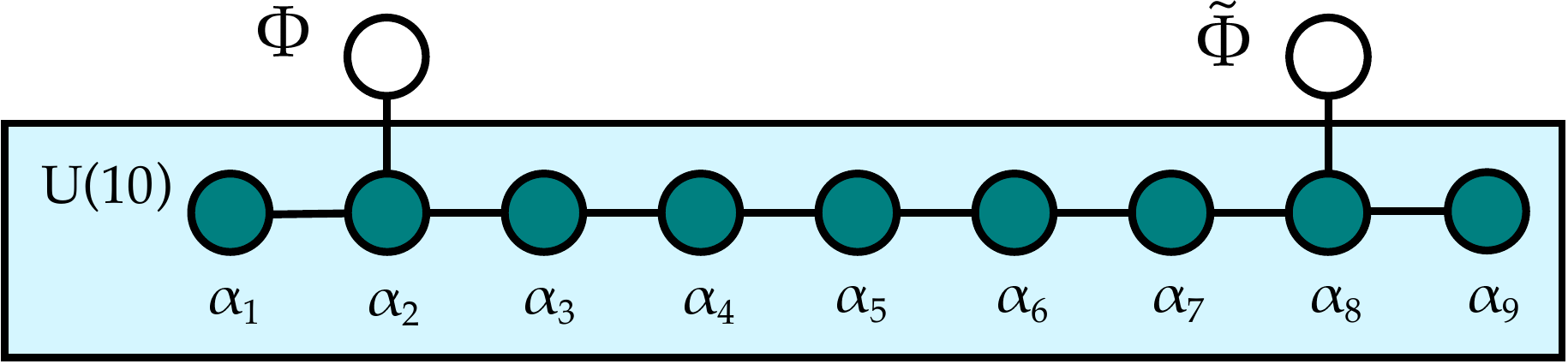} \, .
\ee
Let us denote the curves associated to the ``extra''
nodes of the affine $SO(20)$ codimension one fiber as $F_\Phi \hookrightarrow D_\Phi$ and $F_{\widetilde{\Phi}} \hookrightarrow D_{\widetilde{\Phi}}$, where we fix the ordering by
\begin{equation}
  D_{F_\Phi} \cdot F_2 = D_{2} \cdot F_\Phi = 1 \,, \quad D_{\widetilde{\Phi}} \cdot F_8 = D_8 \cdot F_{\widetilde{\Phi}} = 1 \,,
\end{equation}
and zero intersection with all other components of the affine $SO(20)$ fiber.
These corresponds to roots $\Phi, \widetilde\Phi$ that do not feature in the flavor group $U(10)_\text{BG}$ of the marginal $SU(3)_\text{gauge}$ description.
From a concrete resolution, one can determine the corresponding coroots having the following pairings with the $({\bf 3, 10})$ weights $L_{i,j}$ in the top and bottom row of a decorated $SU(3)_\text{gauge} \times U(10)_\text{BG}$ box graphs:
\begin{equation}
  \begin{aligned}
    \begin{pmatrix} \langle \Phi^\vee, L_{1,j}^\pm \rangle \cr
      \langle {\widetilde\Phi}^\vee, L_{1,j}^\pm \rangle 
    \end{pmatrix} = \mp \begin{pmatrix} 1 & 1 & 0 & 0 & 0 &
      0 & 0 & 0 & 0 & 0 \cr
      1 & 1 & 1 & 1 & 1 & 1 & 1 & 1 & 0 & 0
    \end{pmatrix}
  \end{aligned} \,,
\end{equation}
and
\begin{equation}
  \begin{aligned}
    \begin{pmatrix} \langle \Phi^\vee, L_{3,j}^\pm \rangle \cr
      \langle {\widetilde\Phi}^\vee, L_{3,j}^\pm \rangle 
    \end{pmatrix} = \mp \begin{pmatrix} 0 & 0 & -1 & -1 & -1 &
      -1 & -1 & -1 & -1 & -1 \cr
      0 & 0 & 0 & 0 & 0 & 0 & 0 & 0 & -1 & -1
    \end{pmatrix}
  \end{aligned} \,.
\end{equation}
Using the same methods as for the $SU(3)_\text{gauge} \times U(9)_\text{BG}$ phases above, one can determine with this information whether the fibers $F_\Phi$, $F_{\widetilde{\Phi}}$ are contained in the compact surface for any decorated $SU(3)_\text{gauge} \times U(10)_\text{BG}$ box graph.
An example is shown in figures (g) and (h) of \ref{fig:U9SU3Examples}.

\section{SCFTs/CFDs from Box Graphs}
\label{sec:CFDsfromBG}

As we have already argued in section \ref{sec:somesingle}, the box graphs can be condensed into so-called flavor equivalence classes.
In this section, we show explicitly how this reduction of redundant information is mimicked in the (BG-)CFD representation of the geometry.

To begin with, let us first recall from section \ref{sec:somesingle} that the flavor equivalence classes are characterized by a set $\alpha_j^\text{BG}$ of flavor roots which are contained in the \emph{combined} splitting \eqref{SplitBG} of gauge roots in this equivalence class.
This splitting can be inferred from the reduced box graphs.
Geometrically, these flavor roots correspond to the set of codimension one curves $F_i$ which are contained in the \emph{sum} of all the rulings $f_j$ which split according to \eqref{eq:fiber_splitting_general}.
Since flops between flavor equivalent phases by definition do not change this overall splitting, the particular set of flavor roots remain parts of the rulings on ${\cal S}$ in all phases, and hence collapse in the non-abelian gauge enhancing limit, giving rise to the classical flavor symmetry $G_\text{F,cl}$ of this equivalence class.
Moreover, the set of $(-1)$-curves corresponding to the F-extremal weights also remain, by definition, invariants within a flavor equivalence class.
Note that in some phases, the curve may be reducible; however --- again by definition --- there always exists a phase in which the curve does not split.
Thus, the BG-CFDs precisely correspond to the intersection pattern of the curves associated with the F-extremal weights and the flavor roots contained in the splitting of gauge roots.

The embedding of the BG-CFDs into the full CFD also played an important role in our discussion of constraining possible gauge descriptions of SCFTs.
Again, this is based on the underlying geometry and intersection theory.
In the following, we will show how the reduced box graphs, that specify the BG-CFDs, also determine superconformal flavor symmetry once we specify the attachments of the ``missing'' codimension one components $F_{\Phi_l} \hookrightarrow D_{\Phi_l}$.

\subsection{Box Graphs to Superconformal Flavor Symmetry and CFDs}


The argument follows the same logic as in the previous section, where we have discussed in detail how to determine the explicit fiber structure from a decorated box graph.
Namely, we need to clarify if the extra roots $\Phi_l$ are part of the CFD or not.
This requires the minimal geometric input in form of the pairings
\begin{align}
	{\cal E}_{l,n} = \langle \Phi_l^\vee , L_n \rangle 
\end{align}
between the extra coroots $\Phi_l^\vee$ and the F-extremal weights $L_n^{\epsilon_n}$, which translates into geometric intersections between F-extremal curves $C_n^{\epsilon_n}$ and divisors $D_{\Phi_l}$.
At the level of flavor equivalence classes, consistency of intersection numbers and the reduced box graphs themselves implies the following rules on how to attach the node $\Phi_l$ to the BG-CFD:

\begin{enumerate}
\item If $- \epsilon_n \, \mathcal{E}_{l, n} < 0$ for an F-extremal weight $L_{n}^{\epsilon_n}$, that is \emph{not} contained in the BG-CFD, then $\Phi_l$ is not part of the CFD. This is because the inequality implies $\Phi_l \rightarrow L_n^{\epsilon_n} + \cdots$, and $L_n^{\epsilon_n}$ is not in the BG-CFD. 
  
\item If $- \epsilon_n \, \mathcal{E}_{l, n} \geq 0$ for all $L_n^{\epsilon_n}$ that are not part of the BG-CFD, but there exists one 
$L_m^{\epsilon_m}$ that is a $(-1)$-vertex in the BG-CFD, such that $- \epsilon_m \, {\cal E}_{l,m} > 0$, then $F_{\Phi_l}$ has a non-empty intersection with a curve in the CFD. 
Because it does not split into anything outside the BG-CFD, ${\Phi_l}$ is a $(-2)$-vertex inside the full CFD. 

  \item If $ -\epsilon_n \, {\cal E}_{l,n} \geq 0$ for all F-extremal weights $L_n^{\epsilon_n}$ that are not in the BG-CFD, and $- \epsilon_m \, {\cal E}_{l,m} \leq 0$ for all F-extremal weights $L_m^{\epsilon_m}$ that are part of the BG-CFD, with at least one $m$ such that $- \epsilon_m \, {\cal E}_{l,m} < 0$, then 
  $\Phi_l$ becomes reducible and splits into weights/roots in the BG-CFD, and thereby has to also be contained as a $(-2)$ vertex. 
  
  \item Finally, if ${\cal E}_{l,m} = 0$ for all F-extremal weights $L_m^{\epsilon_m}$ in the BG-CFD, then $\Phi_l$ is a $(-2)$-vertex of the CFD if and only if there is a $(-2)$-vertex $F_{\ell}$ in the BG-CFD $\ell$ such that $D_{\Phi_l} \cdot F_{\ell}>0$.

\end{enumerate}
With these rules we can reconstruct a CFD, which captures the non-abelian part of the superconformal flavor symmetry, from which it is then easy to infer the abelian factors from the classical flavor symmetry and the number of instanton $U(1)$s, see \eqref{eq:rank_counting_G_F}.

Note however, that the CFDs we construct in this way --- which we will refer to as reduced CFDs ---   are generically sub-graphs of the full CFDs that we defined in \cite{Apruzzi:2019vpe} and derived from the geometry in \cite{Apruzzi:2019opn}.
There are $(-1)$-vertices corresponding to non-perturbative states of the gauge description associated with the BG-CFD, as well as unmarked vertices with $n_i \geq 0$, which cannot be reconstructed in this way. 
What we capture using the present gauge theoretic approach is the superconformal flavor symmetry (encoded in the $(-2)$-vertices) and tree-structure (captured by the $(-1)$-vertices, upon which we can apply the standard CFD-transitions) of the reduced CFDs, which have the gauge description of the chosen BG-CFD.
However, as we have discussed in section \ref{sec:descquiv}, one can access other branches of the descendant tree by passing to a dual gauge description, and consider the BG-CFDs of such theories. 

In summary: the reduced CFDs are constructed from the flavor-equivalence classes of $G_\text{gauge} \times G_\text{BG}$ box graphs, in conjunction with minimal input from the geometry, which specifies how the BG-CFD of the marginal $G_\text{gauge}$-theory is embedded into the fiber of the elliptic model that describes the marginal theory. 
The reduced CFDs contain
\begin{itemize}
\item marked subgraph (and thereby the Dynkin diagram of the superconformal flavor symmetry) of the full CFD,
\item $(-1)$ vertices, which in the full CFD have an interpretation as matter hypermultiplets charged under $G_\text{gauge}$.
\end{itemize}
It does not contain those $(-1)$ curves (and higher self-intersection curves), which transition, when flopped, to a geometry without a compatible ruling, and thus to an SCFT that does not have a weakly coupled $G_\text{gauge}$ description.

\begin{figure}
\centering
\includegraphics*[width=11cm]{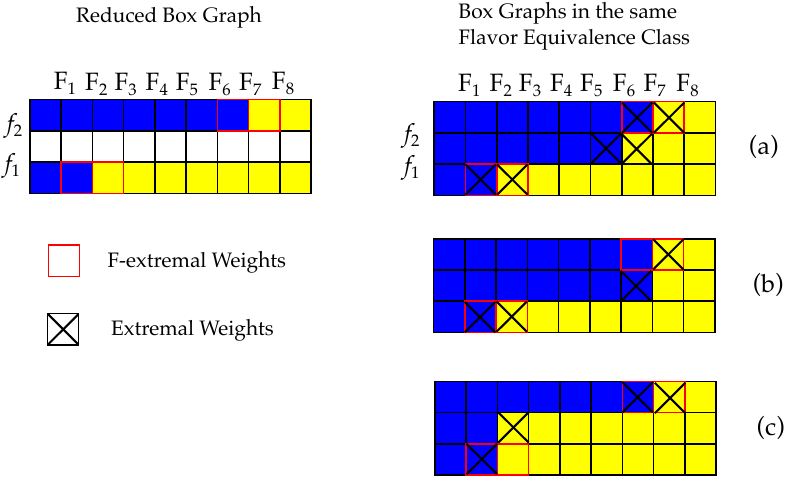}
\caption{An example of a reduced box graph (RHS) and elements in the same flavor equivalence class (box graphs shown on the right) for $SU(3) \times U(9)$. The F-extremal curves are marked in red, the extremal curves of the box graphs are marked by a cross. This shows that F-extremal curves are not necessarily always extremal for every box graph of the flavor equivalence class. The key is that changing the sign assignment of the F-extremal curves results in a different SCFT and superconformal flavor symmetry. The box graphs in a given flavor equivalence class are all distinct gauge theory descriptions that have the same UV fixed point. \label{fig:F-Ex}}
\end{figure}

\subsection{Rank Two CFDs from Box Graphs}

We now use the above approach to determine for all rank two theories the superconformal flavor symmetry, starting from the box graphs and the minimal information from the fibers. The results are summarized in appendix \ref{app:RankTwoBG}.

To start with, let us again consider the $SU(3)_\text{gauge} \times U(9)_\text{BG}$ example we studied in section \ref{sec:AnEx}, and shown in figure \ref{fig:F-Ex} (see also \ref{fig:U9SU3Examples}, (a) and (c)).
On the right hand side of figure \ref{fig:F-Ex}, there are three box graphs that are in the same flavor equivalence class. 
The reduced box graph for this equivalence class is given by simply deleting the middle row, as shown on the left hand side of figure \ref{fig:F-Ex}.
The splitting dictated for the whole flavor equivalence class is given by the F-extremal weights\footnote{We use the same labeling for weights as in figure \ref{fig:SUnUmBG}.}
$L^+_{1,7}, L^-_{1,8}, L^+_{3,2}, L^-_{3,3}$, and the following roots become reducible
\be
\ba
    F_2 & \rightarrow L^+_{3,2}  + L^-_{3,3}\cr 
    F_7 & \rightarrow  L^+_{1,7} + L^-_{1,8}\, .
\ea
\ee
Note that in the full box graphs on the RHS of figure \ref{fig:F-Ex}, the F-extremal weights are not always extremal (can be flopped in the box graph). 
An example is in model (b) the weight $L^+_{1,7}$ and in (c) the weight $L^-_{3,3}$.
We will see momentarily that this is however immaterial in determining the flavor symmetry.

Recall that the $SU(3)_\text{gauge} \times U(9)_\text{BG}$ box graph
provide a gauge theory description of the rank two E-string (realized by a $(E_8,SU(2))$ collision of singularities), where 
the $SU(9)_\text{BG} \subset U(9)_\text{BG}$ fully embedded inside the affine $E_8$ Dynkin diagram as in \eqref{U9IntoE8}. 
 
The BG-CFD is already determined for this flavor equivalence class in figure \ref{fig:CFDInBG}. 
In particular the CFD contains the roots $\alpha_{i}$, $i=3, \ldots, 6$ as $(-2)$-vertices and, 
since $\langle \alpha_3^\vee , L_{3,3} \rangle = - \langle \alpha_6^\vee, L_{1,7} \rangle = -1$, the two weights $L_{3,3}^+$ and $L_{1,7}^-$ as $(-1)$-vertices.
We next need to determine whether there are any additional vertices $\Phi_l$ from the geometry. 

For this note that we determined already the pairings $\mathcal{E}_{l,n}$ between 
the weights $L_n$ and the additional roots $\Phi_l$ for this case in figure 
\ref{fig:intersection_rep_graph_SU9xSU3}. These were determined from the marginal resolution geometry. 
Denoting by ${\Phi}$ and $\Phi_{l}$, $l=1,2$ the roots associated with the three extra nodes, first observe that 
\be
  \langle \Phi^\vee , L_{i,j} \rangle  =0 \,,
\ee
for all F-extremal weights $L_{i,j}$ in the flavor equivalence class. 
Furthermore, $D_{\Phi} \cdot F_6 = 1$. As  $F_6$ is part of the BG-CFD, 
$\Phi$ is a $(-2)$-vertex in the CFD as well, and will contribute thereby to $G_\text{F}$.

To determine whether the roots $\Phi_{1,2}$ of the affine $SU(2)$ are part of the CFD, note that 
\begin{align}
  \langle \Phi_1^\vee , L_{3,2} \rangle = 1 \, .
\end{align}
Since in this reduced box graph, the sign of $L_{3,2}$ (which is not part of the BG-CFD) is $+$, rule 1.~above implies that $\Phi_1$ not part of the CFD either. 
Likewise,
\begin{align}
  \langle \Phi_2^\vee,  L_{1,8} \rangle = -1 \, ,
\end{align}
which, together with the sign ($-$) and the fact that $L_{1,8}$ is not in the BG-CFD, implies that $\Phi_2$ does not contribute, either.
What we obtain is the CFD shown in figure \ref{fig:EnhancedBGCFD}.

\begin{figure}
\centering
\includegraphics*[width=7cm]{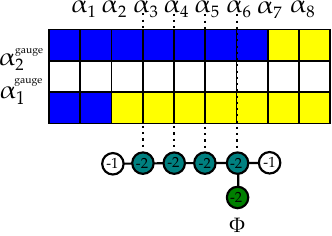}
\caption{The BG-CFD of the $SU(3)_\text{gauge}$ theory presented in figure \ref{fig:CFDInBG}, including the additional node, inferred from the geometric data. The non-abelian part of the superconformal flavor symmetry is determined by the $(-2)$ (marked/green) vertices, and is $SU(6)$. 
The mass deformations that lead to new SCFTs with effective $SU(3)_\text{gauge}$ descriptions are indicated in terms of the $(-1)$ vertices, to which we can apply CFD-transitions. While it is not the full CFD, this reduced CFD,  contains the same marked subgraph, and the $(-1)$ vertices of the reduced graph are also present in the full CFD, which are the central ingredients for the flavor symmetry and the descendant tree.
 \label{fig:EnhancedBGCFD}
}
\end{figure}

To conclude this example, the non-abelian part of the superconformal flavor symmetry is  $G_{\text{F,na}} = SU(6)$.
For this particular flavor equivalence class, the associated 5d effective gauge theory is $SU(3)_\text{gauge} + 5 {\bm F}$, so its total global symmetry (at weak coupling) is $U(5) \times U(1)_I$.
To match the total rank, the superconformal flavor symmetry then must be
\be
  G_\text{F} = SU(6) \times U(1) \, .
\ee

We can repeat this process for all rank two theories and determine the flavor equivalence classes, associated BG-CFDs, and the reduced CFDs.
The results are presented in appendix \ref{app:RankTwoBG}. 
Note that the reduced CFDs contain the full non-abelian part of the flavor symmetry, as above, but may not contain information about additional non-flavor curves. These are therefore sub-graphs of the CFDs, which however contain the complete marked subgraph of the CFD. 
Furthermore, the $(-1)$ vertices in the reduced CFD are also part of the full CFD, and the resulting descendant tree is therefore a subtree.


\section{Conclusions and Outlook}
\label{sec:ConcOut}

In this series of papers, the ``appetizer'' \cite{Apruzzi:2019vpe}, Part I
\cite{Apruzzi:2019opn}, and the present Part II, we made the case that 5d SCFTs
which descend from 6d SCFTs by circle compactifications plus mass
deformations have a concise description in terms of graphs, the so-called
CFDs. These graphs
encode some of the salient physical properties of these superconformal field theories: 
\begin{itemize}
  \item They tell us about the network structure of descendant SCFTs from a given 5d marginal theory.
  \item The marked vertices of a CFD form subgraphs that are Dynkin diagrams, which encode the strongly-coupled flavor symmetry $G_\text{F}$ of the UV fixed point that the CFD characterizes. 
  \item The spectrum of BPS states (Gopakumar--Vafa invariants) is computed by taking suitable combinations of the vertices (i.e., curves). Intersections with divisors associated to the marked subgraph graphs determines the representation under $G_\text{F}$. 
  \item They constrain the possible weakly coupled gauge theory  or quiver descriptions for the associated SCFT. 
  \item They predict dualities among these weakly coupled descriptions. 
\end{itemize}
In summary, the CFDs seem to crystallize some of the important features of 5d SCFTs! 

This approach is particularly powerful when applied to theories that descend
from 6d models whose geometric description has a known construction in terms
of a fully singular Tate or Weierstrass model. This was important in order to compute the CFD
of the marginal theory --- from which all the above properties of the
descendants can be determined in a combinatorial fashion. Examples of such
theories are all the minimal conformal matter theories, as well as some lower
rank theories with automorphisms (e.g., those that occur in the rank two
classification \cite{Jefferson:2018irk,Hayashi:2018lyv}). It is in these instances that we can derive the CFDs and
substantiate all claims regarding weakly coupled descriptions, and dualities
among these, by performing a complementary geometric computation --- a class of
examples where these the details were worked out are the $(E_6, E_6)$ and
$(E_7, SO(7))$ minimal conformal matter theories.  This geometric confirmation
provides backing for other setups, where the geometric computations become
less feasible. Some dualities appeared very recently in  \cite{Bhardwaj:2019ngx} and it would be interesting to 
study the relation with the dualities presented in the present paper.

In particular, there are 6d SCFTs which do not have a known description in
terms of a singular Weierstrass model. 
Specifically, 5d marginal theories with known 6d tensor branch descriptions, e.g., $SU(N)+ 2\bm{AS}+ 8\bm{F}$ for $N>5$, as well as their descendants, were studied using
five-brane webs. In these
cases we can ``bootstrap'' the marginal CFD by using the constraints of known
gauge theory descriptions including their superconformal flavor enhancements.
Perhaps most interestingly, the resulting marginal
CFD in turn predicts new branches of the descendant tree, which would indicate
a yet unknown sequence of SCFT
descendants with different gauge descriptions. Thus, combining the requirement of 
the embedding of the gauge theoretic BG-CFDs,
the known flavor enhancements for parts of the descendant tree, and the
constraint that these all descend from a single marginal CFD by applying the
CFD-transition rules, results in new predictions for these theories.  It would
clearly be very exciting to test these predictions either by constructing the
associated Weierstrass models or by alternative methods such as a five-brane
web realization. 

As already alluded to in the conclusions to Part I \cite{Apruzzi:2019opn}, the
next step in the program to determine all 5d SCFTs that descend from 6d is to
develop a gluing procedure for CFDs. Similar to the classification in 6d,
where the most general theory is built out of a generalized quiver based on a
small set of building blocks (the flavor nodes as well as non-Higgsable
clusters), a similar gluing is expected to exist in 5d. Given the fundamental role that the CFDs seem to play, it is very natural to expect them to be (part of) the building
blocks from which the most general 5d SCFT is glued.  This will be
investigated in the future.

\subsection*{Acknowledgements}

We thank C.~Closset, H.~Hayashi, J.~J.~Heckman, M.~Weidner, K.~Yonekura, and G.~Zafrir for discussions. 
FA, SSN and YNW are supported by the ERC Consolidator Grant number 682608 
``Higgs bundles: Supersymmetric Gauge Theories and Geometry (HIGGSBNDL)'',
CL by NSF CAREER grant PHY-1756996, 
LL by DOE Award DE-SC0013528Y.
FA also thanks the 2019 Summer Workshop at the Simons Center for Geometry and
Physics for hospitality during the completion of this work.
LL thanks the ITP Heidelberg for hospitality during the completion of this
work. SSN is grateful to the CERN Theory Group for hospitality during the course of this work.

\appendix

\section{Coulomb Branch and Reduced CFDs for Rank Two 5d SCFTs}
\label{app:RankTwoBG}

This appendix summarizes the box graphs (and flavor-equivalence classes of box graphs) for the rank two 5d SCFTs , which have a weakly coupled gauge theory description. We also note down in each case the reduced CFDs, which were obtained by using minimal input from the resolution geometry to reconstruct the fiber from the box graphs, as explained in section \ref{sec:CFDsfromBG}.

\subsection[Descendants of Rank 2 E-string and \texorpdfstring{$(D_5, D_5)$}{(D5,D5)} Conformal Matter]{Descendants of Rank 2 E-string and \boldmath{$(D_5, D_5)$} Conformal Matter}

For the rank two E-string on $S^1$, these are shown in figure \ref{fig:EstringSU3}, \ref{fig:EstringSp2}, \ref{fig:EstringSU2SU2}.
For minimal $(D_5, D_5)$ conformal matter, they are in figure \ref{fig:D5D5SU3}, \ref{fig:D5D5Sp2}, \ref{fig:D5D5SU2SU2}.

Each of these theories will be specified by a gauge theory/quiver with a rank two gauge
group together with some matter transforming under some flavor group.
In the main text we provide a detailed description of the $SU(3)$ gauge theory description of the rank two E-string in section \ref{sec:AnEx}.
Furthermore we will specify the reduced CFD: as explained in section \ref{sec:CFDsfromBG}, we can reconstruct the relevant parts of the fiber from the box graphs and flavor-equivalence classes, once we specify the embedding into the fiber of the marginal theory. From this we can further derive the marked vertices of the CFD, i.e., the subgraph that encodes that flavor symmetry at the UV fixed point, as well as the $(-1)$-vertices, which correspond to decoupling hypermultiplet matter. Note that in general this determines only a sub-graph of the full CFD, and may miss curves with self-intersection number $\geq 0$ or $(-1)$-curves, which do not correspond to hypermultiplets of the chosen gauge description.
An example is for instance the rank one CFDs, which have 10 descendants from the marginal theory, from which only 9 have an $SU(2)$ gauge description. 
The complete CFDs that are derived from the geometry capture all these descendants, irrespective of whether they admit a weakly coupled description. 
However, the reduced CFDs that are constructed based on a given gauge theory description, only capture in general a subset. Using the methods in section \ref{sec:CFDsfromBG} we can however determine the full superconformal flavor symmetry.

\begin{figure}
  \centering
  \includegraphics[height= 0.7\textheight]{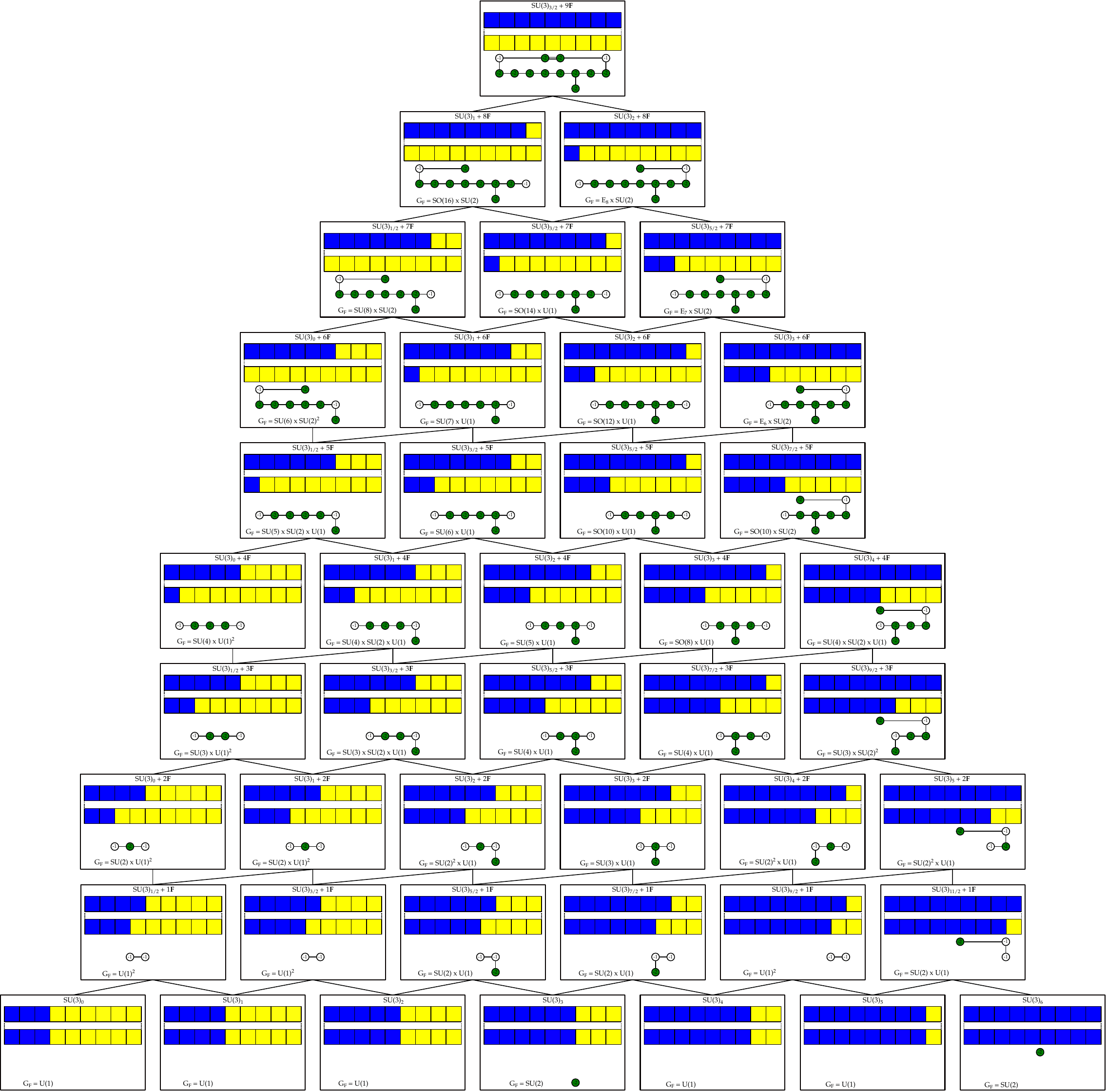}
  \caption{Rank two E-string: the marginal 5d gauge theory description as $SU(3)_{3/2} +
    9\bm{F}$. The flavor-equivalence classes of this marginal theory can be
  written in terms of box graphs, and each box graph corresponds to a
descendant 5d gauge theory. All of the descendant theories have 5d
superconformal fixed points. Below the box graph equivalence classes we show the reduced CFDs which 
encode the superconformal flavor symmetry. 
\label{fig:EstringSU3}}
\end{figure}

\begin{figure}
  \centering
  \includegraphics[height= 0.89 \textheight]{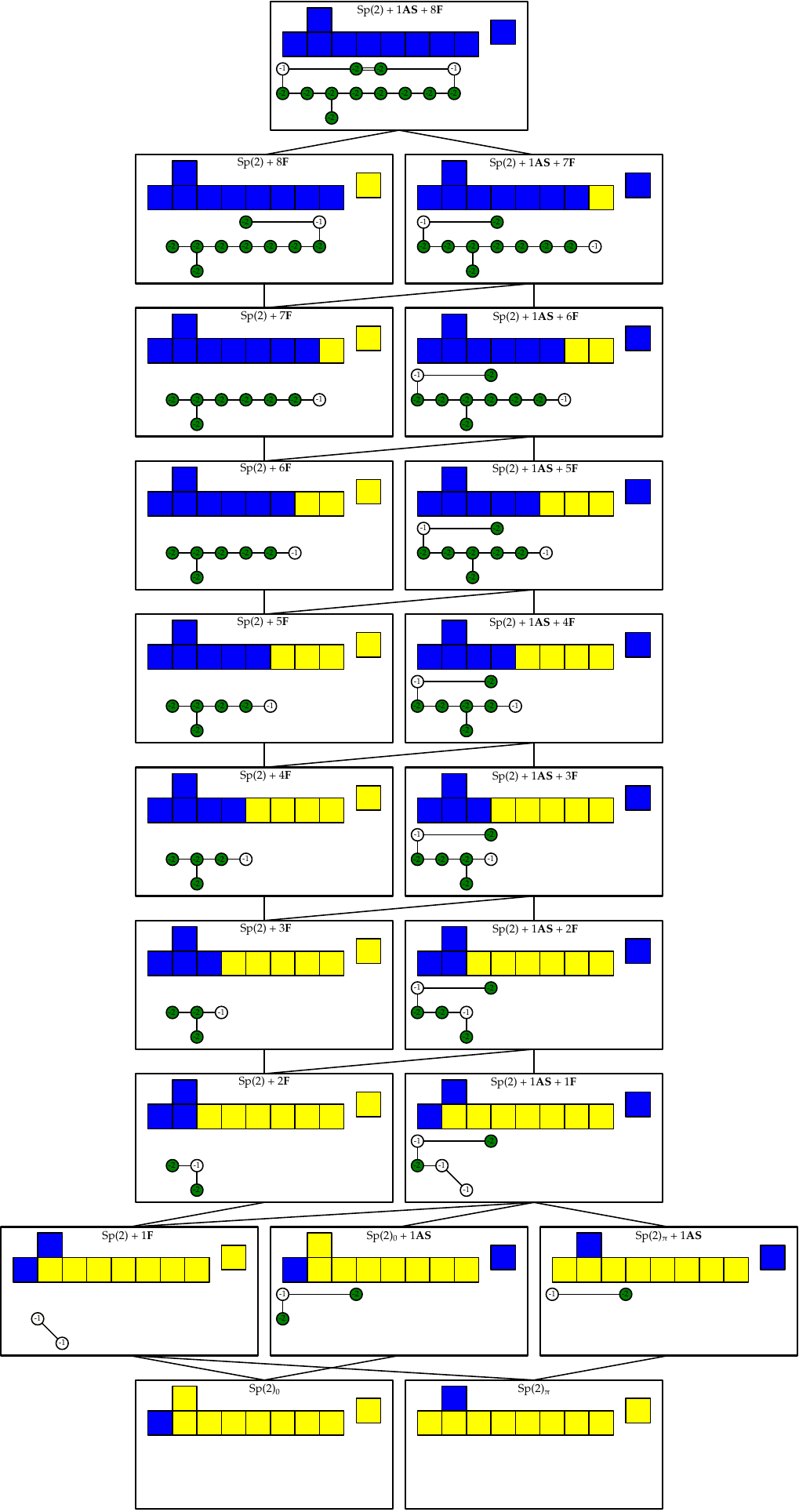}
  \caption{Rank two E-string with the marginal gauge theory described as $Sp(2) +
    8\bm{F} + 1\bm{AS}$. The figure shows the tree of descendant theories, together with their superconformal
    flavor symmetry, $G_\text{F}$. We furthermore specify the reduced CFDs, which encode the flavor symmetry and the $(-1)$ vertices that correspond to hypermultiplet matter in the $Sp(2)$ description.
    \label{fig:EstringSp2}
    }
\end{figure}

\begin{figure}
  \centering
  \includegraphics[width=0.78\textwidth]{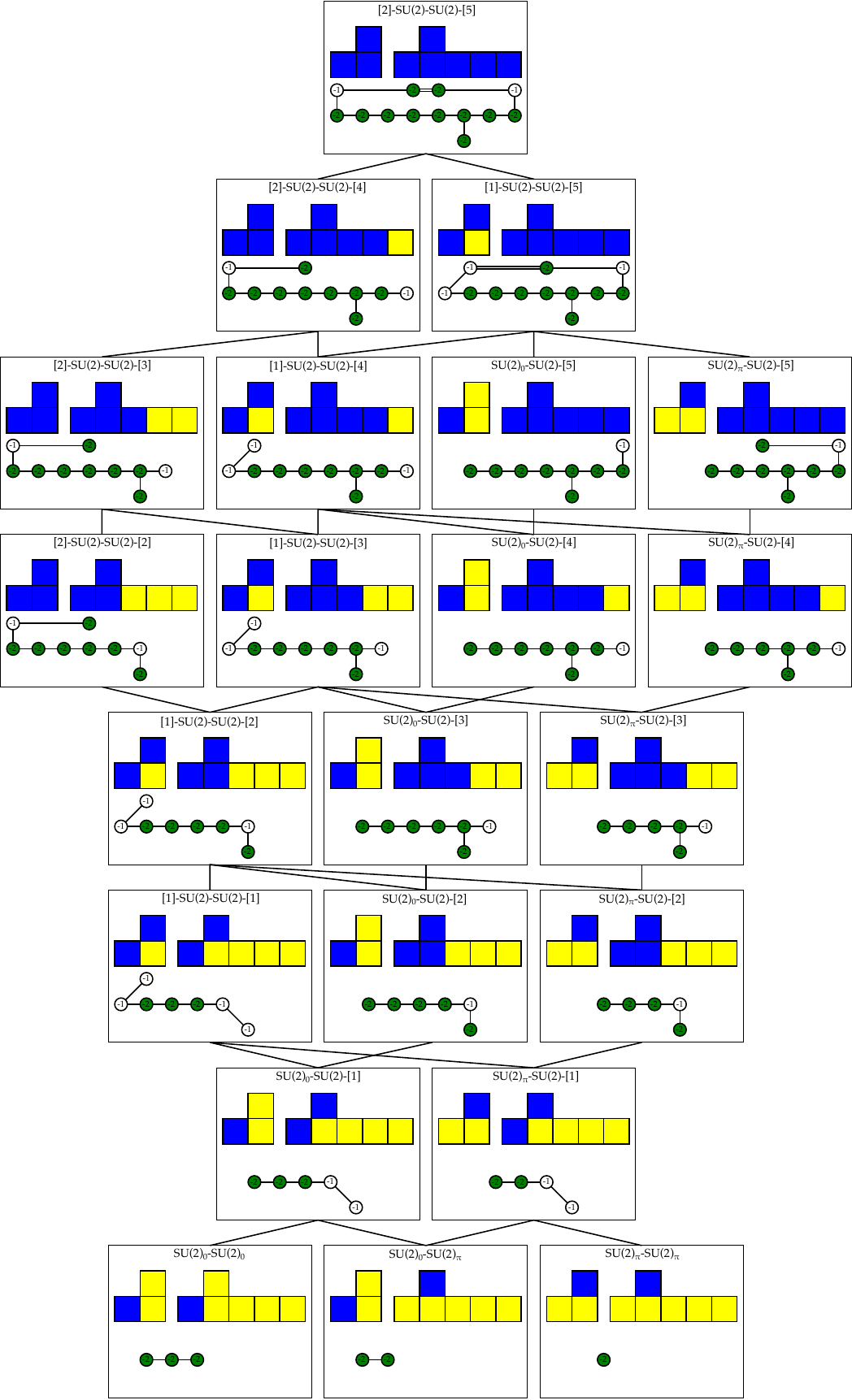}
  \caption{Rank two E-string: Descendants of the marginal theory $5\bm{F} - SU(2) - SU(2) - 2\bm{F}$, with their
flavor-equivalence class of Coulomb phases of $(SU(2) \times SU(2))_\text{gauge} \times (SO(10)
\times SO(4))_\text{BG}$. In addition we also note the reduced CFDs.
\label{fig:EstringSU2SU2}
}
\end{figure}

\begin{figure}
  \centering
  \includegraphics[width=\textwidth]{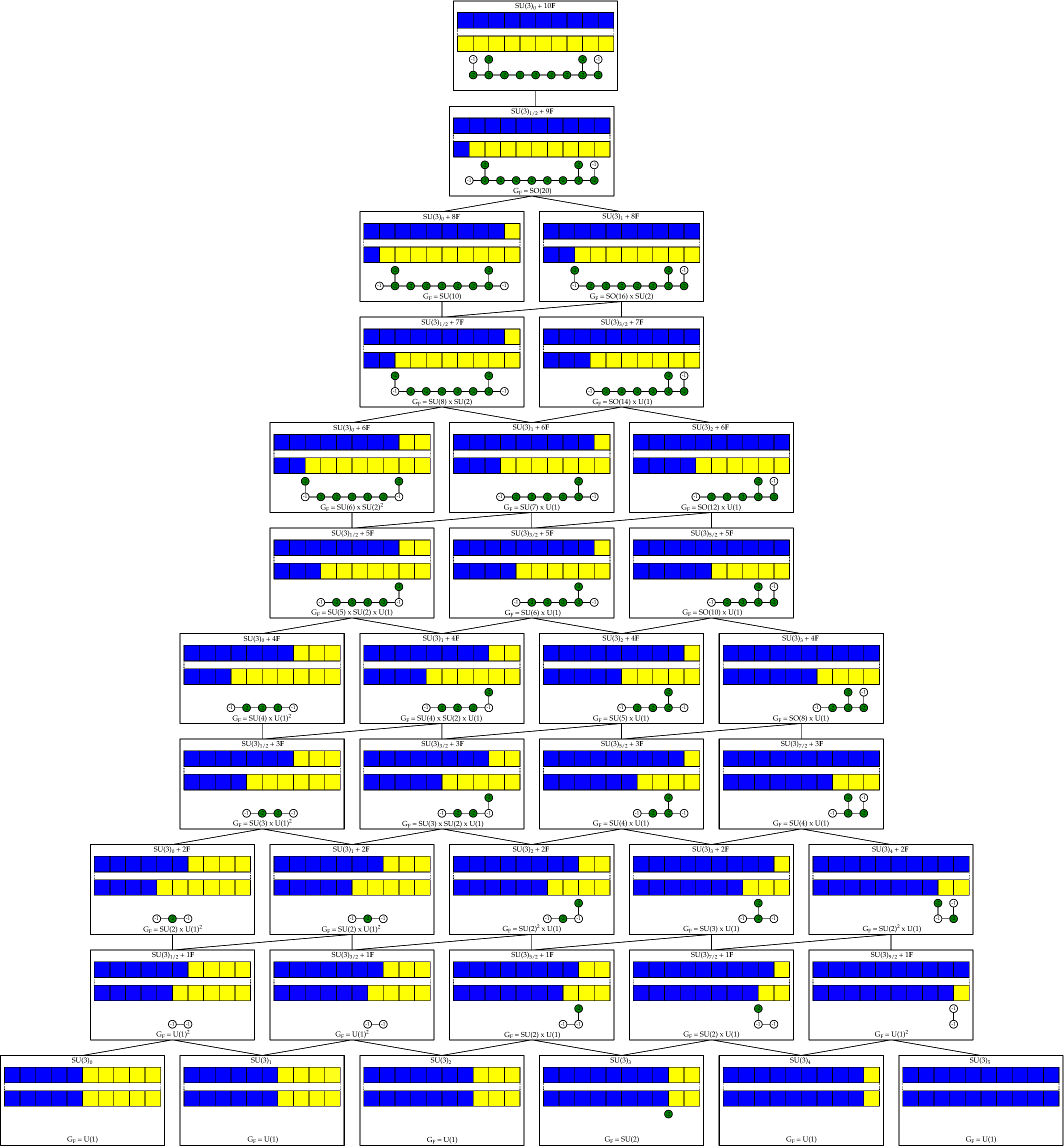}
  \caption{$(D_5, D_5)$ Conformal Matter: 5d marginal gauge theory description 
as  $SU(3)_0 + 10\bm{F}$. The (flavor-equivalence classes of )
    box graphs encode the $SU(3)$ gauge theory description
    of the descendant theory --- this is written at the top of each box in the
    above graph. Knowing that the marginal theory enhances in the UV to the 6d
    theory of $(SO(10),SO(10))$ minimal conformal matter, which is described
    by an affine $SO(20)$ fiber, one can determine which of the curves
    corresponding to the weights of the $(\bm{3,N})$ representation (where $N$
    is the number of flavors in that flavor-equivalence class), and which of
    the $F_i$ associated to the roots of the affine $SO(20)$ are contained
    inside of the non-flat surfaces. The intersection pattern of these curves
    is depicted via a dual graph in the lower half of each box. The superconformal flavor
    symmetry, $G_\text{F}$ is obtained from the reduced CFDs, shown below the box graphs. 
      \label{fig:D5D5SU3}}
\end{figure}

\begin{figure}
  \centering
  \includegraphics[width=\textwidth]{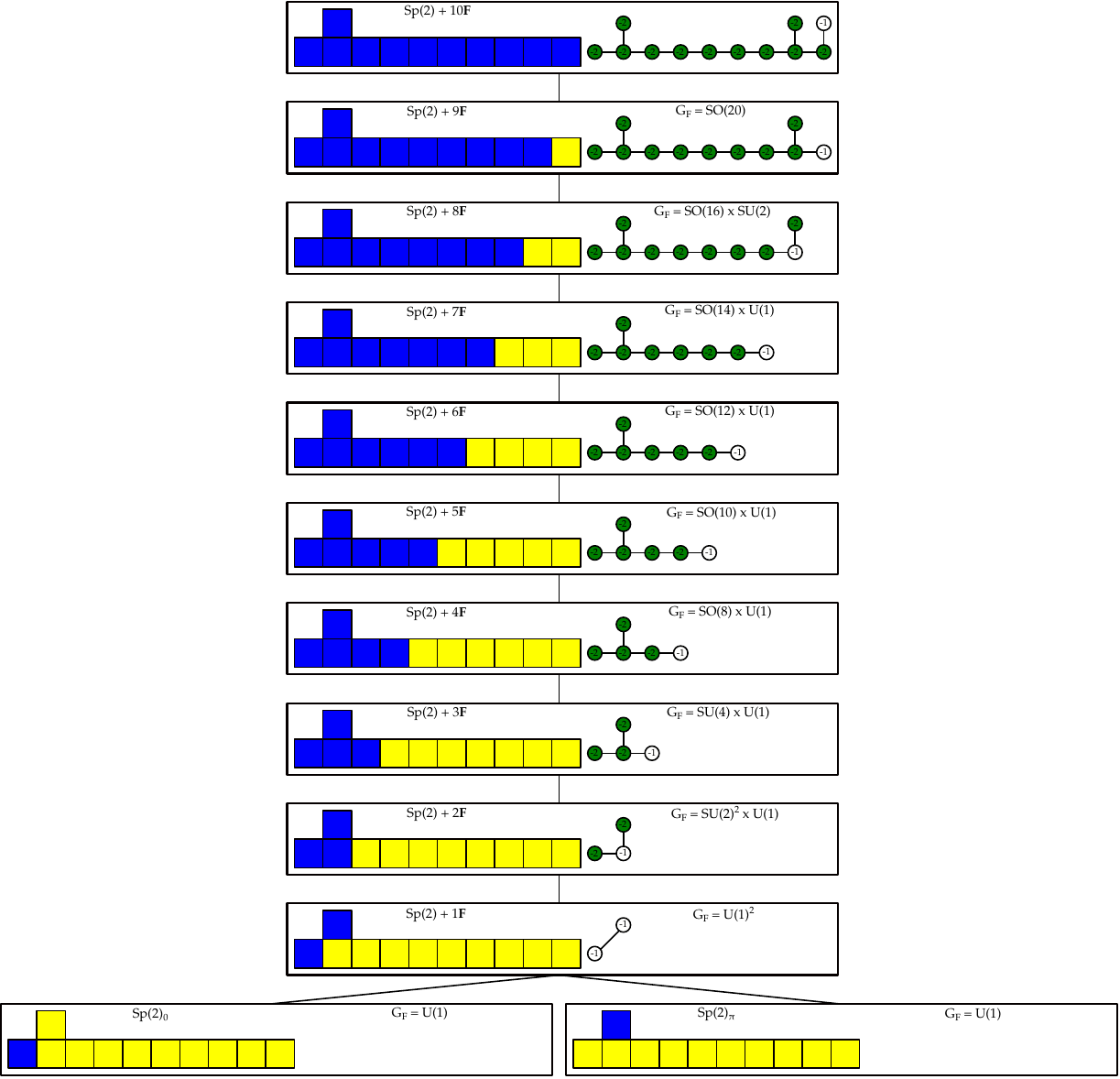}
  \caption{$(D_5, D_5)$ Conformal Matter:  marginal theory description as $Sp(2) + 10\bm{F}$.
  The figure shows the tree of descendant theories, together with their superconformal
    flavor symmetry, $G_\text{F}$.
    \label{fig:D5D5Sp2}}
\end{figure}

\begin{figure}
  \centering
  \includegraphics[width=0.7\textwidth]{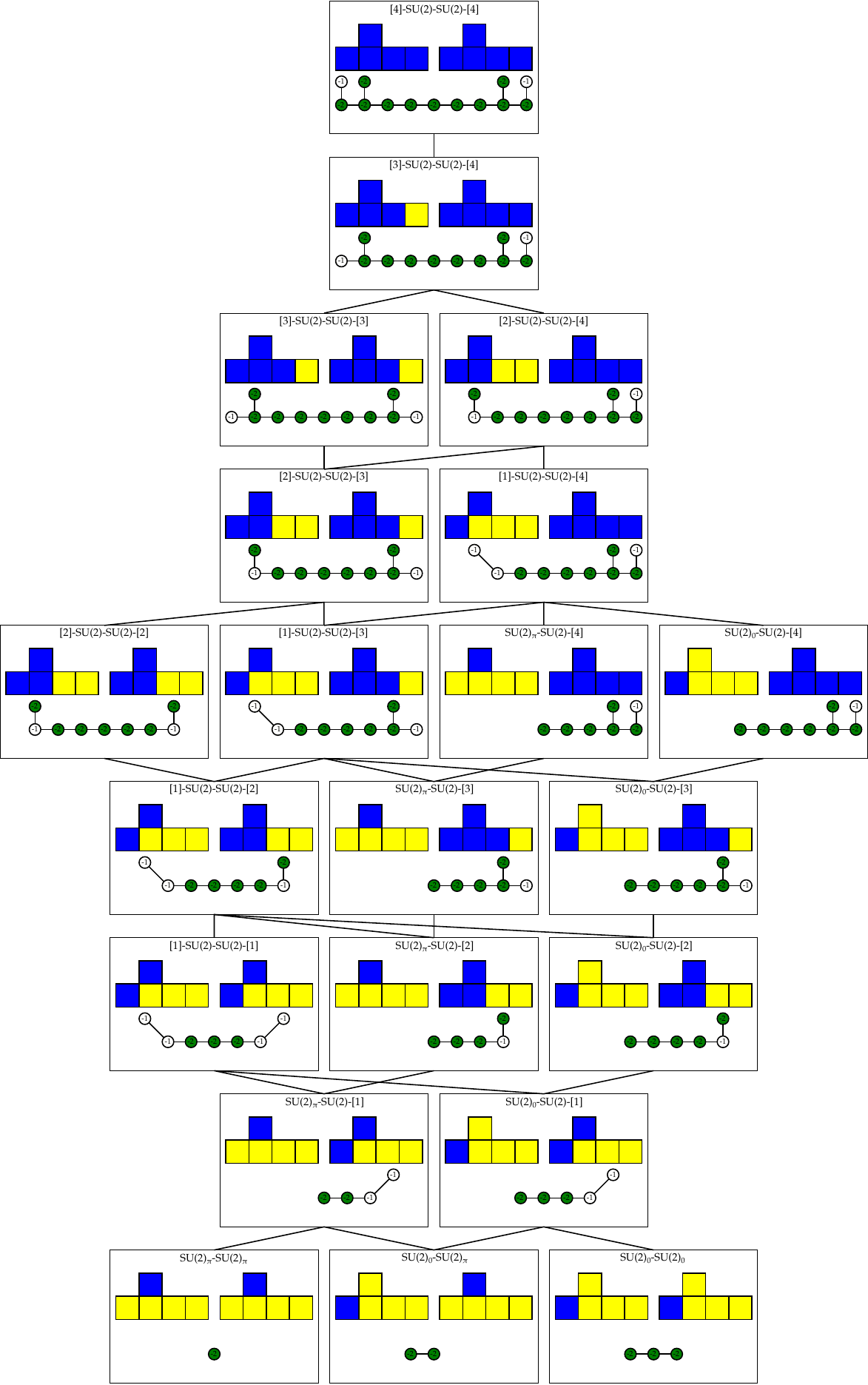}
  \caption{$(D_5, D_5)$ Conformal Matter: the marginal theory in the description as 
  the quiver $4\bm{F}-SU(2)-SU(2)-4\bm{F}$. All of the
descendant 5d theories of such a 5d marginal theory are given in the tree
above, as determined through the flavor-equivalence classes of box graphs.
The codimension two fiber, which is a splitting of affine $SO(20)$, can be
reconstructed in each case, and we draw the curves inside the fiber that are
also contained inside of the compact surfaces in the lower half of each
box. The reduced CFDs are shown from, which we determine the superconformal flavor symmetry. 
\label{fig:D5D5SU2SU2}
}
\end{figure}

\subsection[\texorpdfstring{$SU(3)$}{SU(3)} on a \texorpdfstring{$(-1)$}{(-1)}-curve with 12 Hypermultiplets]{\boldmath{$SU(3)$} on a \boldmath{$(-1)$}-curve with 12 Hypermultiplets}

The marginal theory has three gauge theory descriptions 
\begin{itemize}
\item  $SU(3)_4 + 6 \bm{F}$
\item $Sp(2)+ 2 \bm{AS} + 4 \bm{F}$
\item  $G_2 + 6 \bm{F}$.
\end{itemize}
The box graphs and descendants for these are shown in figures \ref{SU36FTree},
\ref{Sp24F2ASTree}, and \ref{G26FTree}.

\begin{figure}
  \centering
  \includegraphics[scale=0.6]{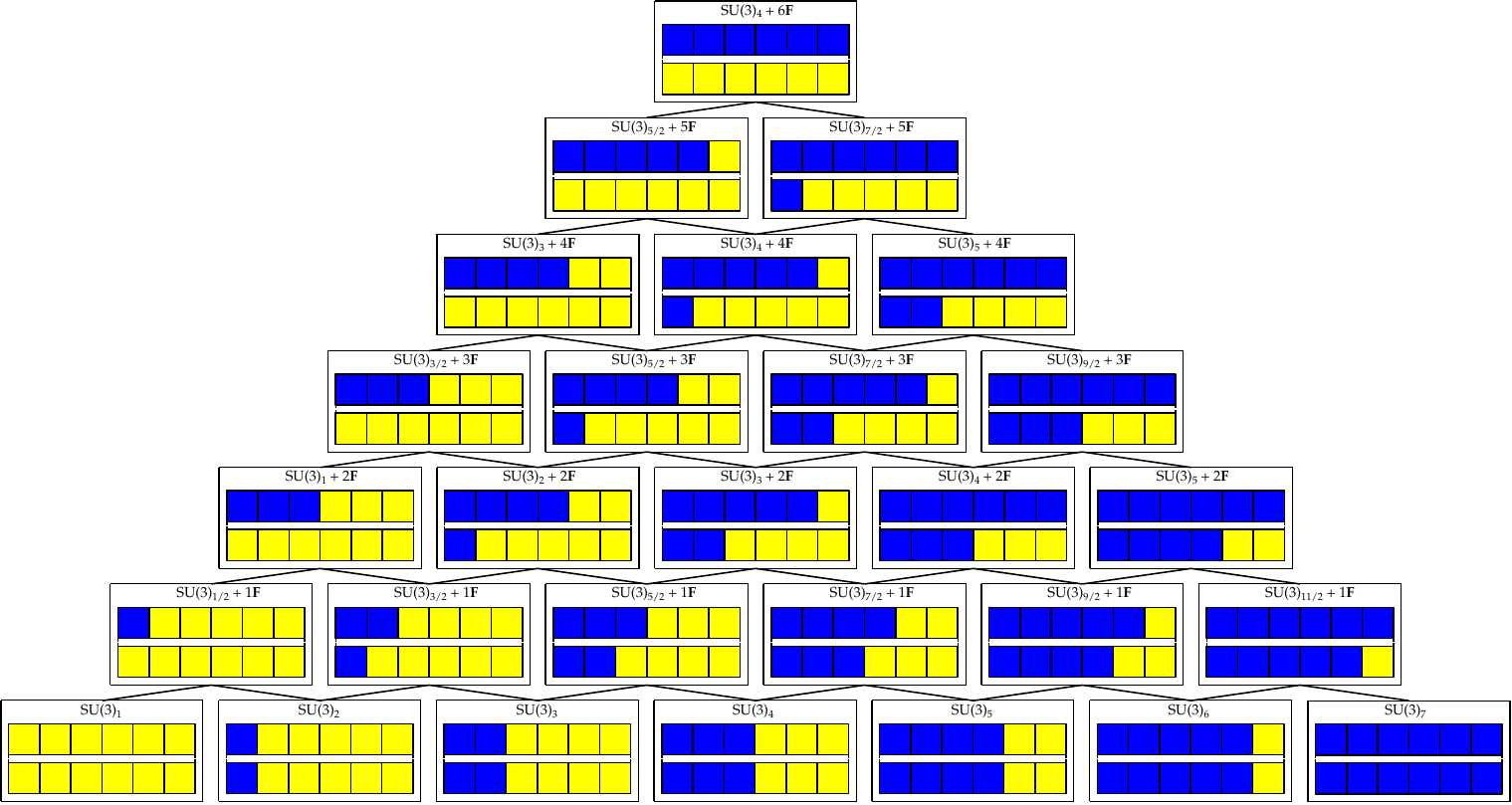}
  \caption{The tree of descendants for $SU(3)_4 + 6 \bm{F}$, which is gauge theory description for the marginal theory obtained from  $SU(3)$ on a $(-1)$-curve with 12 hypermultiplets.
  }\label{SU36FTree}
\end{figure}

\begin{figure}
  \centering
  \includegraphics[scale=0.5]{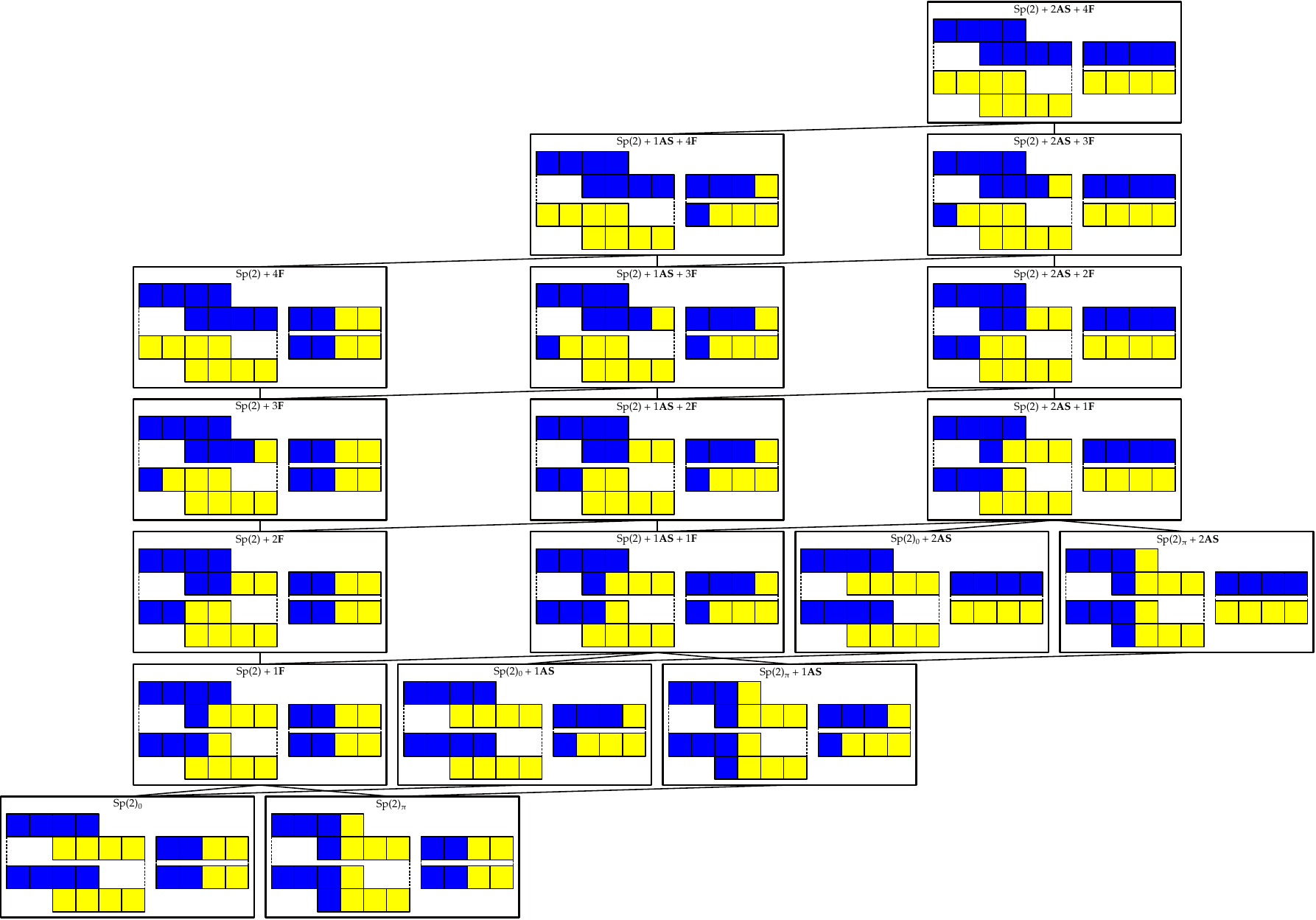}
  \caption{The tree of descendants for $Sp(2) + 6 \bm{F} +
  2\bm{AS}$,  which is gauge theory description for the marginal theory obtained from  $SU(3)$ on a $(-1)$-curve with 12 hypermultiplets.}\label{Sp24F2ASTree}
\end{figure}

\begin{figure}
  \centering
  \includegraphics[scale=0.6]{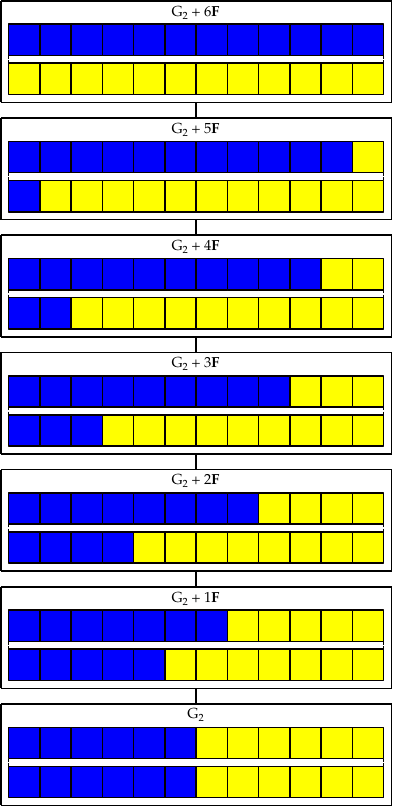}
  \caption{The tree of descendants for $G_2 + 6 \bm{F}$,  which is gauge theory description for the marginal theory obtained from  $SU(3)$ on a $(-1)$-curve with 12 hypermultiplets.}\label{G26FTree}
\end{figure}

\subsection[\texorpdfstring{$SU(3)$}{SU(3)} on a \texorpdfstring{$(-2)$}{(-2)}-curve with 6 Hypermultiplets]{\boldmath{$SU(3)$} on a \boldmath{$(-2)$}-curve with 6 Hypermultiplets}

This theory has a marginal description in terms of 
\be
Sp(2)+3\bm{AS} \,.
\ee
In this case we are considering the phase structure for a theory with gauge
group
\begin{equation}
  Sp(2)_{\text{gauge}}\times Sp(3)_{\text{BG}} \,,
\end{equation}
with matter transforming in the $(\bm{5,6})$ representation. The highest weight of the $(\bm{5,6})$ can be written as
\begin{equation}
  L_{1,1} = (0,1; 1,0,0) \,.
\end{equation}
The five simple roots of this semi-simple Lie algebra are, in the Cartan--Weyl
basis, 
\begin{equation}
  \begin{gathered}
    \alpha_1^2 = (2,-1;0,0,0) \,, \quad \alpha_2^2 = (-2,2;0,0,0) \cr
    \alpha_1^3 = (0,0;2,-1,0) \,, \quad \alpha_2^3 = (0,0;-1,2,-1) \,, \quad
    \alpha_3^3 = (0,0;0,-2,2) \,,
  \end{gathered}
\end{equation}
where the superscript indicates which $Sp(n)$ factor that it acts as the
simple root of. The undecorated box graph, or the weight diagram, of this
representation is depicted in figure \ref{fig:Sp2Sp3BareBG}. Furthermore,
we can see directly from the self-conjugacy of the representation that
determining all of the phases corresponds to determining the different
consistent ways that signs can be assigned to the weights of the subgraph
marked in red on figure \ref{fig:Sp2Sp3BareBG}. 

\begin{figure}
  \centering
  \subfloat[]{{\includegraphics[scale=1.5]{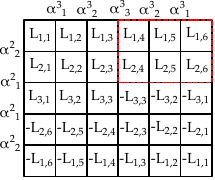}}}
  \qquad\qquad\qquad
  \subfloat[]{{\includegraphics[scale=1.5]{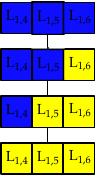}}}
  \caption{In (a) is the weight diagram of the $(\bm{5,6})$ representation of $Sp(2)
  \times Sp(3)$. We can see that, because the representation is self-conjugate,
  that the signs associated to $L_{i,j}$ for $i,j=1,\cdots,3$ are required to
  be all plus. The subdiagram enclosed in red is then the only weights for
whom the sign needs to be specified to fix the phase. The equivalence class of
phases relevant in the limit where the gauge coupling of the $Sp(3)$ is taking
to zero is specified by the signs of the weights $L_{1,4}$, $L_{1,5}$, and
$L_{1,6}$. In (b) is depicted the flop graph of these equivalence classes of
phases.}\label{fig:Sp2Sp3BareBG}
\end{figure}

The consistent phases can be determined by the application of the flow rules
to the decoration of the red-boxed subgraph. The total number of phases can be
seen to be
\begin{equation}
  N_\text{phases} = 10 \,,
\end{equation}
and the adjacency graph can be determined in the usual manner. We are
interested in the equivalence class of the phases where the same set of
$Sp(3)$ simple roots are contained inside of the splitting of the $Sp(2)$
simple roots. Since the weight $-L_{3,3}$ is always associated to a minus sign
we can see that this equivalence class is entirely specified by the signs of
$L_{1,4}$, $L_{1,5}$, and $L_{1,6}$. The flop chain of these equivalence
classes is drawn on the right in figure \ref{fig:Sp2Sp3BareBG}.


\section[Flavor Symmetry Enhancements for \texorpdfstring{\boldmath{$SU(N)_k + 1 AS + N_f F$}}{SU(N)k + 1AS + Nf F}]{Flavor Symmetry Enhancements for \boldmath{$SU(N)_k+ 1 AS + N_{f} F$}}
\label{app:FlavorEnha}

In this appendix we summary some of the known flavor symmetry enhacements of
5d gauge theories of the type $SU(N)_k+ 1  \bm{AS} + N_{f} \bm{F}$ at their UV
fixed points. These were determined from field theoretic methods, are
contained implicitly in \cite{Yonekura:2015ksa}. In particular, we focus on
the cases, which descend from $SU(2n)_0+ 2  \bm{AS} + 8 \bm{F}$ and
$SU(2n+1)_0+ 2  \bm{AS} + 8 \bm{F}$. We spell these out here, in order to
facilitate the comparison with the CFDs. These flavor symmetry
enhancements are summarized in tables \ref{tab:GFsu2n1AS1},
\ref{tab:GFsu2n1AS2}, \ref{tab:GFsu(2n+1)1AS1}, \ref{tab:GFsu(2n+1)1AS2}, and
\ref{tab:GFsu2n1AS3}. 

\paragraph{For $N$ even,}
 i.e. $N=2n$ and $n>2$ we always have that $N_f < N+4$, since $N_f \leq 8$. The superconformal flavor symmetry is related to the one of the theory where $ 1 \bm{AS}$ is decoupled in the following way
\begin{equation}
G_\text{F}= G_\text{F}(SU(2n)_{k+n-2} +N_f \bm{F}) \times U(1),
\end{equation}
where the extra $U(1)$ is the classical symmetry acting on antisymmetric hypermultiplet, and $k$ shifts due to this decoupling, $k\rightarrow k' =  k + n-2$. The flavor symmetries for $SU(2n)_{k'=k+n-2} +N_f \bm{F}$ can be obtained from \cite{Apruzzi:2019vpe, Cabrera:2018jxt}. In fact, in our cases, we have that 
\begin{equation}
 \left|2n-4- \frac{(8-N_f)}{2} \right| \leq |k'| \leq 2n-4+ \frac{(8-N_f)}{2}\, .
\end{equation}
According to \cite{Apruzzi:2019vpe, Cabrera:2018jxt}, we have the following two distinct cases:
\be
\ba
2n-\frac{N_f}{2}>  |k'|\quad & \rightarrow &\quad &G_\text{F}(SU(2n)_{k'=k+n-2})= SU(N_f) \times U(1) \times U(1) \\
2n-\frac{N_f}{2}= |k'|>\frac{1}{2} \quad & \rightarrow &\quad &G_\text{F}(SU(2n)_{k'=k+n-2})= SU(N_f) \times SU(2) \times U(1)\, .
\ea
\ee

\paragraph{For $N$ odd,} i.e. $N=2n+1$ and $n \geq 2$, $N_f < N+4$, since again we have that $N_f \leq 8$. The superconformal flavor symmetry is 
\begin{equation}
G_\text{F}= G_\text{F}(SU(2n+1)_{k+n-\frac{3}{2}} +N_f \bm{F}) \times U(1),
\end{equation}
$k$ shifts, $k\rightarrow k' =  k + n-\frac{3}{2}$ when decoupling an antisymmetric hypermultiplet. The flavor symmetries for $SU(2n+1)_{k'=k+n-\frac{3}{2}} +N_f \bm{F}$ can read from \cite{Apruzzi:2019vpe, Cabrera:2018jxt}. In fact, we have that 
\begin{equation}
 \left|2n-3- \frac{(8-N_f)}{2} \right| \leq |k'| \leq 2n-3+ \frac{(8-N_f)}{2}\, .
\end{equation}
From \cite{Apruzzi:2019vpe, Cabrera:2018jxt}, we have 
\be
\ba
2n+1-\frac{N_f}{2} >  |k'|\quad & \rightarrow &\quad &G_\text{F}(SU(2n)_{k'=k+n-\frac{3}{2}})= SU(N_f) \times U(1) \times U(1) \\
2n+1-\frac{N_f}{2}= |k'|>\frac{1}{2} \quad & \rightarrow &\quad &G_\text{F}(SU(2n)_{k'=k+n-\frac{3}{2}})= SU(N_f) \times SU(2) \times U(1)\, .
\ea
\ee

\begin{sidewaystable}
  \centering
  \begin{small}
  \begin{tabular}{|l||c|c|c|c|c|c|c|c|c|} 
    \hline 
\diagbox{$N_f$}{$|k|$} & $|n-6|$ & $|n-5|$& $|n-4|$ &$n-3$ &  $n-2$ &$n-1$ & $n$  &$n+1$&$n+2$ 
   \cr\hline \hline
  \multirow{2}{*}{8} &  &  & & & $U(8) \times SU(2)$ & & &  &  \\&  &  & & & $\times U(1)$ & & &  & 
   \cr\hline
     \multirow{2}{*}{6} &  &  & &$U(6) \times U(1)$ &$U(6) \times U(1)$   &$U(6) \times SU(2)$ &  & & \\&  &  & &$\times U(1)$  &$\times U(1)$   & $\times U(1)$ &  & &
        \cr\hline
       \multirow{2}{*}{4} &  &  &$U(4) \times U(1)$ &$U(4) \times U(1)$ &$U(4) \times U(1)$   &$U(4) \times U(1)$ &$U(4) \times SU(2)$ & & \\&  &  & $\times U(1)$  &$\times U(1)$  &$\times U(1)$   & $\times U(1)$  &$\times U(1)$   &  &
          \cr\hline
                 \multirow{2}{*}{2} &  &$U(2) \times U(1)$  &$U(2) \times U(1)$ &$U(2) \times U(1)$ &$U(2) \times U(1)$   &$U(2) \times U(1)$ &$U(2) \times U(1)$ & $U(2) \times SU(2)$ & \\&  &$\times U(1)$  & $\times U(1)$  &$\times U(1)$  &$\times U(1)$   & $\times U(1)$  &$\times U(1)$   &$\times U(1)$  &
          \cr\hline
                     0 & $U(1) \times U(1)$  &$U(1) \times U(1)$  &$U(1) \times U(1)$ &$U(1) \times U(1)$ &$U(1) \times U(1)$   &$U(1) \times U(1)$ &$U(1) \times U(1)$ & $U(1) \times U(1)$ & $SU(2) \times U(1)$
          \cr\hline
  \end{tabular}
  \caption{Flavor symmetries at strong coupling, $G_\text{F}$, for $SU(2n)_k + 1 \bm{AS} + N_f \bm{F}$ ($N_f$ even) gauge theories, which are descendants of $SU(2n)_0 + 2 \bm{AS} + 8\bm{F}$.}  \label{tab:GFsu2n1AS1}
   \end{small}
    \centering
  \begin{small}
  \vspace{1cm}
  \begin{tabular}{|l||c|c|c|c|c|c|c|c|}  
    \hline 
\diagbox{$N_f$}{$|k|$}  & $\left|n-\frac{11}{2}\right|$ & $\left|n-\frac{9}{2}\right|$ & $\left|n-\frac{7}{2}\right|$ & $n-\frac{5}{2}$&$n-\frac{3}{2}$  & $n-\frac{1}{2}$  &$n+\frac{1}{2}$ &$n+\frac{3}{2}$
          \cr\hline \hline
  \multirow{2}{*}{7} &  &  & &$U(7) \times U(1)$  &$U(7) \times SU(2)$ & &  &  \\&  &  & & $\times U(1)$&   $\times U(1)$ & &  & 
            \cr\hline
     \multirow{2}{*}{5} &  &  & $U(5) \times U(1)$  &$U(5) \times U(1)$   & $U(5) \times U(1)$ &$U(5) \times SU(2)$  & & \\&  &  &  $\times U(1)$ &$\times U(1)$  &  $\times U(1)$  &$\times U(1)$  & &
   \cr\hline
        \multirow{2}{*}{3} &  &$U(3) \times U(1)$  & $U(3) \times U(1)$  &$U(3) \times U(1)$   & $U(3) \times U(1)$ &$U(3) \times U(1)$  & $U(3) \times SU(2)$ & \\&  & $\times U(1)$ &  $\times U(1)$ &$\times U(1)$  &  $\times U(1)$  &$\times U(1)$  &$\times U(1)$ &
   \cr\hline
             1 & $U(1)^3$ &$U(1)^3$   & $U(1)^3$   &$U(1)^3$    & $U(1)^3$  &$U(1)^3$   &  $U(1)^3$& $U(1)^2 \times SU(2)$
   \cr\hline
  \end{tabular}
  \caption{Flavor symmetries at strong coupling, $G_\text{F}$, for $SU(2n)_k + 1 \bm{AS} + N_f \bm{F}$ ($N_f$ odd) gauge theories, which are descendants of $SU(2n)_0 + 2 \bm{AS} + 8\bm{F}$.}\label{tab:GFsu2n1AS2}
   \end{small}  
\end{sidewaystable}

\begin{sidewaystable}
  \centering
  \begin{small}
  \begin{tabular}{|l||c|c|c|c|c|c|c|c|c|} 
    \hline 
\diagbox{$N_f$}{$|k|$} & $\left|n-\frac{11}{2}\right|$ & $\left|n-\frac{9}{2}\right|$& $\left|n-\frac{7}{2}\right|$ &$\left|n-\frac{5}{2}\right|$ &  $n-\frac{3}{2}$ &$n-\frac{1}{2}$ & $n+\frac{1}{2}$  &$n+\frac{3}{2}$&$n+\frac{5}{2}$ 
   \cr\hline \hline
  \multirow{2}{*}{8} &  &  & & & $U(8) \times SU(2)$ & & &  &  \\&  &  & & & $\times U(1)$ & & &  & 
   \cr\hline
     \multirow{2}{*}{6} &  &  & &$U(6) \times U(1)$ &$U(6) \times U(1)$   &$U(6) \times SU(2)$ &  & & \\&  &  & &$\times U(1)$  &$\times U(1)$   & $\times U(1)$ &  & &
        \cr\hline
       \multirow{2}{*}{4} &  &  &$U(4) \times U(1)$ &$U(4) \times U(1)$ &$U(4) \times U(1)$   &$U(4) \times U(1)$ &$U(4) \times SU(2)$ & & \\&  &  & $\times U(1)$  &$\times U(1)$  &$\times U(1)$   & $\times U(1)$  &$\times U(1)$   &  &
          \cr\hline
                 \multirow{2}{*}{2} &  &$U(2) \times U(1)$  &$U(2) \times U(1)$ &$U(2) \times U(1)$ &$U(2) \times U(1)$   &$U(2) \times U(1)$ &$U(2) \times U(1)$ & $U(2) \times SU(2)$ & \\&  &$\times U(1)$  & $\times U(1)$  &$\times U(1)$  &$\times U(1)$   & $\times U(1)$  &$\times U(1)$   &$\times U(1)$  &
          \cr\hline
                     0 & $U(1) \times U(1)$  &$U(1) \times U(1)$  &$U(1) \times U(1)$ &$U(1) \times U(1)$ &$U(1) \times U(1)$   &$U(1) \times U(1)$ &$U(1) \times U(1)$ & $U(1) \times U(1)$ & $SU(2) \times U(1)$
          \cr\hline
  \end{tabular}
  \caption{Flavor symmetries at strong coupling, $G_\text{F}$, for $SU(2n+1)_k
    + 1 \bm{AS} + N_f \bm{F}$ ($N_f$ even) gauge theories, which are
    descendants of $SU(2n+1)_0 + 2 \bm{AS} +
    8\bm{F}$.}\label{tab:GFsu(2n+1)1AS1}
   \end{small}
    \centering
  \begin{small}
  \vspace{1cm}
  \begin{tabular}{|l||c|c|c|c|c|c|c|c|}  
    \hline 
\diagbox{$N_f$}{$|k|$}  & $\left|n-5\right|$ & $|n-4|$ & $|n-3|$ & $n-2$&$n-1$  & $n$  &$n+1$ &$n+2$
          \cr\hline \hline
  \multirow{2}{*}{7} &  &  & &$U(7) \times U(1)$  &$U(7) \times SU(2)$ & &  &  \\&  &  & & $\times U(1)$&   $\times U(1)$ & &  & 
            \cr\hline
     \multirow{2}{*}{5} &  &  & $U(5) \times U(1)$  &$U(5) \times U(1)$   & $U(5) \times U(1)$ &$U(5) \times SU(2)$  & & \\&  &  &  $\times U(1)$ &$\times U(1)$  &  $\times U(1)$  &$\times U(1)$  & &
   \cr\hline
        \multirow{2}{*}{3} &  &$U(3) \times U(1)$  & $U(3) \times U(1)$  &$U(3) \times U(1)$   & $U(3) \times U(1)$ &$U(3) \times U(1)$  & $U(3) \times SU(2)$ & \\&  & $\times U(1)$ &  $\times U(1)$ &$\times U(1)$  &  $\times U(1)$  &$\times U(1)$  &$\times U(1)$ &
   \cr\hline
             1 & $U(1)^3$ &$U(1)^3$   & $U(1)^3$   &$U(1)^3$    & $U(1)^3$  &$U(1)^3$   &  $U(1)^3$& $U(1)^2 \times SU(2)$
   \cr\hline
  \end{tabular}
  \caption{Flavor symmetries at strong coupling, $G_\text{F}$, for $SU(2n+1)_k + 1 \bm{AS} + N_f \bm{F}$ ($N_f$ odd) gauge theories, which are descendants of $SU(2n+1)_0 + 2 \bm{AS} + 8\bm{F}$.}\label{tab:GFsu(2n+1)1AS2}
   \end{small}  
\end{sidewaystable}

\begin{sidewaystable}
   \begin{small}
   \centering
  \begin{tabular}{|l||c|c|c|c|c|c|c|c|c|}  
    \hline 
\diagbox{$N_f$}{$|k|$} & $0$ & $\frac{1}{2}$& $1$ &$\frac{3}{2}$ &  $2$ &$\frac{5}{2}$ & $3$  &$\frac{7}{2}$&$4$ 
   \cr\hline \hline
8 &$U(8) \times U(1)^2$  &  & & &  & & &  & 
   \cr\hline
7 &  &$U(7) \times U(1)^2$  & & &  & & &  & 
   \cr\hline
6 & $U(6) \times U(1)^2$ &  & $U(6) \times U(1)^2$& &  & & &  & 
   \cr\hline
 5 &  &$U(5) \times U(1)^2$  & &$U(5) \times U(1)^2$ &  & & &  & 
   \cr\hline
 4 &$U(4) \times U(1)^2$  &  &$U(4) \times U(1)^2$ & &$U(4) \times U(1)^2$  & & &  & 
   \cr\hline
 3 &  &$U(3) \times U(1)^2$  & &$U(3) \times U(1)^2$ & &$U(3) \times U(1)^2$ & &  & 
   \cr\hline
 2 &$U(2) \times U(1)^2$  &  & $U(2) \times U(1)^2$& &$U(2) \times U(1)^2$ & &$U(2) \times U(1)^2$ &  & 
   \cr\hline
1 &  & $U(1)^3$  & &$U(1)^3$ & &$U(1)^3$ & & $U(1)^3$ & 
   \cr\hline
0 &$U(1)^2$  &   &$U(1)^2$ & &$U(1)^2$ &&$U(1)^2$ &  & $U(1)^2$
   \cr\hline
  \end{tabular}
  \caption{Flavor symmetries at strong coupling, $G_\text{F}$, for $SU(2n)_k + 1 \bm{AS} + N_f \bm{F}$ ($N_f$ even) gauge theories, which are descendants of $SU(2n)_0 + 2 \bm{AS} + 8\bm{F}$.}\label{tab:GFsu2n1AS3}
   \end{small}
\end{sidewaystable}

\section{Details for Geometric Resolutions}
\label{app:Resolutions}

\subsection{Rank Two E-string}\label{sec:explicit_res_E8SU2}

Here we present an explicit fiber geometry that resolves the non-minimal singularities of the $(E_8 ,SU(2))$ collision.
It is obtained from a flop transition of the marginal geometry in figure 22 of \cite{Apruzzi:2019opn}. Namely, the $(-1)$-curve  $u_8\cdot S_2$ on $S_2$ is flopped into $S_1$.
The resulting non-flat surfaces $S_i = \{\delta_i = 0 \}$ are shown in figure \ref{fig:KK-geometry_E8SU2}.
Here, affine $E_8$ and $SU(2)$ fiber components resolving the codimension one $E_8$ and $SU(2)$ singularities, respectively, are
\begin{align}
  \begin{split}
    & \left( F^{E_8}_0 , \, F^{E_8}_1 , \, F^{E_8}_2 , \, F^{E_8}_3 , \, F^{E_8}_4 , \, F^{E_8}_5 , \, F^{E_8}_6 , \, F^{E_8}_7 , \, F^{E_8}_8 \right) \\
    \longleftrightarrow \quad & \left( U , \, u_8 , \, u_7  , \, u_{11} , \, u_{13} , \, u_{14} , \, u_{15} , \, u_9 , \, u_5 \right) \, , \\
    & \left(F_0^{SU(2)} \, , F_1^{SU(2)} \right) \quad \longleftrightarrow \quad \left( V , v_1 \right) \, .
  \end{split}
\end{align}

\begin{figure}[h]
  \centering
  \includegraphics*[width=.7\hsize]{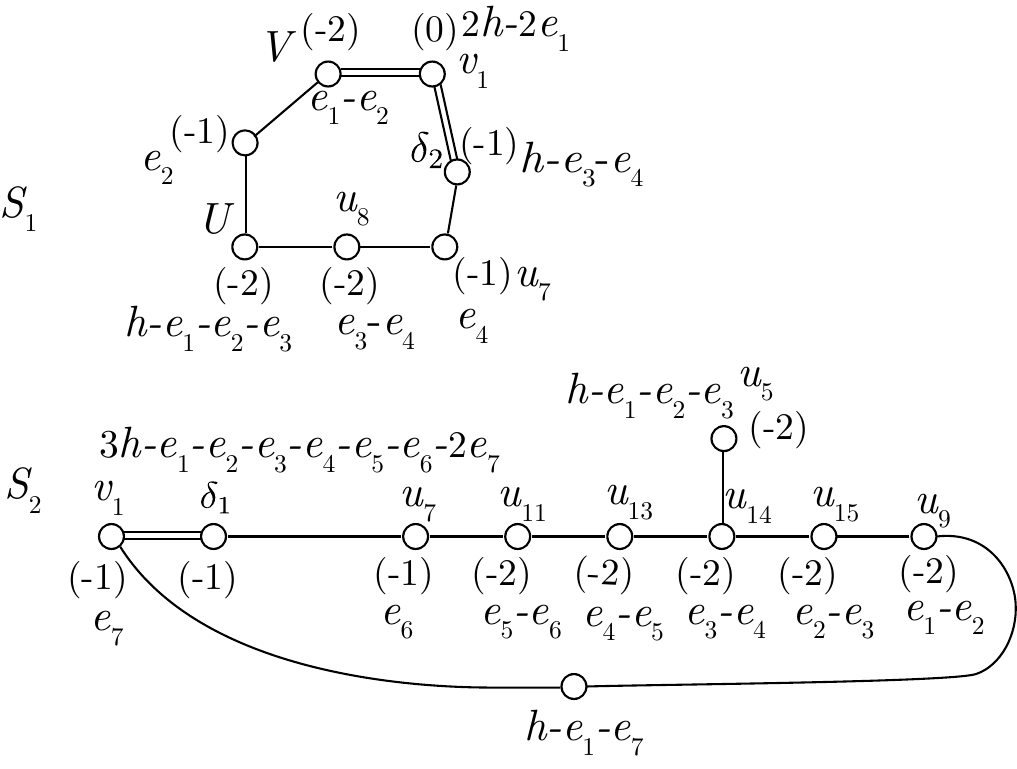}
  \caption{One concrete resolution of the $(E_8, SU(2))$ model that corresponds to a marginal theory in 5d.
  This geometry is obtained by a sequence of flops from the tensor branch resolution of the 6d rank two E-string.
  One of these flops change the genus of the gluing curve $S_1 \cap S_2 = \{\delta_1\} \cap \{\delta_2\}$ from 1 to 0.
  Each depicted node is rational curve generating the Mori cone of each the corresponding surface.}\label{fig:KK-geometry_E8SU2}
\end{figure}

As one can see from figure \ref{fig:KK-geometry_E8SU2}, the codimension one nodes $u_7$ and $v_1$ split into two components, each contained in one of the two surfaces $S_{1,2}$.
The intersection numbers can be inferred from the homology classes, which we choose to represent in the basis of del Pezzo surfaces, i.e., $h^2 = 1, h \cdot e_i = 0, e_i \cdot e_j = -\delta_{ij}$.
In order to determine the gauge phases, it is important to remember that displayed curves are rational, and so we can compute for a curve $C \subset S_i$ the following intersection numbers in the three-fold:
\begin{align}
  C \cdot S_i = -2 - (C \cdot C)|_{S_i} \, , \quad C \cdot S_j = (C \cdot \delta_j)|_{S_i} \, , \quad \text{where } \, i\neq j \, .
\end{align}

Note that the surfaces $S_1$ and $S_2$ contain the $(-1)$ curves labeled $e_2$ and $h-e_1-e_7$, respectively, which do not arise from intersections with exceptional codimension one divisors.
Together with the split products of $F^{SU(2)}_1 \leftrightarrow v_1$ and $F^{E_8}_2 \leftrightarrow u_7$, and the gluing curve $S_1 \cap S_2$, they form the extremal curves in this phase.
For convenience, we list their intersection numbers with all divisors:
\begin{align}\label{eq:int_numbers_extremal_curves_E8SU2_res}
  \renewcommand{\arraystretch}{1.2}
  \begin{array}{c|c c|c @{} c @{} c @{} c @{} c @{} c @{} c @{} c @{} c | c @{} c}
    & S_1 & S_2 & D^{E_8}_0 & D^{E_8}_1 & D^{E_8}_2 & D^{E_8}_3 & D^{E_8}_4 & D^{E_8}_5 & D^{E_8}_6 & D^{E_8}_7 & D^{E_8}_8 & D^{SU(2)}_0 & D^{SU(2)}_1\\ \hline\hline
    e_2|_{S_1} & -1 & 0 & 1 & 0 & 0 & 0 & 0 & 0 & 0 & 0 & 0 & 1 & 0 \\ \hline
    (h - e_1 - e_7)|_{S_2} & 0 & -1 & 0 & 0 & 0 & 0 & 0 & 0 & 0 & 1 & 0 & 0 & 1 \\ \hline
    %
    %
    v_1|_{S_2} & 2 & -1 & 0 & 0 & 0 & 0 & 0 & 0 & 0 & 0 & 0 & 0 & -1 \\ \hline
    u_7|_{S_1} & -1 & 1 & 0 & 1 & -1 & 0 & 0 & 0 & 0 & 0 & 0 & 0 & 0 \\ \hline
    u_7|_{S_2} & 1 & -1 & 0 & 0 & -1 & 1 & 0 & 0 & 0 & 0 & 0 & 0 & 0 \\ \hline
    S_1 \cap S_2 & -1 & -1 & 0 & 0 & 1 & 0 & 0 & 0 & 0 & 0 & 0 & 0 & 2
  \end{array}
\end{align}

\subsubsection*{\texorpdfstring{\boldmath{$SU(3)$}}{\textit{SU(3)}} Gauge Description}

The rulings that give the $SU(3)_\text{gauge}$ gauge theory are
\begin{align}\label{eq:SU3_ruling_for_E8SU2_res}
  \begin{split}
    & S_1 \hookleftarrow f_1 = (h-e_1)|_{S_1} = (e_2 + U + u_8 + u_7)|_{S_1} \, ,\\
    & S_2 \hookleftarrow f_2 = (h-e_7)|_{S_2}=(u_7 + u_{11} + u_{13} + u_{14} + u_{15} + u_9 + (h - e_1 - e_7))|_{S_2} \, .
  \end{split}
\end{align}
The codimension one fibers that are part of these rulings are the $E_8$ roots $(F_0^{E_8} , \cdots , F_7^{E_8})$, which give rise to the non-abelian $SU(9)_\text{BG}$ part of the flavor symmetry $U(9)_\text{BG}$.
We order the Cartan generators of $SU(3)_\text{gauge}$ as $(S_2, S_1)$, and those of $SU(9)_\text{BG}$ as $(F_1 , \cdots, F_8) \leftrightarrow (F_0^{E_8}, \cdots, F_7^{E_8})$, such that the geometry realizes hypermultiplets in the $({\bf 3, 9})$ representation, and not $(\bar{\bf 3}, \bf 9)$.
Furthermore, the extremal curves $e_2|_{S_1}$, $u_7|_{S_i}$, and $(h-e_1-e_7)|_{S_2}$ are special fiber components which shrink when we collapse $f_{1,2}$, and give rise to the bifundamental matter.

\subsubsection*{\texorpdfstring{\boldmath{$Sp(2)$}}{\textit{Sp(2)}} Gauge Description}

We obtain an $Sp(2)_\text{gauge}$ gauge theory with the rulings
\begin{align}\label{eq:Sp2_ruling_for_E8SU2}
  \begin{split}
    & S_1 \hookleftarrow f_1 = (h-e_1)|_{S_1} = (e_2 + U + u_8 + u_7)|_{S_1} \, ,\\
    & S_2 \hookleftarrow f_2 = (v_1 + (h-e_1-e_7))|_{S_2} = (u_5 + u_{15} + 2\,(u_{14} + u_{13} + u_{11} + u_7))|_{S_2} \, .
  \end{split}
\end{align}
Here, the codimension one fibers $(F^{E_8}_0, \cdots, F^{E_8}_6, F^{E_8}_8)$ and $F^{SU(2)}_1$ are parts of the ruling, giving rise to the flavor symmetry $SO(16)_\text{BG} \times SU(2)_\text{BG}$.
Under the total symmetry group $Sp(2)_\text{gauge} \times SO(16)_\text{BG} \times SU(2)_\text{BG}$, the extremal curves $e_2|_{S_1}$ and $u_7|_{S_i}$ give rise to states in the $({\bf 4, 16, 1})$, while $v_1|_{S_2}$ and $(h-e_1-e_7)|_{S_2}$ support $({\bf 5, 1, 2})$ states.

\subsubsection*{\boldmath{$SU(2)^2$} Gauge Description}

The rulings for this gauge theory are
\begin{align}\label{eq:SU2_ruling_for_E8SU2}
  \begin{split}
    & S_1 \hookleftarrow f_1 = (\delta_2 + u_7)|_{S_1} = (U + 2e_2 + V)|_{S_1} \, ,\\
    & S_2 \hookleftarrow f_2 = (\delta_1 + u_7)|_{S_2} = (u_5 + u_{13} + 2\,(u_{14} + u_{15} + u_{9} + (h-e_1-e_7)))|_{S_2} \, .
  \end{split}
\end{align}
The codimension one fibers contained in these rulings are $(F^{E_8}_0, F^{E_8}_2, F^{E_8}_4, F^{E_8}_5, \cdots, F^{E_8}_8)$ and $F^{SU(2)}_0$, which span the flavor symmetry $(SO(4) \times SU(2) \times SO(10))_\text{BG}$.
The extremal curves that give rise to matter are:
\begin{align}
  \renewcommand{\arraystretch}{1.2}
  \begin{array}{c|c|ccc}
    \text{curves} & SU(2)_{\text{gauge}_1} \times SU(2)_{\text{gauge}_2} & SO(4)_\text{BG} \cong (SU(2)^2)_\text{BG} & SU(2)_\text{BG} & SO(10)_\text{BG} \\ \hline \hline
    \delta_2|_{S_1} = \delta_1|_{S_2} \, , & \multirow{2}{*}{({\bf 2 , 2})} & \multirow{2}{*}{(\bf 1,1)} & \multirow{2}{*}{\bf 2} & \multirow{2}{*}{\bf 1} \\ 
    u_7|_{S_1} \, , u_7|_{S_2} & \\ \hline
    e_2|_{S_1} & ({\bf 2, 1}) & ({\bf 2,2}) & {\bf 1} & {\bf 1} \\
    (h-e_1-e_7)|_{S_2} & ({\bf 1,2}) & ({\bf 1,1}) & {\bf 1} & {\bf 10} 
  \end{array}
\end{align}

\subsection[Resolutions with Different Rulings for \texorpdfstring{$E_6 \times E_6$}{E6xE6} Conformal Matter]{Resolutions with Different Rulings for \boldmath{$(E_6, E_6)$} Conformal Matter}
\label{app:E6E6resol}

The singular Tate model of $(E_6,E_6)$ conformal matter is
\be
\label{E7SO7:Tate}
y^2+b_1 UVxy+b_3 U^2 V^2 y=x^3+b_2 U^2 V^2+b_4 U^3 V^3+b_6 U^5 V^5\,.
\ee

Here we present two example resolutions, which have different ruling and quiver gauge theory descriptions. The first one is
\be
\ba
\label{E6E6BU1}
&BU1_{(E_6,E_6)}=\cr
&\{\left\{x,y,U,u_1\right\},\left\{x,y,V,v_1\right\},\left\{x,y,u_1,u_2\right\},\left\{y,u_1,u_2,u_3\right\},\left\{y,u_1,u_4\right\},
\left\{y   ,u_2,u_5\right\},\left\{v_1,u_5,\delta _1\right\},\cr 
&\left\{\delta _1,y,\delta _2\right\},\left\{v_1,u_4,\delta _3\right\},\left\{v_1,\delta_2,\delta_4\right\},\left\{u_3,u_4,u_6\right\},\left\{y,u_3,u_7\right\},\left\{x,y,v_1,v_2\right\},\left\{y,v_1,v_2,v_3\right\},\cr 
& 
\left\{y,v_1,v_4\right\}
   ,   \left\{y,v_2,v_5\right\},\left\{v_3,v_4,v_6\right\},\left\{y,v_3,v_7\right\},\left\{u_7,v_1,\delta _5\right\}\}. \ea
\ee  

The following exceptional divisors form the Dynkin diagram of $(E_6,E_6)$:
\be
\label{f:E6E6-Dynkin}
\centering
\includegraphics[width=8cm]{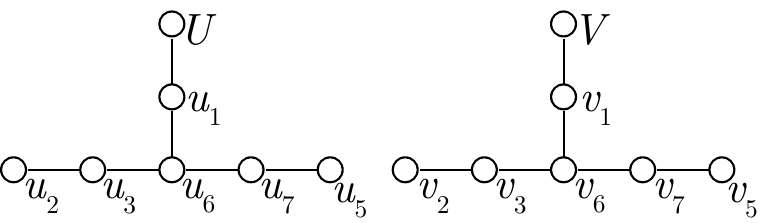}.
\ee

We list some of the triple intersection numbers here:

\be
\label{tripleint-E6E6BU1}
\small
\begin{array}{c|ccccccccccccccccccc}
S_i \cdot D_j^2 & U & u_1 & u_6 & u_3 & u_2 & u_7 & u_5 & V & v_1 & v_6 & v_3 & v_2 & v_7 & v_5
   & \delta _1 & \delta _2 & \delta _3 & \delta _4 & \delta _5 \\ \hline
 S_1 & 0 & 0 & 0 & 0 & -2 & 0 & 0 & 0 & 0 & 0 & 0 & 0 & 0 & 0 & 8 & 0 & 0 & 2 & 0 \\
 S_2 & 0 & 0 & 0 & 0 & 0 & 0 & -2 & 0 & 0 & 0 & 0 & 0 & 0 & 0 & 0 & 8 & 0 & 2 & 0 \\
 S_3 & 0 & -1 & -2 & -1 & 0 & -1 & 0 & 0 & 0 & 0 & 0 & 0 & 0 & 0 & 0 & 0 & 6 & 0 & -2 \\
 S_4 & 0 & 0 & 0 & -1 & 0 & -1 & 0 & 0 & 0 & -2 & -2 & -2 & -2 & -2 & -4 & -4 & 0 & 0 & -2 \\
 S_5 & 0 & 0 & 0 & 1 & 0 & 1 & 0 & 0 & -1 & 0 & 0 & 0 & 0 & 0 & 0 & 0 & 0 & 0 & 8 \\ \hline
  n(F_j)     & 0 & -1 & -2 & -2 & -2 & -2 & -2 & 0 & -2 & -2 & -2 & -2 & -2 & -2 & - & - & - & - & -\\
 \end{array}
\ee

$n(F_j)$ is the ``wrapping number'' of each Cartan node inside the non-flat fiber, introduced in ~\cite{Apruzzi:2019opn}. If $n(F_j)=-2$, then such node is fully wrapped and regarded as a flavor curve. In this case, the actual wrapping number $n(F_j)$ are computed with the following non-trivial multiplicities~\cite{Apruzzi:2019opn}:
\be
\xi_i^{(u)}=(1,1,1,1,2)\,,\ \xi_i^{(v)}=(1,1,1,2,1).\label{E6E6-multiplicity}
\ee
As a reminder, any intersection numbers $S_i\cdot u_j^2$ needs to be multiplied by $\xi_i^{(v)}$, while any intersection numbers $S_i\cdot v_j^2$ needs to be multiplied by $\xi_i^{(u)}$ (including the affine nodes $U$ and $V$).

We draw the configuration of curves on the five surface components in figure~\ref{f:E6E6BU1}. The corresponding CFD is read off as
\be
\label{f:E6E6CFD-03}
\centering
\includegraphics[width=4cm]{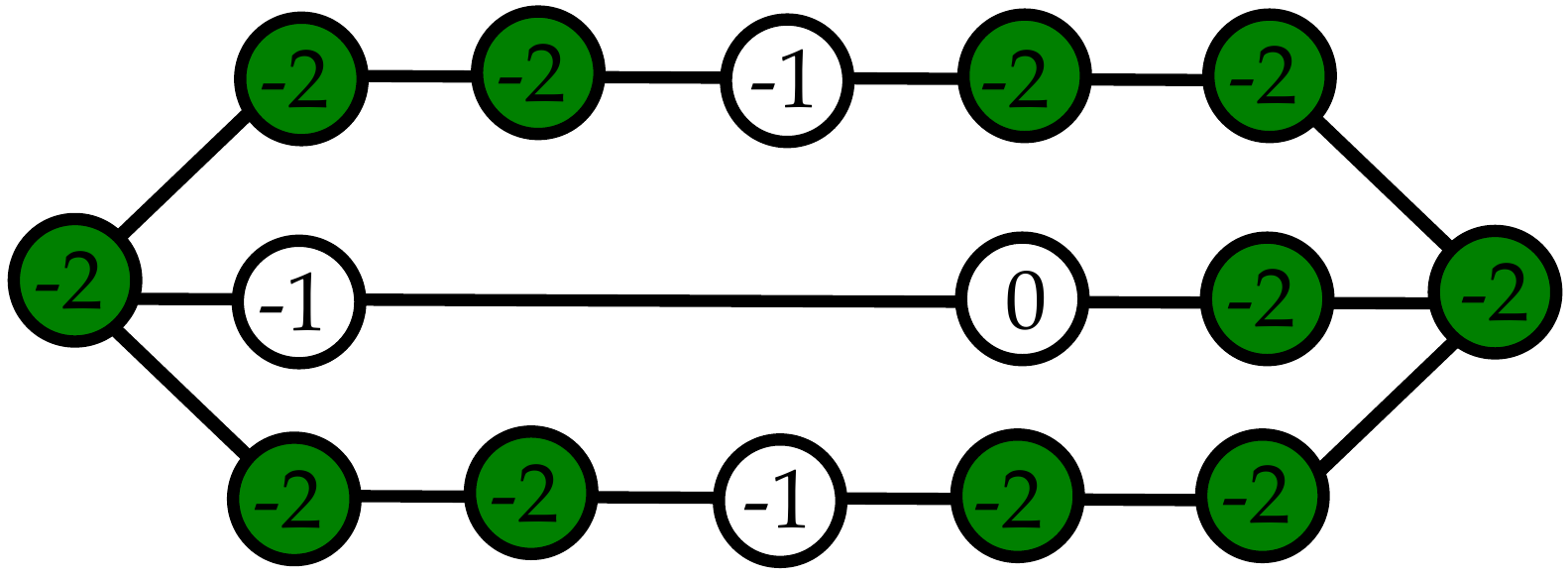}.
\ee

In this case, the assignment of ruling and section for each surface and each curve is uniquely determined. Recall that the ruling curve on each surface needs to be a linear combination of curves with self-intersection number 0 and genus 0. Since $S_1$, $S_2$ and $S_5$ are Hirzebruch surfaces, the 0-curves on them have to be ruling curves. Then we can conclude that $S_1\cdot S_4$, $S_2\cdot S_4$ are section curves, while $S_3\cdot S_5$ and $S_4\cdot S_5$ are a part of ruling curves. If the geometry has a quiver gauge theory description, then the assignment of section and ruling needs to be identical for a curve $S_i\cdot S_j$ on both $S_i$ and $S_j$. With these requirements, the only consistent assignment of section/ruling is shown in figure~\ref{f:E6E6BU1}. We also list the linear combinations of curves on each surface component that correspond to the ruling:
\be
\ba
\label{E6E6BU1:rulings}
&S_1: f^{(1)}\equiv x=u_3\,,\cr
&S_2: f^{(2)}\equiv x=u_7\,,\cr
&S_3: f^{(3)}\equiv V=\delta_5+2u_3+u_6\,,\cr
&S_4: f^{(4)}\equiv u_3+\delta_5+u_7=C_1+v_2+v_3+v_6+v_7+v_5+C_2\,,\cr
&S_5: f^{(5)}\equiv \delta_4=\delta_3\,.
\ea
\ee

\begin{figure}
\centering
\includegraphics[height=6cm]{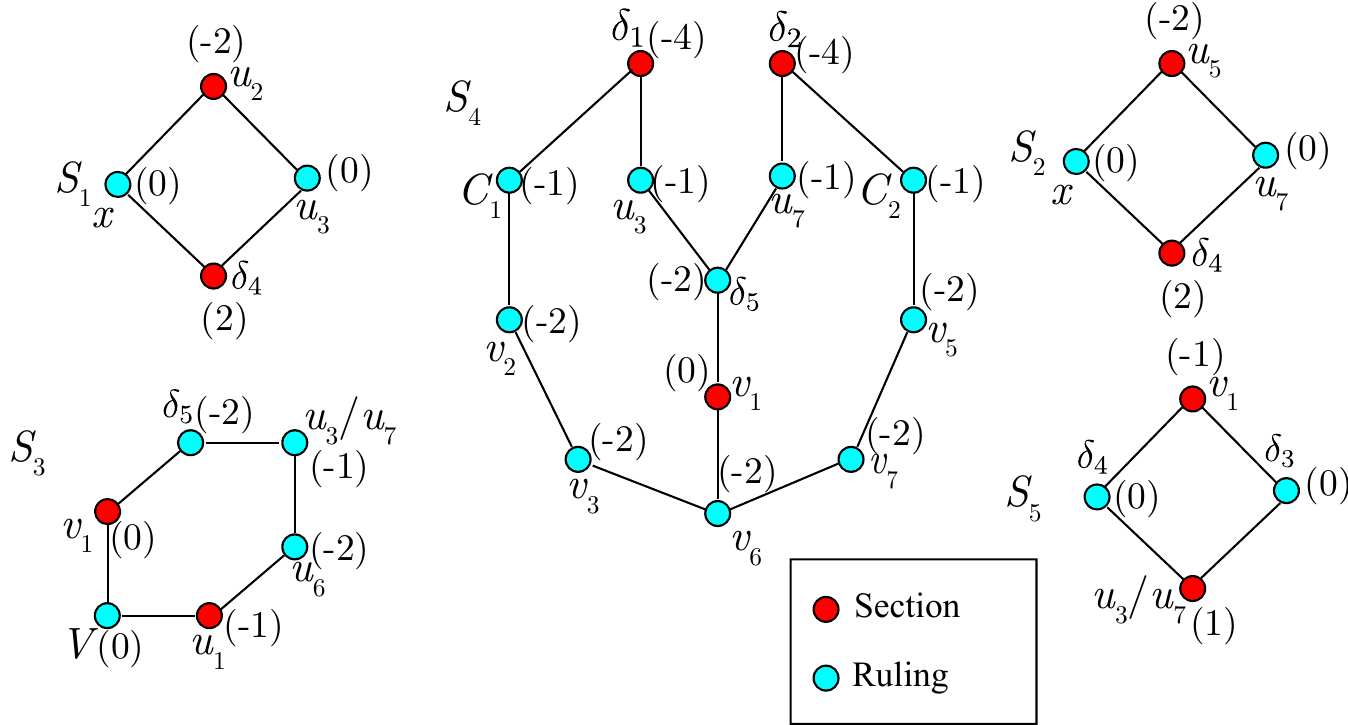}
\caption{The configuration of curves on $S_i(i=1,\dots,5)$ in the geometry $BU1_{(E_6,E_6)}$. The number in the bracket denotes the self-intersection number of the curve. The letter denotes an intersection curve with the corresponding divisor. The ``/'' symbol means that the curves are in the same homology class. The assignment of section/ruling of each curve is marked by red/blue colors.}\label{f:E6E6BU1}
\end{figure}

From this information, we see that the quiver gauge theory has a $SU(4)\times SU(2)^{(1)}\times SU(2)^{(2)}$ gauge group. The three Cartans generators of $SU(4)$ correspond to $S_1$, $S_4$ and $S_2$, while the Cartan generators of  $SU(2)^{(1)}$ and $SU(2)^{(2)}$ correspond to $S_5$ and $S_3$ respectively. 

The massless matter fields are generated by M2 brane wrapping over $(-1)$-curves that are apart of the rulings, which shrinks to zero size in the gauge theory limit. The two unlabeled $(-1)$-curves on $S_2$, along with their linear combinations with the string of five $(-2)$-curves, in total give six copies of $(\mbf{4},\mbf{1},\mbf{1})$ and $(\bar{\mbf{4}},\mbf{1},\mbf{1})$ under $SU(4)\times SU(2)^{(1)}\times SU(2)^{(2)}$. Additionaly, the  curves $u_3\cdot S_2$ and $u_7\cdot S_2$ give rise to bifundamentals $(\bar{\mbf{4}},\mbf{1},\mbf{2})$ and $(\mbf{4},\mbf{1},\mbf{2})$. Moreover, the  curve $u_3\cdot S_3$ gives the bifundamental $(\mbf{1},\mbf{2},\mbf{2})$.

In conclusion, the quiver gauge theory description of the geometry $BU1_{(E_6,E_6)}$ is
\be
[6]-SU(4)-SU(2)^{(1)}-SU(2)^{(2)}\,.
\ee

From the geometry in figure~\ref{f:E6E6BU1}, we can blow up the surface component $S_3$ twice and get the following configuration:
\be
\label{f:E6E6BU1-top}
\centering
\includegraphics[width=12cm]{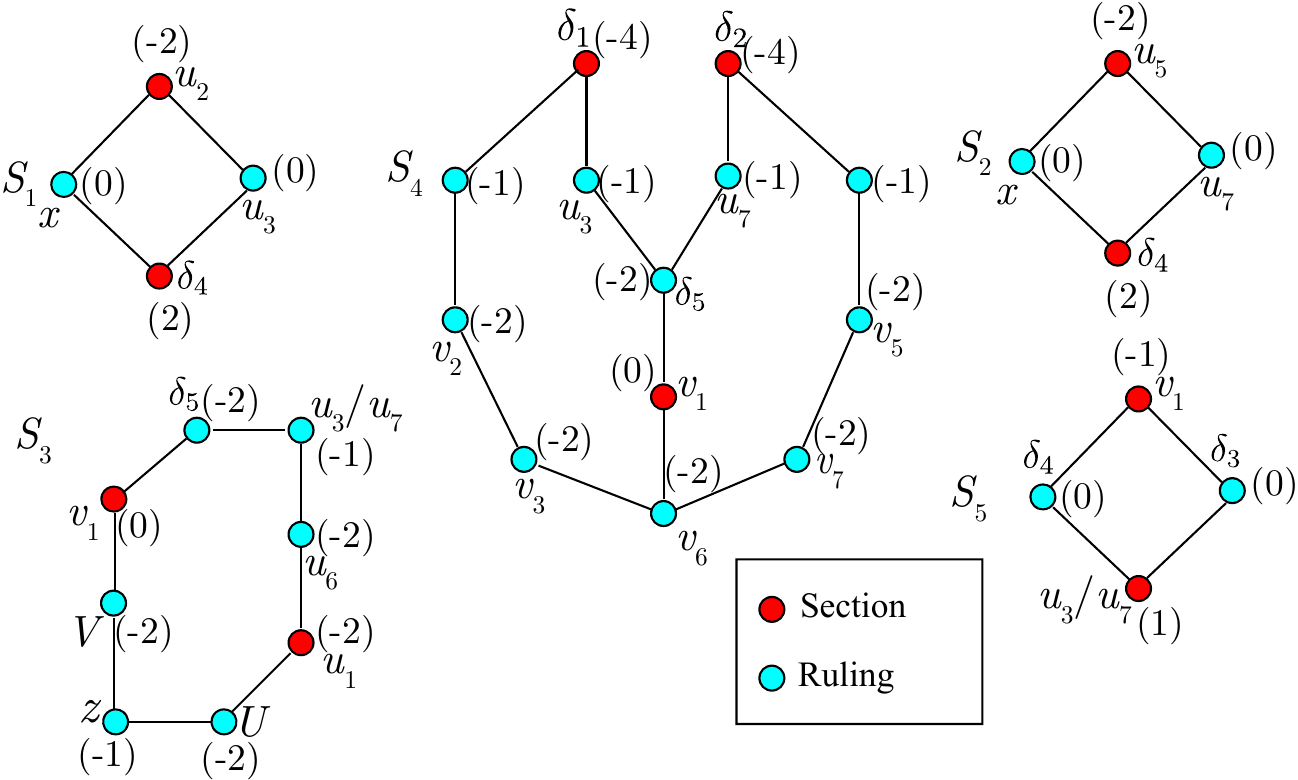}.
\ee
Because of the appearance of new fibral $(-1)$-curve $z\cdot S_3$ on $S_3$ and its two adjacent $(-2)$-curves, the quiver gauge theory description is now
\be
\label{E6E6-quiver-4-2}
[6]-SU(4)-SU(2)^{(1)}-SU(2)^{(2)}-[2]\,,
\ee
with two fundamental flavors on the $SU(2)^{(2)}$ gauge node. Moreover, the corresponding CFD of (\ref{f:E6E6BU1-top}) is exactly the marginal CFD of $(E_6,E_6)$ conformal matter theory:
\be
\label{f:E6E6-CFD-01}
\centering
\includegraphics[width=4cm]{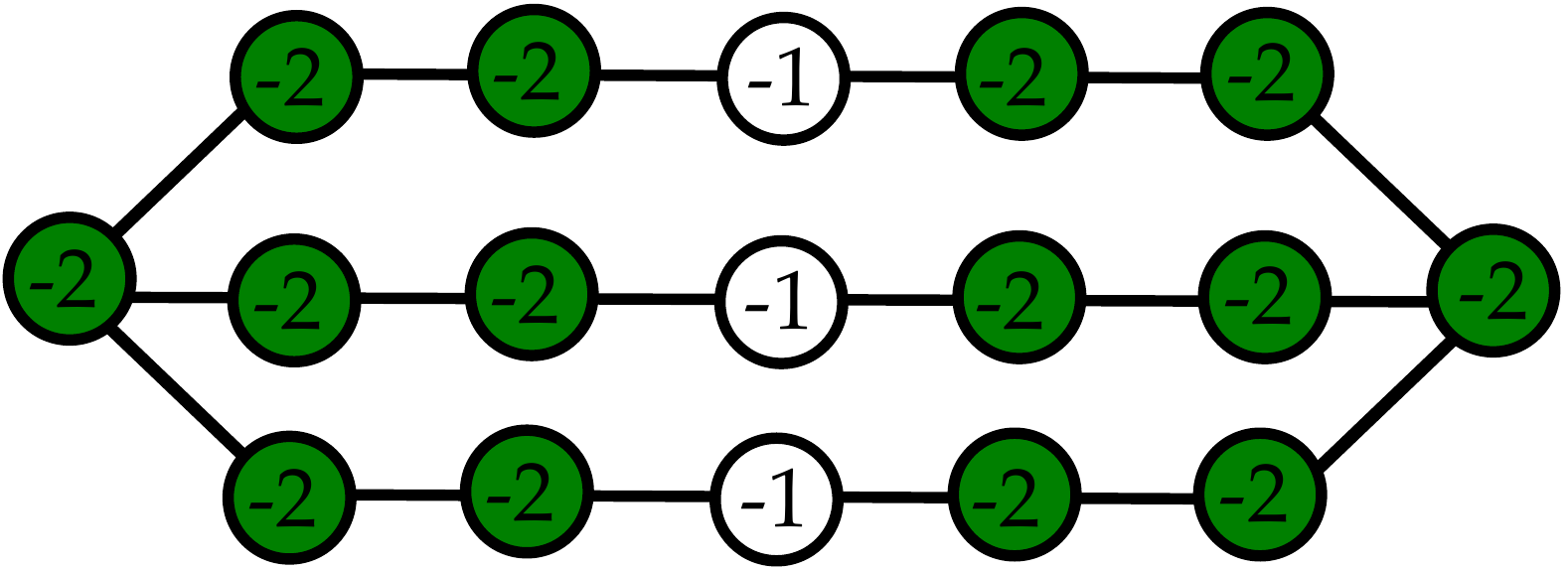}.
\ee

Hence we conclude that the geometry (\ref{f:E6E6BU1-top}) describes the $(E_6,E_6)$ marginal theory, and the quiver description  (\ref{eqn:diego}) in section~\ref{sec:E6asym} indeed appears.

From the geometry (\ref{f:E6E6BU1-top}), there are two ways to flop a $(-1)$-curve outside of these surfaces. One can shrink the $(-1)$-curve $z\cdot S_3$ on $S_3$, and get the following geometry:
\be
\label{f:E6E6BU1-6-1}
\centering
\includegraphics[width=12cm]{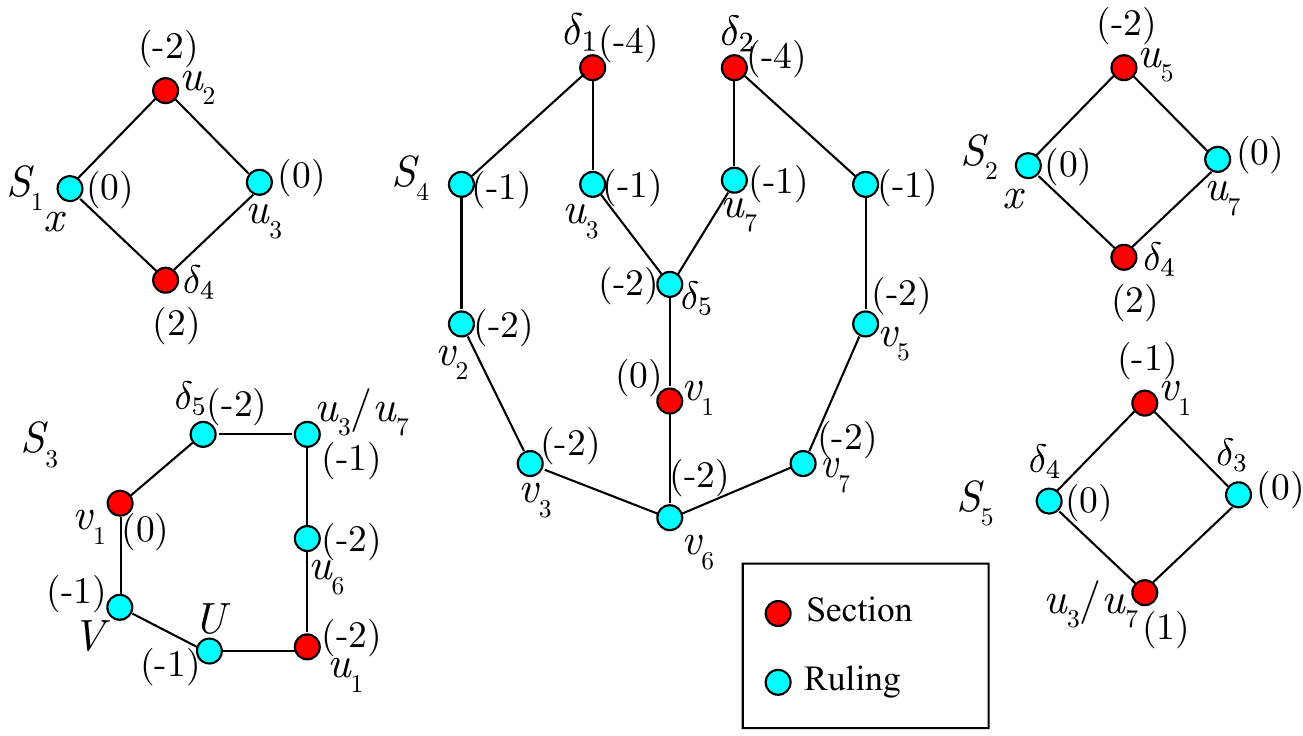}.
\ee
It has quiver gauge theory description
\be
[6]-SU(4)-SU(2)^{(1)}-SU(2)^{(2)}-[1]\,.
\ee

Alternatively, one can flop the $(-1)$-curve connected to $v_2\cdot S_4$ on $S_4$ into $S_1$ and then shrink it. After this birational transformation, the surface geometry becomes
\be
\label{f:E6E6BU1-5-2}
\centering
\includegraphics[width=12cm]{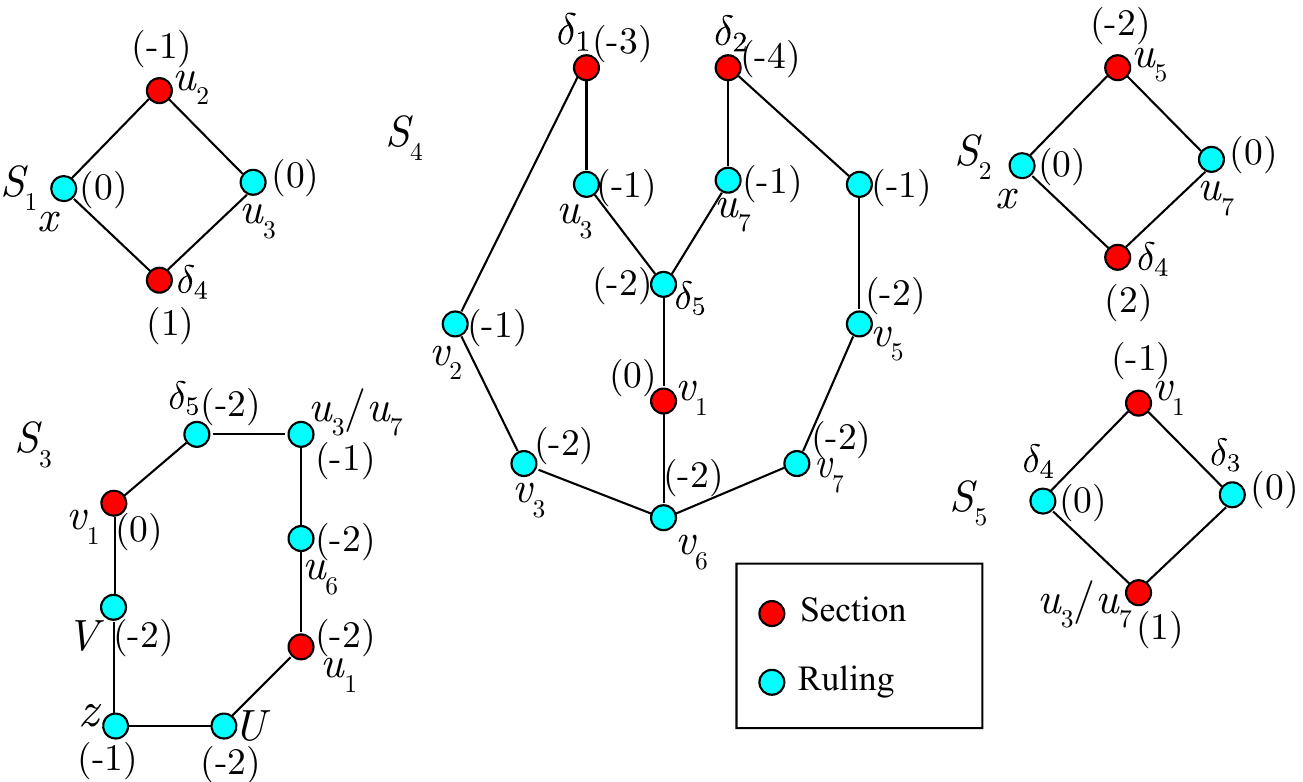}.
\ee
Comparing to (\ref{f:E6E6BU1-top}), the number of fundamental flavors for the $SU(4)$ gauge group is decreased by one, and we have the quiver gauge theory description
\be
[5]-SU(4)-SU(2)^{(1)}-SU(2)^{(2)}-[2]\,.
\ee

Notably, the two different geometries (\ref{f:E6E6BU1-6-1}) and (\ref{f:E6E6BU1-5-2}) have the same CFD:
\be
\label{f:E6E6CFD-02}
\centering
\includegraphics[width=4cm]{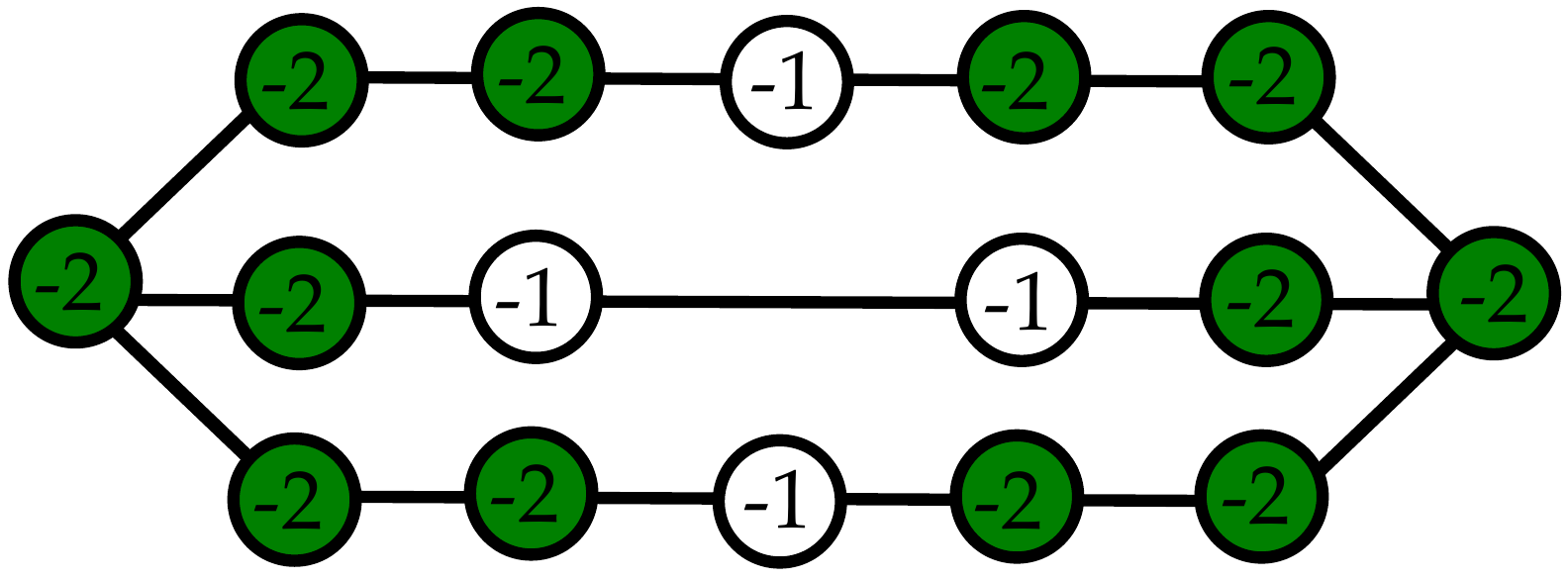}.
\ee
with superconformal flavor symmetry $G_\text{F}=E_6\times E_6$.

This confirms the non-trivial UV duality between the two quiver gauge theory descriptions $[6]-SU(4)-SU(2)-SU(2)-[1]$ and $[5]-SU(4)-SU(2)-SU(2)-[2]$ in section~\ref{sec:E6asym}, from the geometric perspective.

Furthermore, we can flop the $(-1)$-curve connected to $v_2\cdot S_4$ on $S_4$ into $S_1$ and then shrink it, resulting in the geometry 
\be
\label{f:E6E6BU1-5-1}
\centering
\includegraphics[width=12cm]{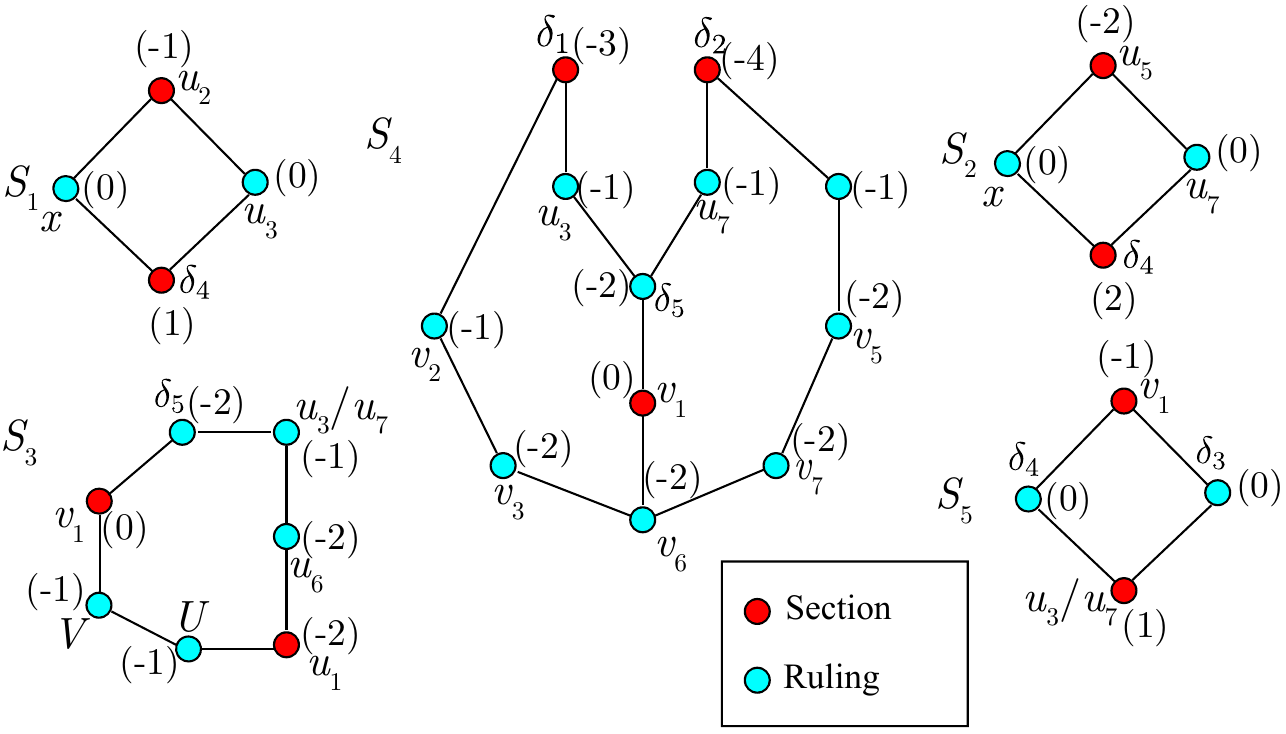}
\ee
with the quiver gauge theory description
\be
[5]-SU(4)-SU(2)^{(1)}-SU(2)^{(2)}-[1]\,.
\ee

From (\ref{f:E6E6BU1-5-2}), we can shrink the $(-1)$-curve $U\cdot S_3$ on $S_3$ to get the geometry (\ref{f:E6E6BU1-5-1}). Alternatively, we can flop the $(-1)$-curve $v_2\cdot S_4$ on $S_4$ into $S_1$ and then shrink it, resulting in the geometry
\be
\label{f:E6E6BU1-4-2}
\centering
\includegraphics[width=12cm]{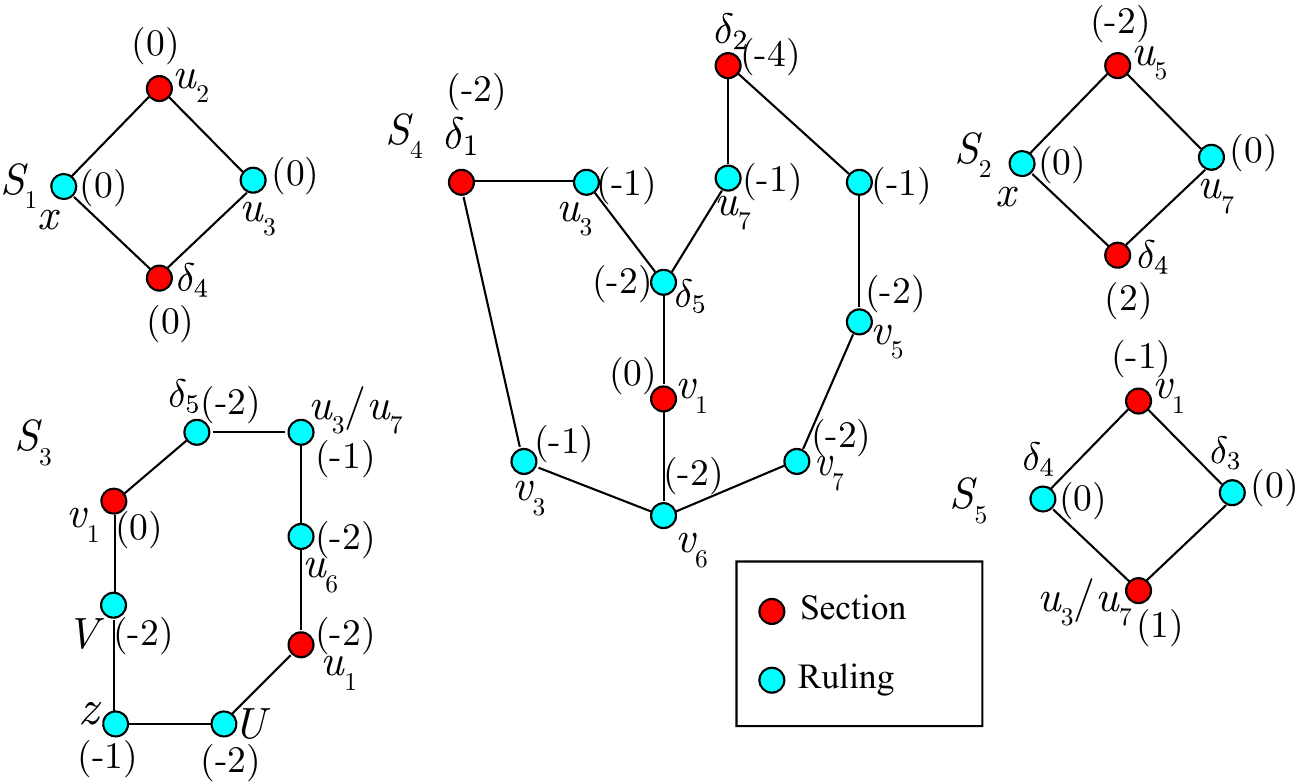}
\ee
with the quiver gauge theory description
\be
[4]-SU(4)-SU(2)^{(1)}-SU(2)^{(2)}-[2]\,.
\ee

The geometries with quiver gauge theory descriptions $[6]-SU(4)-SU(2)-SU(2)$ in figure~\ref{f:E6E6BU1} and $[4]-SU(4)-SU(2)-SU(2)-[2]$ in (\ref{f:E6E6BU1-4-2}) correspond to the same CFD:
\be
\centering
\includegraphics[width=4cm]{E6E6-CFD-03.pdf},
\ee
with the same $G_\text{F}=E_6\times SU(6)$. Hence we can perceive the UV duality between these two quiver gauge theories.

One the other hand, the geometry (\ref{f:E6E6BU1-5-1}) with quiver gauge theory description $[5]-SU(4)-SU(2)-SU(2)-[1]$ corresponds to a different CFD:
\be
\label{f:E6E6CFD-04}
\centering
\includegraphics[width=4cm]{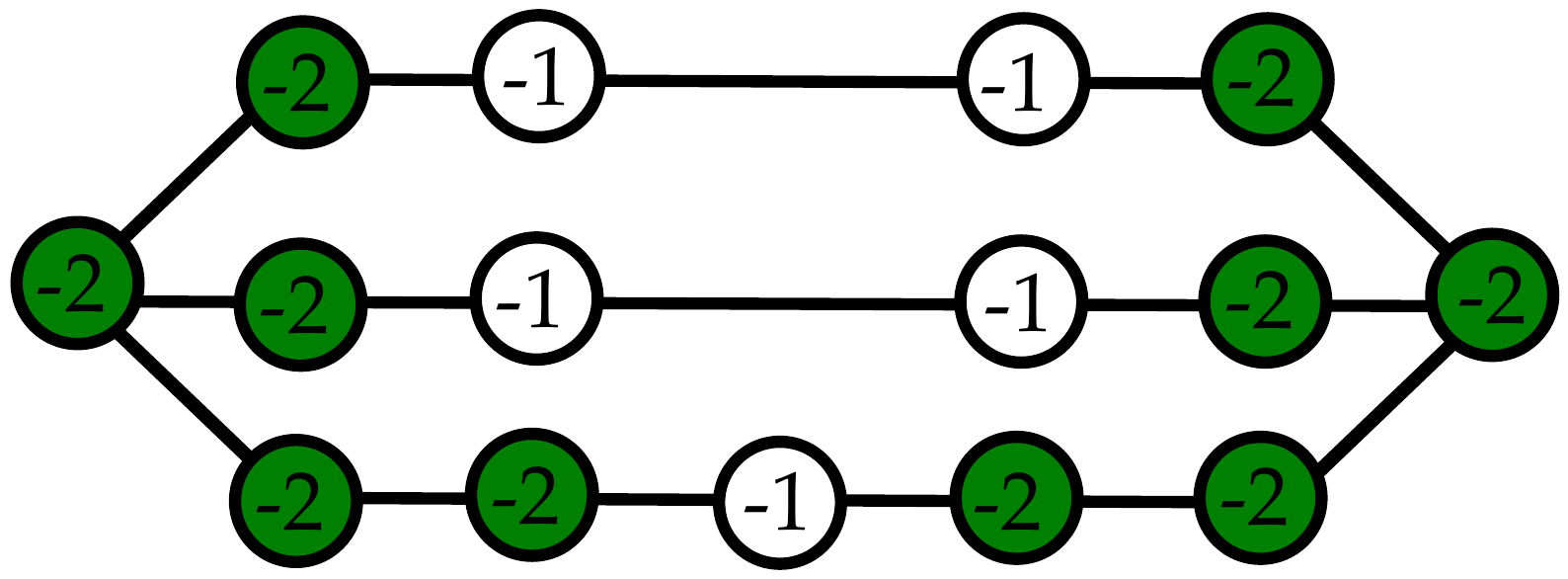}.
\ee
with $G_\text{F}=SO(10)^2\times U(1)$. 

Apart from this class of resolution geometries, we can also choose another resolution sequence: 
\be
\ba
\label{E6E6BU2}
&BU2_{(E_6,E_6)}=\cr
&\{\left\{x,y,V,v_1\right\},\left\{x,y,U,u_1\right\},\left\{x,y,u_1,u_2\right\},\left\{x,y,v_1,v_2\right\},\left\{y,u_1,u_2,u_3\right\},
\left\{y,v_1,v_2,v_3\right\},\cr 
&\left\{y,v_1,v_4\right\},\left\{u_2,v_4,\delta _1\right\},\left\{y,\delta _1,\delta _2\right\},\left\{\delta_1,v_4,\delta _4\right\},\left\{v_4,u_3,\delta _5\right\},\left\{v_4,u_1,\delta_3\right\},\cr 
&\left\{y,u_1,u_4\right\},\left\{y,u_2,u_5\right\},\left\{u_3,u_4,u_6\right\},\left\{y,u_3,u_7\right\},\left\{y,v_2,v_5\right\},\left
   \{v_3,v_4,v_6\right\},\left\{y,v_3,v_7\right\}\}.
\ea
\ee

The multiplicities are the same as (\ref{E6E6-multiplicity}), and we have the following intersection numbers:

\be
\small
\begin{array}{c|ccccccccccccccccccc}
S_i \cdot D_j^2 & U & u_1 & u_6 & u_3 & u_2 & u_7 & u_5 & V & v_1 & v_6 & v_3 & v_2 & v_7 & v_5
   & \delta _1 & \delta _2 & \delta _3 & \delta _4 & \delta _5 \\ \hline
 S_1 & 0 & 0 & 0 & 0 & 0 & 0 & 0 & 0 & 0 & 0 & 0 & 0 & 0 & 0 & 8 & 0 & 0 & 0 & 0 \\
 S_2 & 0 & 0 & 0 & 0 & 0 & -1 & 0 & 0 & 0 & -1 & 0 & 0 & -1 & 0 & 0 & 6 & 0 & -2 & -1 \\
 S_3 & 0 & -1 & 0 & 0 & 0 & 0 & 0 & 0 & 1 & 0 & 0 & 0 & 0 & 0 & 0 & 0 & 8 & 0 & 0  \\
 S _4 & 0 & 0 & 0 & -1 & 0 & 0 & 0 & 0 & 0 & -1 & -1 & 0 & 0 & 0 & -2 & 0 & 0 & 6 & -1 \\
 S_5 & 0 & -1 & -2 & 0 & 0 & -1 & 0 & 0 & -1 & 0 & 0 & 0 & 0 & 0 & 0 & -1 & -2 & -1 & 5 \\
 \hline
  n(F_j)     & 0 & -2 & -2 & -2 & 0 & -2 & 0 & 0 & -1 & -2 & -1 & 0 & -1 & 0 & - & - & - & - & -\\
  
  \end{array}
\ee

The corresponding CFD is
\be
\label{f:E6E6CFD-31}
\centering
\includegraphics[width=4cm]{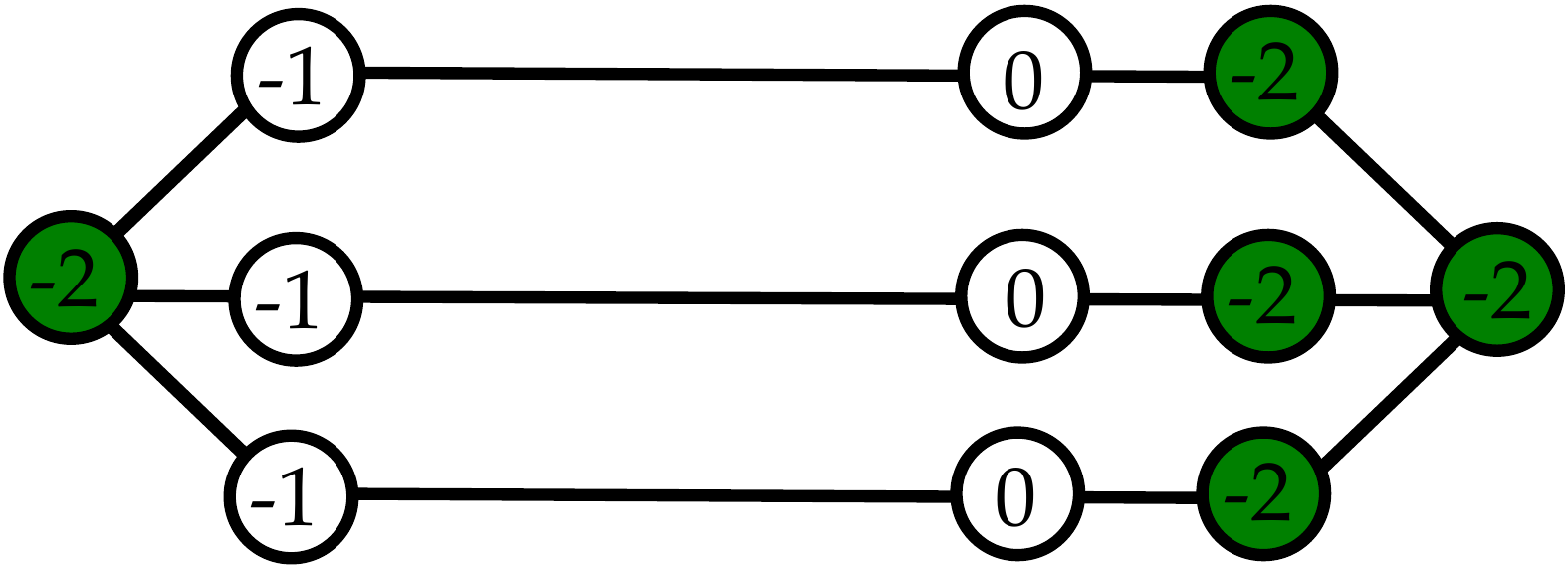}.
\ee

\begin{figure}
\centering
\includegraphics[height=7cm]{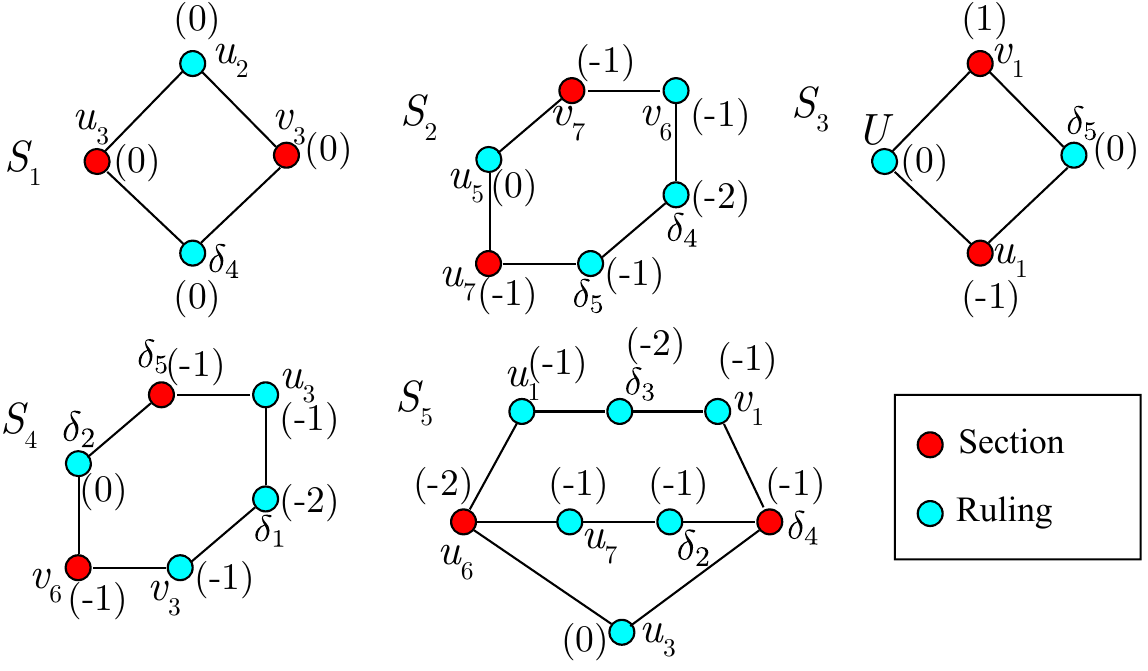}
\caption{ The configuration of curves on $S_i(i=1,\dots,5)$ in the resolution geometry $BU2_{(E_6,E_6)}$. The number in the bracket denotes the self-intersection number of the curve. The letter denotes an intersection curve with the corresponding divisor. The ``/'' symbol means that the curves are in the same homology class. The assignment of ruling/section on each surface component is denoted by blue/red colors.}\label{f:E6E6BU2}
\end{figure}

We plot the configuration of curves on the five non-flat surface components in figure~\ref{f:E6E6BU2}. For this geometry, the assignment of ruling on each surface component is uniquely fixed by the requirement that each intersection curve $S_i\cdot S_j$ is a ruling/section on both $S_i$ and $S_j$. We list the rational ruling curves with self-intersection 0 on each surface component:
\be
\ba
\label{E6E6BU2:rulings}
&S_1: f^{(1)}\equiv u_2=\delta_4\,,\cr
&S_2: f^{(2)}\equiv u_5=v_6+\delta_4+\delta_5\,,\cr
&S_3: f^{(3)}\equiv U=\delta_5\,,\cr
&S_4: f^{(4)}\equiv \delta_2=u_3+\delta_1+v_3\,,\cr
&S_5: f^{(5)}\equiv u_3=u_7+\delta_2=u_1+\delta_3+v_1\,.\cr
\ea
\ee

Hence we conclude that this geometry describes a quiver gauge theory with gauge groups $SU(3)\times SU(2)^{(1)}\times SU(2)^{(2)}\times SU(2)^{(3)}$. The Cartan divisors of the $SU(3)$ factor correspond to the surface components $S_4$ and $S_5$, while the Cartan divisors of $SU(2)^{(1)}$, $SU(2)^{(2)}$ and $SU(2)^{(3)}$ correspond to the surface components $S_1$, $S_2$ and $S_3$ respectively.

The matter fields of this quiver gauge theory can be read off from the $(-1)$-curves in figure~\ref{f:E6E6BU2} that are a part of ruling (colored by blue). Their representations under $SU(3)\times SU(2)^{(1)}\times SU(2)^{(2)}\times SU(2)^{(3)}$ are:
\be
\ba
\label{E6E6BU2-reps}
&v_6\cdot S_2:\quad (\mbf{3},\mbf{1},\mbf{2},\mbf{1})\cr
&S_2\cdot S_5:\quad (\bar{\mbf{3}},\mbf{1},\mbf{2},\mbf{1})\cr
&u_3\cdot S_4:\quad (\mbf{3},\mbf{2},\mbf{1},\mbf{1})\cr
&v_3\cdot S_4:\quad (\bar{\mbf{3}},\mbf{2},\mbf{1},\mbf{1})\cr
&u_1\cdot S_5:\quad (\mbf{3},\mbf{1},\mbf{1},\mbf{2})\cr
&v_1\cdot S_5:\quad (\bar{\mbf{3}},\mbf{1},\mbf{1},\mbf{2})\cr
&u_7\cdot S_5:\quad (\mbf{3},\mbf{1},\mbf{2},\mbf{1})\cr
\ea
\ee

Thus we conclude that the quiver description is:

\begin{equation}
 SU(2)-\overset{%
\begin{array}
[c]{c}%
SU(2)\\
|
\end{array}
}{SU(3)}-SU(2) \,,
\end{equation}

since all the matter fields in (\ref{E6E6BU2-reps}) are bifundamentals of this quiver gauge theory.

From the geometry in figure~\ref{f:E6E6BU2}, we can blow up the surfaces $S_1$, $S_2$ and $S_3$ to get the geometry corresponding to the marginal CFD:

\be
\label{f:E6E6BU2-topresol}
\centering
\includegraphics[width=12cm]{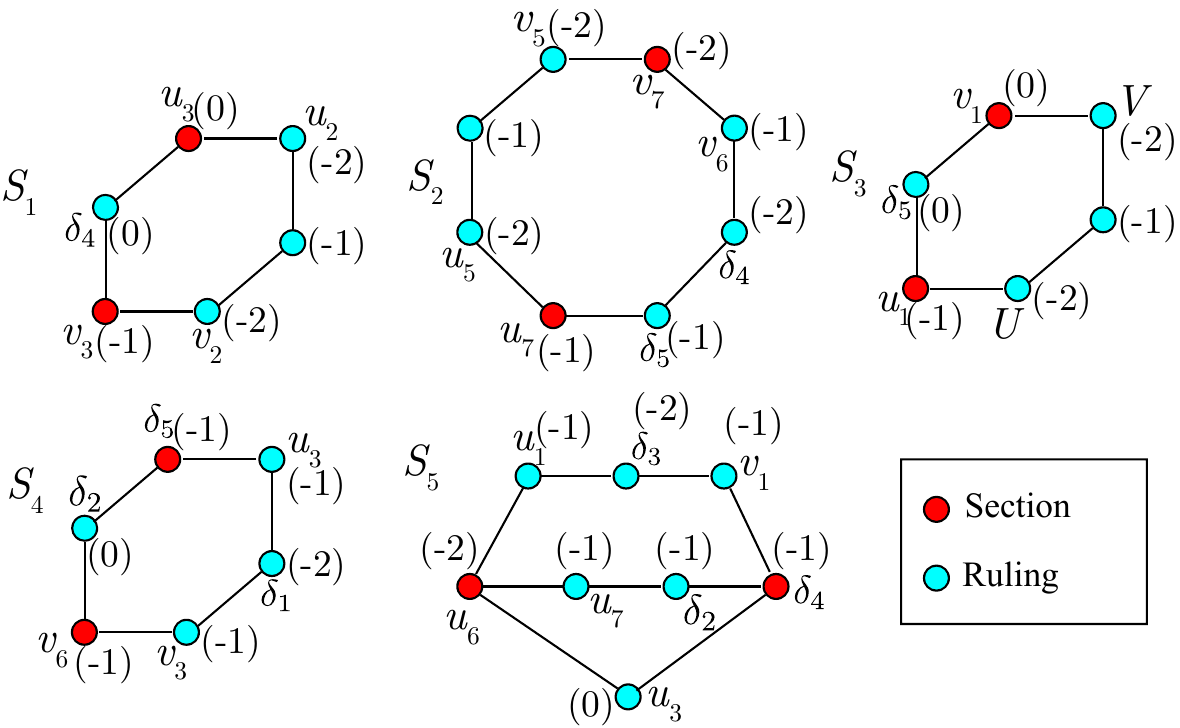}.
\ee

The quiver gauge theory description of this geometry is
\begin{equation}
 [2]-SU(2)-\overset{%
\begin{array}
[c]{c}%
[2]\\
|\\
SU(2)\\
|
\end{array}
}{SU(3)}-SU(2)-[2]\,,
\end{equation}
and the removal of fundamental flavors charged under the three $SU(2)$s will exactly correspond to shrinking the unlabeled $(-1)$-curves on $S_1$, $S_2$ and $S_3$ in (\ref{f:E6E6BU2-topresol}). 

As a summary, we confirmed that the proposed CFD tree with the star-shaped $SU(3)\times SU(2)\times SU(2)\times SU(2)$ quiver in~\cite{Apruzzi:2019vpe,Apruzzi:2019opn} is indeed backed up with a solid Calabi-Yau threefold geometry.

\subsection[Marginal Geometry for \texorpdfstring{$(E_7, SO(7))$}{(E7, SO(7))} Minimal Conformal Matter]{Marginal Geometry for \boldmath{$(E_7, SO(7))$} Minimal Conformal Matter}
\label{app:E7SO7}

In this section we present a marginal geometry of minimal $(E_7, SO(7))$ conformal matter that manifestly has the two dual gauge descriptions $SU(4)_0+2\bm{AS}+8\bm{F}$ and $6\bm{F}-Sp(2)-Sp(1)-2\bm{F}$.
The codimension one affine fibers of $E_7$ and $SO(7)$ are labelled as:

\be
\includegraphics[width=10cm]{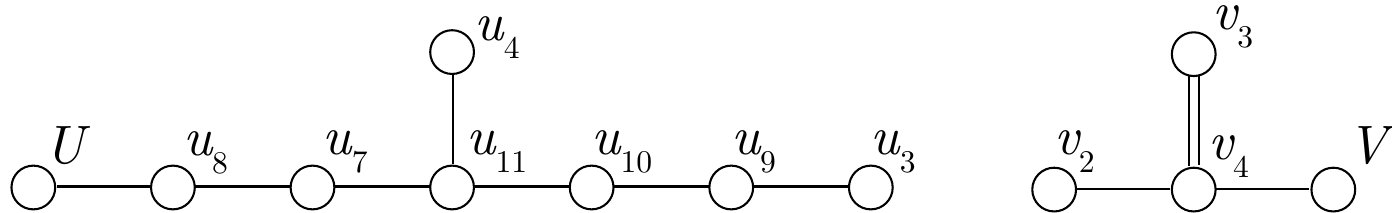}.
\ee

Note that locally, the fiber of $\{v_3 = 0\}$ is formed by two disconnected $\bbP^1$s, which are identified via monodromy effects and reflect the folding of $SO(8)$ to $SO(7)$.
Thus, the condition for $\{v_3\}$ to be fully wrapped requires the self-intersection inside the three compact surfaces ${\cal S} = \bigcup_{j=1}^3 S_j$ to be $-4$, rather than $-2$.
Concretely, the curves on the $S_j$'s are show in figure \ref{f:E7SO7-topresol}.
\begin{figure}
\centering
\includegraphics[width=12cm]{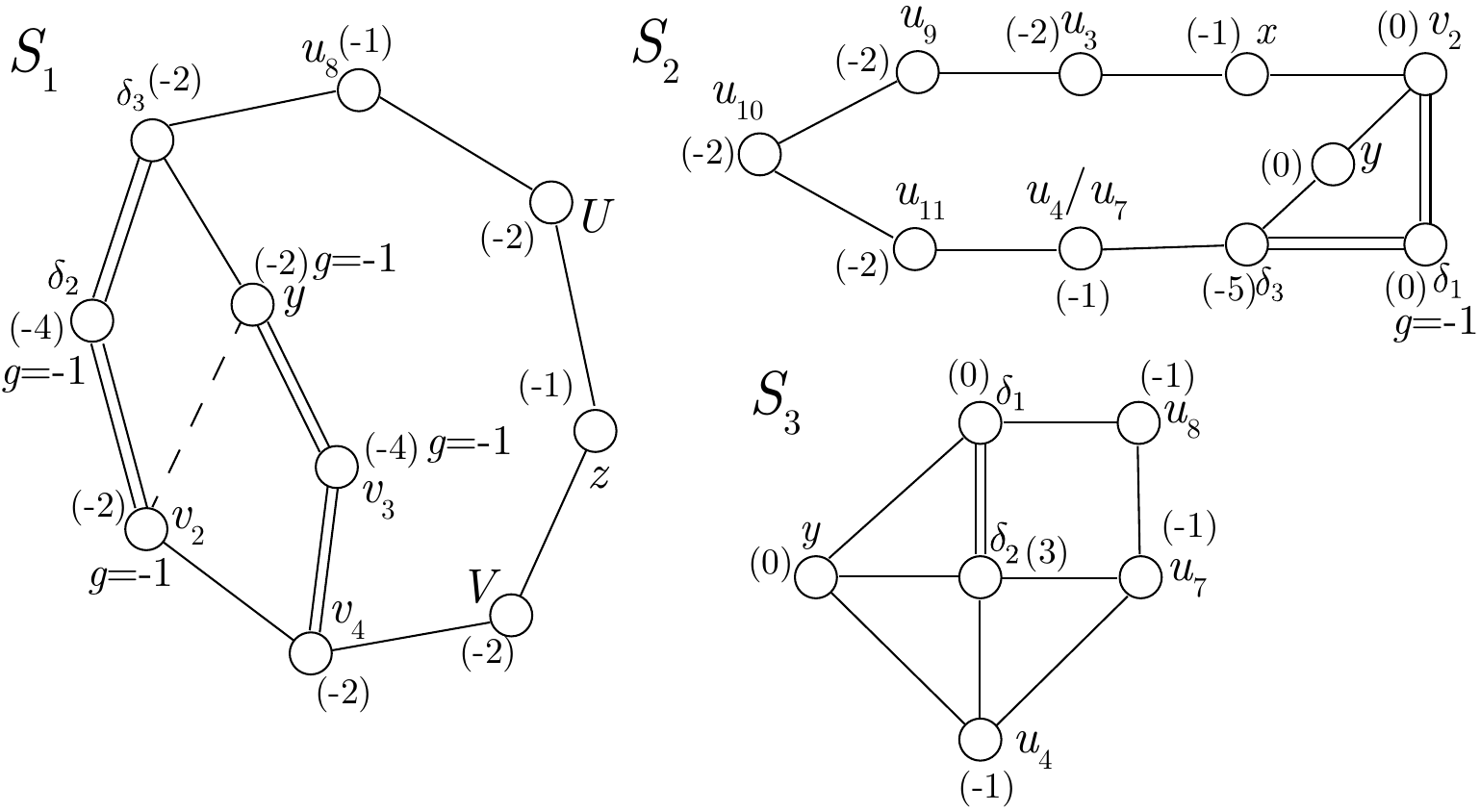}
\caption{
The configuration of curves $S_j \cap \{d=0\} \equiv d$ on the compact surfaces $S_j = \{\delta_j=0\} \ (j=1,\dots,3)$ in the marginal $(E_7, SO(7))$ geometry, where $\{d=0\}$ are (possibly non-compact) divisors in the resolved Calabi--Yau threefold.
The number in the bracket denotes the self-intersection number of the curve.
There are reducible curves $C \subset S$ that satisfy $C \cdot (K_S + C) = -4$, which we formally label as curves with genus $g= -1$.
All other curves have genus $0$.
The dashed line in $S_1$ indicates an intersection number of $-1$; this reflects the fact that the two involved curves, which are reducible, share irreducible components. The precise structure of these irreducible components are, however, immaterial to our discussion here.}\label{f:E7SO7-topresol}
\end{figure}

The rulings $f^{SU(4)}_j \hookrightarrow S_j$ giving rise to the $SU(4)$ gauge description are given by the curves
\begin{align}\label{eq:E7SO7_ruling_for_SU4}
	\begin{split}
		& f_1^{SU(4)} \equiv z + U + u_8 = y + v_3 + v_4 \, ,\\
		& f_2^{SU(4)} \equiv y = u_4 + u_{11} + u_{10} + u_9 + u_3 + x \, ,\\
		& f_3^{SU(4)} \equiv y = u_8 + u_7 \, ,
	\end{split}
\end{align}
where the equalities are understood as rational equivalence relations on each surface.
To realize the $SU(4)$, the surfaces are glued as $S_1 - S_3 - S_2$.
Each gluing curve is a 1-section with respect to the rulings \eqref{eq:E7SO7_ruling_for_SU4}: we have 
\begin{align}
	\left. \left(\delta_3 \cdot f_1^{SU(4)} \right) \right|_{S_1} = \left. \left(\delta_1 \cdot f_3^{SU(4)} \right) \right|_{S_3} = \left. \left(\delta_2 \cdot f_3^{SU(4)} \right) \right|_{S_3} = \left. \left(\delta_3 \cdot f_2^{SU(4)} \right) \right|_{S_2} = 1 \, .
\end{align}

The rulings $f^\text{quiver}_j \hookrightarrow S_j$ realizing the $Sp(2) \times Sp(1)$ quiver are
\begin{align}\label{eq:E7SO7_ruling_for_quiver}
	\begin{split}
		& f_1^\text{quiver} \equiv U+ 2z + V = \delta_3 + \delta_2 + v_2 \, ,\\
		& f_2^\text{quiver} \equiv y = u_4 + u_{11} + u_{10} + u_9 + u_3 + x = f_2^{SU(4)} \, ,\\
		& f_3^\text{quiver} \equiv \delta_1 = u_4 + u_7 \, .
	\end{split}
\end{align}
The $Sp(2)$ is supported on $S_2 - S_3$.
The non-simply laced nature is reflected by the intersection numbers of the gluing curves with these rulings:
\begin{align}
	\left. \left( \delta_3 \cdot f_2^\text{quiver} \right) \right|_{S_2} = 1 \, , \quad \left. \left( \delta_2 \cdot f_3^\text{quiver} \right) \right|_{S_3} = 2 \, .
\end{align}
To have an consistent, independent $Sp(1)$ factor on $S_1$, the gluing curves $S_1 \cap S_2$ and $S_1 \cap S_3$ need to be fibers on all three surfaces, which indeed is the case.

These two gauge descriptions can be verified by matching the prepotentials with the cubic intersection numbers,
\begin{align}
\begin{split}
	& S_1^3 = 2 \, , \quad S_1^2 \cdot S_2 = 0 \, \quad S_1 \cdot S_2^2 = -4 \, , \quad S_1^2 \cdot S_3 = 0 \, , \quad S_1 \cdot S_3^2 = -2 \, , \\
	& S_2^3 = 3 \, , \quad S_2^2 \cdot S_3 = 3 \, , \quad S_2 \cdot S_3^2 = -5 \, , \quad S_3^3 = 7 \, , \quad S_1 \cdot S_2 \cdot S_3 = 2 \, .
\end{split}
\end{align}

%
%
\providecommand{\href}[2]{#2}\begingroup\raggedright\endgroup

\end{document}